\documentclass[halfspacing]{ip3thesis}
\usepackage[utf8]{inputenc} % utf8 input good for non-ascii characters
\usepackage[T1]{fontenc} % Good for hyphenating non-ascii characters
\usepackage{lmodern}
\usepackage[british]{babel}
\usepackage[nottoc]{tocbibind} % Includes bibliography in table of contents, but not the table of contents itself
\usepackage[figuresright]{rotating} % Allows for sideways figures and tables
\usepackage{exscale}
\usepackage{relsize}

\usepackage{xcolor}
\definecolor{solarized-base3}{RGB}{253, 246, 227}

\usepackage[frozencache=true,cachedir=minted-cache]{minted} % syntax highlighting and formatting for code blocks
\setminted{
    breaklines=true,
    breakindentnchars=4,
    linenos=true,
    fontsize=\small,
    bgcolor=solarized-base3,
}

\usepackage{cite} % Combine multiple citations in range

\usepackage[shrink=10,stretch=10]{microtype}
\makeatletter
\@ifpackageloaded{microtype}{%
	\providecommand{\disableprotrusion}{\microtypesetup{protrusion=false}}
	\providecommand{\enableprotrusion}{\microtypesetup{protrusion=true}}
}{%
	\providecommand{\disableprotrusion}{}
	\providecommand{\enableprotrusion}{}
}
\makeatother

\usepackage{booktabs} % Better tables

\usepackage[margin=2em]{caption}
\usepackage{subcaption}
\usepackage{perpage} % Allows to reset counters on each page
\MakePerPage{footnote} % Resets footnote counters on each page

\graphicspath{{img/}}

\newenvironment{itemize*}
{\begin{itemize}
	\setlength{\itemsep}{0pt}
	\setlength{\parskip}{0pt}}
{\end{itemize}}
\newenvironment{enumerate*}
{\begin{enumerate}
	\setlength{\itemsep}{0pt}
	\setlength{\parskip}{0pt}}
{\end{enumerate}}

\usepackage{amsmath,amssymb,mathtools}
\numberwithin{equation}{section}
\allowdisplaybreaks % Allow display equations to break over different pages
\usepackage{amssymb}
\usepackage{bbm} % More blackboard bold characters, via \mathbbm. Mostly for blackboard bold 1
\usepackage{xfrac} % Nice small fractions
\usepackage{array} % Allows us to define custom column types for tables
\newcolumntype{L}{>{$}l<{$}} % a left aligned maths column type

\usepackage[italic]{hepnames} % Add particle name macros (e.g. \PBs)
\usepackage{braket} % Adds Dirac bra-ket notation
\usepackage{slashed} % Adds Dirac slash notation

\usepackage{siunitx}[=v2] % 2022-02-21: Fedora 35 doesn't have v3
\DeclareSIUnit\fb{\femto\barn}
\DeclareSIUnit\pb{\pico\barn}

\usepackage{physics}
\usepackage{xspace}
\usepackage{multirow}
\usepackage{mleftright}
\usepackage[printonlyused,withpage]{acronym}
\usepackage{imakeidx}

\makeindex[columns=2,intoc]

\newcommand{\textbi}[1]{\textit{\textbf{#1}}}

\newcommand{\ie}{\textit{i.e.}\xspace}
\newcommand{\cf}{\textit{c.f.}\xspace}

\newcommand{\br}[1]{\left( #1 \right)}
\newcommand{\sq}[1]{\left[ #1 \right]}
\newcommand{\cu}[1]{\left\{ #1 \right\}}
\newcommand{\ab}[1]{\left| #1 \right|}
\newcommand{\msq}[1]{\ab{#1}^2}

\newcommand{\brf}[1]{\mleft( #1 \mright)}
\newcommand{\brm}[1]{\begin{pmatrix} #1 \end{pmatrix}}

\newcommand{\sqf}[1]{\mleft[ #1 \mright]}

\newcommand{\aem}{{\alpha}}
\newcommand{\astr}{{\alpha_\mathrm{s}}}
\newcommand{\cs}{{\mathrm{g}_\mathrm{s}}}
\newcommand{\cem}{{\mathrm{e}}}

\newcommand{\am}{\mathcal{A}}

\newcommand{\lls}[3]{{{#1}^{\mathrm{(#2)}}_{#3}}}
\newcommand{\amp}[2]{\lls{\am}{#1}{#2}}
\newcommand{\ant}{\amp{0}{n}}
\newcommand{\pa}{A}
\newcommand{\pamp}[2]{\lls{\pa}{#1}{#2}}
\newcommand{\pat}[2]{\pa_{#2}^{\mathrm{(0)}}\brf{#1}}
\newcommand{\pant}[1]{\pat{#1}{n}}

\newcommand{\nc}{{N_{c}}}
\newcommand{\nf}{{N_{f}}}
\newcommand{\gU}[1]{\mathrm{U}(#1)}
\newcommand{\SU}[1]{\mathrm{S}\gU{#1}}
\newcommand{\SUt}[1]{$\SU{#1}$}
\newcommand{\cgf}{t}
\newcommand{\cga}{F}
\newcommand{\cdi}{{\mathrm{T}_\mathrm{F}}}
\newcommand{\casimir}{\mathrm{C}}
\newcommand{\caf}{{\casimir_\mathrm{F}}}
\newcommand{\caa}{{\casimir_\mathrm{A}}}

\newcommand{\cof}{c}

\newcommand{\cm}{\mathcal{C}}

\newcommand{\ff}{F}
\newcommand{\stc}{f}
\newcommand{\covcon}[3]{{#1}_{#2}^{\phantom{#2}#3}}
\newcommand{\concov}[3]{{#1}^{#2}_{\phantom{#2}#3}}
\newcommand{\lws}{\lambda}
\newcommand{\rws}{\widetilde{\lws}}
\newcommand{\lwsu}[2]{\covcon{\lws}{#1}{\dot{#2}}}
\newcommand{\lwsug}[1]{\lws^{\dot{#1}}}
\newcommand{\rwsu}[2]{\covcon{\rws}{#1}{#2}}
\newcommand{\rwsug}[1]{\rws^{#1}}
\newcommand{\lwsd}[2]{{\br{\lws_{#1}}}_{\dot{#2}}}
\newcommand{\lwsdg}[1]{\lws_{\dot{#1}}}
\newcommand{\rwsd}[2]{\br{\rws_{#1}}_{#2}}
\newcommand{\rwsdg}[1]{\rws_{#1}}
\newcommand{\lc}{\varepsilon}
\newcommand{\llc}[3]{\lc#1{\dot{#2}\dot{#3}}}
\newcommand{\llcu}[2]{\llc{^}{#1}{#2}}
\newcommand{\llcd}[2]{\llc{_}{#1}{#2}}
\newcommand{\rlc}[3]{\lc#1{#2#3}}
\newcommand{\rlcu}[2]{\rlc{^}{#1}{#2}}
\newcommand{\rlcd}[2]{\rlc{_}{#1}{#2}}
\newcommand{\ptr}[1]{\tr(#1)}
\newcommand{\ptrs}[2]{\tr_{#1}(#2)}
\newcommand{\ptrf}[1]{\ptrs{5}{#1}}
\newcommand{\pdet}[1]{\det(#1)}
\newcommand{\imi}{{\mathrm{i}\mkern1mu}}
\newcommand{\s}[1]{{{s}_{#1}}}

\newcommand{\trf}{\text{tr}_5}
\newcommand{\sumsquare}[1]{\overline{\ab{#1}^2}}
\newcommand{\deltaplus}[1]{\delta^{(+)}\brf{#1}}

\newcommand{\pov}{\varepsilon}
\newcommand{\cn}{\mathbb{C}}
\newcommand{\integers}{\mathbb{Z}}
\newcommand{\nonnegativeintegers}{\integers^{\ge}}
\newcommand{\positiveintegers}{\integers^{>}}
\newcommand{\oddintegers}{\integers^\text{odd}}

\newcommand{\rationals}{\mathbb{Q}}
\newcommand{\ffield}[1]{{\mathbb{F}_{#1}}}

\newcommand{\rmat}[1]{\begin{pmatrix}#1\end{pmatrix}}
\newcommand{\lm}{\eta}
\newcommand{\sigmabar}{\overline{\sigma}}
\newcommand{\pmin}[1]{\min\brf{#1}}
\newcommand{\pmax}[1]{\max\brf{#1}}

\newcommand{\code}[1]{\texttt{#1}\xspace}
\newcommand{\python}{\code{Python}}
\newcommand{\cpp}{\code{C++}}
\newcommand{\njet}{\code{NJet3}}
\newcommand{\eigen}{\code{Eigen3}}
\newcommand{\openloops}{\code{OpenLoops2}}
\newcommand{\nnlojet}{\code{NNLOjet}}
\newcommand{\oneloop}{\code{OneL0op}}
\newcommand{\collier}{\code{Collier}}
\newcommand{\lhapdf}{\code{LHAPDF}}
\newcommand{\nnpdf}{\code{NNPDF3.1}}
\newcommand{\rivet}{\code{Rivet}}
\newcommand{\otter}{\code{Otter}}
\newcommand{\pentagonfunctions}{\code{PentagonFunctions++}}
\newcommand{\finiteflow}{\code{FiniteFlow}}
\newcommand{\qd}{\code{QD}}
\newcommand{\mathematica}{\code{Mathematica}}
\newcommand{\form}{\code{FORM}}
\newcommand{\qgraf}{\code{QGRAF}}
\newcommand{\spinney}{\code{Spinney}}
\newcommand{\fastjet}{\code{FastJet}}

\newcommand{\incite}[1]{Ref.~\cite{#1}}
\newcommand{\incites}[1]{Refs.~\cite{#1}}

\newcommand{\sAr}[1]{| #1 \rangle}
\newcommand{\sAl}[1]{\langle #1 |}
\newcommand{\sBr}[1]{| #1 ]}
\newcommand{\sBl}[1]{[ #1 |}

\newcommand{\spAa}[1]{\langle#1\rangle}
\newcommand{\spBa}[1]{[#1]}

\newcommand{\spA}[2]{\spAa{#1 \, #2}}
\newcommand{\spB}[2]{\spBa{#1 \, #2}}
\newcommand{\spAB}[3]{\langle #1 \, #2 \, #3 ]}
\newcommand{\spBA}[3]{[ #1 \, #2 \, #3 \rangle}

\newcommand{\lra}{\xrightarrow{\hspace{2em}}}

\newcommand{\disc}[0]{\text{Disc}}
\newcommand{\discb}[1]{\disc\brf{#1}}
\newcommand{\discbb}[2]{\disc_{#1}\brf{#2}}
\newcommand{\li}[2]{\text{Li}_{#1}\brf{#2}}

\newcommand{\eps}{\epsilon}
\newcommand{\ord}{\mathcal{O}}

\newcommand{\mom}[2]{{#1}_{[#2]}}

\newcommand{\anti}[1]{\overline{#1}}

\newcommand{\bigint}{\mathlarger{\int}}
\newcommand{\bigsum}{\mathlarger{\sum}}

\usepackage[hidelinks, pdfusetitle]{hyperref} % hidelinks removes the boxes around links when viewed on screen, pdfusetitle option includes the thesis title and author in the PDF metadata, pagebackref includes links with pages numbers in bibliography to citations in body

\usepackage[capitalise]{cleveref}
\makeatletter
\newcommand{\crefext}[2]{\csname cref@#1@format\endcsname{#2}{}{}}
\newcommand{\Crefext}[2]{\csname Cref@#1@format\endcsname{#2}{}{}}
\newcommand{\crefextp}[2]{\csname cref@#1@name@plural\endcsname{~#2}{}{}}
\newcommand{\Crefextp}[2]{\csname Cref@#1@name@plural\endcsname{~#2}{}{}}
\makeatother

\begin{document}

\title{Precision QCD corrections to gluon-initiated diphoton-plus-jet production at the LHC}
%\subtitle{An optional subtitle}
\author{Ryan Iain Moodie}
\researchgroup{Institute for Particle Physics Phenomenology}
\date{April 2022}
\maketitlepage*

\begin{abstract}
    In this thesis, we present recent advances at the precision frontier of higher-order \ac*{QCD} calculations.
    We consider massless two-loop five-point amplitudes, with a particular focus on diphoton-plus-jet production through gluon fusion.
    We build a library of \acl{IR} functions up to at most \ac*{NNLO} in \acs*{QCD}, which can be used to validate amplitudes and construct counterterms in subtraction schemes at \acs*{NNLO}.
    We review progress in the novel use of \acl{ML} technology to optimise the evaluation of amplitudes in hadron collider simulations.
    We present the \acl{FC} virtual \acs*{QCD} corrections to diphoton-plus-jet production through gluon fusion, discussing the new techniques developed to calculate these non-planar two-loop amplitudes.
    We use these amplitudes to compute the \acl{NLO} \acs*{QCD} corrections to the differential cross sections of diphoton-plus-jet production through gluon fusion at the \acl{LHC}.
    We also present the \acl{LC} \acl{VV} corrections to hadronic trijet production.
    All derived amplitudes are made available in a public implementation that is ready for further phenomenological application.
\end{abstract}

\disableprotrusion
\tableofcontents*
% \listoffigures
% \listoftables
\enableprotrusion

\chapter*{List of Abbreviations}
\addcontentsline{toc}{chapter}{List of Abbreviations}

\begin{acronym}[DGLAP]
    \acro{API}{application programming interface}
    \acro{ATLAS}{\textit{A Toroidal \acs*{LHC} Apparatus}}
    \acro{BCFW}{Britto-Cachazo-Feng-Witten}
    \acro{BLHA}[\code{BLHA}]{\textit{Binoth Les Houches Accord}}
    \acro{CMS}{\textit{Compact Muon Solenoid}}
    \acro{CPU}{central processing unit}
    \acro{CS}{Catani-Seymour}
    \acro{DGLAP}{Dokshitser-Gribov-Lipatov-Altarelli-Parisi}
    \acro{DR}{dimensional regularisation}
    \acro{EM}{electromagnetic}
    \acro{ESSB}{electroweak spontaneous symmetry breaking}
    \acro{f32}[\texttt{f32}]{32-bit floating-point number\acroextra{ (single precision)}}
    \acro{f64}[\texttt{f64}]{64-bit floating-point number\acroextra{ (double precision)}}
    \acro{f128}[\texttt{f128}]{128-bit floating-point number\acroextra{ (quadruple precision)}}
    \acro{f256}[\texttt{f256}]{256-bit floating-point number\acroextra{ (octuple precision)}}
    \acro{FC}{full-colour}
    \acro{FF}{finite field}
    \acro{FKS}{Frixione-Kunszt-Signer}
    \acro{FR}{finite remainder}
    \acro{FS}{full-spin}
    \acro{IBP}{integration-by-parts}
    \acro{IR}{infrared}
    \acro{ISP}{irreducible scalar product}
    \acro{KK}{Kleiss-Kuijf}
    \acro{KLN}{Kinoshita-Lee-Nauenberg}
    \acro{LC}{leading-colour}
    \acro{LHC}{Large Hadron Collider}
    \acro{LI}{linearly independent}
    \acro{LO}{leading order}
    \acro{LS}{leading-spin}
    \acro{MC}{Monte Carlo}
    \acro{ME}{matrix element}
    \acro{MHVb}[$\overline{\text{\acs*{MHV}}}$]{anti-maximally-helicity-violating}
    \acro{MHV}{maximally-helicity-violating}
    \acro{MI}{master integral}
    \acro{ML}{machine learning}
    \acro{MSb}[$\overline{\text{MS}}$]{modified minimal subtraction}
    \acro{MSE}{mean squared error}
    \acro{MTV}{momentum twistor variable}
    \acro{N3LO}[N\textsuperscript{3}\acs*{LO}]{(next-to-)\textsuperscript{3}leading order}
    \acro{N4LO}[N\textsuperscript{4}\acs*{LO}]{(next-to-)\textsuperscript{4}leading order}
    \acro{NkLL}[N\textsuperscript{k}LL]{(next-to-)\textsuperscript{k}leading log}
    \acro{NkLO}[N\textsuperscript{k}\acs*{LO}]{(next-to-)\textsuperscript{k}leading order}
    \acro{NkLP}[N\textsuperscript{k}LP]{(next-to-)\textsuperscript{k}leading power}
    \acro{NLO}{next-to-leading order}
    \acro{NMHV}{next-to-maximally-helicity-violating}
    \acro{NNLO}{next-to-next-to-leading order}
    \acro{NN}{neural network}
    \acro{OPP}{Ossola-Pittau-Papadopoulos}
    \acro{PDF}{parton distribution function}
    \acro{PFD}{partial fraction decomposition}
    \acro{QCD}{quantum chromodynamics}
    \acro{QED}{quantum electrodynamics}
    \acro{QFT}{quantum field theory}
    \acro{RMSE}{root mean squared error}
    \acro{RRR}[R\textsuperscript{3}]{triple-real}
    \acro{RRV}{real-real-virtual}
    \acro{RR}{double-real}
    \acro{RVV}{real-virtual-virtual}
    \acro{RV}{real-virtual}
    \acro{SCET}{Soft-Collinear Effective Theory}
    \acro{SHERPA}[\code{Sherpa}]{\textit{Simulation of High Energy Reactions of Particles}}
    \acro{SLC}{subleading-colour}
    \acro{SM}{Standard Model}
    \acro{tHV}{'t Hooft-Veltman}
    \acro{UHV}{ultra-helicity-violating}
    \acro{UV}{ultraviolet}
    \acro{VVV}[V\textsuperscript{3}]{triple-virtual}
    \acro{VV}{double-virtual}

    \acroindefinite{API}{an}{an}
    \acroindefinite{IR}{an}{an}
    \acroindefinite{ISP}{an}{an}
    \acroindefinite{UHV}{a}{an}
    \acroindefinite{UV}{a}{an}
    \acroindefinite{MTV}{an}{a}
    \acroindefinite{MC}{an}{a}
    \acroindefinite{FF}{an}{a}
    \acroindefinite{MSE}{an}{a}
    \acroindefinite{MI}{an}{a}
    \acroindefinite{FR}{an}{a}
    \acroindefinite{LI}{an}{a}
    \acroindefinite{FKS}{an}{a}
\end{acronym}
\pagebreak

\begin{declaration*}
    The work in this thesis is based on research carried out in the Institute for Particle Physics Phenomenology at Durham University.
    No part of this thesis has been submitted elsewhere for any degree or qualification.
    This thesis is based on joint research: \cref{ch:intro} is a review of the literature; \cref{ch:ir} discusses contributions to the \cpp amplitude library \njet~\cite{njet}; \cref{ch:ann} is based on \incite{moodie:2022flt}; \cref{ch:yy-amp} on \incite{badger:2021imn}; \cref{ch:yy-xs} on \incite{badger:2021ohm}; and \cref{ch:3j} on \incite{3j}.
\end{declaration*}

\begin{acknowledgements*}
    Foremost thanks to my supervisor, Simon Badger, for his mentorship over the past few years.
    Thanks to all those I've had the pleasure of collaborating with and to everyone in the IPPP community that I've enjoyed being a part of so much.
    For proofreading, helpful comments, and discussions on the draft, a huge thanks to Oscar Braun-White, Lucy Budge, Lois Flower, Hitham Hassan, Sebastian Jaskiewicz, Stephen Jones, Matteo Marcoli, Francesco Sarandrea, Henry Truong, Yannick Ulrich, Mia West, and Simone Zoia.

    The work presented in this thesis was funded by UKRI-STFC grant numbers ST/S505365/1 and ST/P001246/1.
\end{acknowledgements*}

\pagenumbering{arabic}

\chapter{Introduction}
\label{ch:intro}

The \ac{SM} is one of the most successful scientific theories of history, including the most precise agreements of prediction and experiment ever achieved~\cite{aoyama:2017uqe,morel:2020dww}.
However, it fails to describe some observed phenomena of the universe and contains empirical parameters~\cite{particledatagroup:2020ssz}.
The subject of this thesis is precision \ac{QCD} phenomenology\index{Precision!phenomenology}\index{Precision!\acs*{QCD}}, which is an indirect search for physics beyond the \ac{SM} by comparing high-precision \ac{SM} predictions to measurements at hadron colliders\index{Hadron collider}~\cite{shiltsev:2012zzc}.
New physics would appear as small deviations from \ac{SM} expectations~\cite{gehrmann:2021qex}.

In this chapter, we introduce the basic toolkit required to calculate \ac{QCD} corrections to amplitudes and use them to construct high-precision predictions for observables at colliders.
We begin in \cref{sec:sm} with a brief review of the relevant sectors of the \ac{SM}.
We discuss how amplitudes can be used to construct collider observables in \cref{sec:observables}, touching on some details of amplitudes in \cref{sec:amplitudes}.
In \cref{sec:colour}, we look at how we can simplify the computation of \ac{QCD} amplitudes by factorising into colour and kinematic contributions, treating the former with colour decomposition.
In \cref{sec:kin}, we cover various representations of the kinematics and some methods for computing partial amplitudes.
In \cref{sec:loops}, we discuss how on-shell methods can be used to calculate loop-level amplitudes.
In \cref{sec:ff}, we show how finite field arithmetic can aid the computation of amplitudes.
We motivate the phenomenology of diphoton production in \cref{sec:pheno}, before summarising the contents of this thesis in \cref{sec:structure}.

For further reading on these topics, see the textbooks~\cite{halzen:1984mc,bagger:1990qh,peskin:1995ev,weinberg:1995mt,srednicki:2007qs,ellis:1996mzs,collins:2011zzd,henn:2014yza,schwartz:2014sze,elvang:2015rqa,campbell:2286381}, lectures~\cite{dixon:1996wi,dixon:2013uaa,skands:2012ts}, reviews~\cite{mangano:1990by,dreiner:2008tw,elvang:2013cua,caola:2022ayt}, and theses~\cite{delaurentis:2020xar,chawdhry:2020xlk,whitehead:2021lzn}.

\section{The Standard Model}
\label{sec:sm}

\begin{figure}
    \begin{center}
        \includegraphics[width=\textwidth]{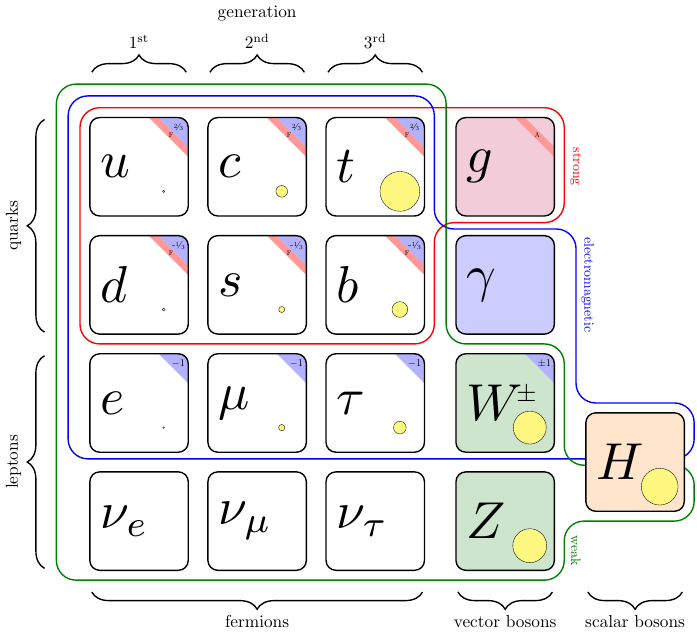}
        \caption{
            Table of the particle content of the \ac{SM}.
            Outlines of the corresponding colour contain particles carrying strong (red), \acl*{EM} (blue), or weak (green) gauge charge.
            The electric charge (blue) and colour representation (red) are shown in the upper right corner of appropriate particle boxes.
            For massive particles, there is a yellow circle with area proportional to the mass on a square-root scale (data from \incite{particledatagroup:2020ssz}).
        }
        \label{fig:sm}
    \end{center}
\end{figure}

The \ac{SM}\index{Standard Model} is our current best description of the fundamental structure of the universe.
It is \iac{QFT} with a direct product gauge group,
\begin{align}
    \label{eq:sm}
    \text{SU}_C(3) \times \text{SU}_L(2) \times \text{U}_Y(1) \to \text{SU}_C(3) \times \text{U}_{\text{\acs*{EM}}}(1) \,.
\end{align}
The first Lie group is the strong interaction, with subscript $C$ for colour, which is described by \ac{QCD}\index{Quantum!chromodynamics}.
The next two comprise the electroweak sector, with subscript $L$ for left-handed chirality and $Y$ for hypercharge.
The transition in \cref{eq:sm} is \ac{ESSB} through the Higgs mechanism~\cite{englert:1964et,higgs:1964pj}, which gives rise to the weak and \ac{EM} interactions we observe.
The resultant \ac{EM} group is described by \ac{QED}\index{Quantum!electrodynamics}.
The \ac{SM} does not describe gravity.

The forces are mediated by vector bosons\index{Boson} (spin-1)\index{Boson!vector}, also called gauge bosons\index{Boson!gauge}: the massless gluon\index{Boson!gluon} $g$ for the strong force, the massive $W^\pm$\index{Boson!$W^\pm$} and $Z$\index{Boson!$Z$} bosons for the weak force, and the massless photon\index{Boson!photon} $\gamma$ for the \ac{EM} force.
The matter particles of the \ac{SM} are fermions (spin-1/2)\index{Fermion}.
This includes the quarks\index{Fermion!quark}---the up $u$, down $d$, charm $c$, strange $s$, top $t$, and bottom $b$---which are massive and carry colour, weak charge, and electric charge.
The other fermions are the leptons, including: charged leptons\index{Fermion!lepton}\index{Fermion!lepton!charged}---the electron $e$, muon $\mu$, and tau $\tau$---which are massive and carry weak and electric charge; and the neutrinos\index{Fermion!lepton!neutrino}---the electron neutrino $\nu_e$, muon neutrino $\nu_\mu$, and tau neutrino $\nu_\tau$---which are massless\footnote{While neutrinos are massless in the \ac{SM}, this is a shortcoming of the model as the experimental evidence of neutrino oscillations indicates that they have a small yet non-zero mass.} and carry only weak charge.
The different types of quarks and leptons are referred to as flavours\index{Fermion!flavour}.
Each fermion comes with an antiparticle, which we denote with an overline.
There is also the Higgs boson\index{Boson!Higgs} $H$, which is a scalar boson (spin-0)\index{Boson!scalar}.
It generates the masses of the $W^\pm$ and $Z$ bosons, the charged leptons, and the quarks through \ac{ESSB}.
The particles of the \ac{SM} are tabulated in \cref{fig:sm}.

As \ac{QED} and \ac{QCD} are the relevant sectors for this thesis, we will further discuss them in the following sections.
The complete \ac{SM} Lagrangian, along with all Feynman rules, is presented in \incite{romao:2012pq}.

\subsection{Quantum electrodynamics}

\Ac{QED}\index{Quantum!electrodynamics} is \iac{QFT} describing the \ac{EM} interaction.
It is an abelian gauge theory with gauge group $\text{U}(1)$.
The gauge charge is electric charge.
The photon field $A_\mu$ couples to (anti)fermion fields ($\anti{\psi_f}$) $\psi_f$ with mass $m_f$, which includes the quarks and charged leptons.
With the imaginary unit,
\begin{align}
    \imi^2=-1\,,
\end{align}
and the Dirac adjoint and Dirac slash,
\begin{align}
    \begin{aligned}
        \label{eq:slash}
        \anti{\psi} &= \psi^\dagger \gamma^0 \,, \\
        \slashed{D} &= D_\mu \gamma^\mu \,,
    \end{aligned}
\end{align}
where $\gamma^\mu$ with $\mu\in\cu{0,1,2,3}$ in four dimensions are the gamma matrices\index{Matrix!gamma}, the classical part of the \ac{QED} Lagrangian is
\begin{align}
    \mathcal{L}_\text{QED}^\text{classical} &= -\frac{1}{4} F_{\mu\nu} F^{\mu\nu} + \sum_f \anti{\psi_f}\br{\imi \slashed{D} - m_f}\psi_f \,,
\end{align}
with the field strength tensor,
\begin{align}
    F_{\mu\nu} &= \partial_\mu A_\nu - \partial_\nu A_\mu \,,
\end{align}
and the covariant derivative,
\begin{align}
    D_\mu &= \partial_\mu - \imi Q_f e A_\mu \,.
\end{align}
The \ac{EM} coupling $Q_f e$ is equal to the electric charge of the fermion $f$, with
\begin{align}
    Q_f = \left\{
        \begin{array}{r l l}
            -1 & \text{charged lepton} & \text{($e$, $\mu$, $\tau$)} \,, \\
            \frac{2}{3} & \text{up-type quark} & \text{($u$, $c$, $t$)} \,, \\
            -\frac{1}{3} & \text{down-type quark} & \text{($d$, $s$, $b$)} \,.
        \end{array} \right.
\end{align}
The coupling is often expressed in terms of the dimensionless quantity,\index{Coupling!electromagnetic}
\begin{align}
    \aem  & =\frac{\cem^2}{4 \pi} \,,
\end{align}
called the fine structure constant.
The coupling depends on the energy scale, as discussed in \cref{sec:qcd}.

The quantisation of the Lagrangian involves adding a gauge fixing term\index{Gauge!fixing}.
This allows a propagator to be defined for the photon, which depends on the gauge choice, such as $R_\xi$ gauge\index{Gauge!covariant!$R_\xi$} (\crefext{chapter}{62} of \incite{srednicki:2007qs}).
Because the gauge degrees of freedom are unphysical, any physical predictions of the theory are independent of the choice of gauge.
They are also unchanged by gauge transformations, which is a property we call gauge invariance\index{Gauge!invariance}.
We discuss gauges in more detail for the non-abelian case of \ac{QCD}.

\subsection{Quantum chromodynamics}
\label{sec:qcd}

\Ac{QCD}\index{Quantum!chromodynamics} is \iac{QFT} describing the strong interaction.
It is a non-abelian gauge theory with gauge group \SUt{3}.
The gauge charge is called colour charge\index{Colour}.
The gluon fields $A^a_\mu$ lie in the adjoint representation with $a\in\cu{1,\ldots,8}$, and are coupled to (anti)quark fields ($\anti{\psi^i_q}$) $\psi^i_q$ lying in the (anti)fundamental representation with $i\in\cu{1,2,3}$.
The classical part of the \ac{QCD} Lagrangian is
\begin{align}
    \begin{aligned}
        \mathcal{L}_\text{QCD}^\text{classical} &= -\frac{1}{4} F^{a}_{\mu\nu} F^{a,\mu\nu} + \sum_q \anti{\psi_q^i} \br{\imi \slashed{D}_{ij} - \delta_{ij} m_q} \psi_q^j \,, \\
        F^a_{\mu\nu} &= \partial_\mu A^a_\nu - \partial_\nu A^a_\mu + g_s \stc^{abc}A^b_\mu A^c_\nu \,,\\
        \br{D_\mu}_{ij} &= \delta_{ij} \partial_\mu - \imi g_s t^a_{ij} A^a_\mu \,,
    \end{aligned}
\end{align}
where $t^a_{ij}$ are the \SUt{3} generators\index{Colour!generator} in a fundamental representation.
There exist various possible matrix representations; one is through the proportionality to the Gell-Mann matrices~\cite{gell-mann:1962yej} $\lambda^a_{ij}$,
\begin{align}
    t^a_{ij}=\frac{1}{2} \lambda^{a}_{ij} \,.
\end{align}
The structure constants\index{Colour!structure constant} $\stc^{abc}$ are defined by the commutator of the fundamental generators,
\begin{align}
    \label{eq:colour-structure-constant}
    \sq{\cgf^a,\cgf^b}=\imi \stc^{abc} \cgf^c \,.
\end{align}
The strong coupling\index{Coupling!strong} $g_s$ is often expressed in terms of the dimensionless parameter,
\begin{align}
    \label{eq:astr}
    \astr & =\frac{\cs^2}{4 \pi}.
\end{align}

Again, a gauge fixing term is introduced as in the abelian case of \ac{QED}.
$R_\xi$ gauge\index{Gauge!covariant!$R_\xi$} adds a covariant gauge fixing term\index{Gauge!covariant},
\begin{align}
    \mathcal{L}_{\text{\ac{QCD}}}^{R_\xi} = -\frac{1}{2\xi}\br{\partial^\mu A^a_\mu}^2\,,
\end{align}
which leads to the gluon propagator,
\begin{align}
    G^{ab}_{\mu\nu}(p)=\frac{\imi\delta^{ab}}{p^2+\imi\varepsilon}\br{-\lm_{\mu\nu}+\br{1-\xi}\frac{p_\mu p_\nu}{p^2+\imi\varepsilon}} \,.
\end{align}
This encompasses Feynman gauge\index{Gauge!covariant!Feynman} for $\xi=1$ and Landau gauge\index{Gauge!covariant!Landau} for $\xi\to0$.

With covariant gauges, the non-abelian theory can also require the introduction of ghost fields \index{Ghosts} to cancel unphysical modes through a procedure such as the Faddeev-Popov method~\cite{faddeev:1967fc}.
The ghost field $c^a$\index{Fermion!\acs*{QCD} ghost} is in the adjoint representation of \SUt{\nc} and is fermionic.
The ghost term of the Lagrangian can take the form,
\begin{subequations}
    \begin{align}
        \mathcal{L}_\text{\ac{QCD}}^{\text{ghost}} &= \br{\partial_\mu \anti{c^a}} D^\mu_{ab} \, c^b \,, \\
        D_{ab}^\mu &= \partial^\mu\delta_{ab}+\imi (F^c)_{ab} (A^{c})^\mu \,, \\
        \label{eq:Fabc}
        (F^a)_{bc} &=-\imi f^{abc} \,,
    \end{align}
\end{subequations}
where $D^\mu_{ab}$ is the covariant derivative in the adjoint representation and $(F^a)_{bc}$ is the \SUt{\nc} generator in an adjoint representation.
These ghosts appear in amplitude loops (\cref{sec:amplitudes}).

Axial gauges\index{Gauge!axial} are an alternative choice which can be ghost-free at the cost of a more complicated gluon propagator.
We fix the gauge field with respect to an arbitrary reference vector $q^\mu$,
\begin{align}
    \mathcal{L}_\text{\ac{QCD}}^\text{axial} = -\frac{1}{2\xi}\br{q^\mu A^a_\mu}^2 \,,
\end{align}
giving the gluon propagator,
\begin{align}
    G^{ab}_{\mu\nu}(p)=\frac{\imi\delta^{ab}}{p^2+\imi\varepsilon}\br{-\lm_{\mu\nu}+\frac{p_\mu q_\nu+p_\nu q_\mu}{p\cdot q}-\br{q^2+\xi p^2}\frac{p_\mu p_\nu}{\br{p\cdot q}^2}} \,.
\end{align}
Light-like axial gauge\index{Gauge!axial!light-like}, or light-cone gauge\index{Gauge!axial!light-cone}, is given in the case of a null (light-like) reference vector $q^2=0$ and $\xi\to0$.
For $\xi\to0$, the ghost fields decouple and can be neglected: the two physical polarisation states of the gluon are explicit.

\begin{figure}
    \begin{center}
        \includegraphics[width=0.8\textwidth]{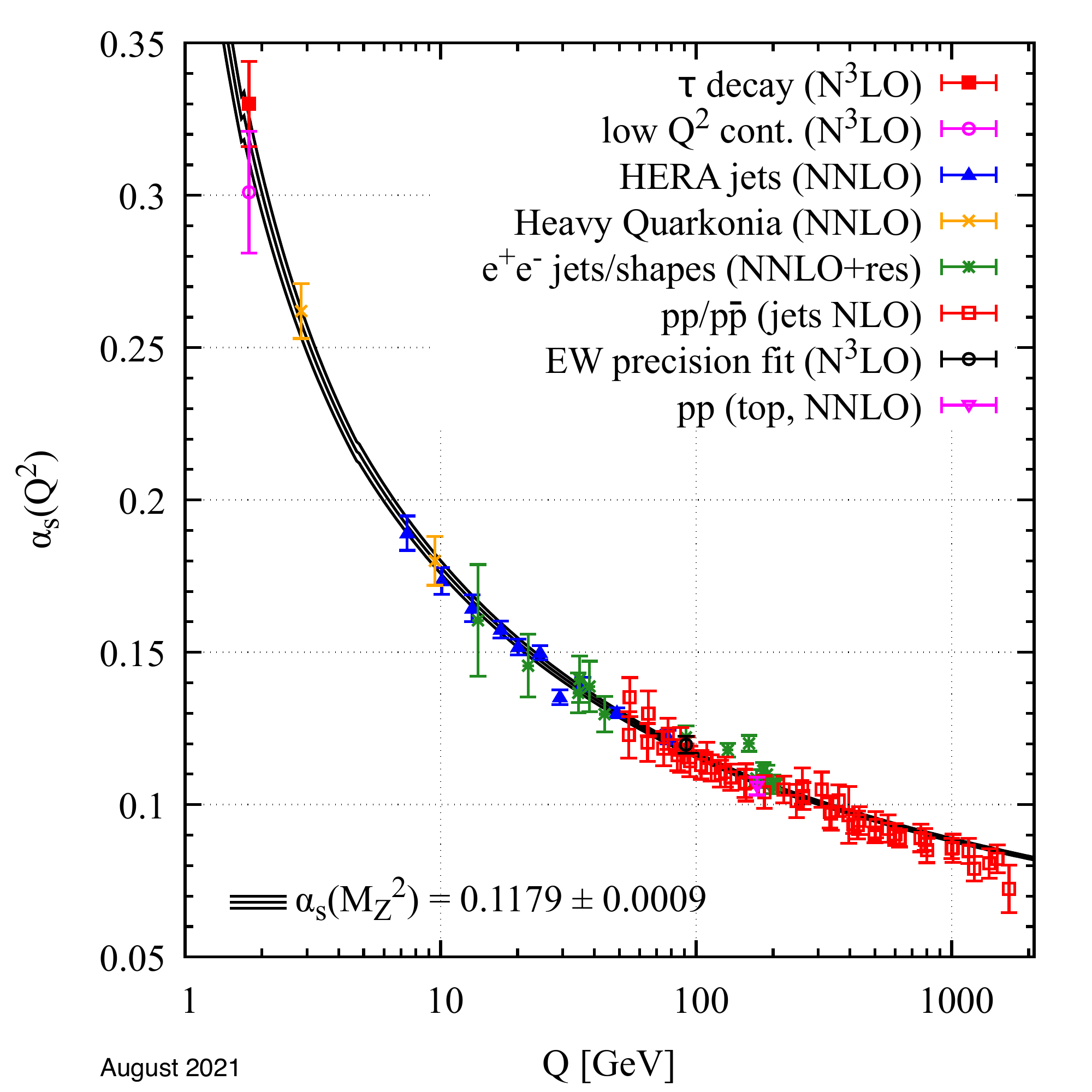}
        \caption{
            Summary of measurements of $\astr$ as a function of the energy scale $Q$.
            The degree of \ac{QCD} perturbation theory (\cref{sec:fixed-order}) used to extract $\astr$ is indicated in brackets, with +res meaning matched to a resummed calculation (\cref{sec:beyond-fixed-order}).
            The theory prediction of $\astr$ running is given by the black line and confidence interval.
            Figure from \incite{particledatagroup:2020ssz}.
        }
        \label{fig:alphas}
    \end{center}
\end{figure}

Couplings depend on the energy scale $\mu$ of the physical process~\cite{wilson:1973jj}.
This phenomenon is called the running of the coupling\index{Coupling!running} and is described by a Callan-Symanzik $\beta$-function~\cite{callan:1970yg,symanzik:1970rt}\index{Quantum!chromodynamics!$\beta$-function}, also called a renormalisation group flow rate.
The \ac{QCD} $\beta$-function is known to five loops~\cite{baikov:2016tgj,pich:2020gzz} and takes the form
\begin{align}
    \label{eq:beta-fn}
    \frac{d\astr}{d\ln(\mu^2)} \eqqcolon \beta(\astr) = -\sum_{n=0}^{\infty} \beta_n \br{\frac{\astr}{4\pi}}^{n+2} \,.
\end{align}
Since it comes with a negative sign, \ac{QCD} exhibits asymptotic freedom\index{Asymptotic freedom}~\cite{gross:1973id,politzer:1973fx}: the strong coupling decreases with increasing energy, as shown in \cref{fig:alphas}.
The scale separating the strongly- and weakly-coupled \ac{QCD} phases is $\Lambda_\text{\ac{QCD}}\sim\SI{200}{\MeV}$, with $\astr<1$ for energy scales $Q\gg\Lambda_\text{\ac{QCD}}$.
This means that hard processes---those at high energy or equivalently small length scale---can be treated perturbatively with $\astr$, \cref{eq:astr}, as the small expansion parameter.
Conversely, soft physics must be treated non-perturbatively, as we discuss in \cref{sec:beyond-fixed-order}.
Composite \ac{QCD} states in the bound regime are called hadrons\index{Hadron}.
This includes mesons\index{Hadron!meson} with an even number of valence quarks such as the pions, and baryons\index{Hadron!baryon} with an odd number of valence quarks such as the proton.

\section{Observables}
\label{sec:observables}

In this section, we discuss how to bridge the gap between perturbative \ac{QFT} calculations, which generally use Feynman diagram technology to compute amplitudes within a theory defined by its Lagrangian, and physical observables, which can be experimentally measured.

\subsection{Factorisation theorem}
\label{sec:factorisation-theorem}

The principles of scale factorisation\index{Scale!factorisation}~\cite{collins:1989gx,collins:2011zzd}---the separation of hard and soft physics---and asymptotic freedom (\cref{sec:qcd}) allow the calculation of observables using \ac{QCD} at hadron colliders.
We introduce the factorisation scale $\mu_F$ as the scale of the interface between soft and hard physics.

\begin{figure}
    \begin{center}
        \includegraphics[width=0.49\textwidth]{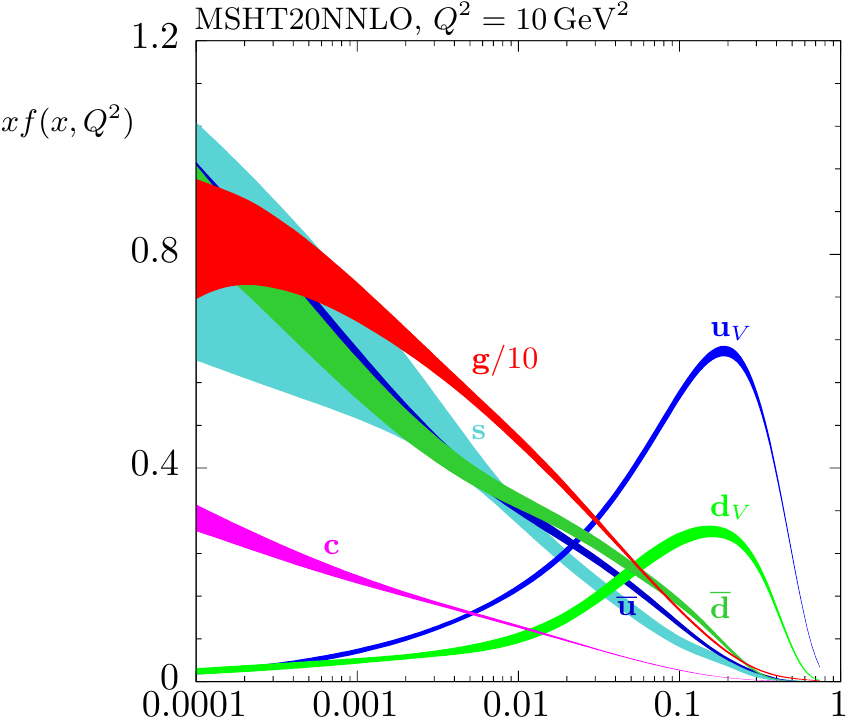}
        \includegraphics[width=0.49\textwidth]{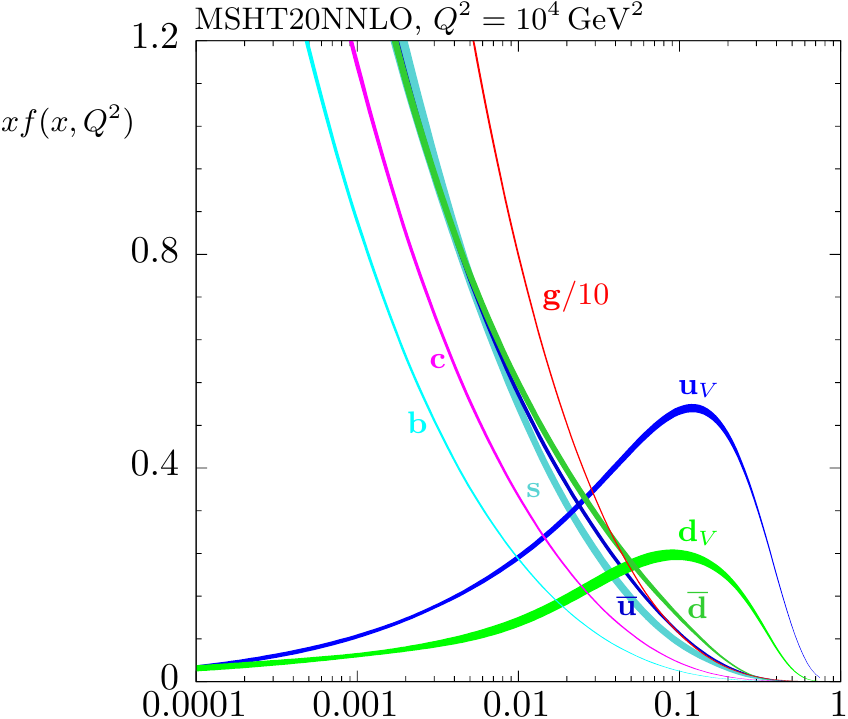}
        \caption{
            \code{MSHT20} \acs*{NNLO} \acsp*{PDF} for the proton at low (left) and high (right) energy scales.
            The $x$-axis shows the momentum fraction $x$.
            Lines are labelled by their particle type, with subscript V for valence.
            The gluon \acs*{PDF} is scaled down by a factor of ten.
            Figure from \incite{bailey:2020ooq}.
        }
        \label{fig:pdfs}
    \end{center}
\end{figure}

The soft physics of the composite initial states can be described by \iac{PDF}\index{Parton!distribution function}~\cite{gao:2017yyd,ethier:2020way,amoroso:2022eow} $f_i(x,\mu_F)$, which is the probability of finding a parton $i$\index{Parton}, meaning a quark or a gluon, with momentum fraction $x$ inside a hadron.
\Acp{PDF} are experimentally determined and evolved between different scales through the \ac{DGLAP} equations, just as the running of $\astr$ is described by the \ac{QCD} $\beta$-function (\cref{sec:qcd}).
Various \ac{PDF} sets are available, including \incites{ball:2021leu,hou:2019efy,h1:2015ubc,alekhin:2017kpj} and those shown in \cref{fig:pdfs}~\cite{bailey:2020ooq}, with recent \ac{PDF} sets achieving percent-level uncertainties~\cite{ball:2021leu}.
While the precision is not currently competitive with empirical techniques, there is also the future possibility to obtain the \acp{PDF} from first principles through lattice \ac{QCD}~\cite{constantinou:2022yye}.

The hard scattering of the partons is treated in fixed-order perturbation theory (\cref{sec:fixed-order}), in which it is given by the squared amplitude (\cref{sec:amplitudes}), which we also refer to as the \ac{ME}\index{Matrix!element}.
To calculate a differential cross section $\dd\sigma$, the \acp{PDF} are convolved with the \ac{ME} and integrated over the phase space by \iac{MC} event generator\index{Monte Carlo event generator}~\cite{campbell:2022qmc,reuschle:2014fya}.
There are several available general-purpose \ac{MC} event generators~\cite{bothmann:2019yzt,gleisberg:2008ta,bahr:2008pv,corcella:2000bw,bellm:2015jjp,sjostrand:2006za,sjostrand:2014zea,alwall:2014hca}.
They additionally simulate the non-perturbative effects discussed in \cref{sec:beyond-fixed-order}.

For a proton-proton collider like the \ac{LHC}, the hadronic cross section\index{Cross section} can be written as\index{Scale!factorisation!equation}
\begin{align}
    \dd\sigma = \sum_{i,j} \int_0^1 \dd x_1 \, \dd x_2 \, f_i(x_1,\mu_F) \, f_j(x_2,\mu_F) \, \dd\hat\sigma_{ij}(Q,\mu_F,\mu_R) + \order{\frac{\Lambda_\text{\ac{QCD}}}{Q}} \,.
\end{align}
The partonic, or hard, cross section for the scattering of the partons $i$ and $j$ in the initial states, $\dd\hat\sigma_{ij}$, depends on the partonic energy scale $Q$, which is given by the partonic centre of mass energy.
The renormalisation scale $\mu_R$ (\cref{sec:amplitudes}) and factorisation scale $\mu_F$ appear, although the scale dependence would vanish in an all-orders expression.
Further non-perturbative effects can be neglected for $Q\gg\Lambda_\text{\ac{QCD}}$~\cite{beneke:1998ui}.
The factorisation of partonic and hadronic physics is depicted in \cref{fig:collision}.

\subsection{Cross sections}
\label{sec-xs}

For a two-particle initial state $\mathcal{I}$ and an $N$-particle final state $\mathcal{F}$, the hard cross section is given by\index{Cross section!hard}
\begin{align}
    \label{eq:cross-section-definition}
    \dd{\hat\sigma} & = \frac{1}{\ff} \sumsquare{\am_{2+N}} \dd{\Phi_N} \,,
\end{align}
where $\sumsquare{\am_{2+N}}$ is the $2\to N$ amplitude squared with the overline denoting that initial spin and colour states are averaged and final spin and colour states are summed, $\dd{\Phi_N}$ is the $N$-particle Lorentz invariant phase space, and $\ff$ is the flux factor, which is
\begin{align}
    \label{eq:flux}
    \ff=2\,\s{12}
\end{align}
for a two-particle initial state ($\s{ij}=(p_i+p_j)^2$, see \cref{sec:dof}).
The treatment of colour derives from the phenomenon of colour confinement, which is that colour is unobservable because only colourless bound hadrons exist at low energy scales $Q<\Lambda_\text{\ac{QCD}}$ (\cref{sec:qcd}).
The treatment of spin is on the assumption that the collider uses an unpolarised input and does not measure the polarisation of the product states.
The phase space\index{Phase space}~\cite{gehrmann-deridder:2003pne} $\dd\Phi_N$ in $d$-dimensions has
\begin{align}
    (d-1)N-d
\end{align}
independent parameters\index{Phase space!degrees of freedom} and is given by
\begin{align}
    \label{eq:ndNpLips}
    \dd{\Phi_N^{(d)}}\brf{\mathcal{I}\rightarrow \mathcal{F}} & = \br{2\pi}^d \, \delta^{(d)}\brf{\sum_{f\in \mathcal{F}} p_f-\sum_{i\in \mathcal{I}} p_i} \prod_{f\in \mathcal{F}} \frac{\dd[d]{p_f}}{\qty(2\pi)^{d-1}} \deltaplus{{p_f}^2-{m_f}^2} \,,
\end{align}
where the first Dirac delta function fixes momentum conservation, and
\begin{align}
    \deltaplus{p^2-m^2} & = \delta\brf{p^2-m^2}\,\theta\brf{E} \,,
\end{align}
is the on-shell condition for each external particle, with $E$ as the temporal component of $p$.
The Heaviside unit step function $\theta(E)$ selects the positive energy solution.

Thus, the quantity we need to compute is the \ac{ME}.
In processes consisting of $D$ diagrams where $D$ is a large number, it is more efficient to evaluate the $D$ diagrams of the amplitude using the toolset built up in \cref{sec:amplitudes,sec:colour,sec:kin,sec:loops} and then modulus square the resulting complex number to calculate the amplitude squared, rather than directly evaluating the $D^2$ terms of the squared amplitude.

\subsection{Beyond fixed order}
\label{sec:beyond-fixed-order}

\begin{figure}
    \begin{center}
        \includegraphics[width=\textwidth]{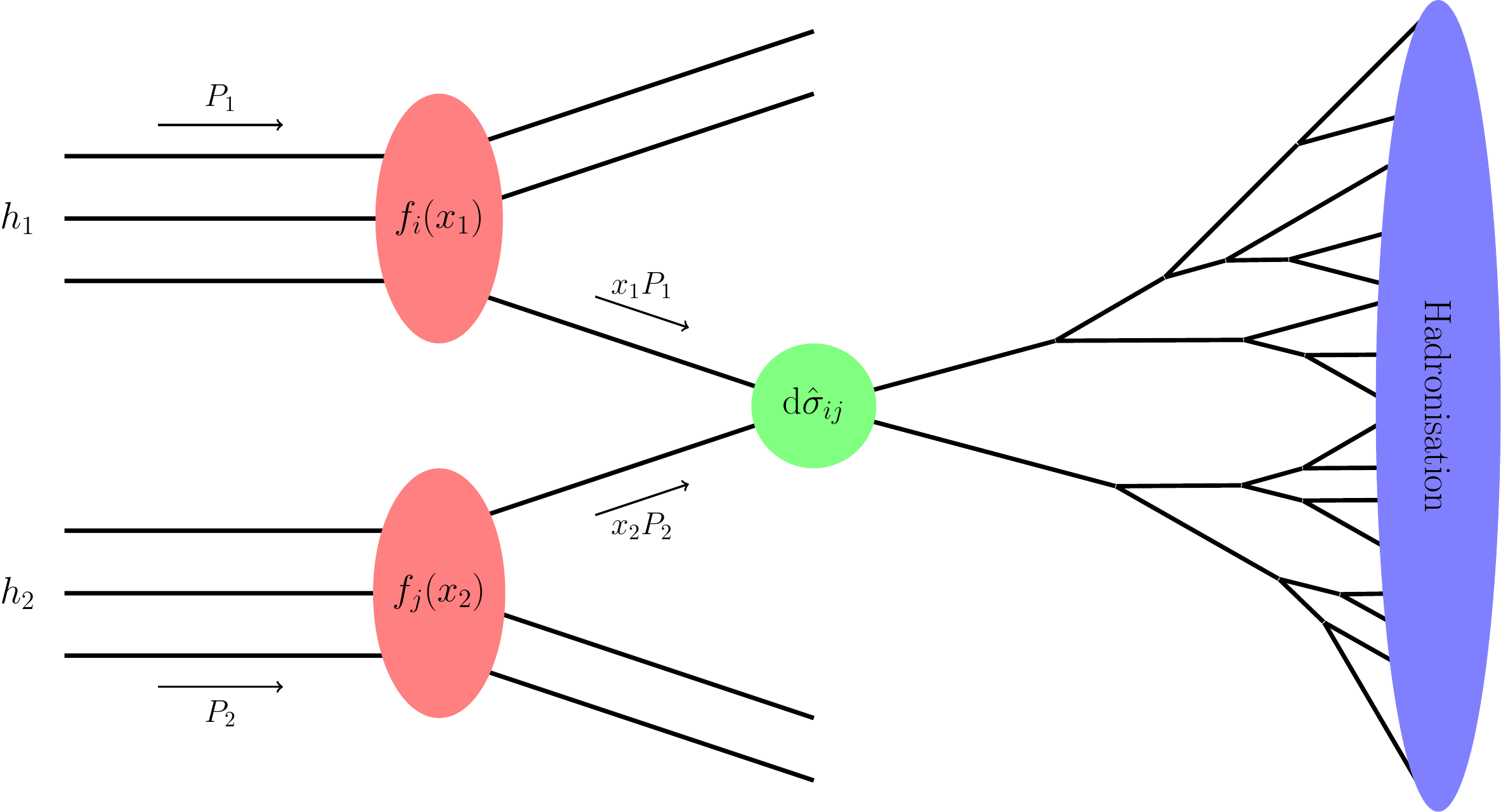}
        \caption{
            Sketch of a hadron collider simulation.
            A hadron $h_1$ with momentum $P_1$ collides with another hadron $h_2$ with momentum $P_2$.
            A parton $i$ with momentum fraction $x_1$ is factorised from $h_1$ through \iac{PDF} $f_i(x_1)$ (red).
            It interacts with a parton $j$ from $h_2$ with momentum fraction $x_2$, factorised by $f_j(x_2)$, in the hard scattering process, described by the hard cross section $\dd\hat\sigma_{ij}$ (green).
            The final states of the hard scattering repeatedly decay, which can be approximated with a parton shower.
            Hadronisation (blue) occurs when the energy scale evolves below $\Lambda_\text{\ac{QCD}}$.
        }
        \label{fig:collision}
    \end{center}
\end{figure}

\begin{figure}
    \begin{center}
        \includegraphics[width=\textwidth]{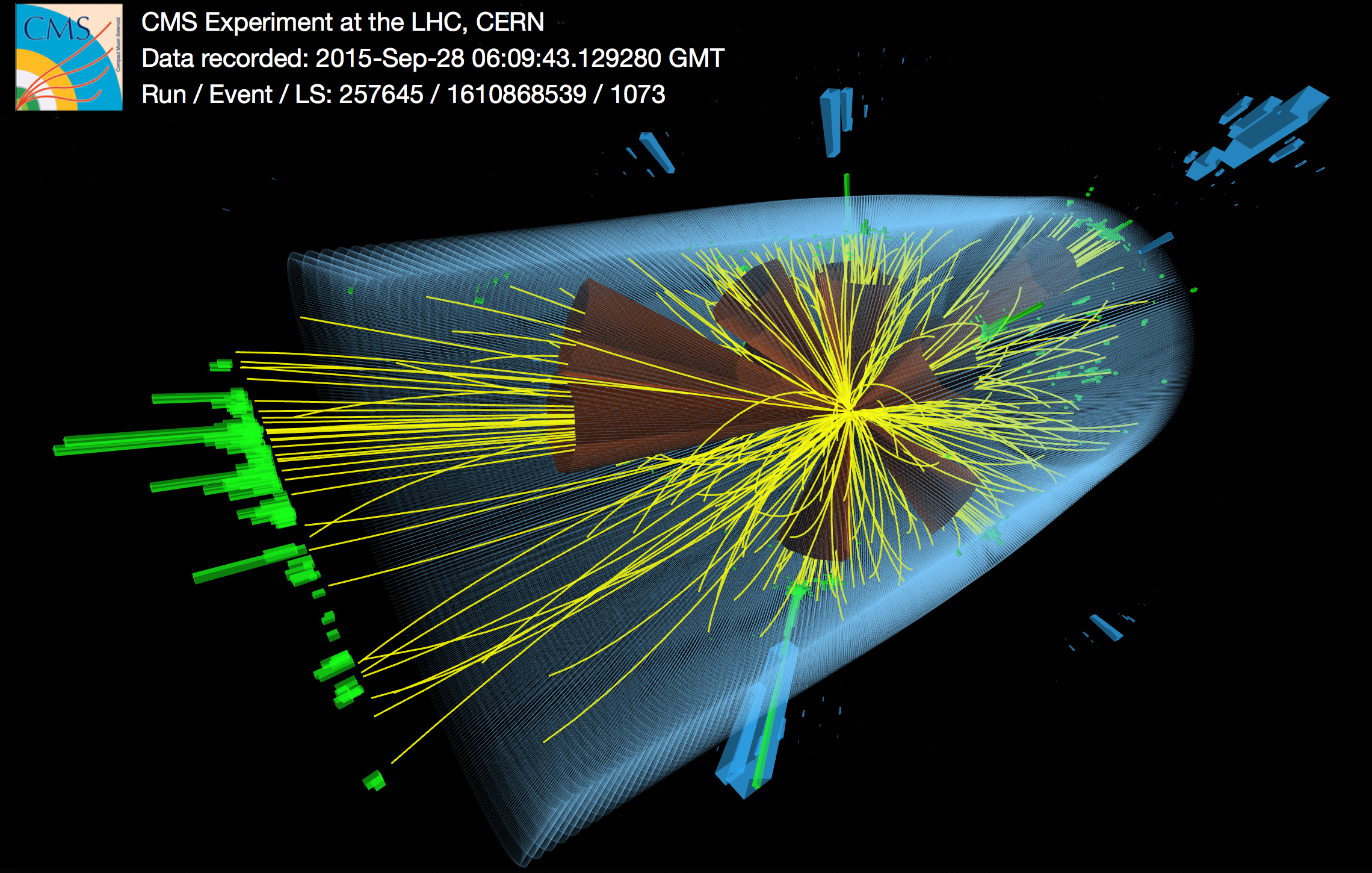}
        \caption{
            A high-multiplicity collision event observed by the \acs*{CMS} detector at the \ac{LHC} from collision data recorded in 2015~\cite{collaboration:2114784}.
        }
        \label{fig:cms}
    \end{center}
\end{figure}

Our discussion so far has assumed that the final states of the scattering are photons or partons.
However, it is not hard partons that are measured at colliders.
The initial and final state particles of the hard scattering process undergo additional radiation as they evolve to lower energy scales, creating a proliferation of extra particles.

The fixed-order prediction is constructed by integrating over the \ac{ME}, so propagators (\cref{sec:amplitudes}) give rise to logarithms.
The presence of large scale separations in the kinematics thus leads to logarithmic corrections to the fixed-order prediction.
Since soft and collinear states (\cref{sec:infrared-limits}) give rise to vanishing propagator momenta, these corrections can be significant in \ac{IR} regions of phase space for massless particles.

Treating these corrections to all orders in $\astr$ is called resummation\index{Resummation}.
It can be performed analytically~\cite{luisoni:2015xha,vanbeekveld:2022blq} in \ac{SCET}~\cite{bauer:2000yr,bauer:2001yt,beneke:2002ph,beneke:2002ni,becher:2014oda}, which systematically expands in powers of the logarithms, giving rise to the \ac{NkLL} language, and kinematic variable, expressed as \ac{NkLP}.
Resummation can also be done numerically below NLL through parton showers~\cite{fox:1979ag,bengtsson:1986et}, which is common with event generators~\cite{hoche:2014rga,sjostrand:2016bif}.
Matching logarithmic corrections to fixed-order results must be done carefully to avoid overcounting contributions~\cite{bertone:2022hig}.

All partons will eventually reach $\Lambda_\text{\ac{QCD}}$ and enter the soft regime, forming bound-state hadrons in a process known as hadronisation\index{Hadronisation}~\cite{andersson:1997xwk,winter:2003tt}.
For exclusive production (\ie~of a single hadron), for example, this can be described by fragmentation functions similarly to \acp{PDF}.
These hadrons decay into child particles, which, along with further emissions such as photons, are what is actually measured.
The parton shower and hadronisation processes within a hadronic simulation are depicted in \cref{fig:collision}.

The hard scattering products are then reconstructed from the measurements using jet clustering algorithms\index{Jet}~\cite{salam:2010nqg}; this is depicted in \cref{fig:cms}.
Care must be taken that additional potentially unresolved emissions at higher orders (see \cref{ch:ir}) do not lead to ambiguities in the jets, which is called \ac{IR} safety.
Specifically, an \ac{IR} safe observable is one for which the \ac{IR} poles in the real- and virtual-type corrections occur in the same bins.

\section{Amplitudes}
\label{sec:amplitudes}

We consider the scattering process of an initial collection of particles, denoted state $\ket{i}$, to a final state $\ket{f}$.
The particles are characterised by properties such as their kinematics and particle type.

The probability of the initial state evolving to the final state is given by
\begin{align}
    P(i\to f) = \ab{\bra{f}S\ket{i}}^2 \,,
\end{align}
where the scattering matrix $S$ is a map between these states.
We impose unitarity\index{Unitarity} on $S$ so that probability is conserved,
\begin{align}
    \label{eq:unitarity}
    S^\dagger S = S S^\dagger = 1 \,,
\end{align}
and explicitly separate the no-scattering case from the interaction, denoted by the transfer matrix $T$, as
\begin{align}
    \label{eq:transfer-matrix}
    S = 1 + \imi T \,.
\end{align}
We define the scattering amplitude $\am$\index{Amplitude}~\cite{dixon:2011xs} by writing the relation
\begin{align}
    \label{eq:T2A}
    \br{2\pi}^4\delta^{(4)}\brf{P_f-P_i}\,\am(i\to f) = \bra{f}T\ket{i} \,,
\end{align}
where ($P_i$) $P_f$ is the total momentum of the (initial) final state and the Dirac delta function imposes momentum conservation.

We calculate scattering amplitudes for $n$ particles as a perturbative expansion in a small dimensionless coupling parameter.
For \ac{QCD}, we use $\astr$,
\begin{align}
    \label{eq:amp-exp}
    \am_n = \astr^a \sum_{\ell=0}^{L} {\astr}^{\ell} \amp{\ell}{n} + \order{{\astr}^{a+L+1}} \,,
\end{align}
where $a$ is the power in the coupling $\astr$ of the \ac{LO} term and the $\amp{\ell}{n}$ are sums over Feynman diagrams with $n$ legs and $\ell$ loops.
We truncate the expansion at some loop order $L$.
The diagrams are composed of external legs\index{Amplitude!leg}, interaction vertices\index{Amplitude!vertex}, and internal lines called propagators\index{Amplitude!propagator}, with the possibility of closed propagators forming loops\index{Amplitude!loop}.
We refer to diagrams with $\ell=0$ as tree-level amplitudes or trees\index{Amplitude!tree}, $\ell=1$ as one-loop amplitudes, and so on.
Diagrams with $\ell\ge1$ are called loop-level amplitudes.
In the following, we are concerned with the efficient computation of the $\amp{\ell}{n}$.
Therefore, these order-by-order contributions are referred to as amplitudes, as well as the perturbative series which comprises the complete amplitude.

\subsection{Properties}
\label{sec:properties}

We can derive the mass dimension\index{Amplitude!mass dimension} $[\am_n]$ of the $2\to n-2$ scattering amplitude by considering the cross section \cref{eq:cross-section-definition}, which has dimensions of area, so mass dimension $[\hat\sigma]=-2$.
The inverse flux factor \cref{eq:flux} has mass dimension $[\ff]=-2$.
The $(n-2)$-particle phase space, \cref{eq:ndNpLips} with $N=n-2$, has mass dimension $[\Phi_{n-2}^{(4)}]=2n-8$ in four dimensions\footnote{
    Recall the usual decomposition and change of variables for the Dirac delta function,
    \begin{align}
        \label{eq:dirac-delta-change-of-variables}
        \dd^dp&=\dd^{d-1}\vec{p}\,\dd E \,, & \delta\brf{f(x)} &= \sum_{x_0\forall f(x_0)=0} \frac{\delta(x-x_0)}{f^\prime(x_0)} \,.
    \end{align}}.
Thus, an amplitude of multiplicity\index{Amplitude!multiplicity} (number of external legs) $n$ must have mass dimension
\begin{align}
    \label{eq:mass-dim}
    [\am_n]=4-n \,.
\end{align}

Amplitudes are gauge-invariant objects.
Thus, they obey the Ward identity\index{Identity!Ward}.
In \ac{QED}, this means that the amplitude vanishes on replacing the polarisation vector of an external gauge boson with its momentum,
\begin{align}
    \amp{\ell}{n}(p_1,\dots,p_n) &\eqqcolon \epsilon_\mu(p_i) \amp{\ell}{n}^\mu(p_1,\dots,p_n) \,, & \br{p_i}_\mu\amp{\ell}{n}^\mu(p_1,\ldots,p_n)&=0 \,.
\end{align}
This also holds in \ac{QCD} for amplitudes involving at most one longitudinal gluon.

Amplitudes exhibit two kinds of complexity~\cite{tancredi:2021oiq}: algebraic, which arises from the increase in kinematic parameters at higher multiplicity; and analytic, which refers to the complicated functions that describe loops (\cref{sec:loops}), such as multiple polylogarithms and their elliptic generalisation~\cite{goncharov:1998kja,duhr:2014woa,bourjaily:2022bwx,travaglini:2022uwo}.

\subsection{Infinities}
\label{sec:infinities}

Tree-level diagrams have all internal line momenta fixed by the external momenta through momentum conservation at each vertex.
However, in loop-level diagrams, there is an undetermined momentum flowing in each closed loop\index{Amplitude!loop}.
Thus, each loop in an amplitude introduces an integral over an unconstrained momentum.
These loop integrals can diverge\index{Divergence}---their value goes to infinity---in large-momentum regimes, where we call them \ac{UV} divergences\index{Divergence!\acl*{UV}}, and low-momentum regimes, which we call \ac{IR} divergences\index{Divergence!\acl*{IR}}.
These points are also called singularities\index{Singularity}, and their treatment is called regularisation.

For instance, the \ac{DR} method\index{Regularisation!dimensional}~\cite{thooft:1972tcz,gnendiger:2017pys} prescribes that we regulate \ac{UV} and \ac{IR} divergences by performing an analytic continuation in the number of spacetime dimensions to
\begin{align}
    d&=4-2\eps \,.
\end{align}
This has the effect of modifying the four-dimensional phase-space integration measure to \cref{eq:ndNpLips}, and replacing each loop integration measure as
\begin{align}
    \astr\,\frac{\dd^4\mom{l}{4}}{(2\pi)^4} &\to \astr\,{\mu}^{2\eps}\,\frac{\dd^d\mom{l}{d}}{(2\pi)^d} \,,
\end{align}
where $\mu$ is the regularisation scale\index{Scale!renormalisation}, an arbitrary scale introduced to fix the mass dimensions of the coupling.
The loop integral is then expressed as a Laurent series in the dimensional regulator\index{Regularisation!dimensional!dimensional regulator} $\eps$, with the singular behaviour cleanly extracted into poles in $\eps$\index{Pole}; the $d=4$ result is reproduced by taking $\eps\to0$.
\Ac{DR} has the advantage over simply introducing \iac{UV} cutoff that it preserves gauge invariance\index{Gauge!invariance}.
Note that scaleless integrals\index{Integral!scaleless}\footnote{
    Scaleless integrals are those that depend only on the regularisation scale $\mu$.
    Since the integral then takes the form $I=\mu^aI^\prime$ and $\mu$ is arbitrary, it must be that $I=0$.
    For example,
    \begin{align}
        &\int\frac{\dd^dk}{(2\pi)^d}\frac{1}{k^2(k+p)^2}\,, \quad p^2=0 \,, & &\int\frac{\dd^dk}{(2\pi)^d}\frac{1}{(k+q)^2}\,.
    \end{align}
} vanish in \ac{DR} due to exact cancellation of the \ac{UV} and \ac{IR} singularities after analytic continuation.

We can also introduce the ``spin dimension''\index{Regularisation!dimensional!spin dimension}, or quasidimension~\cite{goncharov:1998kja}, $d_s$ to regulate the contractions which arise from numerator algebra,
\begin{align}
    \concov{\lm}{\mu}{\mu} = d_s \,,
\end{align}
distinctly from the dimension of the loop integrals $d$.
The conventional \ac{DR} scheme is obtained by setting $d_s = d$.
The \ac{tHV} scheme\index{Regularisation!dimensional!\acl*{tHV}} is given by setting $d_s=d$ only for parts of diagrams that can lead to divergences, and $d_s=4$ otherwise.
Thus, in \ac{tHV}, external momenta are four-dimensional, while divergent loop-momenta are $d$-dimensional.

\Ac{UV} divergences arise due to our definition of the fields, couplings, and masses in the Lagrangian.
These are ``bare'' quantities that must be related to physical quantities before our calculation can return a meaningful answer.
The mechanism for this relationship is the scale dependence of parameters, as discussed for the running coupling in \cref{sec:qcd}.
The process is called renormalisation\index{Renormalisation}.
We express the parameter rescalings as
\begin{align}
    Z_x = 1 + \delta_x \,,
\end{align}
and substitute them into the Lagrangian to introduce counterterms under some scheme, such as the \ac{MSb} scheme.
Note that while it cancels their infinities, renormalisation is necessary regardless of the presence of \ac{UV} divergent integrals.

Renormalised all-orders quantities are naturally independent of the scale $\mu_R$ at which renormalisation is carried out.
However, truncated perturbative expansions in the coupling gain a residual scale dependence due to the omission of higher-order terms.
It is common practice to set the regularisation scale $\mu$ equal to the renormalisation scale $\mu_R$.

\Ac{IR} divergences are further discussed in \cref{ch:ir}.

\section{Colour}
\label{sec:colour}

In this section, the \ac{QCD} gauge group is generalised to \SUt{\nc}, where $\nc$\index{$\nc$} is the number of colours\index{Colour!number of}, to expose the group structure.
For \ac{QCD}, $\nc=3$.

\subsection{Colour decomposition}
\label{sec:colour-decomposition}

The number of Feynman diagrams contributing to an amplitude grows factorially with the number of legs.
It is therefore essential to use techniques that reduce redundant intermediate calculation.
One such method is colour decomposition\index{Colour!decomposition}, which exploits the property that \ac{QCD} amplitudes can be factorised into colour and kinematic parts.
Abstractly,
\begin{align}
    \label{eq:abstract-colour-decomposition}
    \amp{\ell}{n} & = \sum_i \cof^{(\ell)}_i \pamp{\ell}{n,i}
\end{align}
where: $\amp{\ell}{n}$ is the full amplitude at some loop order $\ell$; the colour factors\index{Colour!factor} $\cof^{(\ell)}_i$ contain the colour algebra; and the partial amplitudes\index{Amplitude!partial} $\pamp{\ell}{n,i}$, also called colour-ordered\index{Amplitude!colour-ordered} or primitive\index{Amplitude!primitive} amplitudes, contain the kinematics.
The choice of decomposition is called a colour basis\index{Colour!basis}\footnote{Note that while the colour factors form a basis, the set of partial amplitudes may be linearly dependent.}.
In appropriate colour bases, the partial amplitudes are colour-ordered, meaning that they only receive contributions from diagrams with a particular cyclical ordering of the external partons, and remain gauge invariant\index{Gauge!invariance}.
Thus, only adjacent legs may become collinear (\cref{sec:infrared-limits}).

\subsubsection{Fundamental basis}
\label{sec:fundamental-colour-decomposition}

The fundamental basis, also called the trace basis, is one such colour basis.
It is constructed by taking strings of the \SUt{\nc} generators in the fundamental representation $\cgf^a_{ij}$, which are $\nc^2-1$ traceless hermitian $\nc\times\nc$ matrices~\cite{dixon:1996wi,mangano:1987xk,bern:1990ux,bern:1996je}.
These matrices are normalised, in equation and diagram form, as
\begin{align}
    \label{eq:dynkin-index}
    \ptr{\cgf^a\cgf^b}&=\cdi\,\delta^{ab} \,, &
    \raisebox{-1.2em}{\includegraphics{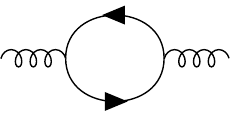}}
    & = \cdi\:
    \raisebox{-0em}{\includegraphics{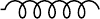}}
    \,.
\end{align}
The trace normalisation $\cdi$, or Dynkin index\index{Colour!Dynkin index}, is $1/2$ for \ac{QCD}, but will be left explicit in symbolic form.
Note that the adjoint colour index is denoted by \mbox{$a\in\cu{1,\ldots,\nc^2-1}$} and the fundamental colour index by \mbox{$i,j\in\cu{1,\ldots,\nc}$}.

Consider what can appear in a colour factor: a generic diagram will receive colour contributions from vertices,
\begin{align}
    \begin{aligned}
        \left.\raisebox{-3em}{\includegraphics{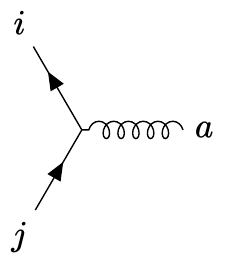}}\right|_\text{colour}
        & = \cgf^a_{ij} \,,
        \\
        \left.\raisebox{-2.5em}{\includegraphics{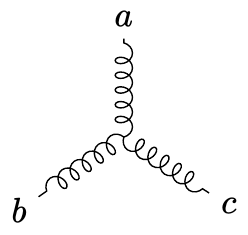}}\right|_\text{colour}
        & = \stc^{abc} \,,
        &
        \left.\raisebox{-2.5em}{\includegraphics{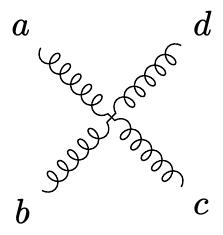}}\right|_\text{colour}
        & = \stc^{abe}\stc^{ecd} \,,
        \label{eq:colour-ordered-feynman-rules}
    \end{aligned}
\end{align}
and propagators,
\begin{align}
    \left.\raisebox{-0.25em}{\includegraphics{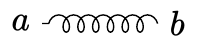}}\right|_\text{colour}
    & = \delta^{ab} \,,
    &
    \left.\raisebox{-0.25em}{\includegraphics{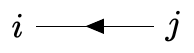}}\right|_\text{colour}
    & = \delta_{ij} \,.
\end{align}
We define the adjoint Casimir operator $\caa$,\index{Colour!Casimir!adjoint}
\begin{align}
    \caa&=\nc\,, &
    \stc^{abc}\stc^{dcb} &= \caa\,\delta^{ad} \,, &
    \raisebox{-1.2em}{\includegraphics{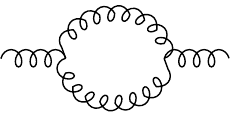}}
    &= \caa\:
    \raisebox{-0em}{\includegraphics{g}}
    \,,
\end{align}
and fundamental Casimir operator $\caf$,\index{Colour!Casimir!fundamental}
\begin{align}
    \caf&=\frac{\nc^2-1}{2\nc}\,, &
    t^a_{ij}t^a_{jk} &= \caf\,\delta_{ik} \,, &
    \raisebox{-0em}{\includegraphics{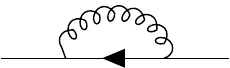}}
    &= \caf\:
    \raisebox{-0em}{\includegraphics{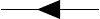}}
    \,.
\end{align}
\Cref{eq:colour-structure-constant} allows us to recast the structure constants in terms of the generators as
\begin{align}
    \begin{aligned}
        \label{eq:structure-function-to-fundamental-generators}
        \stc^{abc}&=-\frac{\imi}{\cdi} \ptr{\cgf^a \sq{\cgf^b,\cgf^c}} \,, \\
        \raisebox{-1.5em}{\includegraphics{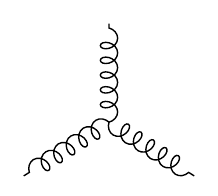}}
        &= -\frac{\imi}{\cdi} \sq{
            \raisebox{-1.5em}{\includegraphics{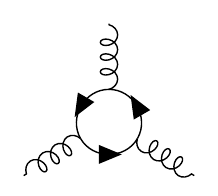}}
            -
            \raisebox{-1.5em}{\includegraphics{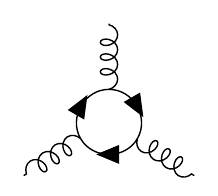}}
        }\,,
    \end{aligned}
\end{align}
which allows all colour factors to be expressed in terms of the generators only.
There also exists a symmetric tensor $d^{\hspace{1pt}abc}$ for the \SUt{\nc} Lie algebra, defined by
\begin{align}
    \cu{t^a,t^b}&=\frac{1}{\nc}\delta^{ab}+d^{\hspace{1pt}abc}t^c \,, &
    d^{\hspace{1pt}abc}=\frac{1}{\cdi}\ptr{t^a\cu{t^b,t^c}} \,,
\end{align}
although we will not make use of it here.

For example, consider $n$-gluon tree-level scattering amplitudes.
After using the Fierz identity\index{Identity!Fierz},
\begin{align}
    \begin{aligned}
        \label{eq:fierz}
        \cgf^a_{ij}\cgf^a_{kl}&=\cdi\br{\delta_{il}\delta_{kj}-\frac{1}{\nc}\delta_{ij}\delta_{kl}}\,, \\
        \raisebox{-1.1em}{\includegraphics{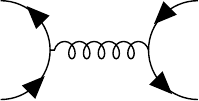}}
        \quad &= \quad \cdi
        \sq{\quad
        \raisebox{-1.2em}{\includegraphics{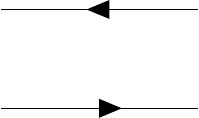}}
        \quad-\quad \frac{1}{\nc}\quad
        \raisebox{-1.0em}{\includegraphics{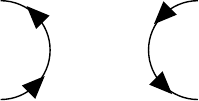}}
        \quad}
        \quad,
    \end{aligned}
\end{align}
which can be recognised as a completeness relation or proved via general tensor decomposition, and the cyclicity of the trace, the amplitude can be written in terms of $(n-1)!$ traces,\index{Colour!basis!fundamental}
\begin{align}
    \label{eq:n-gluon-tree-colour-decomposition-fundamental}
    \ant & =
    \sum_{\sigma\in S_n/\mathbb{Z}_n} \tr\Big(\sigma\brf{\cgf^1,\ldots,\cgf^n}\Big) \pant{\sigma\brf{1,\ldots,n}} .
\end{align}
With $S_n$ as the set of all permutations of $n$ objects and $\mathbb{Z}_n$ as the subset of cyclic permutations, the quotient $S_n/\mathbb{Z}_n$ is the set of all non-cyclic permutations.
This can be generated, for example, by fixing the first element and taking the cyclic permutations of the remaining elements.

The partial amplitudes also inherit a reflection identity from the antisymmetry of the colour-ordered Feynman rules, \cref{eq:colour-ordered-feynman-rules},
\begin{align}
    \label{eq:reflection-identity}
    \pant{1,\ldots,n}=\br{-1}^n\pant{n,\ldots,1} \,,
\end{align}
which reduces the decomposition to $(n-1)!/2$ terms,
\begin{align}
    \label{eq:n-gluon-tree-colour-decomposition-fundamental-with-reflection}
    \ant & =
    \sum_{\sigma\in R_{n}} \lambda\br{\sigma\brf{1,\ldots,n}} \pant{\sigma\brf{1,\ldots,n}} \,,
\end{align}
where
\begin{align}
    \lambda\br{1,\ldots,n} = \ptr{\cgf^1,\ldots,\cgf^n} + \br{-1}^{n} \ptr{\cgf^n,\ldots,\cgf^1} \,,
\end{align}
and the set $R_{n}$ is the reflection-independent subset of $S_n/\mathbb{Z}_n$.

The partial amplitudes also obey the photon decoupling identity,
\begin{multline}
    \label{eq:photon-decoupling-identity}
    0=
    \pant{1,2,3\ldots,n}+
    \pant{2,1,3\ldots,n}+\\
    \pant{2,3,1\ldots,n}
    + \cdots +
    \pant{2,3\ldots,1,n}.
\end{multline}
This arises because \cref{eq:n-gluon-tree-colour-decomposition-fundamental} is also valid for the gauge group $\gU{\nc}=\SU{\nc}\times\gU{1}$, but the extra $\gU{1}$ gauge field, or ``photon'', is colourless and therefore doesn't couple to gluons, so any amplitude containing it must vanish.
Thus, we insert a photon into \cref{eq:n-gluon-tree-colour-decomposition-fundamental-with-reflection} by setting one of the generators to be the generator of $\gU{1}$, which is proportional to the identity matrix, and gather terms to find \cref{eq:photon-decoupling-identity}.
This means only $\br{n-2}!$ of the partial amplitudes are \ac{LI}, so \cref{eq:n-gluon-tree-colour-decomposition-fundamental-with-reflection} is overcomplete.

The fundamental decomposition can be extended for general \ac{QCD} amplitudes containing $q\anti{q}$ pairs and gluons, as well as loop-level partonic amplitudes.

\subsubsection{Adjoint basis}
\label{sec:adjoint-colour-decomposition}

Another colour basis is the adjoint basis~\cite{delduca:1999rs,delduca:1999iql}.
With the \SUt{\nc} generator in the adjoint representation, \cref{eq:Fabc}, the adjoint colour decomposition for the $n$-gluon tree-level amplitude is
\begin{align}
    \label{eq:n-gluon-tree-colour-decomposition-adjoint}
    \ant & =
    \sum_{\sigma\in S_{n-2}} \sq{\sigma\brf{\cga^2,\ldots,\cga^{n-1}}}_{1n} \pant{1,\sigma\brf{2,\ldots,n-1},n}.
\end{align}
This decomposition can be derived from \cref{eq:n-gluon-tree-colour-decomposition-fundamental} using the \ac{KK} relation~\cite{kleiss:1988ne}, which is consistent with the identities of \cref{sec:fundamental-colour-decomposition} and describes the linear relations between partial amplitudes.
It can be stated as
\begin{align}
    \label{eq:kleiss-kuijf-relation}
    \pant{1,\cu{\alpha},n,\cu{\beta}}=\br{-1}^{n_{\beta}} \sum_{\sigma\in\mathrm{OP}} \pant{1,\sigma\brf{\cu{\alpha},\cu{\beta^T}},n} \,,
\end{align}
where
\begin{align}
    \cu{\alpha}\cup\cu{\beta}=\cu{2,\ldots,n-1}\,,
\end{align}
the length of the set $\cu{\beta}$ is $n_{\beta}$, $\{\beta^T\}$ is $\cu{\beta}$ with reversed ordering, and $\mathrm{OP}$ is the set of ordered permutations of $\cu{\alpha}\cup\{\beta^T\}$, \ie~all orderings of the union that preserve the ordering of the elements within the sets $\cu{\alpha}$ and $\{\beta^T\}$.
The adjoint decomposition directly gives the $\br{n-2}!$ partial amplitudes.

The adjoint basis exists only for pure-gluon amplitudes.

\subsubsection{Further bases}

Many further approaches have been explored in the literature~\cite{ochirov:2019mtf,johansson:2015oia,melia:2015ika,reuschle:2013qna,sjodahl:2014opa}.
The Bern-Carrasco-Johansson relations project out $(n-3)!$ partial amplitudes~\cite{bern:2008qj,brown:2016hck}.
The multiplet basis provides a minimal basis~\cite{keppeler:2012ih,du:2015apa,sjodahl:2018cca}.
To approximate two-loop processes, such as trijet production (\cref{ch:3j}), it is common to consider a colour expansion\index{Colour!expansion} in $1/\nc^2$ (in the amplitude squared) and take only the \ac{LC} terms~\cite{thooft:1973alw}.
For calculations involving many particles, \ac{MC} approximation techniques can be used which sample a reduced set of partial amplitudes in the colour sum~\cite{gleisberg:2008fv,caravaglios:1998yr,mangano:2001xp,draggiotis:1998gr,draggiotis:2002hm}.
The colour-flow basis~\cite{maltoni:2002mq,duhr:2006iq} treats gluons as $\nc\times\nc$ matrices $(A_\mu)^i_{\phantom{i}j}$ rather than single-index fields $A_\mu^a$ such that the colour factor is a string of Kronecker delta functions in fundamental colour indices and the colour sum is over $(n-1)!$ terms; it requires the evaluation of fewer partial amplitudes in \ac{MC} colour sums than other techniques.
This has lead to the development of the chirality-flow formalism~\cite{lifson:2020oll,lifson:2020pai,alnefjord:2021yks,alnefjord:2020xqr,lifson:2022ijv} for the treatment of the kinematics, which builds on the spinor-helicity formalism (\cref{sec:shf}).
The colour expansion approach has also been explored at high multiplicity~\cite{frederix:2021wdv}.

\subsection{Colour matrices}
\label{sec:colour-matrices}

The calculation of cross sections requires the squared amplitude.
Due to the phenomenon of colour confinement, colour is unobservable and so we average over initial state colours and sum over final state colours.
Abstractly, we can introduce the colour matrix\index{Colour!matrix},
\begin{align}
    \cm^{(\ell)}_{ij}&\coloneqq\sum_{\mathrm{colours}}{\cof^{(\ell)}_i}^{\dagger}\cof^{(\ell)}_j \,,
\end{align}
which is symmetric and has integer-valued elements,
\begin{align}
    \cm^{(\ell)}_{ij}&=\cm^{(\ell)}_{ji}\,, & \cm^{(\ell)}_{ij}&\in\integers\,,
\end{align}
to organise the computation as a colour sum\index{Colour!sum}, \cf~\cref{eq:abstract-colour-decomposition},
\begin{align}
    \label{eq:abstract-colour-matrix}
    \sum_{\mathrm{colours}} \ab{\amp{\ell}{n}}^2 & = \sum_{i,j} {\pa^{(\ell)}_{n,i}}^{\dagger} \, \cm^{(\ell)}_{ij} \, \pa^{(\ell)}_{n,j} \,.
\end{align}

For example, consider the $n$-gluon tree-level amplitude.
In the fundamental basis, using the decomposition \cref{eq:n-gluon-tree-colour-decomposition-fundamental-with-reflection}, with
\begin{align}
    \begin{aligned}
        l_n&=2\, \cdi^n \nc^{n-6} \br{\nc^2-1} \,, \\
        \gamma_1&=\nc^4-\gamma_2 \,, \\
        \gamma_2&=2\br{\nc^2-3} \,,
    \end{aligned}
\end{align}
the $n=4$ colour matrix with corresponding partial amplitude vector is
\begin{align}
    \cm^{(0)}_{ij} &= l_4
    \begin{pmatrix}
        \gamma_1 & \gamma_2 & \gamma_2 \\
        \gamma_2 & \gamma_1 & \gamma_2 \\
        \gamma_2 & \gamma_2 & \gamma_1 \\
    \end{pmatrix}_{ij} \,,
    &
    \pamp{0}{4,i} &=
    \begin{pmatrix}
        \pat{1,2,3,4}{4} \\
        \pat{1,2,4,3}{4} \\
        \pat{1,3,2,4}{4} \\
    \end{pmatrix}_i \,.
\end{align}
At $n=5$, with
\begin{align}
    \begin{aligned}
        \gamma_3&=\nc^4-4 \nc^2+10 \,, \\
        \gamma_4&=-2\nc^2+4 \,,
    \end{aligned}
\end{align}
we have
\begin{align}
    \cm^{(0)}_{ij} &= l_5
    \br{
        \begin{array}{cccccccccccc}
            \gamma_3  & \gamma_4  & \gamma_4  & 4         & 4         & -\gamma_4 & \gamma_4  & -4       & 4         & 0         & 4         & -\gamma_4 \\
            \gamma_4  & \gamma_3  & 4         & -\gamma_4 & \gamma_4  & 4         & -4        & \gamma_4 & 0         & 4         & \gamma_4  & -4        \\
            \gamma_4  & 4         & \gamma_3  & \gamma_4  & -\gamma_4 & 4         & 4         & 0        & -\gamma_4 & 4         & \gamma_4  & 4         \\
            4         & -\gamma_4 & \gamma_4  & \gamma_3  & 4         & \gamma_4  & \gamma_4  & -4       & -4        & -\gamma_4 & -4        & 0         \\
            4         & \gamma_4  & -\gamma_4 & 4         & \gamma_3  & \gamma_4  & 0         & 4        & 4         & -\gamma_4 & -4        & -\gamma_4 \\
            -\gamma_4 & 4         & 4         & \gamma_4  & \gamma_4  & \gamma_3  & -4        & \gamma_4 & -\gamma_4 & -4        & 0         & 4         \\
            \gamma_4  & -4        & 4         & \gamma_4  & 0         & -4        & \gamma_3  & \gamma_4 & \gamma_4  & 4         & -\gamma_4 & 4         \\
            -4        & \gamma_4  & 0         & -4        & 4         & \gamma_4  & \gamma_4  & \gamma_3 & 4         & \gamma_4  & -4        & \gamma_4  \\
            4         & 0         & -\gamma_4 & -4        & 4         & -\gamma_4 & \gamma_4  & 4        & \gamma_3  & -\gamma_4 & 4         & \gamma_4  \\
            0         & 4         & 4         & -\gamma_4 & -\gamma_4 & -4        & 4         & \gamma_4 & -\gamma_4 & \gamma_3  & -\gamma_4 & -4        \\
            4         & \gamma_4  & \gamma_4  & -4        & -4        & 0         & -\gamma_4 & -4       & 4         & -\gamma_4 & \gamma_3  & \gamma_4  \\
            -\gamma_4 & -4        & 4         & 0         & -\gamma_4 & 4         & 4         & \gamma_4 & \gamma_4  & -4        & \gamma_4  & \gamma_3  \\
        \end{array}
    }_{ij}\,,
\end{align}
and
\begin{align}
    \pamp{0}{5,i} &=
    \begin{pmatrix}
        \pat{1,2,3,4,5}{5} \\
        \pat{1,2,3,5,4}{5} \\
        \pat{1,2,4,3,5}{5} \\
        \pat{1,2,4,5,3}{5} \\
        \pat{1,2,5,3,4}{5} \\
        \pat{1,2,5,4,3}{5} \\
        \pat{1,3,2,4,5}{5} \\
        \pat{1,3,2,5,4}{5} \\
        \pat{1,3,4,2,5}{5} \\
        \pat{1,3,5,2,4}{5} \\
        \pat{1,4,2,3,5}{5} \\
        \pat{1,4,3,2,5}{5} \\
    \end{pmatrix}_i \,.
\end{align}

In the adjoint basis, with
\begin{align}
    k_n=\cdi^{n-2} \nc^{n-2} \br{\nc^2-1},
\end{align}
at $n=4$,
\begin{align}
    \cm^{(0)}_{ij} &= k_4
    \begin{pmatrix}
        4 & 2 \\
        2 & 4 \\
    \end{pmatrix}_{ij} \,,
    &
    \pamp{0}{4,i} &=
    \begin{pmatrix}
        \pat{1,2,3,4}{4} \\
        \pat{1,3,2,4}{4} \\
    \end{pmatrix}_{i} \,,
\end{align}
and at $n=5$,
\begin{align}
    \cm^{(0)}_{ij} &= k_5
    \begin{pmatrix}
        8 & 4 & 4 & 2 & 2 & 0 \\
        4 & 8 & 2 & 0 & 4 & 2 \\
        4 & 2 & 8 & 4 & 0 & 2 \\
        2 & 0 & 4 & 8 & 2 & 4 \\
        2 & 4 & 0 & 2 & 8 & 4 \\
        0 & 2 & 2 & 4 & 4 & 8 \\
    \end{pmatrix}_{ij} \,,
    &
    \pamp{0}{5,i} &=
    \begin{pmatrix}
        \pat{1,2,3,4,5}{5}   \\
        \pat{1,2,4,3,5}{5}   \\
        \pat{1,3,2,4,5}{5}   \\
        \pat{1,3,4,2,5}{5}   \\
        \pat{1,4,2,3,5}{5}   \\
        \pat{1,4,3,2,5}{5}   \\
    \end{pmatrix}_i \,.
\end{align}

The benefit of reducing the overcompleteness of the basis is evident in the reduced size of the colour matrix $\cm^{(0)}_{ij}$ and partial amplitude vector $\pamp{0}{n,i}$ for the adjoint basis compared to the fundamental.
While it does not reduce the size of the colour sum---which becomes increasingly important at high multiplicity---relations between the partial amplitudes like the \ac{KK} relation of \cref{eq:kleiss-kuijf-relation} can, of course, instead be applied in the construction of the partial amplitude vector to optimise its computation in the fundamental basis.

\section{Kinematics}
\label{sec:kin}

Traditionally, the basic kinematic variables used for scattering amplitudes are four-momenta ${p_i}^\mu$.
Correspondingly in squared amplitudes, the momentum invariants and a pseudoscalar are used, introduced in \cref{sec:dof}.
Using Weyl spinors (\cref{sec:weyl-spinors}) to describe massless fermions leads to the spinor-helicity formalism (\cref{sec:shf}), which aims to express amplitudes more compactly.
Momentum twistors (\cref{sec:mtv}) provide another parametrisation that can lead to efficient expression.

Let us begin with some notation.
The mostly-minus Minkowski metric tensor\index{Metric tensor} is used throughout,
\begin{align}
    \lm_{\mu\nu}=\brm{1&0&0&0\\0&-1&0&0\\0&0&-1&0\\0&0&0&-1}_{\mu\nu} \,.
\end{align}
The $n\times n$ identity matrix is denoted by $1_n$.
The Pauli matrices\index{Matrix!Pauli} are denoted by
\begin{align}
    \vec{\sigma}=\br{\sigma_1,\sigma_2,\sigma_3}\,,
\end{align}
with
\begin{align}
    \label{eq:pauli}
    \sigma_1 &= \rmat{0&1\\1&0}\,, & \sigma_2 &= \rmat{0&-i\\i&0}\,, & \sigma_3 &= \rmat{1&0\\0&-1}\,.
\end{align}

We take all particles as massless, giving the on-shell\index{On-shell} constraint,
\begin{align}
    \label{eq:massless}
    p^2=0 \,.
\end{align}
Where quarks are considered, we are concerned only with the light quarks: $u$, $d$, $c$, $s$, and $b$.
With the number of quark flavours denoted as $\nf$\index{Fermion!quark!number of}\index{$\nf$}, this is the $\nf=5$ regime.
As we study hard scattering, this justifies the massless limit.

All particles are taken as outgoing, which reverses the helicity\index{Helicity} (the projection of spin onto the direction of momentum) of physically incoming particles.
Thus, we consider $0\to n$, or $n$-point, scattering; the physical $2\to n-2$ amplitude can be recovered by crossing symmetry\index{Symmetry!crossing}.
Conservation of four-momentum then conveniently reads as\index{Momentum!conservation}
\begin{align}
    \label{eq:mom-cons}
    \sum_{i=1}^n {p_i}^\mu = 0 \,.
\end{align}
For example, we refer to five-gluon scattering as $5g$ in the all-outgoing case and $gg\to ggg$ in the physical case.

\subsection{Degrees of freedom}
\label{sec:dof}

We define the momentum invariants\index{$\s{ij}$} for $n$-point scattering as
\begin{align}
    \s{i\ldots k} &\coloneqq \br{p_{i} + \ldots + p_{k}}^2 \,, & i,\ldots,k&\in\cu{1,\ldots,n} \,.
\end{align}
The two-index case $\s{i j}$ is a generalisation to arbitrary multiplicity of the Mandelstam variables of $2\to2$ scattering.
Note that they are symmetric,
\begin{align}
    \s{ij}=\s{ji}\,,
\end{align}
and for massless particles,
\begin{align}
    \label{eq:sij-massless}
    \s{ij} &= 2\, p_i \cdot p_j \,, & \s{ii} &= 0 \,.
\end{align}
Higher-index invariants decompose to double-index invariants and masses.

The parity-even kinematics can be described by a set of \ac{LI} scalars.
These can be constructed by contracting momenta, leading to a subset of the momentum invariants $\s{ij}$ as a natural choice.
Since they are symmetric, there are $n(n+1)/2$ distinct $\s{ij}$.
For massless particles, we have $n$ constraints---the diagonals $\s{ii}$ vanish---taking us down to $n(n-1)/2$\footnote{This can then be understood as $n$ choose two combinations $\binom{n}{2}$ to generate all $\s{ij}$ with $j>i$.}.
There are a further $n$ constraints,
\begin{align}
    \sum_{j=1}^n\s{ij} =0 \qquad\forall i\in\cu{1,\ldots,n} \,.
\end{align}
given by momentum conservation, \cref{eq:mom-cons}.
This gives the number of \ac{LI} scalar products in $d$-dimensions as\index{Kinematic degrees of freedom}
\begin{align}
    \label{eq:dof}
    \mathcal{D}^{(d)}(n)&=\frac{n(n-3)}{2}\qquad\forall\,n\le (d+1) \,.
\end{align}
We consider the momenta as $d$-dimensional because we are using \ac{DR}.
Giving some explicit cases,
\begin{align}
    \mathcal{D}^{(d)}(4)&=2\,,&
    \mathcal{D}^{(d)}(5)&=5\,,&
    \mathcal{D}^{(d)}(6)&=9\,.
\end{align}
To see why the formula \cref{eq:dof} holds only for multiplicities $n\le (d+1)$, consider the breakdown of the physical case $d=4$ at $n=6$.
We have five ``independent'' momenta by momentum conservation, but as they are in a four-dimensional vector space, one must be dependent.
Thus, we have
\begin{align}
    \mathcal{D}^{(4)}(6)&=8\,.
\end{align}

In the massless case, the scalars, $\s{ij}$ of \cref{eq:sij-massless}, form a Gram matrix\index{Matrix!Gram} of the external momenta when considered as an $m\times m$ matrix, where $m$ gives the range of $i$ and $j$ as $i,j\in\cu{1,\ldots,m}$.
If and only if the set of momenta chosen by the range of $i$ and $j$ are \ac{LI}, then it is invertible and the determinant of the Gram matrix, or Gram determinant\index{Matrix!Gram!determinant}, is nonzero.
Thus, the ($d=4$, $n=6$) Gram determinant with $m=5$ vanishes, giving the additional constraint motivated before.

Considering also parity\index{Symmetry!parity}, which acts by flipping the sign of the spatial momentum components,
\begin{align}
    \label{eq:parity}
    P : \br{{p}^0 , \vec{p}\,} \longrightarrow  \br{{p}^0 , -\vec{p}\,} \, ,
\end{align}
there must also be a parity-odd pseudoscalar invariant\index{Pseudoscalar}.
In $d=4$ with $n$ particles, we can choose it as the parity-odd contraction of four \ac{LI} four-momenta with the fully antisymmetric four-dimensional Levi-Civita symbol\index{Levi-Civita symbol} $\lc_{\mu_1\mu_2\mu_3\mu_4}$.
A common choice is\index{$\trf$}
\begin{align}
    \label{eq:tr5}
    \trf &\coloneqq \ptrs{5}{p_1,p_2,p_3,p_4} \,,
\end{align}
where
\begin{align}
    \label{eq:tr5g}
    \ptrs{5}{p_i,p_j,p_k,p_l} &\coloneqq 4\,\imi\,\lc_{\mu_1\mu_2\mu_3\mu_4}\,{p_i}^{\mu_1}{p_j}^{\mu_2}{p_k}^{\mu_3}{p_l}^{\mu_4}\,.
\end{align}
We can also express this in terms of the spinor brackets defined in \cref{sec:shf},
\begin{align}
    \label{eq:tr5g-sh}
    \ptrs{5}{p_i,p_j,p_k,p_l}&= \spB{i}{j}\,\spA{j}{k}\,\spB{k}{l}\,\spA{l}{i} - \spA{i}{j}\,\spB{j}{k}\,\spA{k}{l}\,\spB{l}{i} \,.
\end{align}
Alternatively, this quantity can be expressed in terms of traces over gamma matrices,
\begin{align}
    \begin{aligned}
        \label{eq:tr+}
        \tr_\pm(p_i,p_j,p_k,p_l) &= \frac{1}{2}\tr[(1\pm\gamma_5)\slashed{p}_i \slashed{p}_j \slashed{p}_k \slashed{p}_l ] \,, \\
        \tr_+(p_i,p_j,p_k,p_l) &= \spB{i}{j} \spA{j}{k} \spB{k}{l} \spA{l}{i} \,, \\
        \tr_-(p_i,p_j,p_k,p_l) &= \spA{i}{j} \spB{j}{k} \spA{k}{l} \spB{l}{i} \,, \\
    \end{aligned}
\end{align}
by
\begin{align}
    \begin{aligned}
        \ptrs{5}{p_i,p_j,p_k,p_l} &= \tr_+(p_i,p_j,p_k,p_l) - \tr_-(p_i,p_j,p_k,p_l) \\
        &= \ptr{\gamma^5\slashed{p}_i\slashed{p}_j\slashed{p}_k\slashed{p}_l} \,. \\
    \end{aligned}
\end{align}
Note that for $n=3$, since we do not have sufficient \ac{LI} four-momenta, the pseudoscalar product \cref{eq:tr5g} vanishes and the kinematics are purely parity-even.
Since the product of two pseudoscalars is a scalar, all contractions of a pseudoscalar with a scalar can be expressed as one pseudoscalar variable, such as $\trf$, multiplied by some algebraic function of scalar variables, such as $\s{ij}$.

The square of $\trf$ can be expressed in terms of the scalars through the Gram determinant with
\begin{align}
    \label{eq:gram}
    {\trf}^2 = \Delta \coloneqq   \det(\s{ij})_{i,j\in\cu{1,2,3,4}} \,,
\end{align}
which is a degree-4 polynomial in the $\s{ij}$.
The pseudoscalar invariant $\trf$ therefore introduces an algebraic dependence on the kinematics, since
\begin{align}
    \label{eq:tr5-sqrt}
    \trf = \pm \sqrt{\Delta}\,.
\end{align}
We emphasise that the sign of $\trf$ changes under parity conjugation, \cref{eq:parity}, and under odd-signature permutations of the external momenta.

For further reading, see \incite{eden:1966dnq} and \crefext{section}{2} of \incite{henn:2021cyv}.

\subsection{Representations}
\label{sec:representations}

The Lorentz\index{Lorentz group} group\footnote{Strictly, ``Lorentz group'' is used here to refer to the identity connected component of the $\mathrm{O}(1,3)$ group.} is the Lie group $\mathrm{SO}(1,3)$.
The algebra of this group, $\mathfrak{so}(1,3)$, maps to two commuting copies of the algebra of \SUt{2}, $\mathfrak{su}(2)$, so its representations can be labelled by the chiral doublet $\br{a,b}$ where $a$ and $b$ are the eigenvalues of the Casimir operators of the two $\mathfrak{su}(2)$ algebras.

Four-momenta\index{Momentum} $p^\mu$, with Lorentz indices $\mu\in\cu{0,1,2,3}$, are Lorentz vectors that transform under the $\br{1/2,1/2}$ representation of the Lorentz group.
In the spinor-helicity formalism, spinors\index{Spinor}\footnote{Strictly, spinors lie in representations of the spin group $\mathrm{SL}\br{2,\cn}$ which is a homomorphism of the Lorentz group.} are instead used as the basic kinematic variable.
In general, Dirac spinors\index{Spinor!Dirac} $\psi^\alpha$, with Dirac spinor indices $\alpha\in\cu{1,2,3,4}$, might be used, which solve the Dirac equation and lie in the $\br{1/2,0}\oplus\br{0,1/2}$ representation.
For massless vectors, however, the Dirac spinor decomposes into two Weyl spinors\index{Spinor!Weyl} and we use these instead.
The first is the left-handed Weyl spinor\index{Spinor!Weyl!left-handed} $\lwsug{\alpha}$, also called the holomorphic spinor\index{Spinor!Weyl!holomorphic}, lying in the $\br{1/2,0}$ representation with left-handed Weyl spinor indices $\dot\alpha\in\{\dot{1},\dot{2}\}$.
The other is the right-handed Weyl spinor\index{Spinor!Weyl!right-handed} $\rwsug{\alpha}$, or antiholomorphic spinor\index{Spinor!Weyl!antiholomorphic}, lying in the $\br{0,1/2}$ representation with right-handed Weyl spinor indices $\alpha\in\cu{1,2}$.
There also exist negative energy solutions to the Weyl equations, but for massless particles they are equal to the positive solutions and hence can be neglected.

\subsection{Weyl spinors}
\label{sec:weyl-spinors}

In analogy to the metric tensor in a Lorentz vector space, the raising and lowering of indices in the Weyl spinor spaces is handled by the two-dimensional Levi-Civita tensor $\lc$ (choosing $\rlcu{1}{2}=\llcu{1}{2}=1$ such that $\lc=\imi\sigma_2$), for example,
\begin{align}
    \lwsdg{\alpha} & = \llcd{\alpha}{\beta} \, \lwsug{\beta} \,,
    &
    \rwsdg{\alpha} & = \rlcd{\alpha}{\beta} \, \rwsug{\beta} \,.
\end{align}
Translations between Lorentz and spinor indices use the sigma matrices\index{Matrix!sigma},
\begin{align}
    \br{\sigma^\mu}_{\alpha\dot\alpha}     & \coloneqq \br{1_2,\,\vec{\sigma}} \,,
    &
    \br{\sigmabar^\mu}^{\dot\alpha\alpha} & \coloneqq \br{1_2,\,-\vec{\sigma}} \,,
\end{align}
noting that
\begin{align}
    \br{\sigmabar_\mu}^{\dot\alpha\alpha} &= \rlcu{\alpha}{\beta} \llcu{\alpha}{\beta} \br{\sigma_\mu}_{\dot\beta\beta} .
\end{align}
The sigma matrices are analogous to the gamma matrices of Dirac spinor algebra.

The four-momentum Lorentz vector ${p}^\mu$ is related to a rank-two spinor (\ie~with two indices), or bispinor\index{Spinor!Weyl!bispinor}, by
\begin{align}
    {p}^{\dot\alpha\alpha} & =\br{\sigmabar_\mu}^{\dot\alpha\alpha} {p}^\mu .
\end{align}
This bispinor is the Weyl spinor analogy to a slashed momentum in the Feynman slash notation, \cref{eq:slash}.
Expressions in Dirac spinors for massless fermions with gamma matrices can always be decomposed into Weyl spinors and sigma matrices by choosing the Weyl, or chiral, representation of the gamma matrices\index{Matrix!gamma},
\begin{align}
    \label{eq:gammas}
    \gamma^\mu &=
    \begin{pmatrix}
        0 & \sigma^\mu \\
        \sigmabar^{\mu} & 0 \\
    \end{pmatrix} \,,
    &
    \gamma^5 &=
    \begin{pmatrix}
        -1_2 & 0 \\
        0 & 1_2 \\
    \end{pmatrix} \,.
\end{align}
Similarly, the massless Dirac equation,
\begin{align}
    \slashed{p}\,\psi(p)=0\,,
\end{align}
decomposes into the left- and right-handed Weyl equations,
\begin{align}
    \label{eq:weyl-eq-mom}
    p_{\alpha\dot\alpha}\,\lwsug{\alpha} &= 0 \,,&
    p^{\dot\alpha\alpha}\,\rwsdg{\alpha} &= 0 \,.
\end{align}

The definition of the sigma matrices leads to the Fierz identities\index{Identity!Fierz},
\begin{align}
    \label{eq:sigma-identities}
    \br{\sigma^\mu}_{\alpha\dot{\alpha}} \br{\sigmabar_\mu}^{\dot{\beta}\beta} & = 2 \, \covcon{\delta}{\alpha}{\beta} \, \covcon{\delta}{\dot{\alpha}}{\dot{\beta}} \,,
    &
    \ptr{\sigma^\mu \sigmabar^\nu}                                                 & = 2 \, \lm^{\mu\nu} \,,
\end{align}
which can be used to express the on-shell constraint\index{On-shell} \cref{eq:massless} as
\begin{align}
    \frac{1}{2} \, {p}^{\dot\alpha\alpha} \, {p}^{\dot\beta\beta} \, \llcd{\alpha}{\beta} \, \rlcd{\alpha}{\beta}
    & =
    \pdet{{p}^{\dot\alpha\alpha}}
    = 0 .
\end{align}
Since the determinant vanishes, the $2\times2$ matrix ${p}^{\dot\alpha\alpha}$ is rank one (\ie~has a single \ac{LI} row or column).
It may be written in bispinor form as
\begin{align}
    {p}^{\dot\alpha\alpha} & = \lwsug{\alpha}\rwsug{\alpha} .
\end{align}
For real momentum, this means that
\begin{align}
    \lwsug{\alpha}=\br{\rwsug{\alpha}}^*\,,
\end{align}
or in other words, complex conjugation is equivalent to a chirality flip.
If the momentum is complex, then the spinors are unrelated.
The bispinor form also demonstrates that the momentum is invariant under little group\index{Little group} transformations\footnote{The little group is the group of transformations for which the four-momentum of an on-shell particle is invariant.}~\cite{wigner:1939cj},
\begin{align}
    \label{eq:little-group}
    \br{\lws,\,\, \rws} \rightarrow \br{\phi\lws,\,\, \frac{\rws}{\phi} } \,,
\end{align}
for a complex phase $\phi\in\cn$, so a four-momentum does not map uniquely to a spinor pair.

\subsection{Spinor-helicity formalism}
\label{sec:shf}

\index{Spinor!helicity formalism}In the spinor-helicity formalism, the scattering amplitude of $n$ massless particles is a function of the set of Weyl spinors $\{\lws_i,\rws_i\}$ with particle label $i\in\cu{1,\ldots,n}$.
Since they describe massless fermions, Weyl spinors with left-handed chirality $\lws$ have negative helicity, and similarly the right-handed spinors $\rws$ have positive helicity.

A bracket notation is used to denote the Weyl spinors, defined by the spinor products,
\begin{align}
    \begin{aligned}
        \label{eq:spinor-product}
        \spB{i}{j} & \coloneqq \lwsd{i}{\alpha}\,\lwsu{j}{\alpha} = \lwsd{i}{\alpha}\,\llcu{\alpha}{\beta}\,\lwsd{j}{\beta} \,, \\
        \spA{i}{j} & \coloneqq \rwsu{i}{\alpha}\,\rwsd{j}{\alpha} = \rwsd{i}{\alpha}\,\rlcu{\alpha}{\beta}\,\rwsd{j}{\beta} \,,
    \end{aligned}
\end{align}
such that,
\begin{align}
    \begin{aligned}
        \label{eq:weyl-spinors}
        \sBr{i} & = \lwsu{i}{\alpha} \,,  &\quad
        \sBl{i} & = \lwsd{i}{\alpha} \,,  \\
        \sAr{i} & = \rwsd{i}{\alpha} \,,  &\quad
        \sAl{i} & = \rwsu{i}{\alpha} \,.
    \end{aligned}
\end{align}
Contractions with matrices are denoted by
\begin{align}
    \spBA{i}{\sigma^\mu}{j} \coloneqq \lwsd{i}{\alpha} \br{\sigma^\mu}^{\dot\alpha\alpha} \rwsd{j}{\alpha} \,,
\end{align}
and contractions with bispinors as
\begin{align}
    \begin{aligned}
        \spBA{i}{j}{k} &\coloneqq {p_j}^{\mu}\,\spBA{i}{\sigma_\mu}{k} = \lwsd{i}{\alpha} \br{p_j}^{\dot\alpha\alpha}\rwsd{k}{\alpha} \,, \\
        \spAB{i}{j}{k} &\coloneqq {p_j}^{\mu}\,\spAB{i}{\sigmabar_\mu}{k} = \rwsu{i}{\alpha} \br{p_j}_{\alpha\dot\alpha}\lwsu{k}{\alpha} \,.
    \end{aligned}
\end{align}

The manipulation of expressions involving these square and angle bracket spinors mainly involves the use of a number of identities.
Some common ones are listed here:
\begin{itemize}
    \item Momentum invariants
        \begin{align}
            \label{eq:sij-spinors}
            \s{ij} & = \spA{i}{j} \spB{j}{i} \,.
        \end{align}

    \item Antisymmetry
        \begin{align}
            \spA{j}{i} & = - \spA{i}{j} \,,&
            \spB{j}{i} & = - \spB{i}{j} \,,&
            \spA{i}{i} &= \spB{i}{i} = 0 \,.
        \end{align}

    \item Projection operators
        \begin{align}
            \label{eq:proj-ops}
            \sAr{i} \sBl{i} & = \br{p_i}_{\alpha\dot\alpha} \,,
            &
            \sBr{i} \sAl{i} & = \br{p_i}^{\dot\alpha\alpha} \,.
        \end{align}

    \item Gordon identity
        \begin{align}
            \label{eq:gordon-id}
            \spBA{i}{\sigma^\mu}{i} & = 2 \, {p_i}^\mu \,.
        \end{align}

    \item Fierz identity\index{Identity!Fierz}
        \begin{align}
            \spBA{i}{\sigma^\mu}{j} \spBA{k}{\sigma_\mu}{l} & = 2 \, \spB{i}{k} \spA{l}{j} \,.
        \end{align}

    \item Charge conjugation
        \begin{align}
            \spBA{i}{\sigma^\mu}{j} & = \spAB{j}{\sigmabar^\mu}{i} \,.
        \end{align}

    \item Complex conjugation (for real momenta ${p_i}^\mu\in\mathbb{R}$ only)
        \begin{align}
            \label{eq:complex-conjugation-spinors}
            \spA{i}{j}^* =
            \begin{cases}
                \phantom{-}\spB{i}{j} & \mathrm{if}\quad \mathrm{sign}({p_i}^0) = \mathrm{sign}({p_j}^0) \,, \\
                -\spB{i}{j} & \mathrm{otherwise} \,.
            \end{cases}
        \end{align}

    \item Schouten identities\index{Identity!Schouten}\footnote{This follows from the fact that a two-component object can be written as the linear combination of two others.}
        \begin{align}
            \label{eq:schouten}
            \spA{i}{j} \spA{k}{l} & = \spA{i}{k} \spA{j}{l} + \spA{i}{l} \spA{k}{j} \,, &
            \spB{i}{j} \spB{k}{l} & = \spB{i}{k} \spB{j}{l} + \spB{i}{l} \spB{k}{j} \,.
        \end{align}

    \item Momentum conservation for an $n$-point amplitude\index{Momentum!conservation}
        \begin{align}
            \label{eq:mom-cons-spinor}
            \sum^n_{j=1} \, \spB{i}{j} \spA{j}{k} &= 0 \qquad\forall\, i,k\in\cu{1,\ldots,n}\,.
        \end{align}
\end{itemize}

To express all the kinematics of a generic amplitude in terms of the spinor variables, the polarisation vectors\index{Polarisation vector} of bosons also require a spinor representation.
This is achieved for massless gauge bosons of definite helicity $\pm1$ by
\begin{align}
    \label{eq:polarisation-weyl}
    \pov_\mu^-\br{p_i,p_r} & = - \frac{\spBA{r}{\sigma_\mu}{i}}{\sqrt{2}\,\spB{r}{i}} \,,
    &
    \pov_\mu^+\br{p_i,p_r} & = \frac{\spAB{r}{\sigmabar_\mu}{i}}{\sqrt{2}\,\spA{r}{i}} \,,
\end{align}
where the superscript $\pm$ refers to the helicity of the boson, $p_i$ is the momentum of the boson, and $p_r$ is a reference momentum due to gauge invariance\index{Momentum!reference}.
We define
\begin{align}
    \pov_i^\lambda\br{q}\coloneqq\pov^\lambda\br{p_i,q}
\end{align}
for convenience.
The amplitude is independent of the choice of reference momentum; judicious choice of its value can greatly simplify algebra.
The Weyl equation, \cf~\cref{eq:weyl-eq-mom},
\begin{align}
    \spBA{r}{i}{i} = 0\,,
\end{align}
implies that the polarisation vector is transverse to its momentum for any reference momentum,
\begin{align}
    \pov_i^\pm(q) \cdot p_i = 0 \,.
\end{align}
Again, there are some useful identities to deal with these expressions:
\begin{itemize}
    \item Complex conjugation
        \begin{align}
            \label{eq:reverse}
            \br{\pov_i^\mp\br{q}}^* & = \pov_i^\pm\br{q} \,.
        \end{align}
    \item Normalisation
        \begin{align}
            \label{eq:normal}
            \pov_i^\pm\br{q} \cdot \br{\pov_i^\pm\br{q}}^* & = -1 \,.
        \end{align}
    \item Orthogonality
        \begin{align}
            \label{eq:orthog}
            \pov_i^\pm\br{q} \cdot \br{\pov_i^\mp\br{q}}^* & = 0 \,.
        \end{align}
    \item Completeness relation\index{Identity!completeness}
        \begin{align}
            \label{eq:polarisation-completeness}
            \sum_{\lambda\in\{-,+\}} \pov_\mu^\lambda\br{p,q} \br{\pov_\nu^\lambda\br{p,q}}^* & = -\eta_{\mu\nu} + \frac{p_\mu q_\nu + p_\nu q_\mu}{q \cdot p} \,.
        \end{align}
    \item Cancellation identities
        \begin{subequations}
            \begin{align}
                \label{eq:polarisation-dot-reference}
                \pov_i^\pm\br{q}\cdot q                                                   & = 0 \,, \\
                \label{eq:polarisation-dotted-same}
                \pov^\pm_i\br{q} \cdot \pov^\pm_j\br{q}                             & = 0 \,, \\
                \label{eq:polarisation-dotted-opposite}
                \pov_i^\pm\br{p_j} \cdot \pov^\mp_j\br{q}                           & = 0 \,, \\
                \label{eq:slashed-polarisation-right}
                \slashed{\pov}_i^{+}\br{p_j}|j\rangle = \slashed{\pov}_i^{-}\br{p_j}|j] & = 0 \,, \\
                \label{eq:slashed-polarisation-left}
                [j|\slashed{\pov}_i^-\br{p_j} = \langle j | \slashed{\pov}_i^+\br{p_j}  & = 0 \,.
            \end{align}
        \end{subequations}
\end{itemize}
With these identities, it is possible to express and simplify any amplitude with massless fermions and vector bosons as the external states in the spinor-helicity formalism.

Recall the little group scaling\index{Little group} of the Weyl fermions, \cref{eq:little-group}, or in the new notation,
\begin{align}
    \begin{aligned}
        \sBr{i} &\to \phi\sBr{i} \,, & \sAr{i} &\to \phi^{-1}\sAr{i} \,, \\
        \sBl{i} &\to \phi\sBl{i} \,, & \sAl{i} &\to \phi^{-1}\sAl{i} \,.
    \end{aligned}
\end{align}
We now scale \cref{eq:polarisation-weyl} to find for the bosons that
\begin{align}
    \pov_i^\pm &\to \phi^{\mp 2} \pov_i^\pm \,.
\end{align}
Thus, all particles transform as $\phi^{-2h}$, where $h$ is the helicity of the particle,
\begin{align}
    h=\begin{cases}
        -\frac{1}{2} & \text{for } \sBr{i} \,,\\
        \phantom{-}\frac{1}{2} & \text{for } \sAr{i} \,,\\
        \pm1 & \text{for } \pov^\pm \,.
    \end{cases}
\end{align}
Therefore, an amplitude of $n$ massless particles scales according to the helicity of each particle,
\begin{align}
    \label{eq:amp-hel-scaling}
    \prod_{i=1}^{n} \phi_i^{-2h_i}\,.
\end{align}
These scaling factors are also referred to as phase or helicity weights\index{Little group!weight}.

\subsection{Helicity amplitudes}
\label{sec:helicity-amplitudes}

Treating spin on the assumption that the collider uses an unpolarised input and does not measure the polarisation of the product states, we construct the full amplitude as a sum over helicity amplitudes\index{Amplitude!helicity},
\begin{align}
    \amp{\ell}{n} = \sum_{h_1,\ldots,h_n} \amp{\ell}{n}(1^{h_1},\ldots,n^{h_n}) \,,
\end{align}
where $h_i\in\cu{+,-}$.
For tree-level helicity amplitudes, the polarisation vector cancellation identity \cref{eq:polarisation-dotted-same} causes all amplitudes of helicity signature\footnote{Helicity signature $\br{a,b}$ refers to an $(a+b)$-particle amplitude that has $a$ positive helicity legs and $b$ negative helicity legs.} $\br{+,-} = \br{0,n}$ and $\br{1,n-1}$, and their parity conjugates, to vanish,
\begin{align}
    \pant{1^\pm,\ldots,n^\pm} = \pant{1^\pm,\ldots,i^\mp,\ldots,n^\pm} &= 0 \,.
    \label{eq:uhv}
\end{align}
Therefore, amplitudes with signature $\br{2,n-2}$ receive the name \ac{MHV} amplitudes\index{Amplitude!\acs*{MHV}}, and their conjugates are called \ac{MHVb} amplitudes.
Signature $(3,n-3)$ amplitudes are called \ac{NMHV} amplitudes, and so on.

To calculate pure-gluon tree-level \ac{MHV} amplitudes, the tools we have now built up come into good use.
The number of diagrams to evaluate is reduced using colour decomposition and relations between partial amplitudes.
Others vanish due to judicious choice of the reference momenta.
Finally, use of the identities of \cref{sec:shf} allow the result to be collected in a single term.
A remarkably simple expression is found as the result: the Parke-Taylor amplitudes~\cite{parke:1986gb,berends:1987me}, which have the form
\begin{align}
    \label{eq:mhv}
    \begin{aligned}
        \pa_{n}^{(0),\text{\acs{MHV}}}\br{1^+,\ldots,j^-,\ldots,k^-,\ldots,n^+} &= \imi \, \frac{\spA{j}{k}^4}{\prod_{i=1}^n \spAa{i|i+1}} \,, \\
        \pa_{n}^{(0),\text{\acs{MHVb}}}\br{1^-,\ldots,j^+,\ldots,k^+,\ldots,n^-} &= \imi \, \frac{\spB{j}{k}^4}{\prod_{i=1}^n \spBa{i|i+1}} \,,
    \end{aligned}
\end{align}
with the indices $i$ defined modulo $n$.

Massless three-particle scattering presents a special case.
By momentum conservation \cref{eq:mom-cons}, $\s{12}=0$, and so by \cref{eq:sij-spinors}, either $\spA{1}{2}$ or $\spB{1}{2}$ must vanish.
Furthermore, by \cref{eq:mom-cons-spinor,eq:weyl-eq-mom}, we find either all square brackets or all angle brackets vanish,
\begin{align}
    \spA12=\spA23=\spA13=0 \qquad\text{or}\qquad \spB12=\spB23=\spB13=0\,.
\end{align}
% and thus all square spinors or all angle spinors are proportional,
% \begin{align}
%     \sAr1 \propto \sAr2 \propto \sAr3 \qquad\text{or}\qquad \sBr1 \propto \sBr2 \propto \sBr3 \,.
% \end{align}
Since a nonzero three-particle amplitude must therefore depend only on angle or square brackets, and for real momenta, these are each other's complex conjugates by \cref{eq:complex-conjugation-spinors}, such amplitudes do not exist for physical kinematics.
However, they are nonvanishing with complex momenta, with values given by \cref{eq:mhv}.

We use curly brackets to denote the exponents of the little group scalings \cref{eq:amp-hel-scaling} for each leg of helicity amplitudes, or in general any object carrying helicity weights.
For example, consider the four-gluon helicity amplitude,
\begin{align}
    \cu{\am_4\br{1_g^+,2_g^+,3_g^-,4_g^-}} = \cu{-2,-2,2,2} \,.
\end{align}

Beyond tree level in the amplitude---or equivalently \ac{LO} in the amplitude squared, assuming the process is not loop induced, as discussed in \cref{sec:fixed-order}---the helicity configurations of \cref{eq:uhv} are finite.
These are referred to as \ac{UHV} amplitudes\index{Amplitude!\acs*{UHV}}.
At \ac{NLO}, the virtual correction is the interference of one-loop amplitudes and tree-level amplitudes.
Therefore, \ac{UHV} amplitudes first appear at \ac{NNLO} in the \ac{VV} corrections as one-loop squared amplitudes.

\subsection{Recursion}

\index{Recursion}In \cref{sec:colour-decomposition}, colour decomposition was used to reduce the number of Feynman diagrams required to calculate an amplitude.
In this spirit, it is also possible to construct higher-multiplicity amplitudes recursively from lower-multiplicity amplitudes.
This is a useful technique in the numerical evaluation of high-multiplicity amplitudes.
One such method is Berends-Giele off-shell recursion\index{Recursion!Berends-Giele} for partial amplitudes~\cite{berends:1987me}.
A partial amplitude is expressed as
\begin{align}
    \label{eq:off-shell-decomposition}
    \pant{1,\ldots,n} &= \pov_\mu\br{p_n} J^\mu\br{1,\ldots,n-1} \,,
\end{align}
where $J^\mu$ is called the off-shell current\index{Current!off-shell}.
In the case of gluon amplitudes, it is defined recursively as
\begin{multline}
    \label{eq:off-shell-recursion-definition}
    J^\mu\br{1,\ldots,n} =
    \frac{-\imi}{\s{1,n}} \Bigg[
        \sum_{i=1}^{n-1} V_3^{\mu\nu\rho}\br{p_{1,i},p_{i+1,n}} \, J_\nu\br{1,\ldots,i} \, J_\rho\br{i+1,\ldots,n} \\
        + \sum_{i=1}^{n-2} \sum_{j=i+1}^{n-1} V_4^{\mu\nu\rho\sigma} \, J_\nu\br{1,\ldots,i} \, J_\rho\br{i+1,\ldots,j} \, J_\sigma\br{j+1,\ldots,n}
        \Bigg] \,,
\end{multline}
with
\begin{align}
    p_{i,k}&\coloneqq p_{i} + \ldots + p_{k}\,, & \s{i,k}&\coloneqq{p_{i,k}}^2\,.
\end{align}
The gluon vertices $V_3^{\mu\nu\rho}$ and $V_4^{\mu\nu\rho\sigma}$ are given in $R_{\xi=1}$ gauge by the colour-ordered Feynman rules,
\begin{align}
    \begin{aligned}
        V_3^{\mu\nu\rho}\br{p,q} &= \imi\br{\lm^{\nu\rho}\br{p-q}^\mu+2\lm^{\rho\mu}q^\nu-2\lm^{\mu\nu}p^\rho} \,,
        \\
        V_4^{\mu\nu\rho\sigma} &= \imi\br{2\lm^{\mu\rho}\lm^{\nu\sigma}-\lm^{\mu\nu}\lm^{\rho\sigma}-\lm^{\mu\sigma}\lm^{\nu\rho}}.
    \end{aligned}
\end{align}
Pictorially, \cref{eq:off-shell-recursion-definition} can be stated as
\begin{align}
    \raisebox{-3.4em}{\includegraphics{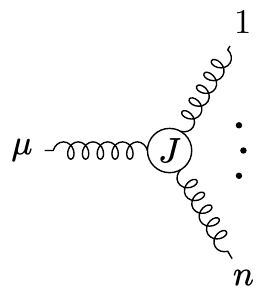}}
    =
    \frac{-\imi}{\s{1,n}}
    \br{
        \sum_i
        \raisebox{-6.0em}{\includegraphics{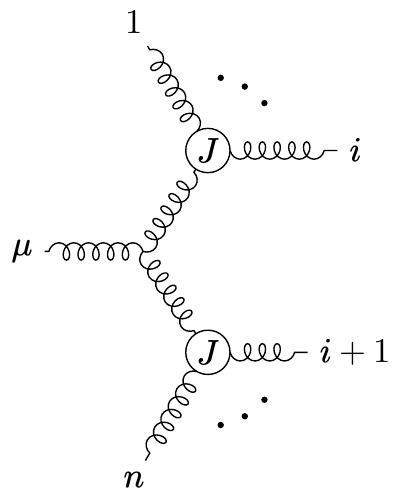}}
        +
        \sum_{i,j}
        \raisebox{-5.3em}{\includegraphics{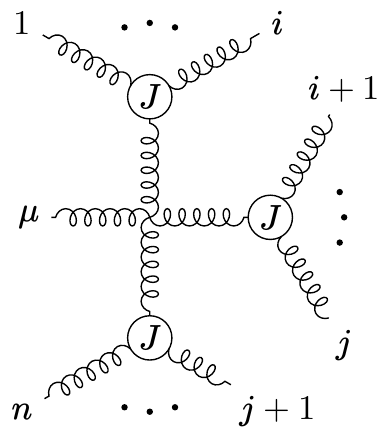}}
    } \,.
\end{align}

The off-shell current $J_\mu$ has one off-shell leg, labelled by $\mu$, and is therefore gauge dependent.
It obeys the photon decoupling equation \cref{eq:photon-decoupling-identity}, the reflection identity \cref{eq:reflection-identity} with an extra sign, and current conservation,
\begin{align}
    {p_{1,n}}^\mu J_{\mu}\br{1,\ldots,n}=0 \,.
\end{align}
An amplitude is recovered on contraction of the off-shell current with a polarisation vector, as in \cref{eq:off-shell-decomposition}.
Off-shell recursion can be extended to additionally include quarks~\cite{badger:2011znk}.

An alternative method is \ac{BCFW} on-shell recursion\index{Recursion!\acs*{BCFW}}~\cite{britto:2004ap,britto:2005fq}, which can be advantageous for helicity expressions, although it requires the introduction of complex momenta.
Recursive techniques allow common building blocks of an amplitude to be reused so that the number of independent diagram evaluations is reduced to polynomial growth in the number of legs~\cite{badger:2012uz}.

\subsection{Momentum twistor variables}
\label{sec:mtv}

Our methods so far have cast amplitudes as analytical expressions of the scalars $\s{ij}$, the pseudoscalar $\trf$, and the spinor products $\spA{i}{j}$ and $\spB{i}{j}$.
One advantage of these variables is that the poles in the kinematics (see \cref{ch:ir}) are manifest in expressions~\cite{laurentis:2019bjh}.
While these variables are highly linearly dependent, it is possible to reduce them to \iac{LI} set.
However, it can be the case that using a particular choice of overcomplete set leads to more compact expressions.
Automating the reduction to a set that gives the shortest expressions is very challenging.
In addition, recall that $\trf$ introduces algebraic dependence on the kinematics with a square root, \cref{eq:tr5-sqrt}.

An alternative representation is provided by momentum twistor\index{Momentum!twistor} parametrisation~\cite{hodges:2009hk,badger:2013gxa}, in which $n$-point kinematics are represented by a momentum twistor matrix,
\begin{align}
    \label{eq:mom-twi-mat}
    Z = \brm{ Z_1 \cdots Z_n } \,.
\end{align}
Each momentum twistor,
\begin{align}
    Z_i = \brm{ \lws_i \\ \mu_i } \, ,
\end{align}
has the first two components as those of the negative-helicity Weyl spinor $\lws_i$, while the last two components $\mu_i$ are defined through their relation to the positive-helicity Weyl spinor $\rws_i$,
\begin{align}
    \rws_i = \frac{\spAa{i | i+1} \, \mu_{i-1} + \spAa{i+1 | i-1} \, \mu_{i} + \spAa{i-1 | i} \, \mu_{i+1}  }{\spAa{i | i+1} \spAa{i-1 | i}} \, ,
\end{align}
with the indices defined modulo $n$.
Thus, $Z$ is an $4\times n$ matrix and so has $4n$ momentum twistor components.

The momentum twistor matrix $Z$ enjoys Poincar\'e symmetry, which has ten generators.
The corresponding physical momentum $p_i$ is invariant under a $\text{U}(1)$ scaling of the components of a momentum twistor,
\begin{align}
    \lws_i &\to \phi_i\,\lws_i \,, & \mu_i &\to \phi_i\,\mu_i \,, & \phi_i&\in\text{U}(1)\,,
\end{align}
which arises from the little group invariance \cref{eq:little-group}.
Consequently, each momentum twistor is defined projectively,
\begin{align}
    p_i\to p_i \qquad\text{under}\qquad Z_i\to\phi_iZ_i\,.
\end{align}
Therefore, $Z$ has the number of independent components,\index{Kinematic degrees of freedom}
\begin{align}
    \mathcal{D}^{(4)}(n)&=4n-10-n=3n-10\,.
\end{align}
This agrees with our ($d=4$, $n\le6$) parity-even analysis in \cref{sec:dof}.
The choice of these free components, which we refer to as \acp{MTV}\index{Momentum!twistor!variables}, is not unique.
They are also referred to as momentum twistor coordinates.

\Acp{MTV} are unconstrained by construction, in that both massless on-shell constraints \cref{eq:massless} and momentum conservation \cref{eq:mom-cons} are automatically satisfied.
Hence, the Schouten identities \cref{eq:schouten} are trivially satisfied when written in terms of \acp{MTV}.
Identifying relations between functions of \acp{MTV} is therefore easy, and the automated simplification of expressions is unambiguous.
It is also straightforward to convert them back to traditional kinematic variables.

Another advantage of \acp{MTV} is that they can provide a rational parametrisation of the kinematics: the spinors $\lws$ and $\rws$,  and momenta $p$, and therefore the spinor products, scalars $\s{ij}$, and pseudoscalar $\trf$, are all rational functions of the \acp{MTV}.
This is necessary for the use of \ac{FF} techniques (\cref{sec:ff}), as we will see in \cref{sec:Reduction}.

A disadvantage of \acp{MTV} is that they do not contain the helicity phase information, which is lost when translating from spinor products.
However, the correct phase weights can easily be restored to \iac{MTV} expression by multiplying by a suitable phase factor.

\Acp{MTV} are invaluable tools to provide rational kinematics and obtain compact expressions in traditional variables, as well as providing efficient expressions themselves.
The geometric origin of momentum twistors is discussed in \crefextp{section}{5.3 and 5.4} of \incite{elvang:2015rqa}.
We discuss an explicit parametrisation in \cref{sec:momtwistors}.

\section{On-shell methods and integral reduction}
\label{sec:loops}

The implementation of on-shell methods and integral reduction led to an ``\ac{NLO} revolution'' in the automated computation of one-loop amplitudes~\cite{giele:2008ve,berger:2008sj,bevilacqua:2011xh,hirschi:2011pa,cascioli:2011va,cullen:2014yla,alwall:2014hca,denner:2017wsf,degrande:2018neu}.
This technology has recently been applied to automate two-loop amplitudes~\cite{pozzorini:2022ohr,pozzorini:2022msb}, but evaluation is relatively costly.
This indicates the need for a mixed analytical and numerical approach at \ac{NNLO}, such as reconstruction of analytic expressions (\cref{sec:ff}).
For further reading on the techniques discussed in this section, see \incites{ellis:2011cr,bern:2011qt,britto:2010xq,berger:2009zb,weinzierl:2022eaz}.

\subsection{Loops}

\index{Amplitude!loop}An $L$-loop partial amplitude can be written as a sum over all contributing $L$-loop diagrams,
\begin{align}
    \label{eq:l-loop}
    \pamp{L}{n} &= \sum_{j} \int \br{\prod_{\ell=1}^{L} \frac{\dd^d{k_\ell}}{\br{2\pi}^d}} \partial{\pamp{L}{n,j}} \,,
\end{align}
where $k_\ell$ are the loop-momenta, and the diagram integrands take the general form,
\begin{align}
    \partial{\pamp{L}{n,j}} &= S_{j} \frac{N_{j}}{\prod_{\alpha_{j}} D_{\alpha_{j}}} \,,
\end{align}
where $S_{j}$ is a symmetry factor, $\alpha_{j}$ index the propagators, $D_{\alpha_{j}}$ are inverse propagators, and $N_{j}$ are polynomials in Lorentz-invariant contractions of loop-momenta, external momenta, and polarisation vectors.

After regulating the divergences (\cref{sec:infinities}), loop integrals are typically not expressible as rational functions.
A common analytic structure in massless amplitudes are polylogarithms\index{Polylogarithm} $\li{i}{z}$ which are defined iteratively by
\begin{align}
    \label{eq:polylog}
    \li{i}{z} &= \int_0^z \dd{y} \frac{\li{i-1}{y}}{y} \,, & \li{1}{z}&=-\ln(1-z) \,,
\end{align}
where $\ln(z)$ is the natural logarithm (with base $e$).
Such functions are referred to as transcendental functions\index{Transcendentality}, and have a transcendental weight, or degree of transcendentality, given by the dimension of the integral(s).
Hence, $\ln(z)$ and $\pi$ have transcendentality one, since
\begin{align}
    \ln(z)&=\int_1^z \frac{\dd{t}}{t}\,, & \imi\pi&=\ln(-1)\,,
\end{align}
and $\li{i}{z}$ has transcendentality $i$.

\subsection{Unitarity cuts}
\label{sec:unitarity-cuts}

Let us take some massless one-loop amplitude and choose the loop-momentum $k$ in each diagram such that we can rewrite \cref{eq:l-loop} under one integral,
\begin{align}
    \label{eq:one-loop}
    A^{(1)}_n &= \int \frac{\dd^d k}{\br{2\pi}^d}\sum_{j} \partial{A^{(1)}_{n,j}} \,, &
    D_{\alpha_{j}}&=\br{k-\sum_{\beta_{j}}^{\alpha_{j}-1} p_{\beta_{j}}}^2 \,, &
    \alpha_{j}&\in\cu{1,\ldots,n} \,,
\end{align}
where $p_i$ are the external momenta.
We can apply cut conditions on the loop-momentum where we impose that certain loop propagators $\alpha_j$ have zero momentum.
In other words, we take these lines on-shell, which we call taking unitarity cuts\index{Unitarity!cut}.
Then, for this particular subplane of loop-momentum, the loop amplitude factorises as a product of on-shell tree-level amplitudes.

We will now arrive at this relationship from unitarity of the $S$-matrix.
We insert \cref{eq:transfer-matrix} into \cref{eq:unitarity},
\begin{align}
    \label{eq:twice-im}
    \begin{aligned}
        T^\dagger T &= \imi \br{T^\dagger-T} \\
        &= 2\Im(T) \,,
    \end{aligned}
\end{align}
and sandwich with states,
\begin{align}
    2\Im(\bra{f}T\ket{i}) = \sum_a \bra{f}T^\dagger\ket{a}\bra{a} T\ket{i} \,,
\end{align}
where the insertion of identity involves a sum over the degrees of freedom of the state $a$.
We can write this in terms of amplitudes using \cref{eq:T2A},
\begin{align}
    \label{eq:optical-theorem}
    2\Im(\am(i\to f)) = \sum_a (2\pi)^4 \delta^{(4)}(P_i-P_a) \am(f\to a)^*\am(i\to a) \,.
\end{align}
This equality is called the optical theorem.
Excluding overall factors of $\imi$, the only place that imaginary parts enter in the Feynman rules is in the $\imi\eps$ prescription for the Feynman propagator.
Therefore, a nonzero $\Im(\am(i\to f))$ can only arise when virtual particles go on-shell and the $\imi\eps$ becomes relevant. 

The discontinuity of a function $f:\cn\to\cn$ at $x_0$ is given by the difference between the function evaluated at the point plus and minus a small imaginary part,
\begin{align}
    \label{eq:discontinuity}
    -\imi \discbb{x_0}{f(x)} \coloneqq \lim_{\eta\to0^+} \Big( f(x_0+\imi\eta)-f(x_0-\imi\eta) \Big)\,,
\end{align}
where we include the $-\imi$ factor for later convenience.
The discontinuity is only nonzero if it crosses a branch cut.

Consider our amplitude $\am(i\to f)$ as an analytic function of a complex variable, $\am(s)$.
If the energy of state $i$ in \cref{eq:optical-theorem} is insufficient to produce on-shell intermediate states $a$, then $\Im(\am(i\to f))$ vanishes.
In this domain,
\begin{align}
    \label{eq:star-star}
    \am(s) = {\am(s^*)}^* \,.
\end{align}
Analytically continuing this to the entire complex $s$ plane, in the domain where the initial energy is high enough to allow on-shell intermediate states and therefore $\Im(\am(s))$ is nonzero,
\begin{align}
    \Re(\am(s+\imi\eps)) &= \Re(\am(s-\imi\eps)) \,,\\
    \Im(\am(s+\imi\eps)) &= -\Im(\am(s-\imi\eps)) \,.
\end{align}
Thus, there is a branch cut across the real axis of $s$, showing that a nonzero $\Im(\am(s))$ requires a branch cut singularity.
The discontinuity is
\begin{align}
    \label{eq:disc}
    \discb{\am(s)} &\coloneqq 2\Im(\am(s)) \,,
\end{align}
which can also be seen by applying \cref{eq:star-star,eq:twice-im} to \cref{eq:discontinuity}.

Examining \cref{eq:optical-theorem} order by order in perturbation theory, we discover relations between the discontinuities of a given order and the products of lower-order terms (dropping factors),
\begingroup
\allowdisplaybreaks[0]
\begin{subequations}
    \begin{align}
        \discb{\amp{0}{n}} &= 0 \,, \\
        \label{eq:disc1}
        \discb{\amp{1}{n}} &= \amp{0}{n}^* \amp{0}{n} \,, \\
        \discb{\amp{2}{n}} &= \amp{0}{n}^* \amp{1}{n} + \amp{1}{n}^* \amp{0}{n} + \amp{0}{n+1}^* \amp{0}{n+1} \,,
    \end{align}
\end{subequations}
\endgroup
and so on.
The constraint \cref{eq:disc1} is none other than the factorisation we found at one loop~\cite{cutkosky:1960sp}.

\subsection{Generalised unitarity}
\label{sec:generalised-unitarity}

Beyond those prescribed by unitarity, we can continue to make further cuts.
This technique is called generalised unitarity~\cite{bern:1994zx,bern:1994cg}\index{Unitarity!generalised}.
The full loop integrand can be reconstructed by analysing the complete set of possible unitarity cuts.
This proceeds by identifying a basis of integrals, then isolating their coefficients by considering the projections provided by the cuts.

Consider $n$-point massless one-loop amplitudes with external states in four dimensions.
They can be decomposed onto a basis of $n$-point scalar integrals\index{Integral!scalar} $I^{(1)}_{k,i}$, labelled by the number of internal lines $k$, and a rational part $R$ as
\begin{align}
    \label{eq:scalar-basis}
    \amp{1}{n} = \sum_{k=2}^4 \sum_i c^{(0)}_{k,i} \, I^{(1)}_{k,i} + R + \order{\eps} \,.
\end{align}
The integrals are called boxes $I^{(1)}_{4,i}$, triangles $I^{(1)}_{3,i}$, and bubbles $I^{(1)}_{2,i}$.
We neglect tadpoles $I^{(1)}_{1,i}$ since for massless propagators, the integrals are scaleless and hence vanish in \ac{DR}.
The size of the integral basis is set by the number of spacetime dimensions, four.
This basis is presented in \incite{ellis:2007qk}.

The coefficients $c^{(0)}_{k,i}$ are functions of the external kinematics and $\eps$.
Having distinct propagator structures, only a subset of the scalar integrals contribute to a given unitarity cut.
Therefore, a single coefficient or set of coefficients may be singled out by performing a certain set of cuts.
Quadruple cuts isolate only the box coefficients $c^{(0)}_{4,i}$~\cite{britto:2004nc}; these are maximal cuts\index{Unitarity!cut!maximal} in $d=4$.
Triple cuts pick out both the box coefficients $c^{(0)}_{4,i}$ and the triangle coefficients $c^{(0)}_{3,i}$.
Similarly, double cuts additionally include the bubble coefficients $c^{(0)}_{2,i}$.
Thus, all coefficients can be extracted by solving a system of linear equations formed from various cuts.
This gives the coefficients as combinations of products of trees~\cite{arkani-hamed:2008owk,britto:2007tt,forde:2007mi}.

We can demonstrate this by considering the integral decomposition,
\begin{align}
    \amp{1}{5}(1_g^-,2_g^-,3_g^+,4_g^+,5_g^+) &= c^{(0)}_{1^+|2^+|3^-|4^-5^-} \br{\raisebox{-3.8em}{\includegraphics{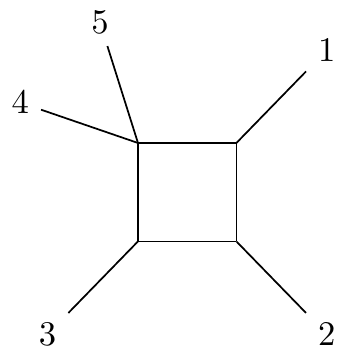}}} + \ldots \,.
\end{align}
We focus on a quadruple cut term to find the scalar box coefficient,
\begin{align}
    \label{eq:scalar-box-coeff}
     c^{(0)}_{1^+|2^+|3^-|4^-5^-} = \raisebox{-4.3em}{\includegraphics{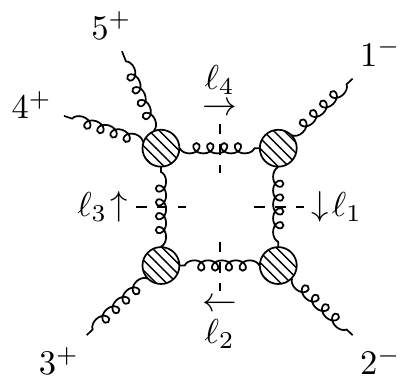}} \,,
\end{align}
following the prescription of \incite{britto:2004nc}.
First, we write the loop-momenta $\ell_i$ in terms of the external momenta $p_i$.
We take as the dependent loop-momenta
\begin{align}
    \label{eq:dep-moms}
    \ell_2 &= \ell_1-p_2\,, & \ell_3&=\ell_1-p_2-p_3 \,,& \ell_4&=\ell_1+p_1\,.
\end{align}
Solving for on-shell loop-momenta,
\begin{align}
    \label{eq:on-shell-loop-moms}
    \ell_1\cdot p_1 &= 0\,, & \ell_1\cdot p_2&=0 \,,& \ell_1\cdot \br{p_{2}+p_{3}}&=\frac{\s{23}}{2} \,.
\end{align}
Since it is four-dimensional, we can write $\ell_1$ as
\begin{align}
    {\ell_1}^\mu &= a_1 \, {p_1}^\mu + a_2 \, {p_2}^\mu + a_3 \, \frac{1}{2} \, \spBA{1}{\sigma^\mu}{2} + a_4 \, \frac{1}{2} \, \spAB{2}{\sigma^\mu}{1} \,.
\end{align}
\Cref{eq:on-shell-loop-moms} sets $a_1=a_2=0$ and solving for $\ell_1$ on-shell has two solutions: either $a_3$ or $a_4$ vanishing,
\begin{align}
    {\ell_1^A}^\mu &= a_3 \, \frac{1}{2} \, \spBA{1}{\sigma^\mu}{2} \,, & {\ell_1^B}^\mu &= a_4 \, \frac{1}{2} \, \spAB{2}{\sigma^\mu}{1} \,.
\end{align}
Inserting these into \cref{eq:on-shell-loop-moms} and solving for $a_{3}$ and $a_4$ gives
\begin{align}
    a_3 &= \frac{\spB{2}{3}}{\spB{1}{3}} \,, & a_4 &= \frac{\spA{2}{3}}{\spA{1}{3}} \,,
\end{align}
so we find that
\begin{align}
    {\ell_1^A}^\dagger=\ell_1^B \,.
\end{align}

Consider solution B.
We can write helicity spinors which satisfy $\ell_1^B$, $\ell_2^B$, and $\ell_4^B$ through \cref{eq:gordon-id},
\begin{align}
    \label{eq:dep-sp-hls}
    \begin{aligned}
        \sAr{\ell_1^B} &= a_4 \sAr{1}\,, & \sBr{\ell_1^B} &= \sBr{2} \,, \\
        \sAr{\ell_2^B} &= a_4 \sAr{1} -\sAr{2} \,, & \sBr{\ell_2^B} &= \sBr{2} \,, \\
        \sAr{\ell_4^B} &= \sAr{1}\,, & \sBr{\ell_4^B} &= a_4 \sBr{2} + \sBr{1} \,.
    \end{aligned}
\end{align}
We can treat solution A similarly.
We now have all the ingredients we need to compute $c^{(0)}_{1^+|2^+|3^-|4^-5^-}$ as a function of the external kinematics.

Now we consider the trees of the cut one-loop diagram depicted in \cref{eq:scalar-box-coeff}.
Recall from \cref{sec:helicity-amplitudes} that three-gluon trees are nonzero for complex momenta.
Since the trees are nonzero only for \ac{MHV} helicity configurations, there is only a single free internal helicity, $h_1$, with the others fixed,
\begin{align}
    h_2 &= - \,, & h_3 &= - \,, & h_4 &= + \,.
\end{align}
Inserting our kinematic solutions into the \ac{MHV} expressions for the trees \cref{eq:mhv}, we find solution A vanishes, while solution B with $h_1=-$ vanishes, leaving solution B with $h_1=+$ as the only nonzero piece,
\begin{align}
    c^{(0)}_{1^+|2^+|3^-|4^-5^-} \Big|_{B, h_1=+} =
    \frac{\spB{1}{\ell_1^B}^3}{\spB{\ell_1^B}{\ell_4^B} \spB{\ell_4^B}{1}}
    \frac{\spA{\ell_2^B}{\ell_1^B}^3}{\spA{\ell_1^B}{2} \spA{2}{\ell_2^B}}
    \frac{\spA{3}{\ell_3^B}^3}{\spA{\ell_3^B}{\ell_2^B} \spA{\ell_2^B}{3}}
    \frac{\spA{4}{5}^3}{\spA{5}{\ell_4^B} \spA{\ell_4^B}{\ell_3^B} \spA{\ell_3^B}{4}}
    \,.
\end{align}
The answer is given by the sum over internal helicities and average of the kinematic solutions, which after substitution of \cref{eq:dep-moms,eq:dep-sp-hls} and subsequent algebra is
\begin{align}
    c^{(0)}_{1^+|2^+|3^-|4^-5^-} = \frac{1}{2} \, \s{12} \, \s{23} \, A^{(0)}(1_g^+,2_g^+,3_g^-,4_g^-,5_g^-) \,.
\end{align}

The rational part $R$ of \cref{eq:scalar-basis} can be the most time-consuming to compute.
While it is possible for the integral coefficients $c^{(0)}_{k,i}$, the rational part is not cut constructable with the loop-momentum in four dimensions.
We can instead use cuts in analytic dimensions of $d=4-2\eps$~\cite{anastasiou:2006jv,anastasiou:2006gt,badger:2008cm}.
As both the coefficients and integrals are Laurent series in $\eps$, with the integrals containing poles, we see the origin of the rational part as terms in the expansion of the form
\begin{align}
    \order{\frac{1}{\eps^a}}&\times\order{\eps^a}\,, & a&\in\nonnegativeintegers\,.
\end{align}

Outside of four dimensions, the trees lose their simple structure.
One way to manage this is to partition the four-dimensional part of the loop $d$-momentum, with the size of the extra dimensions introducing a new scale that we treat as a fictitious mass,
\begin{align}
    \mom{k}{d}^\mu &= \mom{k}{4}^\mu + \mom{k}{-2\eps}^\mu \,, &
    {\mom{k}{d}}^2 &= {\mom{k}{4}}^2 - \mu^2 \,, &
    \mu^2 &= -{\mom{k}{-2\eps}}^2 \,,
\end{align}
so that propagators look like massive four-dimensional propagators,
\begin{align}
    D_{\alpha_{j}}\br{\mom{k}{d},p_i}&=D_{\alpha_{j}}\br{\mom{k}{4},p_i} - \mu^2 \,.
\end{align}

\subsection{Integrand reduction}
\label{sec:integrand-reduction}

We have discussed one-loop amplitudes in the form,
\begin{align}
    \label{eq:one-loop-numerator}
    \pamp{1}{n,j} &= S_{j}\,a_j(p_i)\cdot \int \frac{\dd^d{k}}{\br{2\pi}^d} \frac{\mathcal{N}_{j}(k,p_i)}{\prod_{\alpha_{j}=1}^n D_{\alpha_{j}}(k,p_i)} \,,
\end{align}
with inverse propagators $D_{\alpha_{j}}$ defined in \cref{eq:one-loop}.
We have factorised out from the numerator $N_j$ the part depending only on external four-momenta $a_j$,
\begin{align}
    N_j(k,p_i)=a_j(p_i)\cdot\mathcal{N}_j(k,p_i)\,,
\end{align}
where $\cdot$ represents the contraction of Lorentz indices.
These are generally tensor integrals, as $\mathcal{N}_j$ can carry free indices.

We can reduce to scalar integrals, meaning $\mathcal{N}_j$ is scalar, by rewriting the numerator $N_{j}$ in a basis of inverse propagators $D_{\alpha_{j}}$ and spurious \acp{ISP}.
\Iac{ISP}\index{Irreducible scalar product} is a scalar product that cannot be expressed in terms of only inverse propagators, while a spurious \ac{ISP} vanishes after integration.
A convenient choice is the van Neerven-Vermaseren basis~\cite{vanneerven:1983vr}.

For example, consider a massless box topology\footnote{This example is inspired by the lecture series \textit{Loop Amplitudes in Gauge Theories} given by Simon Badger at Durham University in 2019.}.
By momentum conservation, there are only three \ac{LI} external four-momenta.
The four-dimensional components of the loop-momentum $\mom{k}{4}^\mu$ require one more orthogonal vector to find a basis.
We introduce the vector,
\begin{align}
    \omega_{124}^\mu &\propto \lc^{\mu\nu\sigma\rho}\br{p_1}_{\nu}\br{p_2}_\sigma\br{p_4}_\rho \,,
\end{align}
such that it is orthogonal to \iac{LI} set of external four-momenta,
\begin{align}
    p_i\cdot\omega_{124}&=0\quad\forall\, i\in\cu{1,2,4}\,,
\end{align}
to form the basis,
\begin{align}
    \mom{k}{4}^\mu &= \sum_i\,b_i {v_i}^\mu\,, & v &= \br{p_1,p_2,p_4,\omega_{124}}\,.
\end{align}
We can form the \ac{ISP},
\begin{align}
    \mom{k}{4}&\cdot\omega_{124}\,,
\end{align}
which is spurious because
\begin{align}
    \int\dd{\mom{k}{4}}\br{\mom{k}{4}\cdot\omega_{124}}^x&=0\quad\forall x\in\oddintegers\,,
\end{align}
while other possible scalar products,
\begin{align}
    \mom{k}{4}\cdot p_i\,,
\end{align}
are reducible.
For example,
\begin{align}
    \begin{aligned}
        \mom{k}{4}\cdot p_1 &= \frac{-\br{\mom{k}{4}-p_1}^2+{\mom{k}{4}}^2}{2} \\
        &=\frac{-D_2(k)+D_1(k)}{2}\,,
    \end{aligned}
\end{align}
so can be expressed as a linear combination of inverse propagators.
We can then write an ansatz $\Delta_{4}$ for a general box numerator.
For a general renormalisable gauge theory, the maximum tensor rank for an $n$-propagator integral is $n$, so
\begin{align}
    \label{eq:box-numerator}
    \Delta_{4}&=y_0+y_1\br{\mom{k}{4}\cdot\omega_{124}}+y_2\,\mu^2+y_3\,(\mom{k}{4}\cdot\omega_{124})\mu^2+y_4\,\mu^4\,.
\end{align}
The coefficients $y_1$ and $y_3$ are spurious, $y_0$ is the four-dimensional scalar box coefficient, and $y_2$ and $y_4$ contain $\ord(\eps)$ parts of the $d$-dimensional box cut.
Choosing the index $j=0$, this gives the integrand,
\begin{align}
    \partial\amp{1}{4,0}=\frac{\Delta_4\brf{\mom{k}{4}\cdot\omega_{124},\mu^2}}{\prod_{\alpha_0=1}^{4}D_{\alpha_0}} + \text{sub-topologies} \,.
\end{align}
After similarly treating the triangles and bubbles, we arrive at \cref{eq:scalar-basis}.

The connection to on-shell technology is provided by methods such as those from \ac{OPP}~\cite{ossola:2006us} or Forde~\cite{forde:2007mi}, such that the integrand coefficients are projected out by numerical evaluations of cut diagrams.
% $y_0,\ldots,y_4$
% The coefficients can be extracted by solving a linear system constructed from the $d$-dimensional box cut.
% The triangle and bubble sub-topologies can be extracted by subtracting these numerators, with a denominator formed from the appropriate propagators, from lower-multiplicity cuts.

\subsection{Integration-by-parts identities}
\label{sec:ibp}

Having reduced a loop-level amplitude to scalar integrals, we may be interested in using a basis of \ac{LI} scalar integrals, which we call \acp{MI}\index{Integral!master}~\cite{chetyrkin:1981qh,grozin:2011mt}.
This reduces the number of integral evaluations required for the computation.
Furthermore, the linear independence of the coefficients can be necessary for reconstruction techniques such as those discussed in \cref{sec:ff}.
We can reduce to \iac{MI} basis by using \ac{IBP} identities, also called recurrence relations,\index{Identity!Integration-by-parts} between the scalar integrals~\cite{chetyrkin:1981qh,tkachov:1981wb,grozin:2011mt}.

To illustrate \ac{IBP} identities, consider a prototype one-loop scalar integral of the form,
\begin{align}
    I_a = \int \frac{\dd^d{k}}{\br{2\pi}^d} \frac{1}{\br{k^2-m^2}^a}\,.
\end{align}
It is invariant under shifts of the loop-momentum $k^\mu$, such as
\begin{align}
    k^\mu\to (1+ \zeta) k^\mu\,,
\end{align}
which gives the integral as
\begin{align}
    I_a = (1+\zeta)^d \int \frac{\dd^d{k}}{\br{2\pi}^d} \frac{1}{\br{\sq{k\br{1+\zeta}}^2-m^2}^a}\,.
\end{align}
For infinitesimal shifts $\zeta\ll1$, we can Taylor expand to find
\begin{align}
    \label{eq:exp-int}
    I_a = I_a +\zeta \br{d \, I_a + \int \frac{\dd^d{k}}{\br{2\pi}^d} \frac{-2 a k^2}{\br{k^2-m^2}^{a+1}}}+\order{\zeta^2}\,.
\end{align}
Using
\begin{align}
    \frac{\partial}{\partial k^\mu} k^\mu &= d \,, &
    k^\mu \frac{\partial}{\partial k^\mu} \frac{1}{\br{k^2-m^2}^a} &= \frac{-2ak^2}{\br{k^2-m^2}^{a+1}} \,,
\end{align}
we can rewrite \cref{eq:exp-int} as
\begin{align}
    \frac{\partial}{\partial k^\mu} k^\mu \, I_a + \int \frac{\dd^d{k}}{\br{2\pi}^d} k^\mu \frac{\partial}{\partial k^\mu}\frac{1}{\br{k^2-m^2}^a}&=\order{\zeta^2}\,.
\end{align}
Neglecting higher-order terms, we collect with the chain rule to obtain
\begin{align}
    \int \frac{\dd^d{k}}{\br{2\pi}^d} \frac{\partial}{\partial k^\mu}\br{\frac{k^\mu}{\br{k^2-m^2}^a}}&=0\,.
\end{align}
This construction generalises to multi-loop integrals to give \ac{IBP} identities of the form,
\begin{align}
    \int \br{\prod_{\ell} \frac{\dd^d{k_\ell}}{\br{2\pi}^d}} \frac{\partial}{\partial {k_i}^\mu}\br{\frac{{v_i}^\mu}{\prod_{\alpha}D_{\alpha}}}&=0\,,
\end{align}
where ${v_i}^\mu$ is some external or internal momentum and the $D_\alpha$ are the inverse propagators.
The equation shows that the integral vanishes on the boundary, which it must for the result to be finite.
This provides a method to generate relations between different integrals.

A general method for identifying \iac{MI} basis and applying \ac{IBP} identities to project onto it is the Laporta algorithm~\cite{laporta:2001dd}.

\section{Finite fields}
\label{sec:ff}

We now have the technology to decompose amplitudes into rational coefficients and special functions, and numerically evaluate the coefficients (\cref{sec:loops}).
To obtain an optimal representation of the coefficients, we can reconstruct their analytic form through many numerical evaluations by randomly sampling an appropriate ansatz \cite{klappert:2019emp,peraro:2016wsq}.
Were we to perform this procedure using a floating-point number representation, the precision loss from rounding errors would spoil the computation.
Instead, we sample over \acp{FF}, allowing us to use an integer number representation of fixed size.
Unlike floats, integers have no precision loss\footnote{There is the caveat that the integer size is sufficiently large that overflows are not encountered. Working with operations modulo $n$ for some integer $n$ means that the calculation is safe if $n$ is less than the integer size.} and we obtain an exact answer.
We can extract the rational result from this computation at the end.
This requires a rational parametrisation of the kinematics, which is provided by \acp{MTV} (\cref{sec:momtwistors}).
\Ac{FF} techniques have found much use in reconstruction and other methods in recent works~\cite{vonmanteuffel:2014ixa,peraro:2016wsq,klappert:2019emp,peraro:2019svx,smirnov:2019qkx,klappert:2020aqs,klappert:2020nbg,abreu:2020xvt}.

\subsection{Definition}

\Iac{FF} $\ffield{n}$ is a set of $n$ non-negative integers,
\begin{align}
    \ffield{n}=\cu{0,\ldots,n-1}\,,
\end{align}
together with the arithmetic operations of addition modulo $n$ and multiplication modulo $n$, and their inverses.
The size $n$ of the field must be a prime power.
The modular additive inverse is simply subtraction modulo $n$.
The modular multiplicative inverse of a non-zero integer $a$ is the integer $x$ such that
\begin{align}
    a x &= 1 \mod n \,, & a&\neq0 \,.
\end{align}
We use the following abuse of notation to denote the modular multiplicative inverse,
\begin{align}
    x &= a^{-1} \mod n \,.
\end{align}
This can be efficiently computed using the Extended Euclidean algorithm~\cite{knuthdonaldervin1997taoc}.

The modulus operation provides a non-injective homomorphism (a one-to-many mapping) from the integers to \iac{FF},
\begin{align}
    \text{mod } n : \integers \to \ffield{n}\,.
\end{align}
Similarly, we can map from the rationals $\rationals\to\ffield{n}$ with
\begin{align}
    \frac{a}{b} \,\to\, a\, b^{-1} \mod n \,.
\end{align}

\subsection{Reconstruction}
\label{sec:reconstruction}

Let us demonstrate the idea of \ac{FF} reconstruction with an example\footnote{This example is inspired by a question set by Ben Page at the SAGEX Mathematica and Maple School 2021.}.
Consider a box contribution to the one-loop six-gluon helicity amplitude,
\begin{align}
    \amp{1}{6}(1_g^+,2_g^-,3_g^+,4_g^-,5_g^+,6_g^-) = d^{1^+|2^-|3^+4^-5^+|6^-} \br{\raisebox{-4.3em}{\includegraphics{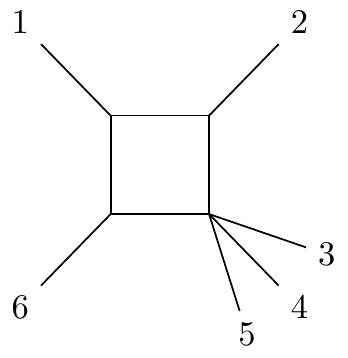}}} + \ldots \,,
\end{align}
where
\begin{align}
    d^{1^+|2^-|3^+4^-5^+|6^-} = \sum_{h_1,h_2,h_3,h_4\in\cu{+,-}} d_{h_1h_2h_3h_4}^{1^+|2^-|3^+4^-5^+|6^-} \,.
\end{align}
The lines in the superscript of the coefficient indicate the cuts, while the subscript shows the helicities in the loop, as shown in the cut diagram,
\begin{align}
    d_{h_1h_2h_3h_4}^{1^+|2^-|3^+4^-5^+|6^-} = \raisebox{-4.9em}{\includegraphics{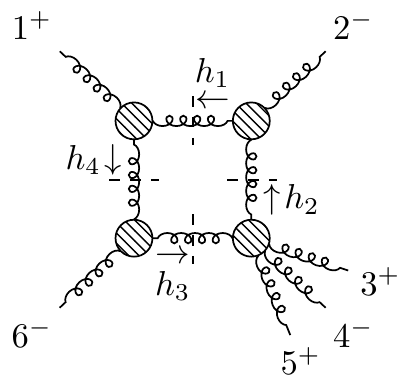}} \,.
\end{align}
In particular, consider the coefficient for $\cu{h_1,h_2,h_3,h_4}=\cu{+,-,+,-}$, which takes the form,
\begin{align}
    d_{+-+-}^{1|2|345|6} = \frac{\mathcal{N}}{\spB{3}{4}\spB{4}{5}\sAl{2}(1+6)\sBr{5}\sAl{6}(1+2)\sBr{3}\s{345}}\,.
\end{align}

The numerator $\mathcal{N}$ can be written in terms of the spinor brackets, so we can write the ansatz,
\begin{align}
    \mathcal{N} &= \sum_i \alpha_i\,g_i\,, & g_i&=\prod_{j=1}^{30}{v_j}^{\beta_{ij}} \,, & \beta_{ij}&\in\positiveintegers\,,
\end{align}
such that the numerator is a linear combination of polynomials $g_i$ in the spinor brackets $v_j$,
\begin{align}
    \vec{v}&=\cu{\spA ij,\spB ij \,:\, i<j \,;\, i,j\in\cu{1,\ldots,6}} \,.
\end{align}
By \cref{eq:mass-dim} and \cref{eq:amp-hel-scaling}, we know the mass dimension and little group scalings of the numerator,
\begin{align}
    \sq{\mathcal{N}} &= 10\,, & \cu{\mathcal{N}} &= \cu{-2,3,-4,0,-4,3}\,.
\end{align}
This provides physical constraints on the ansatz,
\begin{align}
    \label{eq:constraints}
    \sum_{j=1}^{30}\beta_{ij}&=\sq{\mathcal{N}} \,, & \sum_{j=1}^{30} \beta_{ij}\cu{v_j}_k &=\cu{\mathcal{N}}_k & &\forall i,k\,,
\end{align}
since
\begin{align}
    [v_j]&=1\,,&[\alpha_i]&=0\,,&[g_i]&=[\mathcal{N}]\,.
\end{align}

There are 93 solutions to the linear system of constraints \cref{eq:constraints}, so $i\in\cu{1,\ldots,93}$.
However, the solutions are not all \ac{LI}.
To reduce to the \ac{LI} subset, we numerically sample 93 phase-space points $\cu{x_i}_{i=1}^{93}$ to construct the $93\times93$ matrix,
\begin{align}
    M_1=
    \brm{
        g_1(x_1) & \hdots & g_{93}(x_1) \\
        \vdots & \ddots & \vdots \\
        g_1(x_{93}) & \hdots & g_{93}(x_{93}) \\
    }\,.
\end{align}
We randomly generate the points $x_i$ using a momentum twistor parametrisation such that they are rational and map them to a finite field $\mathbb{Q}\to\mathbb{F}_{n_1}$.
We perform row reduction, also called Gaussian elimination, (modulo $n_1$) on the matrix $M_1$ to identify the 24 \ac{LI} polynomials $\tilde g_i$ by the position of the leading number one in the upper 24 rows of the matrix in reduced row echelon form.
This reduces the ansatz to
\begin{align}
    \label{eq:ansatz-24}
    \mathcal{N}(x) =\sum_{i=1}^{24} a_i \, \tilde{g}_i(x)\,.
\end{align}

To solve for the \ac{LI} coefficients $a_i$, we numerically sample the components of
\begin{align}
    0 = -\mathcal{N}(x) +\sum_{i=1}^{24} a_i \, \tilde{g}_i(x)
\end{align}
over 25 points $\cu{y_i}_{i=1}^{25}$ to form the $25\times25$ matrix,
\begin{align}
    M_2=
    \brm{
        \mathcal{N}(y_1) & \tilde g_1(y_1) & \hdots & \tilde g_{25}(y_1) \\
        \vdots & \vdots & \ddots & \vdots \\
        \mathcal{N}(y_{25}) & \tilde g_1(y_{25}) & \hdots & \tilde g_{25}(y_{25}) \\
    }\,,
\end{align}
and find the nullspace, also called the kernel, (modulo $n_1$) of $M_2$.
In other words, we solve
\begin{align}
    M_2 \, \vec{\tilde a} = \vec{0}
\end{align}
for the vector $\vec{\tilde a}\in(\ffield{n_1})^{25}$.
After multiplying modulo $n_1$ the components $\tilde a_i$ by
\begin{align}
    -{\tilde a_0}^{-1}\mod n_1
\end{align}
to fix the coefficient of $\mathcal{N}$ as $-1$, we find the form
\begin{align}
    \label{eq:a}
    \vec{\tilde a} = \cu{-1 , a_1 , \ldots , a_{24}} \,,
\end{align}
with numerical values for the $a_i$ in $\ffield{n_1}$.
Thus, having the values of the coefficients of our ansatz \cref{eq:ansatz-24}, we have found $\mathcal{N}$ in $\ffield{n_1}$.

\subsection{Recovering rationals}

We now want to recover the rational solution for $\mathcal{N}$, but the map $\ffield{n_1}\to\rationals$ is many-to-one.
For each coefficient $a_i$, we could solve
\begin{align}
    \frac{r_i}{s_i}\mod n_1 = a_i
\end{align}
and fix the ambiguity by choosing the rational $r_i/s_i$ with smallest Euclidean norm ${r_i}^2+{s_i}^2$.
However, this introduces a lower bound on $n_1$ as it requires that
\begin{align}
    \label{eq:euc-norm}
    {r_i}^2+{s_i}^2<n_1
\end{align}
and we don't know how large the norm will be a priori.
In addition, if $n_1$ were required to be too large, we would lose computational efficiency by being forced to use non-native integer number representation.

The solution is provided by the Chinese Remainder Theorem, which allows us to construct a result in a larger \ac{FF} from multiple evaluations in smaller \acp{FF}.
Let us demonstrate this by applying it to our reconstruction example from \cref{sec:reconstruction}, labelling the components of the result \cref{eq:a} as $a_i^{\mkern1mu(n_1)}$.
We repeat the steps in \cref{sec:reconstruction} using a different \ac{FF} $\ffield{n_2}$, where $n_1$ and $n_2$ are coprime, to obtain the $a_i^{\mkern1mu(n_2)}$.
These numbers are related through a number $b_i$, in our case the rational result, by
\begin{subequations}
    \begin{align}
        & & a_i^{\mkern1mu(n_1)} &= b_i\mod n_1 & a_i^{\mkern1mu(n_2)}&=b_i\mod n_2 \\
        \label{eq:bis}
        \implies& & b_i&=a_i^{\mkern1mu(n_1)}+m_{1,i}n_1 & b_i&=a_i^{\mkern1mu(n_2)}+m_{2,i}n_2 \,,
    \end{align}
\end{subequations}
for some $m_{1,i},m_{2,i}\in\nonnegativeintegers$.
Since $n_1$ and $n_2$ are coprime, B\'ezout's identity tells us that there exist $q_1,q_2\in\integers$ such that
\begin{align}
    1=q_1n_1+q_2n_2\,.
\end{align}
% Multiplying through by $b_i$,
% \begin{align}
%     b_i=b_iq_1n_1+b_iq_2n_2\,,
% \end{align}
Multiplying through by $b_i$ and inserting from \cref{eq:bis},
\begin{align}
    \begin{aligned}
        b_i&=\br{a_i^{\mkern1mu(n_2)}+m_{2,i}n_2}q_1n_1+\br{a_i^{\mkern1mu(n_1)}+m_{1,i}n_1}q_2n_2 \\
        &=\underbrace{a_i^{\mkern1mu(n_1)}q_2n_2+a_i^{\mkern1mu(n_2)}q_1n_1}_{a_i^{(n_1n_2)}}+\underbrace{\br{q_1m_{2,i}+q_2m_{1,i}}}_{m_{3,i}}n_1n_2\,.
    \end{aligned}
\end{align}
Hence, we have constructed the result in $\ffield{n_1n_2}$,
\begin{align}
    a_i^{(n_1n_2)}=b_i\mod (n_1n_2) \,.
\end{align}

This method generalises to evaluations over an arbitrary number of \acp{FF}.
Choosing the \ac{FF} sizes $n_j$ as prime numbers ensures all $n_j$ are coprime.
This provides a way to construct a result $a_i^{(\prod_j n_j)}$ in a sufficiently large \ac{FF} $\ffield{\prod_j n_j}$ that
\begin{align}
    {r_i}^2+{s_i}^2<\prod_j n_j
\end{align}
is satisfied for the recovery of the rational result $r_i/s_i$, while keeping evaluations over sufficiently small \acp{FF} $\ffield{n_j}$ for computational efficiency.

\section{Phenomenology}
\label{sec:pheno}

Precise theoretical predictions are in high demand for the current \ac{LHC} experiments, which are aiming to look for tiny deviations from the \ac{SM}.
These include the \ac{ATLAS}~\cite{atlas:2008xda} and \ac{CMS}~\cite{cms:2008xjf} experiments.
With experimental bottlenecks like the determination of interaction luminosity at around one percent at \ac{ATLAS}~\cite{atlas:2019pzw} and \ac{CMS}~\cite{cms:2021xjt}, and similarly for the resolution of jet energies~\cite{cms:2016lmd,atlas:2017bje}, the current target for theory is to also achieve one percent precision~\cite{salam:2018rwo}.
In the near future, the High Luminosity upgrade of the \ac{LHC} will also overcome statistical limitations~\cite{dainese:2019rgk}.

Due to the relatively large size of the strong coupling constant, \ac{NNLO} corrections in \ac{QCD} are desirable for a wide variety of final state processes.
In particular, a class of two-to-three scattering processes with many kinematic scales have presented a considerable challenge to the theoretical community and there has been a good deal of activity leading to new methods able of overcoming their algebraic and analytic complexity~\cite{kosower:2011ty,mastrolia:2011pr,badger:2012dp,zhang:2012ce,mastrolia:2012an,mastrolia:2012wf,ita:2015tya,badger:2013gxa,badger:2015lda,abreu:2017xsl,abreu:2020xvt}.
We discuss recent progress for diphoton-plus-jet production in \cref{sec:yy}.

\subsection{Fixed-order perturbation theory}
\label{sec:fixed-order}

\begin{figure}
    \begin{center}
        \includegraphics[width=\textwidth]{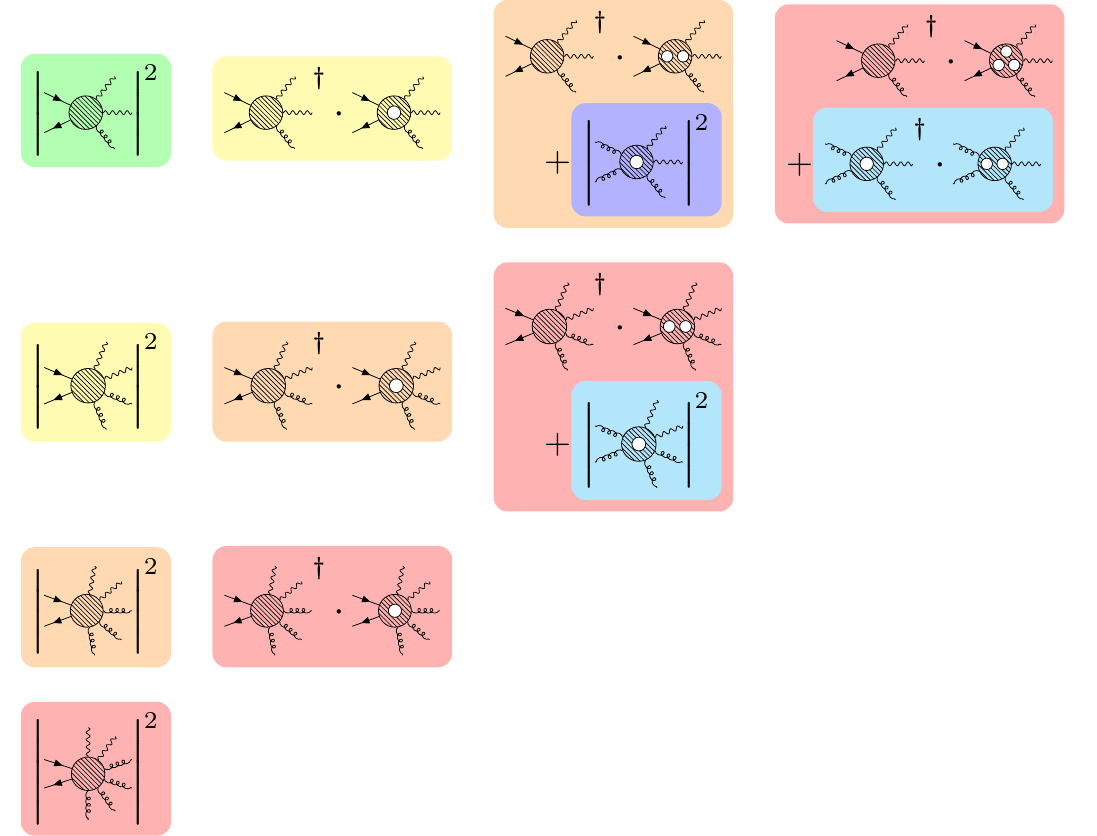}
        \caption{
            Schematic of the perturbative expansion in $\astr$ of $pp\to\gamma\gamma j$.
            Blobs represent sums over possible internal vertex configurations for the external particles, with holes denoting loops.
            Single representative terms are shown to highlight each contribution, with other terms omitted.
            Moving to the right horizontally, we have an increase in the number of loops (virtual corrections).
            Moving downwards vertically, we have an increase in the number of additional emissions (real corrections).
            \Ac{LO} (green) is $\order{\astr}$.
            At \ac{NLO} (yellow), which is $\ord({\astr}^2)$, there are real and virtual contributions.
            At \ac{NNLO} (orange), which is $\ord({\astr}^3)$, there are \ac{RR}, \ac{RV}, and \ac{VV} contributions.
            In the \ac{VV}, we see the first appearance of the gluon-fusion subprocess (blue).
            At \ac{N3LO} (red), which is $\ord({\astr}^4)$, there are \ac{RRR}, \ac{RRV}, \ac{RVV}, and \ac{VVV} contributions.
            The \ac{NLO} contributions to the gluon-fusion subprocess are highlighted here in cyan.
        }
        \label{fig:pert-exp}
    \end{center}
\end{figure}

In fixed-order perturbation theory, in analogy to our expansion of the amplitude, \cref{eq:amp-exp}, we calculate the cross section as an expansion in the coupling.
For example, up to \ac{NNLO},
\begin{align}
    \mathrm{d}\sigma&=\astr^{A}\br{\mathrm{d}\sigma_{\text{\acs{LO}}}+\astr\,\mathrm{d}\sigma_{\text{\acs{NLO}}} + {\astr}^2\,\mathrm{d}\sigma_{\text{\acs{NNLO}}}}+\order{{\astr}^{A+3}} \,,
\end{align}
where $A$ is the coupling power of the \ac{LO} term.
It is important to calculate order-by-order for the regularisation of \ac{IR} divergences, as we will see in \cref{sec:subtraction-methods}.
Since $\astr\approx 0.1$ at the energy scales of modern colliders, we can estimate that at least \ac{NNLO} predictions are necessary in general for percent-level precision.

To evince the anatomy of this series, let us consider the \ac{ME} contributions (up to permutations and antiparticles) to hadronic diphoton-plus-jet production $pp\to\gamma\gamma j$\index{Diphoton-plus-jet production}:
\begin{description}
    \item[\ac{LO}] At \ac{LO}, which is also called Born level, we have $A=1$.
        The contributions are the square of the amplitudes $q\anti{q}\to\gamma\gamma g$ and $qg\to\gamma\gamma q$.
        We do not have the gluon-fusion channel as there is no vertex coupling photons to gluons in the \ac{SM}.
    \item[\ac{NLO}] Moving to \ac{NLO}, we can increase the order in $\astr$ by manipulating the \ac{LO} diagrams in two ways: either by emitting an additional particle in the final state, called a real correction; or by introducing a new internal line to form a closed loop, called a virtual correction.
        Note that both of these corrections are experimentally indistinguishable from the Born as virtual particles are not observed and the final state reconstruction is dependent on the jet clustering algorithm.
        The real correction includes the square of the amplitudes $q\anti{q}\to\gamma\gamma gg$, $qg\to\gamma\gamma qg$, and $gg\to\gamma\gamma q\anti{q}$, while the virtual correction is the interference of the tree-level and one-loop diagrams for $q\anti{q}\to\gamma\gamma g$ and $qg\to\gamma\gamma q$.
        Note that the virtual contribution to the cross section involves an integration over a three-particle phase space, while the real has an extra particle and thus is over a four-particle phase space.
    \item[\acs*{NkLO}] We can continue this pattern to higher orders, \ac{NkLO}, as depicted in \cref{fig:pert-exp}.
        The first place that the gluon-fusion subprocess $gg\to\gamma\gamma g$\index{Diphoton-plus-jet production!gluon fusion} can appear is in the \ac{VV} correction at \ac{NNLO} of the full process $pp\to\gamma\gamma j$.
        Because it has no tree-level diagram and thus contains a loop in the \ac{LO} contribution, we refer to $gg\to\gamma\gamma g$ as a loop-induced\index{Amplitude!loop-induced} process.
\end{description}

\subsection{Estimating uncertainty}
\label{sec:uncertainties}

\subsubsection{Theory uncertainty}

\begin{figure}
    \begin{center}
        \includegraphics[width=0.55\textwidth]{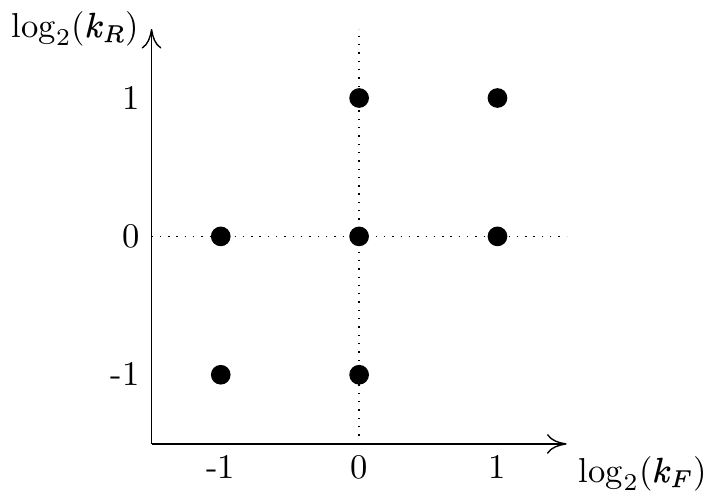}
        \caption{
            Plot of the $\br{k_F,k_R}$ scale factor set for the seven-point scale variation.
        }
        \label{fig:scale-variation}
    \end{center}
\end{figure}

Fixed-order theoretical predictions come with an error due to the missing higher orders.
While the only way to really know this uncertainty is to calculate the next order, in practice we can estimate it.
As the scale dependence is due to missing higher orders (\cref{sec:infinities}), and a scale variation amounts to an $\order{\astr}$ correction, we use it as a measure of the uncertainty due to the missing higher orders.

The de facto choice is an ad hoc method known as the seven-point scale variation\index{Scale!seven-point variation}.
The computation for an observable $\sigma(\mu_F,\mu_R)$ is repeated for a set of different scales varied around a central value.
Typically, the central value is taken as
\begin{align}
    \mu_F=\mu_R\eqqcolon\mu \,,
\end{align}
with $\mu$ set to a fixed scale or calculated dynamically as in \cref{eq:mu}.
We consider a set of values of $\br{k_F,k_R}$ in
\begin{align}
    S_7 = \cu{\br{0.5,0.5},\br{0.5,1},\br{1,0.5},\br{1,1},\br{1,2},\br{2,1},\br{2,2}}\,,
\end{align}
such that we sample the seven cases of $\sigma(k_F\mu,k_R\mu)$ where
\begin{align}
    k_F,k_R&\in\cu{0.5,1,2} \,, & k_F/k_R&\in\cu{0.5,1,2} \,;
\end{align}
this is depicted in \cref{fig:scale-variation}.
The result is then that of the central value $\sigma(\mu,\mu)$, with the error estimated by the interval
\begin{align}
    \sq{\pmin{\sigma\brm{k_F\mu,k_R\mu}:\br{k_F,k_R}\in S_7},\,\pmax{\sigma\brm{k_F\mu,k_R\mu}:\br{k_F,k_R}\in S_7}} \,.
\end{align}

There are also efforts to prescribe more reliable methods of uncertainty estimation, such as Bayesian modelling~\cite{cacciari:2011ze,bonvini:2020xeo,duhr:2021mfd}.

\subsubsection{Monte Carlo uncertainty}
\label{sec:mc}

The phase-space integration involved in the computation of an observable like the cross section in a collider simulation is generally performed using the numerical method of \ac{MC} integration \cite{james:1980yn}.
This technique approximates an integral over a function $f(\vec{x})$,
\begin{align}
    I = \int_V \dd^Dx\,f(\vec{x})\,,
\end{align}
as the sum,
\begin{align}
    I \approx \frac{V}{N} \sum_{i=1}^{N} f(\vec{x}_i) \,,
\end{align}
where $D$ is an integer and the $D$-dimensional phase space of finite volume $V$ is sampled $N$ times at points $\vec{x}_i$ with a uniform random distribution.
The error scales like $1/\sqrt{N}$ and can be estimated by the sample standard error \cite{hughes2010measurements},
\begin{align}
    \alpha &= \frac{\sigma_{N-1}}{\sqrt{N}}\,,
\end{align}
which we define through the sample standard deviation,
\begin{align}
    \sigma_{N-1} &= \sqrt{\frac{1}{N-1}\sum_{i=1}^{N}\br{f(x_i) - \bar{f}\,}^2}\,,
\end{align}
and the mean,
\begin{align}
    \label{eq:mean}
    \bar{f} &= \frac{1}{N}\sum_{i=1}^{N}f(x_i)\,.
\end{align}
% \begin{align}
%     \varepsilon &\approx \frac{1}{N} \sqrt{\frac{1}{N-1}\sum_{i=1}^{N}\br{f(x_i) - \langle f \rangle_x}}\,, &
%     \langle f \rangle_x &= \frac{1}{N}\sum_{i=1}^{N}f(x_i)\,,
% \end{align}
% where the angle brackets denote the mean.
% This can also be written as the square root of the variance,
% \begin{align}
%     \varepsilon^2 \approx \langle f^2 \rangle_x - {\langle f \rangle_x}^2
%     = \frac{1}{N}\sum_{i=1}^{N} f^2(x_i) - \br{\frac{1}{N}\sum_{i=1}^{N}f(x_i)}^2\,.
% \end{align}
By the Central Limit Theorem \cite{hughes2010measurements}, the \ac{MC} result converges on the true result as $N\to\infty$.
Various methods such as importance sampling and stratified sampling are used to accelerate convergence.

For differential observables, the \ac{MC} error on the $i^{th}$ histogram bin, which contains points in the set $b_i$, can be estimated by
\begin{align}
    \alpha_i &= \sqrt{\frac{1}{N\br{N-1}}\sum_{j=1}^{N}\br{\theta(j\in b_i)f(x_{j}) - \bar{f}_i}^2}\,,
\end{align}
with
\begin{align}
    \bar{f}_i &= \frac{1}{N}\sum_{j=1}^{N}\theta(j\in b_i)f(x_{j})\,,
\end{align}
where $N$ is the total number of points in all bins.
To ensure the area under the histogram is equal to the total integrated observable, the bin heights and errors can be divided by the bin width.

Histograms can be rebinned by aggregating adjacent bins.
Assuming the samples to be random and uncorrelated, the original bin errors can be propagated to the new bins by adding errors in quadrature.
The resultant bin error is a lower bound.
To propagate errors exactly in rebinning, the covariance matrix $C$ can be used,
\begin{align}
    C_{ik} = \frac{1}{N-1}\sum_{j=1}^{N}\br{\theta(j\in b_i)f(x_j) - \bar{f}_i}\br{\theta(j\in b_k)f(x_j) - \bar{f}_k}\,
\end{align}
where the indices $i$ and $k$ denote the bins.
The diagonal elements of the covariance matrix give the variances and are related to the bin errors by
\begin{align}
    \alpha_i = \sqrt{\frac{C_{ii}}{N}}\,,
\end{align}
since the variance is the square of the standard deviation, while the off-diagonal terms give the correlations between bins.
When aggregating bins, for example from four to two bins by combining the first and second pair, the covariance matrix is aggregated as
\begin{align}
    \left(\begin{array}{cc|cc}
        c_{11} & c_{12} & c_{13} & c_{14} \\
        c_{21} & c_{22} & c_{23} & c_{24} \\
        \hline
        c_{31} & c_{32} & c_{33} & c_{34} \\
        c_{41} & c_{42} & c_{43} & c_{44} \\
    \end{array}\right)
    \to
    \begin{pmatrix}
        c_{11} + c_{12} + c_{21} + c_{22} & c_{13} + c_{14} + c_{23} + c_{24} \\
        c_{31} + c_{32} + c_{41} + c_{42} & c_{33} + c_{34} + c_{43} + c_{44} \\
    \end{pmatrix}
    \,.
\end{align}
The variances for the new bins are given by the diagonal entries.

\subsection{Diphoton production}
\label{sec:yy}

The production of a pair of high-energy photons is an important experimental signature at hadron colliders and can be used, for example, to study the Higgs boson through its decay to photons~\cite{caola:2015wna} or in beyond the \ac{SM} searches~\cite{franceschini:2015kwy}.
The \ac{SM} background is dominated by \ac{QCD} corrections, and a precise description of the kinematics of these observables requires the theoretical predictions to include perturbative information from the production in association with additional jets.
Diphoton-plus-jet signatures form the largest background to Higgs production at high transverse momenta; the extra jet is necessary to ensure a non-zero transverse momentum in the diphoton system.
In addition, with at least one jet, the Higgs coupling can be probed independently of initial-state effects, thus bypassing additional uncertainties introduced by the \acp{PDF}.

Diphoton production $pp\to\gamma\gamma$ has been known at \ac{NNLO} for some time~\cite{catani:2011qz,campbell:2016yrh,gehrmann:2020oec} and the two-loop scattering amplitudes were among the first complete $2\to 2$ processes to be calculated at this order~\cite{anastasiou:2002zn,bern:2001df}.
A $q_T$-resummed calculation at order\footnote{A prime indicates the inclusion of partial but dominant corrections from the next higher order.} \ac{NNLO}$\,+\,\mathrm{N}^3\mathrm{LL}^\prime$ was presented recently~\cite{neumann:2021zkb}.
Steps towards \ac{N3LO} are being taken with the completion of the first four-point three-loop \ac{QCD} amplitude~\cite{caola:2020dfu}.
For the gluon-fusion subprocess $gg\to\gamma\gamma$, the three-loop amplitudes were recently calculated~\cite{bargiela:2021wuy}.
Along with the five-point two-loop results of \cref{ch:yy-amp}, this makes available the final remaining piece for the \ac{NNLO} corrections to the subprocess, which contribute to the full process at \ac{N4LO}.

\Ac{NNLO} \ac{QCD} corrections to diphoton-plus-jet production\index{Diphoton-plus-jet production} $pp\to \gamma\gamma j$, which is initiated at \ac{LO} by quark-antiquark and quark-gluon processes (\cref{sec:fixed-order}), have been considered a high priority for current and future experiments for several years~\cite{badger:2016bpw,bendavid:2018nar,amoroso:2020lgh}.
Following recent breakthroughs in two-loop amplitude technology, there has been a flurry of activity around this channel, including complete sets of \ac{LC}~\cite{agarwal:2021grm,chawdhry:2021mkw} and now \ac{FC}~\cite{agarwal:2021vdh} two-loop helicity amplitudes, and \ac{LC} \ac{NNLO} distributions~\cite{chawdhry:2021hkp}.
This progress is extremely timely given the continually improving experimental measurements of diphoton signatures~\cite{atlas:2021mbt}.

The loop-induced gluon-fusion channel\index{Diphoton-plus-jet production!gluon fusion} $gg\to \gamma\gamma g$ starts contributing to the cross section only from \ac{NNLO} onwards.
Owing to the large gluon luminosity, \ie~the gluon \ac{PDF} diverges at small momentum fraction $x$ in high-energy scales (see \cref{fig:pdfs}), this channel yields a dominant contribution to the \ac{NNLO} corrections and dominates their scale uncertainty~\cite{chawdhry:2021hkp}.
To improve upon this uncertainty requires the \ac{NLO} corrections, for which we derive the \ac{FC} two-loop virtual amplitudes in \cref{ch:yy-amp} and the \ac{FC} \ac{NLO} distributions in \cref{ch:yy-xs}.

\section{Structure of this thesis}
\label{sec:structure}

In \cref{ch:ir}, we study the \ac{IR} factorisation properties of \ac{QCD} amplitudes and discuss how \ac{IR} divergences are regulated in fixed-order cross section calculations before presenting a library of \ac{IR} singular functions.
In \cref{ch:ann}, we review the use of \acp{NN} to emulate \acp{ME} for processes with two photons and many gluons in hadron collider simulations.
In \cref{ch:yy-amp}, we present the first full set of \ac{FC} five-point helicity amplitudes at two loops in \ac{QCD}, which are those for the process of diphoton-plus-jet production through gluon fusion.
In \cref{ch:yy-xs}, we combine these amplitudes with the real contributions to produce \ac{NLO} \ac{QCD} distributions, the first time that \ac{FC} two-to-three two-loop amplitudes have been integrated to provide fully differential cross section predictions relevant for the \ac{LHC}.
In \cref{ch:3j}, we also present a set of \ac{LC} two-loop helicity amplitudes for trijet production at hadron colliders.
We conclude in \cref{ch:conc}.

\chapter{Infrared behaviour}
\label{ch:ir}

In this chapter, we look into the \ac{IR} behaviour of \ac{QCD} amplitudes and cross sections.
In \cref{sec:infrared-limits}, we discuss the soft and collinear limits that can appear in phase space, and in \cref{sec:infrared-factorisation}, the factorisation properties of amplitudes in these limits.
We motivate the utility of these limits in \cref{sec:utility}, and explain how they are regulated in the calculation of observables in \cref{sec:subtraction-methods}, also listing proposed schemes for \ac{NNLO}.
In \cref{sec:ir-impl}, we present a library of universal \ac{IR} singular helicity functions.
For further reading, see the review~\cite{agarwal:2021ais}.

\section{Infrared limits}
\label{sec:infrared-limits}

We saw in \cref{sec:infinities} that loops contain \ac{IR} divergences when the loop-momentum becomes small, which we regulate with \ac{DR}.
\index{Phase space!\acl*{IR} limit}\index{Divergence!\acl*{IR}}\Ac{IR} divergences also manifest in amplitudes when the phase space of the external momenta contains unresolved particles, which occurs in soft\index{Divergence!\acl*{IR}!soft} and collinear\index{Divergence!\acl*{IR}!collinear} limits.
In the following, we take the massless four-momentum as $p=\br{E, \vec{p}\,}$.
The double-collinear limit, denoted $i\parallel j$, occurs when $\theta_{ij} \rightarrow 0$ where $\theta_{ij}$ is the angle between $\vec{p}_i$ and $\vec{p}_j$.
Similarly, the single-soft limit $\mathcal{E}_i$ of a massless particle $i$ is when $E_i\rightarrow0$, or equivalently when \(|\vec{p}_i|\rightarrow0\).
These configurations cause an internal propagator to go on-shell, giving rise to a singularity.

The kinematics of the collinear limit $i\parallel j$ are characterised by $\s{ij}\rightarrow0$ and the soft limit $\mathcal{E}_i$ by \(\mathrm{s}_{ij} \rightarrow 0 \,\,\forall j\) since
\begin{align}
    \s{ij} = 2 \ab{\vec{p}_i} |\vec{p}_j| \br{1-\cos\br{\theta_{ij}}}\,.
\end{align}
Soft limits and collinear limits can overlap in phase space, so care must be taken to avoid double counting when considering them.

\subsection{Cuts}
\label{sec:cuts}

At \ac{LO}, there exist \ac{IR} phase-space regions.
However, these are not physical states as they cannot be measured: soft particles are invisible and collinear particles are indistinguishable.
Thinking of detectors, there is a finite resolution to which particles can be measured; in theory, our phase space should be similarly constrained.
This is achieved by setting appropriate cuts\index{Phase space!cut}, which are criteria that phase-space points must satisfy.

Some quantities to cut on, which we will make use of later in this thesis, include the following.
\begin{itemize}
    \item Transverse momentum magnitude,\index{$p_T$}
        \begin{align}
            p_T=\sqrt{{p_x}^2+{p_y}^2} \,,
        \end{align}
        with the beam along the $z$-axis.
    \item Pseudorapidity,\index{$\eta$}
        \begin{align}
            \eta=-\ln\brf{\tan(\frac{\theta}{2})} \,,
        \end{align}
        where the polar angle $\theta$ is between the three-momentum of the particle $\vec{p}$ and the positive direction of the beam axis $\hat{z}$.
    \item Azimuthal angle $\phi$.
\end{itemize}

Related to cuts is the concept of jet clustering\index{Jet}.
A common choice for identifying \ac{QCD} jets is the anti-$k_T$ algorithm~\cite{cacciari:2008gp}, which is implemented in \fastjet~\cite{cacciari:2011ma}.
Considering \ac{QED} states, photons $\gamma$ are often selected in \iac{IR} safe fashion by Frixione smooth cone isolation\index{Boson!photon!isolation}~\cite{frixione:1998jh}.
It prescribes that all cones of radius $r_\gamma<\Delta R$ must satisfy
\begin{align*}
    E_{\text{hadronic}}(r_\gamma)\leq\epsilon_\gamma\,E_{T,\gamma}\frac{1-\cos(r_\gamma)}{1-\cos(\Delta R)} \,,
\end{align*}
where $E_{\text{hadronic}}(r_\gamma)$ is the hadronic energy found inside the cone of radius $r_\gamma$, $E_{T,\gamma}$ is the transverse momentum of the photon, the isolation cut cone radius is given by the R-separation,\index{$\Delta R$}
\begin{align}
    \label{eq:r-sep}
    \Delta R=\sqrt{(\Delta\eta)^2+(\Delta\phi)^2} \,,
\end{align}
and $\epsilon_\gamma$ is an arbitrary parameter.

\subsection{Poles}
\label{sec:poles}

\index{Pole}Beyond \ac{LO}, we calculate contributions to the fixed-order result that are not themselves observable and thus really contain \ac{IR} divergences.
For instance, at \ac{NLO}, the real correction contains single-unresolved singularities, and the loop in the amplitude for the virtual correction diverges when the loop-momentum goes soft.
In \ac{DR}, the one-loop amplitude therefore contains poles in the dimensional regulator $\eps$; when expanded as a Laurent series in $\eps$, it contains coefficients of $\eps^{-2}$ and $\eps^{-1}$, and a finite part, with $\order{\eps}$ terms being neglected\footnote{Note that we need up to $\ord(\eps^2)$ for the amplitudes in the one-loop squared contributions to \ac{VV} corrections at \ac{NNLO} as crossterms in the square yield poles and finite parts.} as they vanish in $\eps\to0$.
However, the poles in the real and virtual corrections come with opposite signs and exactly cancel.
In fact, for any \ac{IR} safe observable, when all components at any fixed order are combined, the result is finite, as described by the \ac{KLN} theorem~\cite{kinoshita:1962ur,lee:1964is}.

It can also be the case, at any order, that certain terms within an amplitude diverge in the \ac{IR} limit of some kinematic variable or combination thereof, but do not correspond to a physical \ac{IR} singularity.
These crop up depending on the particular algebraic expression of the amplitude and cancel out in the full finite result.
As they are unphysical, they are called spurious poles\index{Pole!spurious}.
Performing \ac{PFD} can introduce spurious poles; it can be necessary to take care with them as they can introduce large intermediate cancellations that reduce numerical stability.
An example is considered in \cref{sec:Apart}.

\section{Infrared factorisation}
\label{sec:infrared-factorisation}

An all-orders $(n+m)$-particle amplitude factorises in the $m$-unresolved \ac{IR} limit as
\begin{align}
    \am_{n+m} \rightarrow \widetilde{\am}_{n} \otimes \mathcal{S}_m \,,
\end{align}
where: \(\widetilde{\am}_{n}\) is the reduced amplitude, which is none other than the on-shell $n$-particle amplitude; \(\otimes\) indicates summation over spin and colour states of correlated particles\index{Correlation}; and \(\mathcal{S}_m\) is a universal singular function for the $m$-unresolved limit.
The double-collinear and single-soft limits described in \cref{sec:infrared-limits} are single-unresolved limits\index{Phase space!\acl*{IR} limit!single-unresolved}.
If there is no summation over correlated states, \ie~the \(\otimes\) reduces to an ordinary multiplication \(\times\), this is called an exact factorisation.
For \ac{QCD} amplitudes, it is the partial amplitudes that exhibit this factorisation property, for which only adjacent legs can become singular.

\subsection{Collinear limits}
\label{sec:collinear}

Consider the tree-level single-unresolved \ac{IR} limit when two partons go collinear\index{Divergence!\acl*{IR}!collinear}.
For an \((n+1)\)-parton amplitude, there is a factorisation into an \(n\)-parton reduced amplitude and a three-legged splitting amplitude\index{Amplitude!splitting},
\begin{align}
    \label{eq:double-collinear}
    \amp{0}{n+1} \rightarrow \lls{\widetilde{\am}}{0}{n} \otimes \lls{\mathcal{P}}{0}{1} \,.
\end{align}
The $i\parallel j$ phase-space configuration can be constructed by the parametrisation,
\begin{subequations}
    \begin{align}
        \label{eq:col-par-one}
        {p_i}^\mu    & = z\,\tilde{p}^{\hspace{1pt}\mu} + {p_\perp}^\mu - \frac{{p_\perp}^2}{z \, 2 \, \tilde p \cdot q} \, q^\mu \,,
        \\
        \label{eq:col-par-two}
        {p_j}^\mu    & = \br{1-z} \tilde{p}^{\hspace{1pt}\mu} - {p_\perp}^\mu - \frac{{p_\perp}^2}{\br{1-z} 2 \, \tilde p \cdot q} \, q^\mu \,,
        \\
        \label{eq:col-par-three}
        \tilde{p}^{\hspace{1pt}\mu} & = {p_{ij}}^\mu - \frac{\s{ij}}{2\,p_{ij} \cdot q} q^\mu \,,
        \\
        \label{eq:col-par-four}
        z      & = \frac{p_i \cdot q}{p_{ij} \cdot q} \,, \qquad \s{ij} =-\frac{{p_\perp}^2}{z(1-z)}\,,
    \end{align}
\end{subequations}
where:
the transverse momentum $p_\perp$ is timelike (${p_\perp}^2<0$) and orthogonal to $q$, $\tilde p$, and $p_{ij}$;
$\tilde p$ is the massless projection of $p_{ij}$;
and $q$ is a reference null (light-like) four-vector ($q^2=0$).
The parameter $z$ controls the momentum fraction of the unresolved pair in leg $i$, and $\ab{p_\perp}\to0$ gives the collinear limit.
We can write this as
\begin{align}
    \label{eq:col-ps}
    \cu{\ldots,p_i,p_j,\ldots,p_k,\ldots}_{n+1}\overset{i\parallel j}{\longrightarrow}\cu{\ldots,\tilde p,\ldots,\tilde{p}_k,\ldots}_{n} \,,
\end{align}
where $p_i$ and $p_j$ live in the full $(n+1)$-particle phase space, and map to the on-shell momentum $p$ in the reduced $n$-particle phase space.
When not in the exact limit, the reduced phase space must be provided through a mapping of the full phase space that imposes on-shellness and momentum conservation.
We also select an arbitrary momentum $p_k$ from the full phase space that we recoil off in the mapping to obtain $\tilde p_k$ in the reduced phase space.
One such mapping is described in the \ac{CS} dipole subtraction scheme (\cref{sec:subtraction-methods}), in which they call the correlated leg $\tilde p$ the emitter and the recoiling leg $\tilde p_k$ the spectator.
It can be stated as
\begin{align}
    \label{eq:cs-map}
    \tilde{p}_k^{\phantom{k}\mu} &= \br{1+\frac{\s{ij}}{2\,p_{ij}\cdot q}} {p_k}^\mu \,, & q&=p_k \,.
\end{align}
Another mapping is provided by Kosower's antenna scheme~\cite{kosower:1997zr}.

A helicity amplitude in the \(i\parallel j\) limit factorises as
\begin{multline}
    \am_n^{c_1 \ldots c_{i-1} c_i c_j c_{j+1} \ldots c_{n}}(1^{h_1},\ldots,(i-1)^{h_{i-1}},i^{h_i},j^{h_j},(j+1)^{h_{j+1}},\ldots,n^{h_n}) \overset{i\parallel j}{\longrightarrow} \\*
    \sum_{h_\rho,c_\rho}\widetilde{\am}_{n-1}^{c_1 \ldots c_{i-1} c_{\rho} c_{j+1} \ldots c_{n}}(1^{h_1},\ldots,(i-1)^{h_{i-1}},\rho^{h_{\rho}},(j+1)^{h_{j+1}},\ldots,n^{h_n}) \,
    \mathcal{P}_1^{c_{i} c_{\rho} c_{j}}(i^{h_i},\rho^{-h_{\rho}},j^{h_j}) \,,
    \label{eq:collinear-factorisation}
\end{multline}
where \(c_i\) is the abstract colour index (\ie~we have not specified the representation) associated with leg \(i\), \(h_i\in\cu{-,+}\) is the helicity of leg \(i\), and \(\rho\) labels the correlated leg.
There remains a sum over the helicity \(h_\rho\) and the colour \(c_{\rho}\).
Pictorially, drawing gluons in lieu of any parton and with colour implicit, this looks like
\begin{align}
    \raisebox{-3em}{\includegraphics{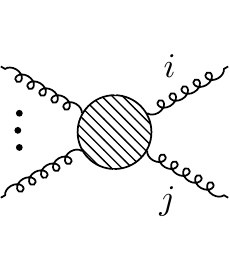}}
    \overset{i\parallel j}{\longrightarrow}
    \,\bigsum_h\,
    \raisebox{-3em}{\includegraphics{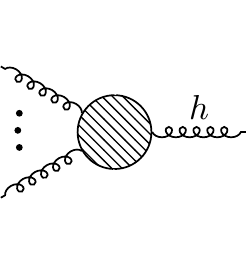}}
    \otimes
    \raisebox{-3em}{\includegraphics{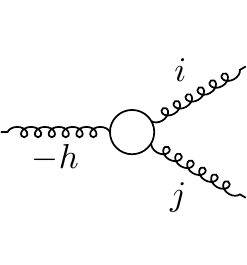}}.
\end{align}
With this picture in mind, we see that the double-collinear partonic splitting amplitudes, up to antiparticles and permutations, are $g\to gg$ and $q\to qg$ (giving $g\to \anti{q}q$ by crossing symmetry).
The splitting amplitudes can be algebraically derived by constructing these configuration with \ac{MHV} amplitudes~\cite{birthwright:2005vi}, for example.

Turning to the squared helicity amplitude, we have
\begin{align}
    \ab{\am}^2 \overset{i\parallel j}{\longrightarrow}
    \br{\sum_{h_1,c_\rho}\widetilde{\am}_{h_1}^{c_1 \ldots c_{\rho} \ldots c_{n_g}}
    \mathcal{P}_{-h_1}^{c_{i} c_{\rho} c_{j}}}^\dagger
    \sum_{h_2,c_\sigma}\widetilde{\am}_{h_2}^{c_1 \ldots c_{\sigma} \ldots c_{n_g}}
    \mathcal{P}_{-h_2}^{c_{i} c_{\sigma} c_{j}} \,,
    \label{eq:collinear-square-gluon}
\end{align}
where subscripts now denote the helicity of the correlated leg.
We can represent this diagrammatically as
\begin{align}
    \ab{\am}^2 \overset{i\parallel j}{\longrightarrow}
    \raisebox{-4.75em}{\includegraphics[height=10em]{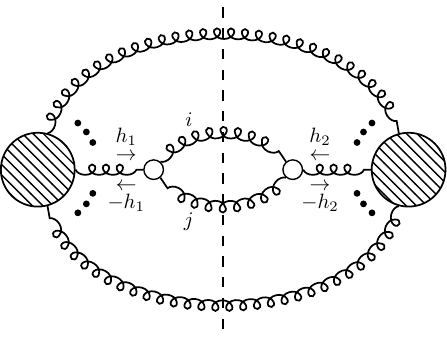}} \,.
\end{align}

Consider the colour of the squared splitting amplitude,
\begin{align}
    \left.\br{\mathcal{P}_{-h_1}^{a_{i} a_{\rho} a_{j}}}^\dagger \mathcal{P}_{-h_2}^{a_{i} a_{\sigma} a_{j}}\right|_\text{colour} =
    \left\{
        \begin{array}{r l}
            f^{a_i a_{\rho} a_j} f^{a_i a_{\sigma} a_j} = \caa\,\delta^{a_\rho a_\sigma} & \mathrm{for\ } g\to gg \,, \\
            t_{a_i a_j}^{a_{\rho}} t_{a_i a_j}^{a_{\sigma}} = \cdi\,\delta^{a_\rho a_\sigma} & \mathrm{for\ } g\to \anti{q}q \,, \\
            t_{a_i a_{\rho}}^{a_j} t_{a_i a_{\sigma}}^{a_j} = \caf\,\delta_{a_\rho a_\sigma} & \mathrm{for\ } q\to qg \,. \\
        \end{array}
        \right.
        \label{eq:delta-colour}
\end{align}
Thus, contracting the colour indices of the splitting amplitudes always results in a Kronecker delta of the collinear colour indices (\cref{sec:fundamental-colour-decomposition}).
This breaks the colour correlation that we observed at amplitude level to give exact factorisation of colour in the squared amplitude.
Armed with this observation, \cref{eq:collinear-square-gluon} can be rearranged as
\begin{align}
    \ab{\am}^2 \overset{i\parallel j}{\longrightarrow}\sum_{h_1,h_2}
    {\widetilde{\mathcal{M}}}_{h_1 h_2}
    \mathcal{P}_{-h_1 -h_2} \,,
    \label{eq:collinear-square-two}
\end{align}
where we have introduced the \(2\times2\) spin matrices,
\begin{align}
    \mathcal{P}_{h_1 h_2} &\coloneqq {\mathcal{P}_{h_1}}^\dagger\cdot\mathcal{P}_{h_2} \,, &
    {\widetilde{\mathcal{M}}}_{h_1 h_2} &\coloneqq {\widetilde{\am}_{h_1}}^\dagger\cdot\widetilde{\am}_{h_2} \,,
\end{align}
with \(\cdot\) denoting contraction of colour indices.
The off-diagonal terms (\(h_1\neq h_2\)) are called spin correlations\index{Correlation!spin} and make subdominant contributions to the sum.
They are omitted in the \ac{LS} approximation.
Particularly for high-multiplicity processes, the spin correlations can become significant deep in the \ac{IR} limit, in which case a \ac{FS} evaluation is necessary.

\subsection{Soft limits}
\label{single-soft}

Next, we consider the tree-level limit when a single parton goes soft\index{Divergence!\acl*{IR}!soft}.
This is only possible with a gluon, \ie~the limit \(\mathcal{E}_g\), as taking a quark in the exact soft limit would result in a violation of quark number conservation in the reduced amplitude.
The full \((n+1)\)-parton amplitude factorises to an \(n\)-parton reduced amplitude and an eikonal amplitude\index{Amplitude!Eikonal} with dependence on three legs of the full amplitude,
\begin{align}
    \amp{0}{n+1} \rightarrow \lls{\widetilde{\am}}{0}{n} \otimes \lls{\mathcal{S}}{0}{1}\,.
\end{align}
In the limit, the $(n+1)$-parton phase space reduces to an $n$-parton phase space,
\begin{align}
    \label{eq:soft-ps}
    \cu{\ldots,p_i,\ldots,p_j,\ldots,p_k,\ldots}_{n+1}
    \overset{E_j\rightarrow\,0}{\lra}
    \cu{\ldots,\tilde p_i,\ldots,\tilde{p}_k,\ldots}_{n} \,,
\end{align}
where $j$ is the soft leg.
Near the limit, we can map the full momenta to the reduced momenta while recoiling to enforce momentum conservation  and on-shellness in the reduced phase space, for example,
\begin{align}
    \label{eq:soft-map}
    \begin{aligned}
        \tilde{p}_i &= p_i + p_j - \beta p_k \,, \\
        \tilde{p}_k &= \br{1+\beta}p_k \,,
    \end{aligned}
    &\qquad
    \beta = \frac{\s{ij}}{\s{ik}+\s{jk}} \,.
\end{align}

The kinematic part of the eikonal amplitude takes a simple form,
\begin{align}
    \begin{aligned}
        S_+(i,j,k)&=\frac{\langle i k \rangle}{\langle i j \rangle \langle j k \rangle}\,,
        &\qquad
        S_-(i,j,k)&=-\frac{[ i k ]}{[ i j ][ j k ]}\,,
        \label{eq:eikonal}
    \end{aligned}
\end{align}
where the subscript denotes the helicity of the soft leg $j$.
These expressions are universal: they are independent of the parton type and helicity of partons \(i\) and \(k\).
This limit can be pictorially represented for a partial amplitude as
\begin{align}
    \raisebox{-3em}{\includegraphics{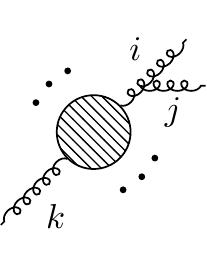}}
    \;\overset{E_j\rightarrow\,0}{\lra}\;
    \raisebox{-3em}{\includegraphics{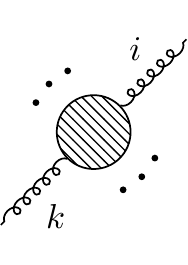}}
    \times
    \raisebox{-3em}{\includegraphics{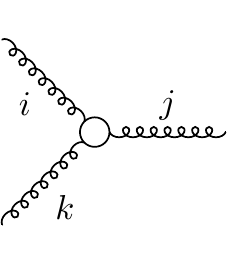}} \,,
    \label{eq:soft-partial-diagram}
\end{align}
where $p_i$ emits the soft $p_j$, with $p_k$ as a reference momentum.

While kinematics factorise exactly at amplitude level, no matter how soft the gluon is, it still carries the same colour charge.
Thus, colour correlations\index{Correlation!colour} are produced.
In fact, to reproduce the full colour structure of an \((n+1)\)-parton colour-summed amplitude from the reduced and eikonal amplitudes, it is necessary to
sum over emissions of the soft gluon from all the external legs of the reduced amplitude,
\begin{align}
    \label{eq:soft-amp}
    \am \overset{E_j\rightarrow\,0}{\lra}&
    \,\sqrt{2} \, \sum_{\substack{i=1\\i\neq j}}^{n+1} \, \widetilde{\mathcal{C}}^{(i)} \, S_{h_j}(i,j,k)
    \,,
    &
    \widetilde{\mathcal{C}}^{(i)} \coloneqq& \,\widetilde{R} \,\, \widetilde{C} \cdot T^{(i)} \,,
\end{align}
where we adopt a brief notation to show the colour-correlated amplitude as \(\widetilde{\mathcal{C}}^{(i)}\), the reduced amplitude decomposed into its kinematics $\widetilde{R}$ and colour $\widetilde{C}$, and the eikonal colour contribution as $T^{(i)}$.
The $\cdot$ represents the partial colour contraction of ``attaching'' the soft gluon $j$ to leg $i$.
The parton type of leg \(i\) determines the form of \(T^{(i)}\):
\begin{align}
    T^{(i)} = \begin{cases}
        f^{a_i a_j a_k} & \mathrm{gluon,} \\
        t^{a_j}_{f_i f_k} & \mathrm{quark,} \\
        \bar{t}^{a_j}_{\bar{f}_i \bar{f}_k} = -t^{a_j}_{f_k f_i} & \mathrm{antiquark.}
    \end{cases}
\end{align}
Note that the sum in \cref{eq:soft-amp} runs over the indices of the reduced $n$-parton phase space.
Since for each term in the sum $i$ is different and $k$ may be different, a local mapping \cref{eq:soft-ps} is required for each term.
This is in contrast to the global mapping \cref{eq:col-ps} we found for the collinear limit.

For the squared helicity amplitude,
\begin{align}
    \ab{\am}^2 \overset{E_j\rightarrow\,0}{\lra}
    \sum_{i,k}
    \widetilde{\mathrm{C}}^{(i,k)} \, \mathrm{S}_{j^{h_j}}^{(i,k)} \,,
    \label{eq:soft-square-zero}
\end{align}
with
\begin{align}
    \widetilde{\mathrm{C}}^{(i,k)} =& \;
    \widetilde{\mathcal{C}}^{(i)\dagger}\cdot{\widetilde{\mathcal{C}}^{(k)}} \,,
    &
    \mathrm{S}_{j^{h_j}}^{(i,k)} =& \;
    {S_{h_j}(i,j,k)}^\dagger\,S_{h_j}(k,j,i)
    = -2 \frac{\s{ik}}{\s{ij} \s{jk}}
    \,,
\end{align}
where $i$ and $k$ run over the indices of the reduced phase space.
The colour-correlated reduced \ac{ME} \(\widetilde{\mathrm{C}}^{(i,k)}\) can be constructed from the reduced partial amplitude vector and the colour correlations as
\begin{align}
    \begin{aligned}
        \widetilde{\mathrm{C}}^{(i,k)} =&
        {\widetilde{R}_a}^\dagger \, \br{\widetilde{\mathrm{D}}^{(i,k)}}^{ab} \, \widetilde{R}_b \,, \\
        \br{\widetilde{\mathrm{D}}^{(i,k)}}^{ab} \coloneqq&
        \widetilde{C}^a_{\ldots c_i \ldots c_{\bar{k}} \ldots}
        \, T^{(i)}_{c_i c_j c_{\bar{i}}} \, T^{(k)}_{c_k c_j c_{\bar{k}}} \,
        \widetilde{C}^b_{\ldots c_k \ldots c_{\bar{i}} \ldots} \,,
    \end{aligned}
\end{align}
where $a$ and $b$ run over the colour basis.
In the definition of the colour-correlation matrix $(\widetilde{\mathrm{D}}^{(i,k)})^{ab}$ we explicitly show abstract colour indices, with remaining contractions occurring over the omitted indices.
Expressed in diagram form, the colour-correlation matrix is
\begin{align}
    \widetilde{\mathrm{D}}^{(i,k)}=
    \raisebox{-4.75em}{\includegraphics[height=10em]{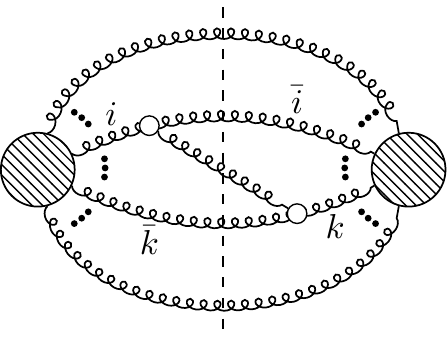}} \,.
\end{align}

As the degree of colour correlation\index{Correlation!colour} grows with the number of legs on the amplitude, construction of the reduced \ac{ME} $\widetilde{\mathrm{C}}^{(i,k)}$ can be computationally intensive at high multiplicities.
Notice that the diagonal of the eikonal matrix vanishes by masslessness and the reduced \ac{ME} is symmetric,
\begin{align}
    \mathrm{S}_{j^{h_j}}^{(i,i)}&=0 \,, &
    \widetilde{\mathrm{C}}^{(i,k)}&=\widetilde{\mathrm{C}}^{(k,i)} \,,
\end{align}
so \cref{eq:soft-square-zero} can be optimised as
\begin{align}
    \ab{\am}^2 \overset{E_j\rightarrow\,0}{\lra}
    2 \sum_{\substack{i,k\\k>i}}
    \widetilde{\mathrm{C}}^{(i,k)} \, \mathrm{S}_{j^{h_j}}^{(i,k)}\,.
    \label{eq:soft-square}
\end{align}

In contrast to the notation used here, the literature generally treats the colour with a notation described in \crefext{section}{3.1} of \incite{catani:1999ss}.
This notation uses an abstract colour charge \(\boldsymbol{T}_i\) which comes with the emission of a gluon from the \(i^\mathrm{th}\) parton.
The limit expression is constructed from an \(n\)-parton reduced \ac{ME} and an eikonal current\index{Current!eikonal},
\begin{align}
    \boldsymbol{J}^{\mu}(q) = \sum_{i=1}^{n} \boldsymbol{T}_i \frac{{p_i}^{\mu}}{p_i \cdot q} \,.
\end{align}
To see the connection with our notation, the eikonal amplitudes, \cref{eq:eikonal}, can be obtained from the kinematic part of the eikonal current by contracting with a gluon polarisation vector of the appropriate helicity, \cref{eq:polarisation-weyl}, and using \cref{eq:sij-spinors}.

\subsection{Beyond \acs{NLO}}

The previously considered limits, double-collinear and single-soft, are single-unresolved configurations and first appear in the real corrections at \ac{NLO}.
For \ac{RR} contributions at \ac{NNLO}, we also have double-unresolved configurations\index{Phase space!\acl*{IR} limit!double-unresolved}, which factorise $\am_{n+2}$ to $\am_{n}$~\cite{campbell:1997hg,catani:1999ss}.
This involves multiple soft or collinear pairs of particles, or a mixture of the two, which includes: correlated limits, which introduce new singular functions; and uncorrelated limits, where the reduced amplitude is iteratively factorised by taking multiple single-unresolved limits.

One correlated double-unresolved limit is the triple-collinear limit, which up to antiparticles and permutations includes $g\to ggg$, $q\to qgg$ (with $g\to \anti{q}qg$ by crossing symmetry), and $q\to q\anti{Q}Q$, where $Q$ may be a quark of different flavour to $q$.
The amplitude factorises as
\begin{align}
    \amp{0}{n+2} \rightarrow \lls{\widetilde{\am}}{0}{n} \otimes \lls{\mathcal{P}}{0}{2} \,.
\end{align}
For instance, $g\to ggg$ can be drawn as
\begin{align}
    \raisebox{-3em}{\includegraphics{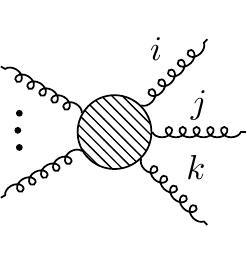}}
    \overset{i\parallel j\parallel k}{\lra}
    \,\bigsum_h\,
    \raisebox{-3em}{\includegraphics{r_c2}}
    \otimes
    \raisebox{-3em}{\includegraphics{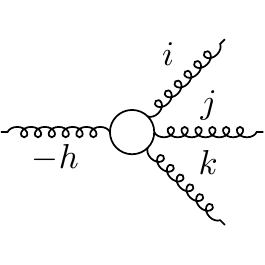}} \,.
\end{align}
This is treated similarly to the double-collinear limit.
In analogy to the parametrisation of \cref{eq:col-par-one,eq:col-par-two,eq:col-par-three,eq:col-par-four}, we can define a multi-collinear parametrisation of the kinematics~\cite{badger:2015cxa}.

When considering the \ac{RV} contribution at \ac{NNLO}, we need to calculate single-unresolved limits of loop-level amplitudes.
With loops, we must take care to include all contributions at the fixed order.
For instance, the one-loop single-unresolved limit factorises as
\begin{align}
    \label{eq:one-loop-double-collinear}
    \amp{1}{n+1} \rightarrow \widetilde{\am}^{(1)}_{n} \otimes \mathcal{S}^{(0)}_1 + \widetilde{\am}^{(0)}_{n} \otimes \mathcal{S}^{(1)}_1 \,.
\end{align}
Explicitly for the one-loop double-collinear limit~\cite{bern:1995ix}, we have
\begin{align}
    \begin{aligned}
        \raisebox{-3em}{\includegraphics{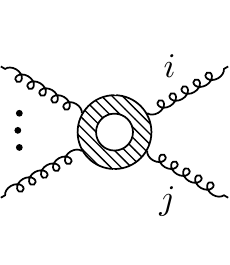}}
        \overset{i\parallel j}{\longrightarrow}\,
        &\,\bigsum_h\,
        \raisebox{-3em}{\includegraphics{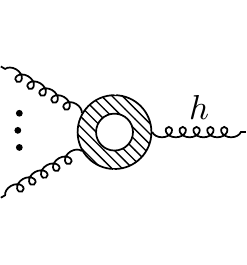}}
        \otimes
        \raisebox{-3em}{\includegraphics{split_g2gg}}\\
        +
        &\,\bigsum_h\,
        \raisebox{-3em}{\includegraphics{r_c2}}
        \otimes
        \raisebox{-3em}{\includegraphics{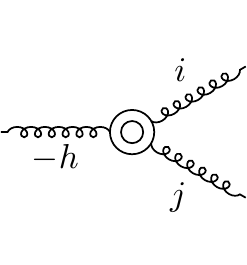}} \,.
    \end{aligned}
\end{align}

\subsection{Utility}
\label{sec:utility}

These limits offer a convenient way to validate new higher-order expressions.
In \ac{IR} regions of phase space, a numerical evaluation of the full amplitude is approximated by the product of the appropriate singular function and lower-multiplicity reduced amplitude.
A momentum mapping scheme may be necessary to relate the full and reduced phase space, as discussed for the double-collinear limit (\cref{sec:collinear}).
Generally, the reduced amplitude is available, or more easily calculated than the full amplitude, and the singular functions are well known in the literature.
Thus, by numerically evaluating both the full and factorised expressions at points in \ac{IR} limits, we can check that a new higher-order amplitude correctly displays the expected \ac{IR} behaviour.

Factorised expressions can also be used to provide an implementation of amplitudes that, while valid only near their \ac{IR} limit, is more numerically stable than the full amplitude~\cite{banerjee:2021mty}.
The limit functions will also provide the building blocks of counterterms to regulate \ac{IR} divergences at \ac{NNLO}, as discussed in \cref{sec:subtraction-methods}.

\section{Infrared subtraction}
\label{sec:subtraction-methods}

Recall the perturbative expansion in $\astr$ of the cross section discussed in \cref{sec:fixed-order}.
The \ac{NLO} correction to the cross section for a process with $n$ final state partons is given by
\begin{align}
    \dd\sigma_{\text{\ac{NLO}}}=\int_{n}\dd\sigma_{\text{\ac{NLO}}}^{\text{V}}+\int_{n+1}\dd\sigma_{\text{\ac{NLO}}}^{\text{R}} \,,
\end{align}
where $\dd\sigma_{\text{\ac{NLO}}}^{\text{V}}$ and $\dd\sigma_{\text{\ac{NLO}}}^{\text{R}}$ indicate respectively the renormalised virtual and the real correction, with the real integration over a higher-dimensional phase space due to the additional emission.
The virtual amplitude contains \ac{IR} divergent loops (\cref{sec:infinities}), while the real cross section diverges in single-unresolved corners of phase space.

While possible in principle, analytical integration in $4-2\eps$ dimensions of real-type corrections to cross sections for modern phenomenology is computationally intractable.
In addition, non-perturbative effects in simulations are generally computed numerically, so analytic integration is not possible.
Instead, we use the numerical technique of \ac{MC} integration (\cref{sec:mc}), which requires an integer number of dimensions.
The method of \ac{IR} subtraction\index{Regularisation!infrared subtraction} is used to regulate \ac{IR} divergences while allowing efficient integration.

The aim of \ac{IR} subtraction is to separately cancel the divergences in the virtual and real contributions.
We introduce a real subtraction term $\dd\sigma_{\text{\ac{NLO}}}^{\text{S}}$ which locally isolates the singular behaviour of the real correction.
This term can be constructed by exploiting the previously discussed factorisation properties of \ac{QCD} in \ac{IR} limits.
It is important for computational efficiency that the subtraction term is as simple as possible while still reproducing the correct \ac{IR} behaviour.
The divergences of the virtual correction are subtracted at the level of the amplitude to define the \acp{FR}, as in \cref{eq:finremdef}; we represent this schematically as a virtual subtraction term $\dd\sigma_{\text{\ac{NLO}}}^{\text{T}}$.
The \ac{NLO} correction can then be rewritten as
\begin{align}
    \label{eq:sigNLO}
    \dd\sigma_{\text{\ac{NLO}}}=\int_{n}\left(\dd\sigma_{\text{\ac{NLO}}}^{\text{V}}-\dd\sigma_{\text{\ac{NLO}}}^{\text{T}}\right)+\int_{n+1}\left(\dd\sigma_{\text{\ac{NLO}}}^{\text{R}}-\dd\sigma_{\text{\ac{NLO}}}^{\text{S}}\right) \,,
\end{align}
with
\begin{align}
    \label{eq:sigT}
    \dd\sigma_{\text{\ac{NLO}}}^{\text{T}}+\int_{1}\dd\sigma_{\text{\ac{NLO}}}^{\text{S}} = \order{\eps} \,,
\end{align}
such that both integrals in \cref{eq:sigNLO} are \ac{IR} finite in $\eps\to0$ and can therefore be numerically evaluated through \ac{MC} integration.
Various definitions of the subtraction terms give rise to different subtraction schemes.
Depending on the scheme, the real subtraction term can be required to be analytically integrable over the phase space of the unresolved parton, such that the matching of the singularities of the virtual corrections in \cref{eq:sigT} occurs in a fully analytical fashion.

\index{Regularisation!infrared subtraction!scheme}At \ac{NLO}, general algorithms are firmly established for \ac{IR} subtraction, including \ac{FKS} subtraction~\cite{frixione:1995ms,frederix:2009yq} and \ac{CS} subtraction~\cite{catani:1996vz,catani:2002hc}.
The implementation of the subtraction terms at \ac{NNLO} is not yet automated to this degree~\cite{torresbobadilla:2020ekr,magnea:2020trj,heinrich:2020ybq}.
However, many regularisation schemes are proposed, including:
\begin{itemize}
    \item antenna subtraction~\cite{gehrmann-deridder:2005btv,daleo:2006xa,currie:2013vh,chen:2022ktf} (we use this scheme to regulate the cross section in \cref{sec:setup}),
    \item sector-improved residue subtraction~\cite{czakon:2010td,boughezal:2011jf,czakon:2014oma},
    \item nested soft-collinear subtraction~\cite{caola:2017dug},
    \item CoLoRFulNNLO subtraction~\cite{delduca:2016ily} (closest to \ac{CS} subtraction),
    \item projection-to-Born subtraction~\cite{cacciari:2015jma},
    \item local analytic sector subtraction (Torino scheme)~\cite{magnea:2018hab,magnea:2018ebr},
    \item $q_T$-slicing~\cite{catani:2007vq},
    \item n-jettiness slicing~\cite{gaunt:2015pea},
    \item geometric \ac{IR} slicing~\cite{herzog:2018ily}.
\end{itemize}

\section{Implementation}
\label{sec:ir-impl}

In version 3.0.0 of \njet~\cite{njet}, we present an analytic library of the various soft and collinear limit helicity functions of \ac{QCD} amplitudes.
We use amplitude-level factorisation~\cite{kosower:1999xi} for efficient construction of factorised \acp{ME} including spin correlations.

It is possible that \acp{f64} do not provide sufficient precision when probing deep in the limit.
For this case, all \njet classes are templated to support higher fixed precisions of \ac{f128} and \ac{f256} provided by the \qd library~\cite{qdweb}.

We make available the splitting amplitudes \(\mathcal{P}_{h}\), spin matrices \(\mathcal{P}_{h_1 h_2}\), and colour-correlation matrices ${\widetilde{\mathrm{D}}^{(i,j)}}$ for the following partonic limits:
\begin{itemize}
    \item the single-soft tree-level limit, $\mathcal{E}_g$,
    \item all independent double-collinear limits up to one-loop, $g\to gg$ and $q\to qg$, from~\cite{bern:1999ry},
    \item the triple-collinear tree-level limits, $g\to ggg$ and $g\to\anti{q}qg$, from~\cite{badger:2015cxa}.
\end{itemize}
As amplitudes with photons and gluons can be constructed from pure-gluon amplitudes by summing over permutations~\cite{dicus:1987fk,deflorian:1999tp}, the loop-induced mixed \ac{QCD}$+$\ac{QED} limit functions, $g\to\gamma g$ and $g\to\gamma\gamma$, can also be generated from this set.

This includes all \ac{NLO} limits, and all limits for \ac{RV} at \ac{NNLO}, but omits the $q\to q\anti{Q}Q$~\cite{catani:2003vu} and double-soft limits that appear in \ac{RR}.
While therefore currently incomplete for \ac{NNLO}, this library can be used for validation of new amplitudes and extended to provide a full library of the limits necessary to build counterterms at \ac{NNLO} within a consistent framework.

\subsection{Validation}
\label{sec:valid}

To demonstrate the stability of the limit functions, we show the behaviour of the factorised expression compared to the full \ac{ME} over a phase space path driving into the collinear limit for several cases.
The full and reduced amplitudes are provided by the existing processes in the \njet library.
We use colour- and helicity-summed \acp{ME} for these tests.

% \subsubsection{Collinear limits}
% \label{sec:valid-collinear}

\begin{figure}
    \begin{center}
        \includegraphics[width=\textwidth]{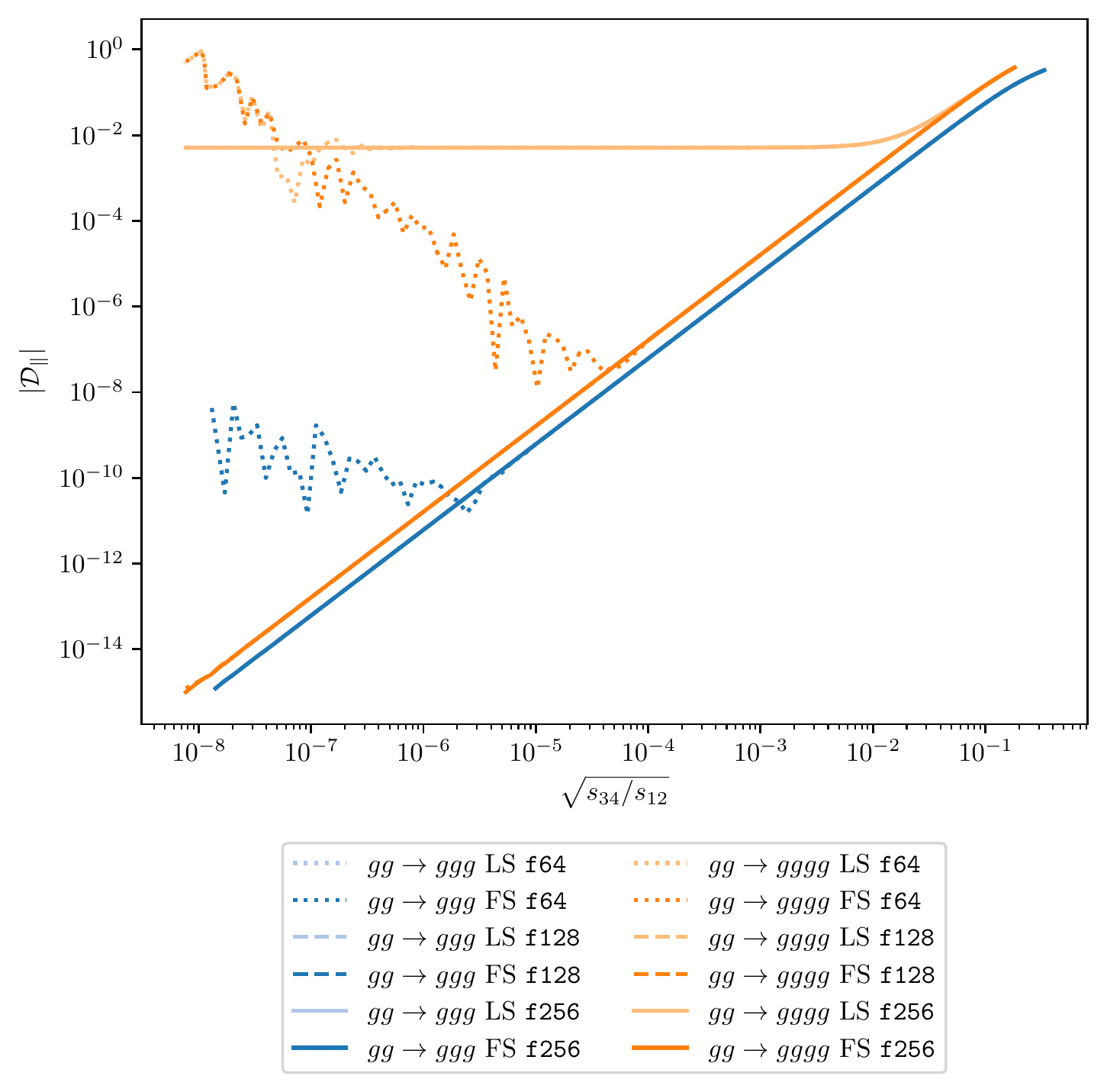}
        \caption{
            The scaling behaviour of the relative difference \cref{eq:rel-diff-coll} between the factorised expression and the full \ac{ME} in a collinear limit at tree level.
            We consider two multiplicities of gluon scattering at \ac{LS} and \ac{FS} with three numerical precisions.
        }
        \label{fig:collinear-scaling}
    \end{center}
\end{figure}

We begin with the $3\parallel4$ double-collinear limit \cref{eq:double-collinear}, considering tree-level five- and six-gluon scattering.

We parametrise phase space using the prescription of \incite{badger:2015cxa} and use this to generate a \num{100}-point slice that approaches a collinear limit.
The prescription uplifts an $n$-point phase space to an on-shell mass-conserving $(n+1)$-point phase space that is parametrised in a collinear limit.
For the double-collinear limit, this is exactly \cref{eq:col-ps} in reverse, with ${p_i}^\mu$ and ${p_j}^\mu$ given by \cref{eq:col-par-one,eq:col-par-two} respectively along with
\begin{align}
    q^\mu&=\covcon{\tilde{p}}{k}{\mu} \,, &
    {p_\perp}^\mu &= \delta \Im\brf{\spBA{\tilde{p}}{\sigma^\mu}{\tilde{p}_k}-\spBA{\tilde{p}_k}{\sigma^\mu}{\tilde{p}}} \,,
\end{align}
and
\begin{align}
    {p_k}^\mu &= \br{1+\frac{{p_\perp}^2}{z(1-z)\,2\,\tilde{p}\cdot \tilde{p}_k}} \covcon{\tilde{p}}{k}{\mu} \,,
\end{align}
such that $\delta$ controls the degree of collinearity of the point.
The exact limit is given by $\delta=0$.
We set $z=0.5$ and vary $\delta$ over the slice as shown in \cref{fig:collinear-scaling}.
The four-point seed phase space is generated randomly using the algorithm from~\incite{byckling:1971vca}, which is provided in the \njet library.

For each point in the slice, we construct the reduced phase space using the \ac{CS} mapping defined by \cref{eq:cs-map,eq:col-ps,eq:col-par-three}.
We evaluate both sides of \cref{eq:collinear-square-two} in \ac{f64}, \ac{f128}, and \ac{f256} and plot their relative differences,
\begin{align}
    \label{eq:rel-diff-coll}
    \mathcal{D}_\parallel = \frac{\ab{\am}^2-\sum_{h_1,h_2}{\widetilde{\mathcal{M}}}_{h_1 h_2}\mathcal{P}_{-h_1 -h_2}}{\ab{\am}^2} \,,
\end{align}
in \cref{fig:collinear-scaling}.
The values of the factorised and full \acp{ME} smoothly become more similar as we probe deeper into the collinear limit.
The five-gluon amplitude \ac{f64} evaluation becomes numerically unstable below $\s{34}/\s{12}\approx10^{-10}$, where we see $\mathcal{D}_\parallel$ deviate from the \ac{f128} evaluation.
While the five-gluon amplitude in \njet is implemented as a hard-coded analytic expression, the six-gluon amplitude is implemented numerically and becomes unstable with \ac{FS} \ac{f64} evaluation at the higher value of $\s{34}/\s{12}\approx10^{-8}$.
The \ac{FS} six-gluon \ac{f128} evaluation begins to destabilise as it approaches $\s{34}/\s{12}\approx10^{-16}$.

The five-gluon \ac{ME} does not carry any spin correlations, so the \ac{LS} evaluations match the \ac{FS} ones.
At six-gluons, however, the spin correlations are nonzero.
We see the \ac{LS} approximation level off at $\s{34}/\s{12}\approx10^{-3}$, while the \ac{FS} evaluation continues to approach the value of the full amplitude deeper into the limit.
The \ac{LS} \ac{f64} numerical instability does not become visible until the fluctuations grow to the scale of the \ac{ME}, which is where the \ac{FS} \ac{f64} line approaches it at $\s{34}/\s{12}\approx10^{-13}$.

\begin{figure}
    \begin{center}
        \includegraphics[width=\textwidth]{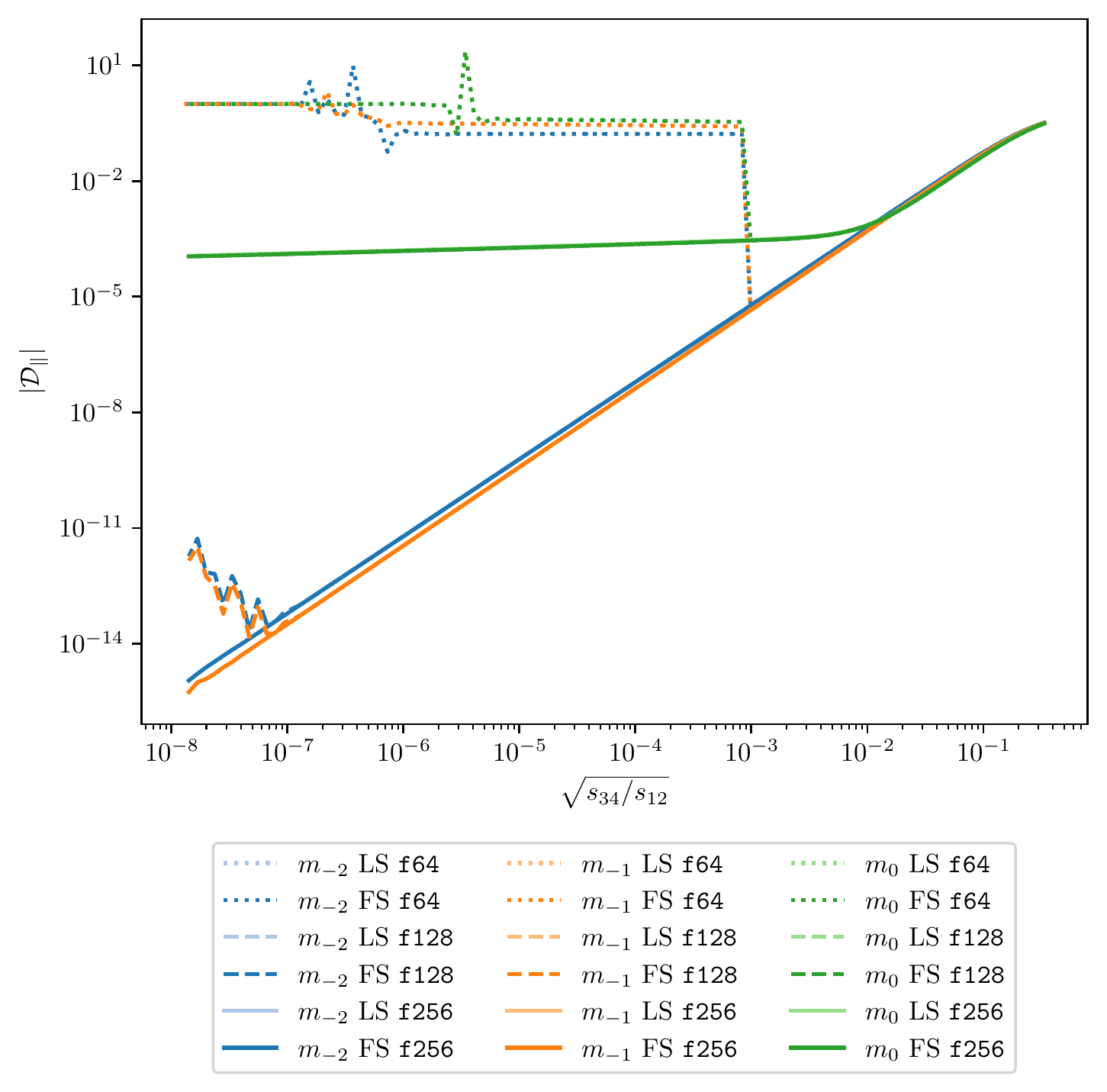}
        \caption{
            The scaling behaviour of the relative difference between the factorised expression and the full \ac{ME} in a collinear limit for the interference between the tree-level and one-loop amplitudes of $gg\to ggg$.
        }
        \label{fig:collinear-scaling-1l}
    \end{center}
\end{figure}

We also consider the $3\parallel4$ double-collinear limit \cref{eq:one-loop-double-collinear} of five-gluon scattering at one-loop level; the full \ac{ME} is the interference of the tree-level and one-loop five-gluon amplitudes.
We use the same phase space parametrisation as for the tree-level case.
\oneloop~\cite{vanhameren:2010cp} is used to provide the one-loop scalar integrals.
Evaluations are performed in \ac{f64}, \ac{f128}, and \ac{f256}.
In \cref{fig:collinear-scaling-1l}, we plot the \ac{ME} coefficients $m_a$ defined by
\begin{align}
    {\mathcal{A}^{(0)}}^*\mathcal{A}^{(1)} &= \sum_{a=-2}^0 m_{a}\,\eps^{a} + \mathcal{O}(\eps) \,.
\end{align}
The one-loop integrals become unstable at $\s{34}/\s{12}\approx10^{-6}$ in \ac{f64}, while the $\eps$ pole coefficients $m_{-2}$ and $m_{-1}$ become unstable near $\s{34}/\s{12}\approx10^{-13}$ in \ac{f128}.
The $\mathcal{D}_\parallel$ for the \ac{f256} evaluations smoothly decreases for all coefficients, although the finite part $m_{0}$ has a much more gradual slope than the poles after $\s{34}/\s{12}\approx10^{-3}$.
Only $m_{0}$ has nonzero spin correlations, which approach from below the magnitude of the diagonal spin contributions towards the left side of the figure and are therefore not easily visible on the logarithmic scale.

% \subsubsection{Soft limits}

% \begin{figure}
%     \begin{center}
%         \includegraphics[width=\textwidth]{soft-scaling.together}
%         \caption{
%             The scaling behaviour of the relative difference between the factorised expression and the full \ac{ME} in a soft limit.
%         }
%         \label{fig:soft-scaling}
%     \end{center}
% \end{figure}

% Mapping \cref{eq:soft-map} for each $(i,k)$ pair in the sum \cref{eq:soft-square}.

% Relative difference
% \begin{align}
%     \label{eq:rel-diff-soft}
%     \mathcal{D}_s &= \frac{\ab{\am}^2-\sum_{i,k}\widetilde{\mathrm{C}}^{(i,k)} \mathrm{S}_{j^{h_j}}^{(i,k)}}{\ab{\am}^2} \,,
% \end{align}

% \cref{fig:soft-scaling}

\subsection{Example code}
\label{sec:example}

\subsubsection{Collinear limit}

Consider the $3\parallel4$ double-collinear limit of a five-gluon tree-level helicity amplitude,
\begin{align}
    \amp{0}{5}\brf{1_g^+,2_g^+,3_g^+,4_g^-,5_g^-} \overset{3\parallel4}{\longrightarrow}
    \lls{\widetilde{\am}}{0}{4}\brf{1_g^+,2_g^+,\rho_g^{h},5_g^-} \otimes
    \lls{\mathcal{P}}{0}{g\rightarrow gg}\brf{3^+,\rho_g^{-h},4^-}\,.
\end{align}
To demonstrate the use of the spin matrices \(\mathcal{P}_{h_1 h_2}\) in \njet, we show how to evaluate \cref{eq:collinear-square-two} for this case in the following \cpp code.

\begin{minted}{c++}
#include <algorithm>
#include <array>
#include <cmath>
#include <complex>
#include <iomanip>
#include <iostream>

#include "analytic/0q4g-analytic.h"
#include "analytic/0q5g-analytic.h"
#include "ir/split_g2gg-analytic.h"
#include "ngluon2/Mom.h"

int main()
{
    // 2 || 3 collinear massless 5-particle phase space (0 indexing)
    const std::array<MOM<double>, 5> full_mom { {
        { -5.0000000000000000e-01, 0.0000000000000000e+00, 0.0000000000000000e+00, 5.0000000000000000e-01 },
        { 4.9999949999999999e-01, 4.3301226887951738e-01, 0.0000000000000000e+00, 2.4999975000000005e-01 },
        { 2.9999999999999999e-01, -2.5960910680884997e-01, 2.8795483728213630e-04, -1.5034303689869646e-01 },
        { 2.0000050000000008e-01, -1.7340316207066742e-01, -2.8795483728213630e-04, -9.9656713101303585e-02 },
        { -5.0000000000000000e-01, 0.0000000000000000e+00, 0.0000000000000000e+00, -5.0000000000000000e-01 },
    } };

    // Catani-Seymour momentum mapping for 2 || 3 (0 indexing)
    const MOM<double> p { full_mom[2] + full_mom[3] };
    const double x { dot(full_mom[2], full_mom[3]) / dot(p, full_mom[4]) };

    // reduced phase space
    std::array<MOM<double>, 4> reduced_mom {};
    std::copy_n(full_mom.cbegin(), 2, reduced_mom.begin());
    reduced_mom[2] = p - x * full_mom[4];
    reduced_mom[3] = (1. + x) * full_mom[4];

    // momenta for splitting amplitude
    const std::array<MOM<double>, 3> splitting_mom { { -reduced_mom[2], full_mom[2], full_mom[3] } };

    // initialise amplitudes
    Amp0q5g_a<double> full_amp;
    Amp0q4g_a<double> reduced_amp;
    Splitg2gg_a<double> splitting_amp;

    // set momenta
    full_amp.setMomenta(full_mom.data());
    reduced_amp.setMomenta(reduced_mom.data());
    // the second argument is a reference momentum
    splitting_amp.setMomenta(splitting_mom.data(), reduced_mom[3]);

    // helicities for amplitudes
    const std::array<int, 5> full_hels { +1, +1, +1, -1, -1 };
    // the helicity values of the correlated legs do not matter
    const std::array<int, 4> reduced_hels { +1, +1, 0, -1 };
    const std::array<int, 3> splitting_hels { 0, +1, -1 };

    // initialise (flattened) 2x2 spin matrices
    // order: ++ +- -+ --
    std::array<std::complex<double>, 4> reduced_spn_mat {}, splitting_spn_mat {};

    // compute spin matrices
    // the first argument is the index of the correlated leg
    reduced_amp.born_spnmatrix(2, reduced_hels.data(), reduced_spn_mat.data());
    splitting_amp.born_spnmatrix(0, splitting_hels.data(), splitting_spn_mat.data());

    // while the final result will be real, the intermediate steps may be complex
    std::complex<double> limit {};

    // compute factorised matrix element including spin correlations
    for (int i { 0 }; i < 4; ++i) {
        limit += reduced_spn_mat[i] * splitting_spn_mat[3 - i];
    }

    // compute full matrix element
    const double amp { full_amp.born(full_hels.data()) };

    // print results (with unity indexing)
    std::cout
        << std::setprecision(1) << std::scientific
        << '\n'
        << std::setw(25) << "|s_{34}/s_{12}| = "
        << std::setw(7) << std::abs(dot(full_mom[2], full_mom[3]) / dot(full_mom[0], full_mom[1]))
        << '\n'
        << std::setw(25) << "|(amp^2-lim^2)/amp^2| = "
        << std::setw(7) << std::abs((amp - limit.real()) / amp)
        << '\n'
        << '\n';
}
\end{minted}

This runs to give the result:
\begin{minted}{c++}
          |s_{34}/s_{12}| = 1.3e-06
    |(amp^2-lim^2)/amp^2| = 7.9e-04
\end{minted}
The small ratio $\s{34}/\s{12}=\SI{1.3e-6}{\relax}$ shows that we are near a $3\parallel4$ collinear limit.
The relative difference \cref{eq:rel-diff-coll} between the factorised and full \acp{ME} $\mathcal{D}_\parallel=\SI{7.9e-4}{\relax}$ is small, showing a similar order of magnitude to the helicity-summed value in \cref{fig:collinear-scaling}.
This code can be used for the helicity-summed \ac{ME} by simply removing the helicity argument in all function calls.

\subsubsection{Soft limit}

Consider the $\mathcal{E}_4$ single-soft limit of a five-gluon tree-level helicity amplitude,
\begin{align}
    \amp{0}{5}\brf{1_g^+,2_g^+,3_g^-,4_g^-,5_g^-} \overset{E_4\rightarrow\,0}{\lra}
    \lls{\widetilde{\am}}{0}{4}\brf{1_g^+,2_g^+,3_g^-,5_g^-} \otimes
    \lls{\mathcal{S}}{0}{1}\brf{4_g^-}\,.
\end{align}
To demonstrate the use of the colour-correlation matrices ${\widetilde{\mathrm{D}}^{(i,j)}}$ in \njet, we show how to evaluate \cref{eq:soft-square} for this case in the following \cpp code.

\begin{minted}{c++}
#include <algorithm>
#include <array>
#include <cmath>
#include <complex>
#include <iomanip>
#include <iostream>

#include "analytic/0q4g-analytic.h"
#include "analytic/0q5g-analytic.h"
#include "ir/soft_gg2g-analytic.h"
#include "ngluon2/Mom.h"

int main()
{
    // massless 5-point phase space with full_mom[3] soft
    const std::array<MOM<double>, 5> full_mom { {
        { -5.0000000000000000e-01, 0.0000000000000000e+00, 0.0000000000000000e+00, -5.0000000000000000e-01 },
        { -5.0000000000000000e-01, 0.0000000000000000e+00, 0.0000000000000000e+00, 5.0000000000000000e-01 },
        { 4.9999949999999999e-01, 4.3301226887951738e-01, 0.0000000000000000e+00, 2.4999975000000005e-01 },
        { 4.9999999999772449e-06, -2.2505887718767004e-06, 1.7633487034433740e-06, -4.1018839000804920e-06 },
        { 4.9999550000000009e-01, -4.3301001829074548e-01, -1.7633487034433740e-06, -2.4999564811609998e-01 },
    } };

    // initialise amplitudes
    Amp0q5g_a<double> full_amp;
    Amp0q4g_a<double> reduced_amp;
    Softgg2g_a<double> soft_amp;

    // set momenta
    full_amp.setMomenta(full_mom.data());

    // helicities for amplitudes
    const std::array<int, 5> full_hels { +1, +1, -1, -1, -1 };
    const std::array<int, 4> reduced_hels { +1, +1, -1, -1 };
    // the eikonal amplitude is independent of the helicities of the correlated legs
    const std::array<int, 3> soft_hels { 0, -1, 0 };

    // perform reduced colour sum
    // indices i,3,k are in the full 5-point phase space
    // indices ii,kk are in the reduced 4-point phase space
    double lim {};
    int ref_index {};
    for (int i { 0 }; i < 5; ++i) {
        if (i != 3) {
            const int ii { i < 3 ? i : (i - 1) % 4 };

            for (int k { i + 1 }; k < 5; ++k) {
                if (k != 3) {
                    const int kk { k < 3 ? k : (k - 1) % 4 };

                    // reduced phase space is just the full phase space without the soft leg
                    std::array<MOM<double>, 4> reduced_mom {};
                    std::copy_n(full_mom.cbegin(), 3, reduced_mom.begin());
                    reduced_mom[3] = full_mom[4];

                    // ensure physical reduced phase space
                    const double b { dot(full_mom[i], full_mom[3]) / (dot(full_mom[i], full_mom[k]) + dot(full_mom[3], full_mom[k])) };
                    reduced_mom[ii] = full_mom[i] + full_mom[3] - b * full_mom[k];
                    reduced_mom[kk] = (1. + b) * full_mom[k];

                    reduced_amp.setMomenta(reduced_mom.data());

                    // ensure reference momentum is not the same as another momentum in use
                    // to avoid division by zero from brackets in the denominator of the form < i i >
                    while ((ref_index == i) || (ref_index == 3) || (ref_index == k)) {
                        ref_index = (ref_index + 1) % 5;
                    }

                    // set momenta for eikonal amplitude
                    // the second argument is the reference momentum
                    soft_amp.setMomenta({ reduced_mom[ii], full_mom[3], reduced_mom[kk] }, full_mom[ref_index]);

                    // compute eikonal and colour-correlated reduced matrix elements
                    const double soft_val { soft_amp.born(soft_hels.data()) },
                        cc_val { reduced_amp.born_ccij(reduced_hels.data(), ii, kk) };

                    lim += cc_val * soft_val;
                }
            }
        }
    }

    // we summed over upper triangle of colour correlation matrix
    // diagonals receive no contribution because eikonal kinematic factor is zero due to antisymmetry of brackets
    // matrix is symmetric, so add lower triangle by doubling result
    lim *= 2.;

    // compute full matrix element
    const double amp_val { full_amp.born(full_hels.data()) };

    // print results (with unity indexing)
    std::cout
        << std::setprecision(1) << std::scientific
        << '\n'
        << std::setw(25) << "|{E_4}^2/s_{12}| = "
        << std::setw(7) << std::abs(pow(full_mom[3].x0, 2) / dot(full_mom[0], full_mom[1])) << '\n'
        << std::setw(25) << "|(amp^2-lim^2)/amp^2| = "
        << std::setw(7) << std::abs((amp_val - lim) / amp_val) << '\n'
        << '\n';
}
\end{minted}

This evaluates to:
\begin{minted}{c++}
         |{E_4}^2/s_{12}| = 5.0e-11
    |(amp^2-lim^2)/amp^2| = 3.1e-05
\end{minted}
The small ratio ${E_{4}}^2/\s{12}=\SI{5.0e-11}{\relax}$ shows that we are in an $\mathcal{E}_4$ soft limit.
The relative difference,
\begin{align}
    \label{eq:rel-diff-soft}
    \mathcal{D}_s &= \frac{\ab{\am}^2-\sum_{i,k}\widetilde{\mathrm{C}}^{(i,k)} \mathrm{S}_{j^{h_j}}^{(i,k)}}{\ab{\am}^2} \,,
\end{align}
between the factorised and full \acp{ME} $\mathcal{D}_s=\SI{3.1e-05}{\relax}$ is small, supporting that the factorisation \cref{eq:soft-square} holds.

\chapter{Matrix element neural networks}
\label{ch:ann}

Precision phenomenological studies of high-multiplicity scattering processes present a substantial theoretical challenge and are vitally important ingredients in measurements at collider experiments.
\Ac{ML} technology has the potential to dramatically optimise simulations for complicated final states.
We investigate the use of \acp{NN} to approximate \acp{ME}, studying the case of loop-induced diphoton-plus-jets production through gluon fusion.
We train \ac{NN} models on one-loop amplitudes from the \njet \cpp library~\cite{njet,badger:2012pg,badger:2013vpa} (see also \cref{ch:yy-amp}) and interface them with the \ac{SHERPA} \ac{MC} event generator~\cite{bothmann:2019yzt,gleisberg:2008ta} to provide the \ac{ME} within a realistic hadron collider simulation.
Computing some standard observables, such as jet transverse momentum, with the models and comparing to conventional techniques, we find excellent agreement in the distributions and a reduced total simulation time by a factor of thirty.

This chapter is organised as follows.
We first motivate the use of \ac{ME} \acp{NN} in the gluon-initiated diphoton-plus-jets sector in \cref{sec:nn-back}.
Then, in \cref{sec:nn-amp}, we discuss the gluon-initiated diphoton-plus-jets amplitudes and their implementations within \njet which form the target distribution for training the \acp{NN}.
We then describe the phase-space partitioning used to handle \ac{IR} divergent regions in \cref{sec:nn-ps}.
Next, in \cref{sec:nn-arch}, we present the architecture of the \acp{NN} used.
In \cref{sec:nn-flow}, we discuss the simulation pipeline and interface of the \ac{NN} model to \ac{SHERPA}.
Finally, in \cref{sec:nn-res}, we study the performance of the model compared to \njet for $gg\to\gamma\gamma gg$ and present some distributions before concluding in \cref{sec:nn-conc}.

Our code is publicly available~\cite{n3jet_diphoton}.

\section{Background}
\label{sec:nn-back}

With the increasing size of the \ac{LHC} dataset driving ever more precise experimental measurements, \ac{SM} predictions for high-multiplicity scattering at hadron colliders form a vital part of precision phenomenology studies.
These calculations mainly rely on automated numerical codes~\cite{degrande:2018neu} to calculate amplitudes up to high multiplicities, including tree-level real corrections at \ac{NLO} and \ac{RR} corrections at \ac{NNLO}, and one-loop \ac{RV} corrections at \ac{NNLO} (\cref{sec:fixed-order}).
These codes have been a theoretical revolution, particularly at one-loop (\cref{sec:loops}).
However, due to the high dimensionality of the phase space, these real-type corrections are often the computational bottleneck in higher-order calculations (for instance, see \cref{sec:cost}).

As discussed in \cref{sec:yy}, the gluon-fusion channel of diphoton-plus-jets production is of high phenomenological relevance.
Therefore, we study the loop-induced class of processes with two photons and many gluons (\cref{sec:fixed-order}).
We stress that because they are loop induced, these amplitudes are finite in $\eps$ (\cref{sec:infinities}).

\Ac{ML} technology has found a wealth of application in high-energy physics: see the reviews~\cite{feickert:2021ajf,butter:2022rso} and references therein.
For an introduction to \ac{ML}, see \incite{goodfellow}.
We employ the ensemble \ac{NN} model of \incite{badger:2020uow}, which studied $e^+e^-$ annihilation to jets, to emulate the gluon-initiated diphoton-plus-jets \acp{ME} within a full \ac{MC} event generator simulation (\cref{sec:factorisation-theorem}).
This tests the methodology against the additional complexity of hadron collider simulations, including \ac{PDF} convolution and variable partonic centre-of-mass scale, complex phase-space cuts and jet clustering (\cref{sec:cuts}), and phase-space sampling optimisation methods of integrators.

\section{Amplitudes}
\label{sec:nn-amp}

\begin{figure}
    \begin{center}
        \includegraphics[width=0.25\textwidth]{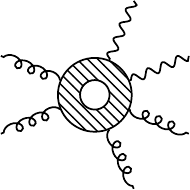}
        \caption{
            Diagram of $gg\to\gamma\gamma gg$ ($N=6$) at \ac{LO}.
            The photons couple to an internal quark loop.
        }
        \label{fig:ggyygg_1l}
    \end{center}
\end{figure}

\begin{figure}
    \begin{center}
        \includegraphics[width=0.67\textwidth]{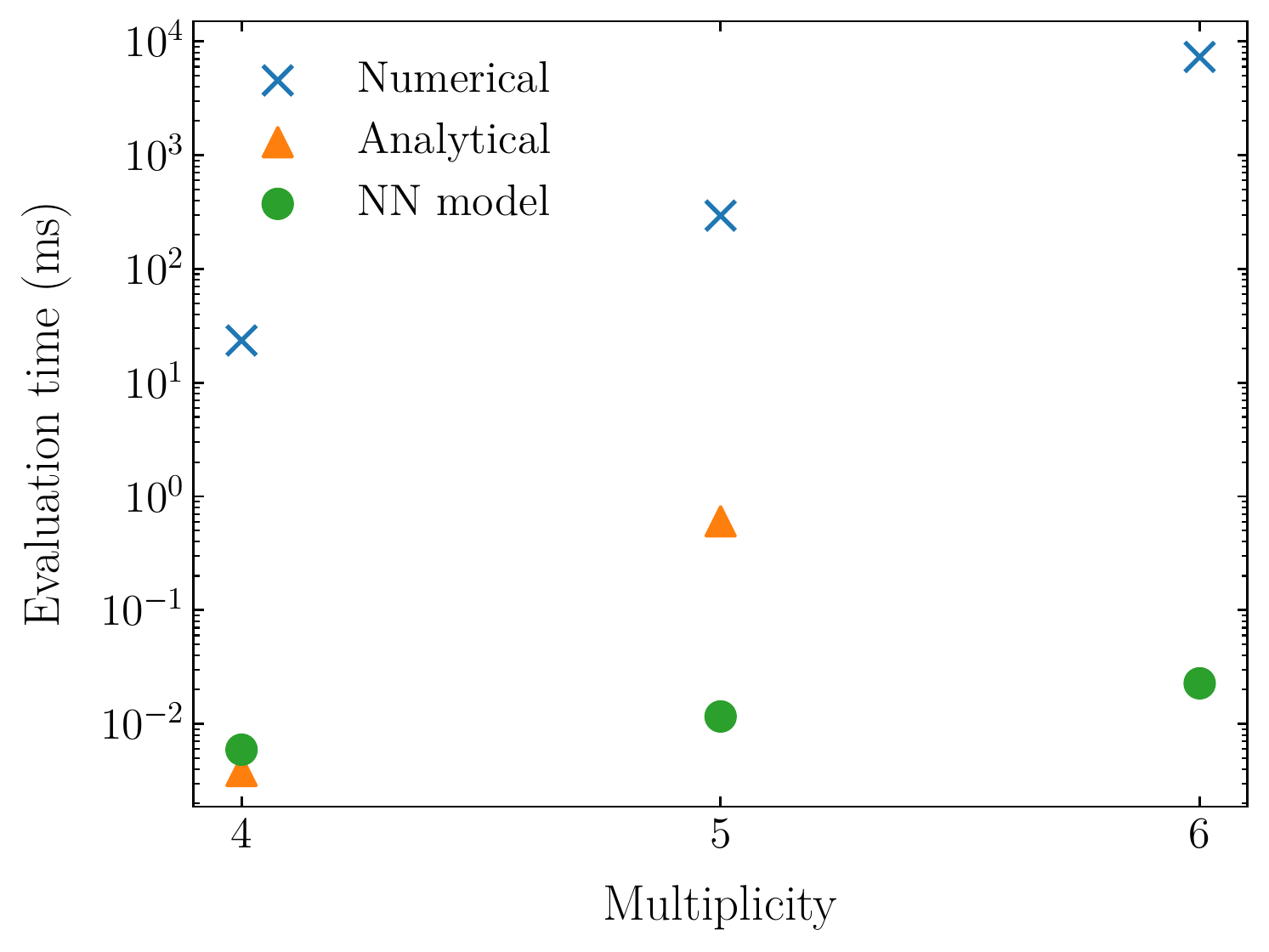}
        \caption{
            Typical evaluation time of the \ac{ME} for a single phase-space point.
            Results are shown for available implementations at various multiplicities, including numerical (blue cross) and analytical (orange triangle) evaluations using \njet and inference on the \ac{NN} model (green circle).
        }
        \label{fig:timing-single}
    \end{center}
\end{figure}

As there is no vertex coupling gluons to photons in the \ac{SM}, diphoton-plus-jets production through gluon fusion is loop induced, as depicted in \cref{fig:ggyygg_1l} and discussed in \cref{sec:fixed-order}.
The \ac{LO} process is $\ord({\astr}^{N-2})$ for multiplicity $N$, appearing at \ac{NNLO} in the perturbative expansion of the combined quark- and gluon-initiated process, as shown in \cref{fig:pert-exp}.
We study the channels with only gluons and photons in the external particles $gg\to\gamma\gamma+n\times g$.
These proceed through a quark loop at \ac{LO}.

Conventional event generator simulations optimise virtual corrections in \ac{NLO} calculations by learning the phase-space distribution of the \ac{LO} process and using this to sample the virtual contribution.
This technique fails for loop-induced processes, where the expensive one-loop amplitude has no tree-level process to optimise the phase space on.
Therefore, new methods are required to improve the efficiency of integrating these channels at high multiplicity.

We use the one-loop-squared \acp{ME} from the \njet library as the targets for our \ac{NN} emulation.
These include two classes of amplitudes: an automated numerical setup for arbitrary multiplicity; and hard-coded analytical expressions for $N\in\{4,5\}$.
The former obtains the diphoton-plus-jets amplitudes by summing permutations of pure-gluon partial amplitudes~\cite{dicus:1987fk,deflorian:1999tp}, which are themselves based on generalised unitarity (\cref{sec:generalised-unitarity}) and integrand reduction (\cref{sec:integrand-reduction}).
While completely automated, evaluation time and numerical stability are increasingly difficult to control.
The hard-coded implementations offer compact analytical expressions with extremely fast and stable evaluation, although they are unavailable for higher multiplicity.
The $N=5$ result is obtained through \iac{FF} reconstruction~\cite{peraro:2019svx} of the permutation-sum result.

The evaluation timings of these methods are compared to the \ac{NN} model in \cref{fig:timing-single}.
Note that this is a single \ac{NN} model, which is comprised of \iac{FKS} ensemble (\cref{sec:nn-ps}), and not the stochastic ensemble (\cref{sec:nn-flow}).
The value is the mean of \num{100} evaluations using random sampling over a uniform phase space~\cite{byckling:1971vca}.
We time single-threaded \ac{CPU} calls as parallelisation is applied at the level of events in event generator simulations.

\section{Phase space partitioning}
\label{sec:nn-ps}

Training a single \ac{NN} over the entire phase space results in a poor fit, especially at higher multiplicity~\cite{badger:2020uow}.
This is caused by regions where the amplitude becomes \ac{IR} divergent, which arise from soft ($\mathcal{E}_i$) and collinear ($i\parallel j$) emissions (\cref{sec:infrared-limits}).
These singularities are regulated at \ac{LO} with cuts (\cref{sec:cuts}), but the amplitude in local regions exhibits extreme curvature which causes problems for the global fit.
Therefore, we train a separate \ac{NN} on each of the \ac{IR} structures of the phase space.

We first partition the phase space into a non-divergent region $\mathcal{R}_{\textrm{non-div}}$ and a divergent region $\mathcal{R}_{\textrm{div}}$.
Phase space points which pass the following cut are included in $\mathcal{R}_{\textrm{div}}$,
\begin{align}
    \mathrm{min}\br{\cu{\frac{\s{ij}}{\s{12}} : i,j\in\cu{1,\ldots,N}}}<y.
\end{align}
The threshold $y$ must be tuned to discriminate points of a similar scale into a single region, while having sufficient points in $\mathcal{R}_{\textrm{div}}$ to train on.

We then sub-divide $\mathcal{R}_{\textrm{div}}$ similarly to the decomposition of the \ac{FKS} subtraction scheme (\cref{sec:subtraction-methods}).
We define a set of \ac{FKS} pairs,
\begin{align}
    \mathcal{P}_{\mathrm{FKS}} = \left\{ (i,j) : \mathcal{E}_i \lor \mathcal{E}_j \lor i\parallel j \right\},
\end{align}
corresponding to the singular configurations, of which there are
\begin{align}
    \binom{N}{2} - 1 = \frac{N^2-N-2}{2} \,.
\end{align}
This includes redundancy as it overcounts soft singularities, which means the model must learn this, but is favoured for its simplicity.
Each pair is assigned a partition function,
\begin{align}
    \mathcal{S}_{ij} &= \frac{1}{ \s{ij} \sum_{k,\ell\in \mathcal{P}_{\mathrm{FKS}}} \frac{1}{\s{k\ell}} }\,, &
    \sum_{i,j\in\mathcal{P}_{\mathrm{FKS}}} \mathcal{S}_{ij} &= 1 \,,
\end{align}
which smoothly isolates that divergence on multiplication with the \ac{ME}.
The set of all partition functions sum to unity.

We train \iac{NN} on $\left|\mathcal{A}(p^\mu)\right|^2$ for $p^\mu\in\mathcal{R}_{\textrm{non-div}}$, and \iac{NN} on each of the partition-function-weighted \acp{ME},
\begin{align}
    \left\{\mathcal{S}_{ij} \left|\mathcal{A}(p^\mu)\right|^2 \,:\, i,j\in\mathcal{P}_{\mathrm{FKS}} \,;\, p^\mu\in\mathcal{R}_{\textrm{div}}\right\}.
\end{align}
We reconstruct the complete \ac{ME} in $\mathcal{R}_{\textrm{div}}$ by summing the weighted \acp{ME},
\begin{align}
    \left|\am\right|^2 = \sum_{i,j\in \mathcal{P}_{\mathrm{FKS}}} \mathcal{S}_{ij} \left|\am\right|^2.
\end{align}
This ensemble of \acp{NN}, referred to as the model, can be used to accurately infer the \ac{ME} over the complete phase space $\mathcal{R}_{\textrm{non-div}}\cup\mathcal{R}_{\textrm{div}}$.

Note that increasing the cut $y$, which increases the proportion of points in $\mathcal{R}_{\textrm{div}}$, incurs a performance penalty due to the higher cost of inferring over several \acp{NN} in $\mathcal{R}_{\textrm{div}}$ compared to the single \ac{NN} in $\mathcal{R}_{\textrm{non-div}}$.

\section{Model architecture}
\label{sec:nn-arch}

\begin{figure}
    \begin{center}
        \includegraphics[width=0.67\textwidth]{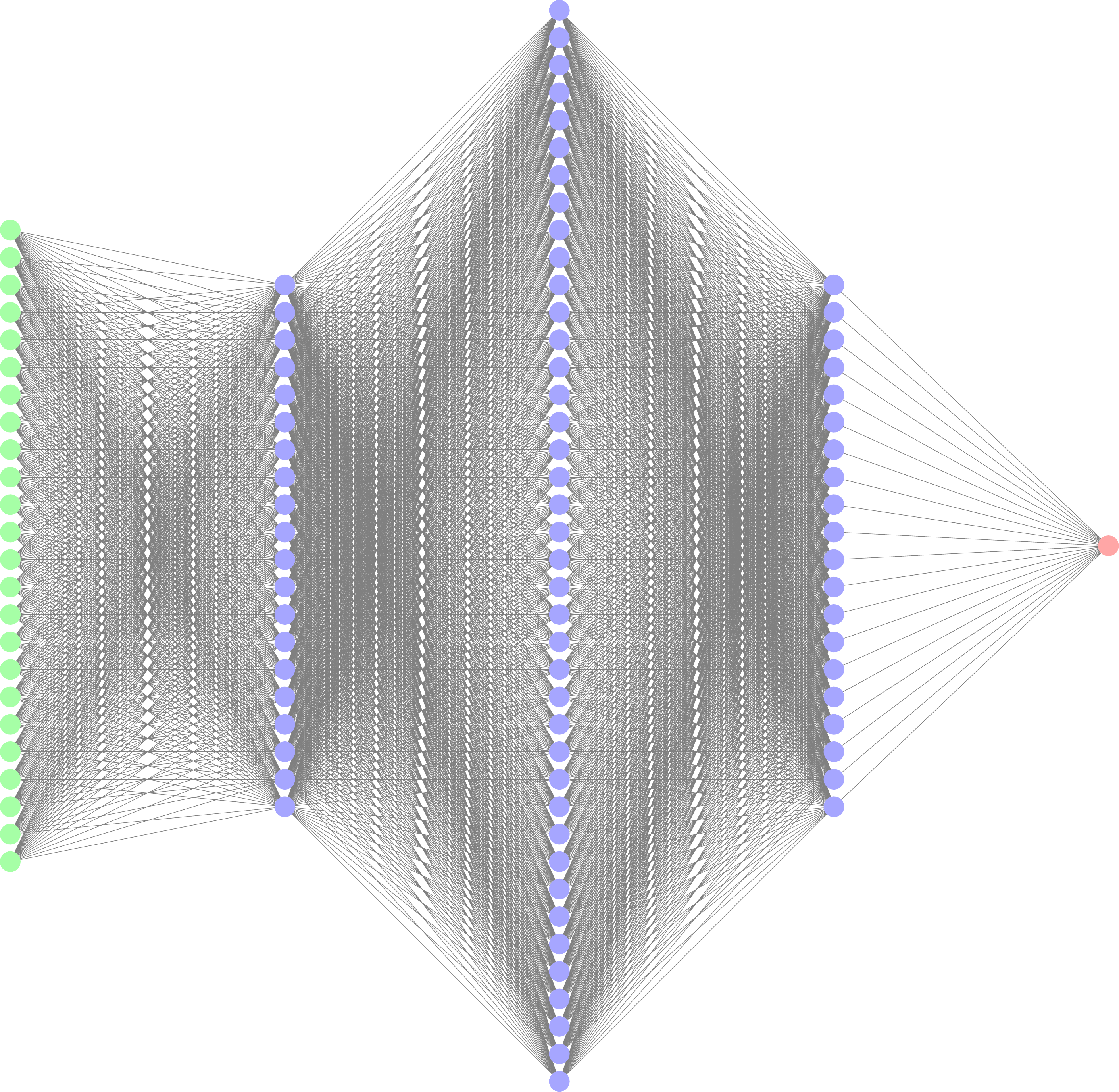}
        \caption{
            Diagram of a single \ac{NN} in the $gg\to\gamma\gamma gg$ model.
            There are \num{24} input nodes (green).
            The hidden layer nodes are shown in blue and the output node in red.
        }
        \label{fig:neural_network}
    \end{center}
\end{figure}

Although using fine-tuned architectures for each configuration (processes, cuts, etc.) would provide optimal performance, this would be prohibitively expensive in terms of personnel resources.
We use a general setup as the gains of specialised \ac{NN} optimisation are beyond the scope of this pioneering work, performing hyperparameter optimisation on the $gg\to\gamma\gamma g$ process (see \crefext{appendix}{A} of \incite{aylett-bullock:2021hmo} for details).

As depicted in \cref{fig:neural_network}, each \ac{NN} uses a fully-connected architecture, a standard choice for a regression problem, parametrised using the \texttt{Keras} \python \ac{API}~\cite{keras} to the \texttt{TensorFlow} \ac{ML} library~\cite{tensorflow}.
There are $4\times N$ input nodes: one for each component of each momentum in the phase-space point.
The three hidden layers are comprised of \num{20}, \num{40}, and \num{20} nodes respectively, all with hyperbolic-tangent activation functions, which we found to outperform standard alternatives such as rectified linear unit.
There is a single output node with a linear activation function, which returns the approximation of the \ac{ME}.
We find that this is a sufficient number of layers and nodes to learn the \acp{ME}, while remaining economical for computational performance.

We train with \iac{MSE} loss function,
\begin{align}
    L=\frac{1}{n}\sum_{i=1}^{n}\left(f(x_i)-y_i\right)^2\,,
\end{align}
where the model is represented by a function $f$ acting on $n$ input data points $x_i$ with targets $y_i$, using Adam-optimised gradient descent~\cite{adam}.
We expect that the model will learn the mean of the target distribution using this loss function (\crefext{appendix}{A} of \incite{badger:2020uow}).
The number of training epochs is determined by Early Stopping regularisation (\crefext{section}{8.1.2} of \incite{goodfellow}), with a patience of \num{100} epochs to mitigate the effects of the limited size of $\mathcal{R}_\text{div}$ that may appear in the validation set.
We use \acp{f32} throughout.

\section{Pipeline}
\label{sec:nn-flow}

\begin{figure}
    \begin{center}
        \includegraphics[width=0.67\textwidth]{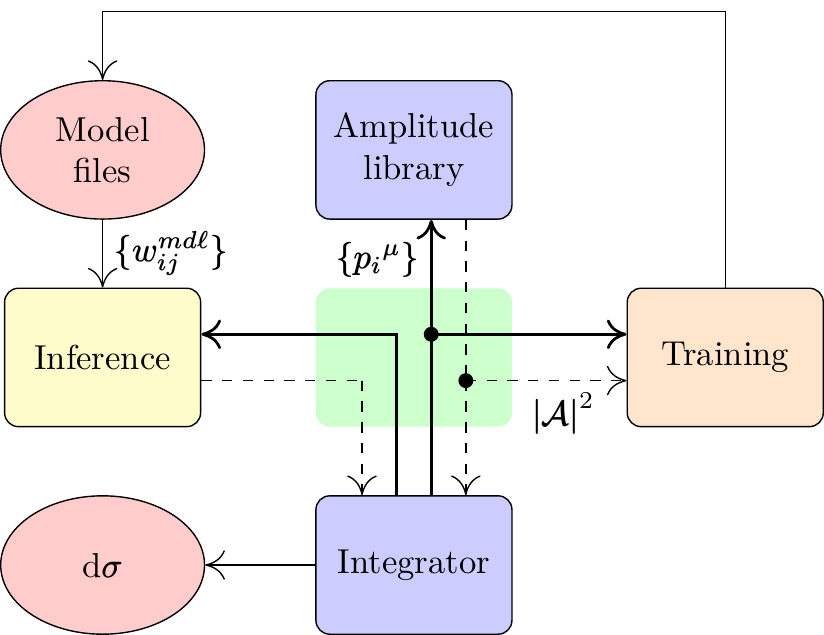}
        \caption{
            Flow chart of our \acs*{ML} pipeline.
            \Iac*{MC} event generator and \ac{ME} library (blue) are used to generate a training dataset of phase-space points (bold line) and \acp{ME} (dashed line).
            Our interface code (green) extracts this data and makes it available for training (orange).
            The \acs*{NN} weights are denoted by $w_{ij}^{mdl}$, where $m$ is the index of the model in the ensemble, $d$ is the \acs*{NN} in the model, $l$ denotes the layers for a link between nodes in layers $l$ and $l+1$, and $i$ and $j$ identify the nodes in these layers.
            They are encoded in model files (red) that are read by our inference code (yellow) to provide an approximation of the \ac{ME} for a given phase-space point, which can be provided by the integrator in a hadronic simulation to efficiently compute a cross section (red).
        }
        \label{fig:pipeline}
    \end{center}
\end{figure}

Our \ac{ML} pipeline used to produce the $gg\to\gamma\gamma gg$ results presented, sketched in \cref{fig:pipeline}, is:
\begin{enumerate}
    \item Generate the training and validation datasets by running \ac{SHERPA} with \njet on a unit integration grid, \ie~uniformly sampling the phase space such that all phase space weights are one. Similarly, generate the testing dataset with a different random seed.
    \item Train the model.
    \item Use the model to estimate the \acp{ME} during event generation with \ac{SHERPA}, using the same integration grid.
\end{enumerate}

Input data consists of a list of phase-space points ${p_i}^{\mu}\in\mathbb{R}^{4N}$ and the corresponding colour- and helicity-summed one-loop-squared \ac{ME} $\left|\am\right|^2\in\mathbb{R}$.
Phase space sampling is determined by the integrator, meaning the training is optimal only for a specific integrator.
The results presented here use the \code{RAMBO} integrator~\cite{kleiss:1985gy}, although we also study \code{VEGAS}~\cite{ohl:1998jn} in \incite{aylett-bullock:2021hmo}.
The data is extracted from a run of the integrator, generating \SI{100}{\kilo\relax} points which are split 4:1 into training and validation datasets.
A \SI{3}{\mega\relax} point testing dataset is produced by a second run of the integrator with a different random number seed and used to evaluate model performance.

We perform inference on an ensemble of twenty models, each of which has different random weight initialisation and shuffled training and validation datasets.
We take as the result the mean of the ensemble,
\begin{align}
    \bar{x} = \frac{\sum_{i=1}^{20} x_i }{20}\,,
\end{align}
where $x_i$ is the result from each model, with the standard error providing the precision/optimality error~\cite{badger:2020uow},
\begin{align}
    \varepsilon_\text{std} = \frac{\sqrt{\sum_{i=1}^{20} (x_i - \bar{x})^2 }}{20}\,.
\end{align}

While training was performed using \python, event generators are generally written in \cpp.
To use the model within a simulation, we wrote a \cpp inference code as well as a bespoke \cpp interface for \ac{SHERPA}.
The weights of the trained models are written to file and read by the inference code at runtime; the library \eigen~\cite{eigenweb} is used to perform efficient linear algebra on the \ac{CPU}.
The interface can also be used to call \cpp amplitude libraries directly instead of the model.
We use this to interface \njet to \ac{SHERPA} to generate the datasets, which is performed with \acp{f64}.
Calls are made though \iac{BLHA} interface~\cite{binoth:2010xt,alioli:2013nda}, which is compatible with all \acs{BLHA}-supporting amplitude libraries with minor modifications.

\Acp{PDF} are provided by \lhapdf~\cite{buckley:2014ana} using the \ac{NLO} \nnpdf set\footnote{\texttt{NNPDF31\_nlo\_as\_0118}} with $\astr\br{m_Z}=0.118$~\cite{nnpdf:2017mvq}.
Cuts are adapted from those in \incite{badger:2013ava}.
Analysis, including all treatment of \ac{MC} errors, is performed using \rivet~\cite{bierlich:2019rhm} with an adapted reference analysis script~\cite{aaboud:2017vol}.

\section{Results}
\label{sec:nn-res}

\begin{figure}
    \begin{center}
        \includegraphics[width=0.67\textwidth]{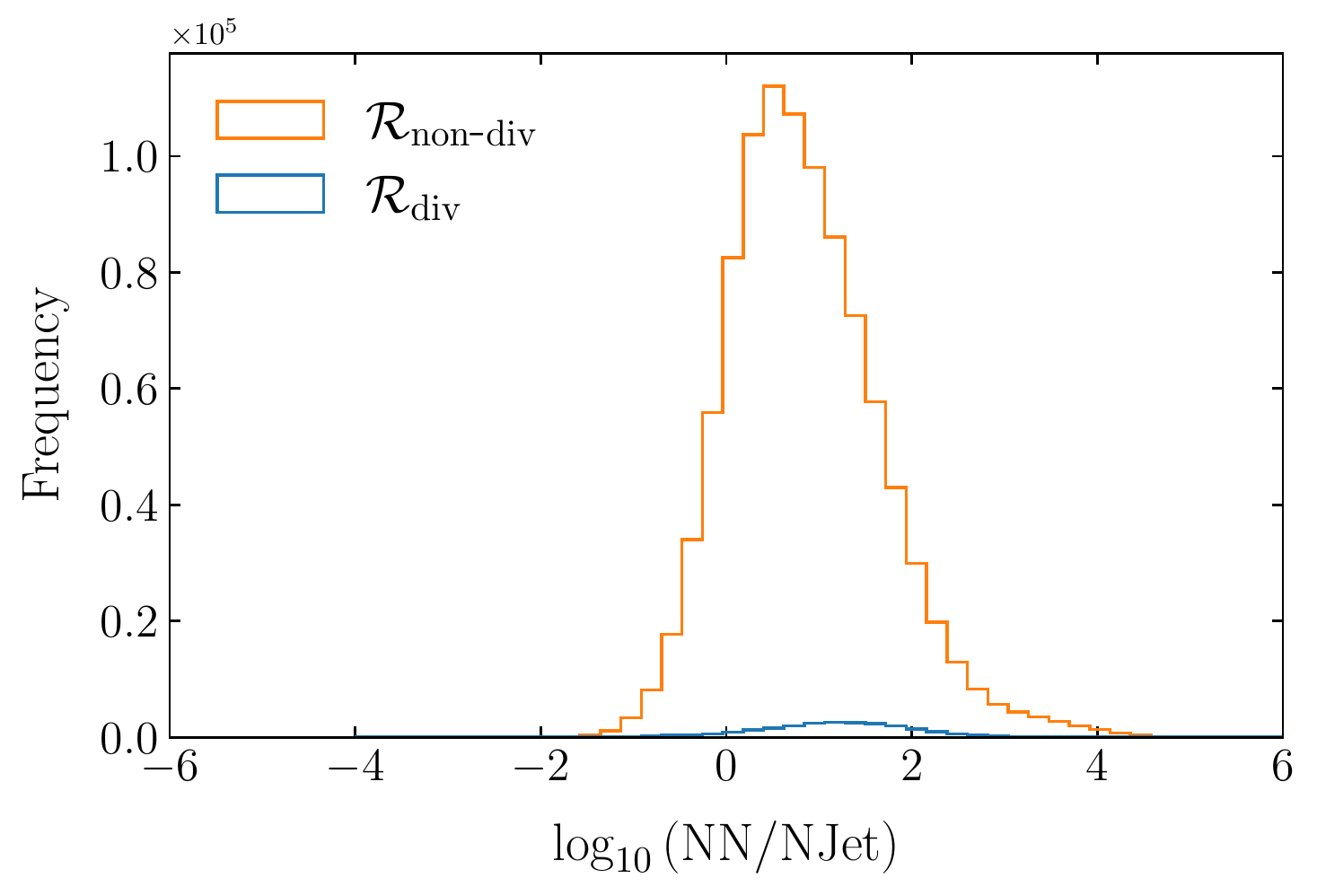}
        \caption{
            Histogram by region of the decimal logarithm of the ratio between the \ac{ME} returned by the model and \njet for each point in a \SI{1}{\mega\relax} subset of the testing data for $gg\to\gamma\gamma gg$.
            The region cut is $y=10^{-3}$ and $\mathcal{R}_{\textrm{div}}$ (blue) contains \SI{2.4}{\percent} of the points.
        }
        \label{fig:error-plot}
    \end{center}
\end{figure}

\begin{figure}
    \begin{center}
        \includegraphics[width=0.67\textwidth]{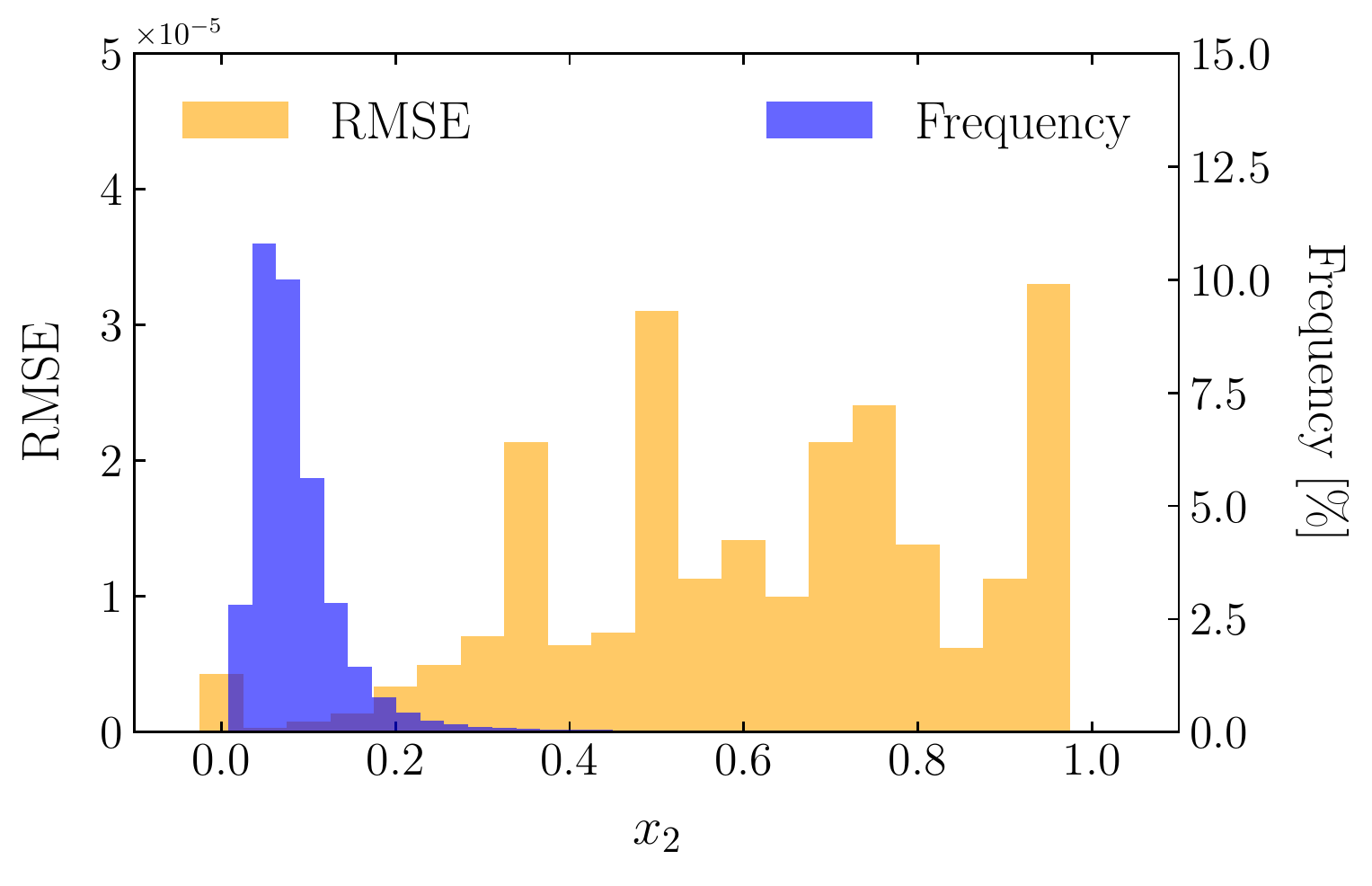}
        \caption{
            Histogram of the \ac{RMSE} of the model compared to \njet for a slice of phase space in $x_2$, the momentum fraction of the second incoming parton (gluon), for $gg\to\gamma\gamma g$ (yellow).
            Also shown are the points in the training dataset, binned in $x_2$ (blue).
        }
        \label{fig:x2-error-plot}
    \end{center}
\end{figure}

\begin{figure}
    \begin{center}
        \includegraphics[width=0.67\textwidth]{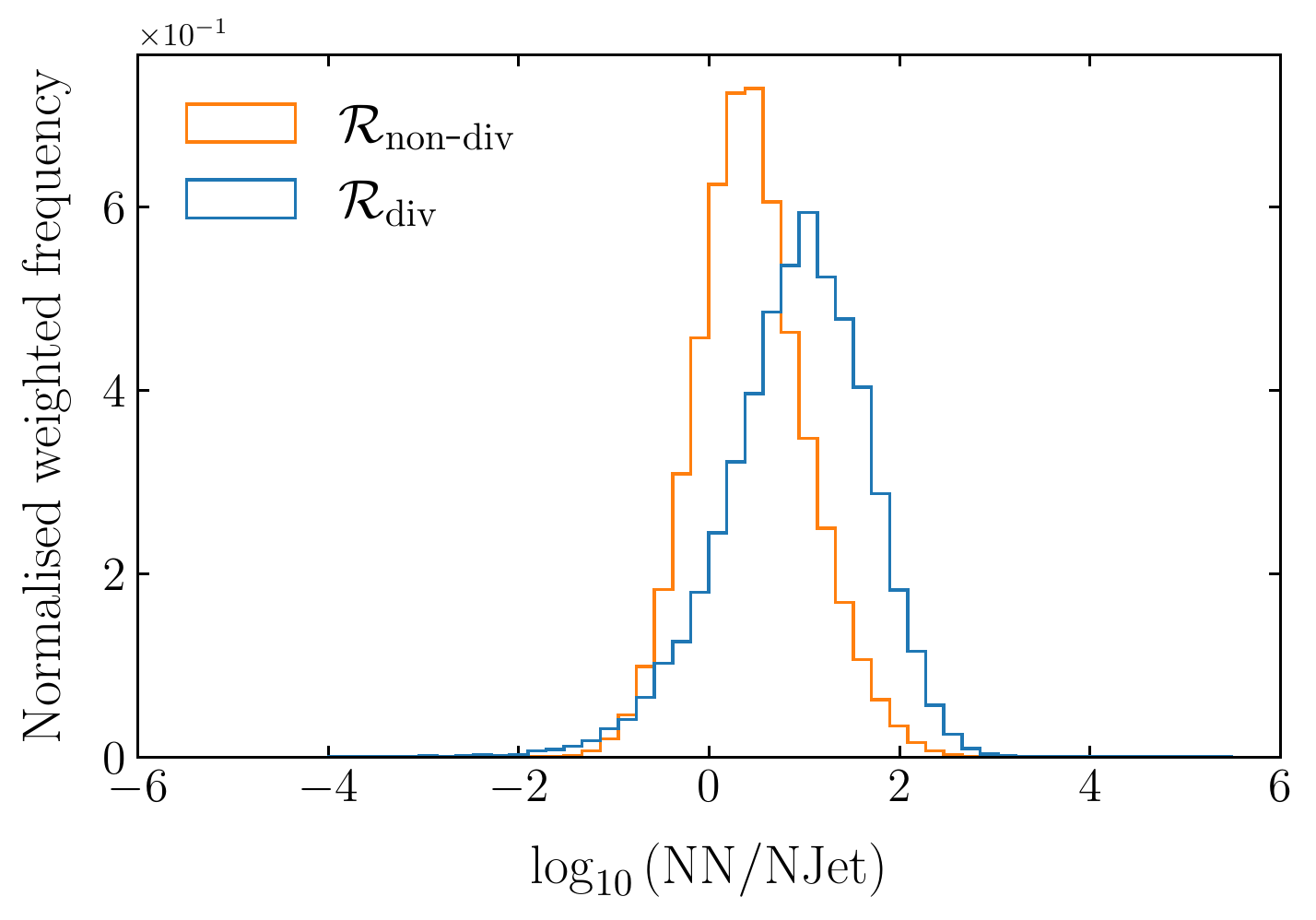}
        \caption{
            As \cref{fig:error-plot}, but weighted by the \acp{PDF}.
        }
        \label{fig:error-plot-weighted}
    \end{center}
\end{figure}

Comparing the output of the trained $gg\to \gamma\gamma gg$ model to the amplitude library value through point-by-point ratio in \cref{fig:error-plot}, we see a peaked and approximately symmetric error distribution with a shifted mean in both regions.
Both region histograms have a similar mean, indicating comparable accuracy, with $\mathcal{R}_{\textrm{non-div}}$ performing slightly better.
The distributions are fairly broad.
$\mathcal{R}_{\textrm{non-div}}$ shows a slight tail on the right, which arises from points near the cutoff $y$.

Despite the per-point agreement being somewhat poor, the total cross section is found to be in agreement, with
\begingroup
\allowdisplaybreaks[0]
\begin{align*}
    \sigma_\text{\njet} &= \SI{49(5)e-7}{\pb} \quad\text{(\ac{MC}\ error)}, \\
    \sigma_\text{\ac{NN}} &= \SI{45(6)e-7}{\pb} \quad\text{(precision/optimality\ error)}.
\end{align*}
\endgroup
\Cref{fig:x2-error-plot} shows the \ac{RMSE} $\varepsilon_{\text{\ac{RMSE}}}$ of the model compared to \njet against the frequency of points appearing on a univariate slice of phase space in the training dataset.
For a bin with $n$ points, the \ac{RMSE} is calculated as
\begin{align}
    \varepsilon_{\text{\ac{RMSE}}}=\sqrt{\frac{1}{n}\sum_{i=1}^{n}\left(r_{\text{\ac{NN}}}-r_{\text{\njet}}\right)^2}\,,
\end{align}
for target result $r_{\text{\njet}}$ and model result $r_{\text{\ac{NN}}}$.
Sampling frequency during unit-grid integration is determined by the gluon \ac{PDF}.
The figure shows that the regions that are sampled the most due to the shape of the gluon \ac{PDF} are those that have the lowest error, which is why the agreement in the total cross section is much better than for point-by-point.
To validate this observation, in \cref{fig:error-plot-weighted} we weight the per-point ratios of the model result compared to the target by the \acp{PDF}, using bin contributions for each point of
\begin{align}
    f_g(x_1,\mu_F) \, f_g(x_2,\mu_F) \,,
\end{align}
where $f_g(x_i,\mu_F)$ is the gluon \ac{PDF} for the momentum fraction $x_i$ of gluon $i$ and factorisation scale $\mu_F$.
We plot the normalised weighted histograms.
The result is indeed more narrowly peaked and closer to being unit-centred, further suggesting that poorly performing points fall in \ac{PDF}-suppressed regions.
This indicates that the accuracy of distributions inferred with the model is dependent on the choice of process, cuts, and observable.

\begin{figure}
    \includegraphics[width=0.6\textwidth]{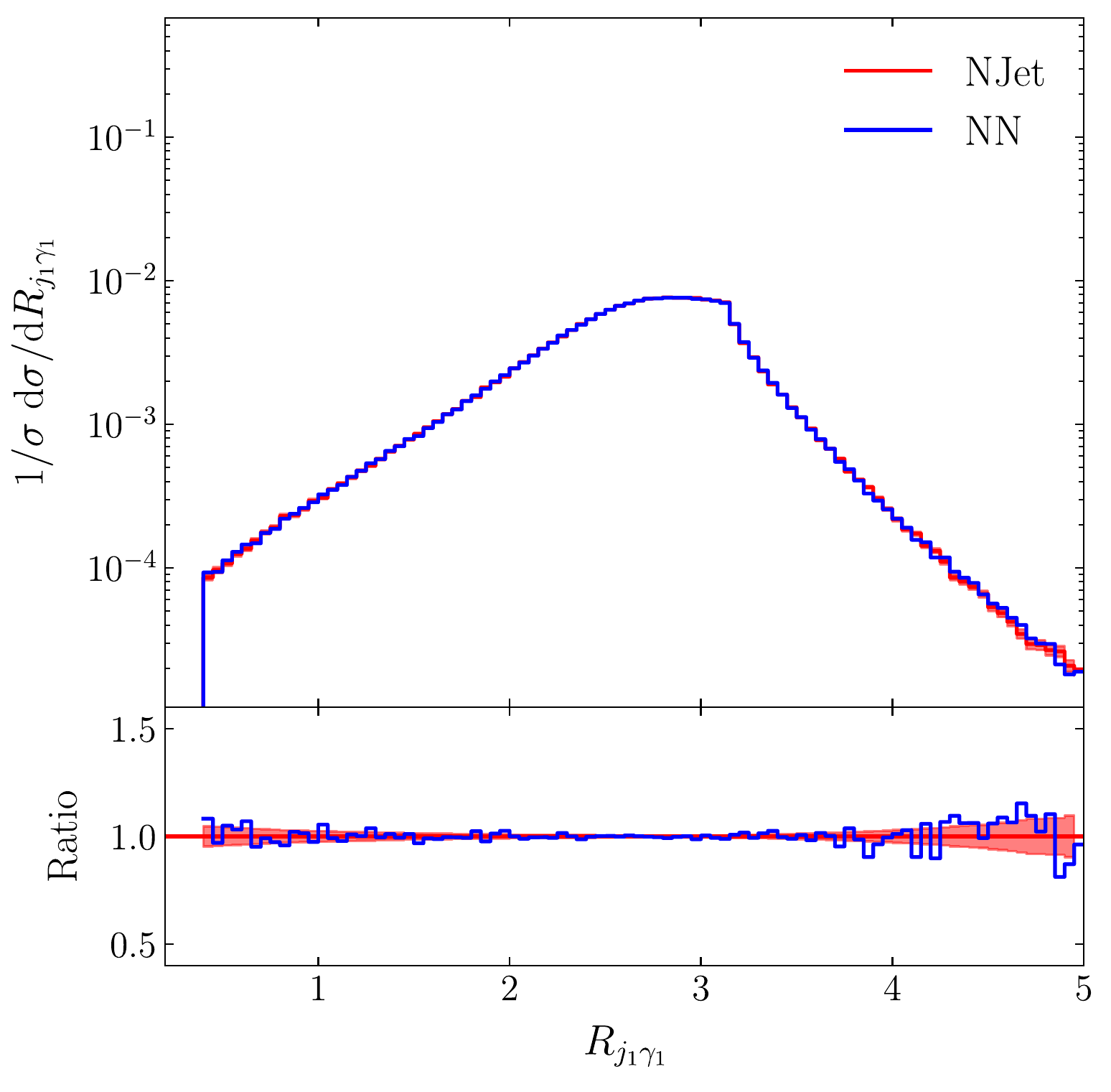}
    \includegraphics[width=0.6\textwidth]{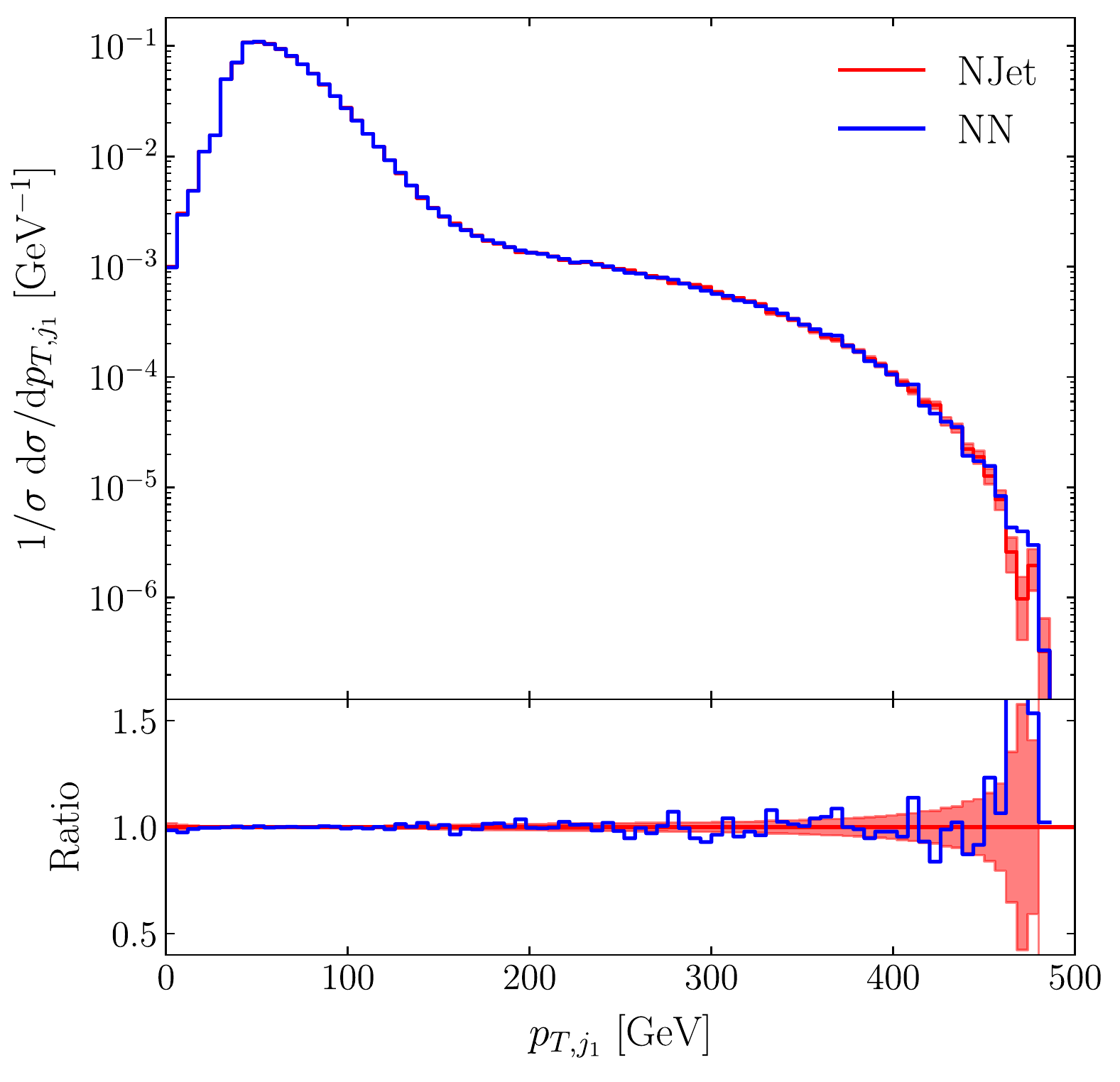}
    \caption{
        Differential normalised cross sections for $gg\to\gamma\gamma gg$, comparing \njet (red; \acs*{MC} error) to the model (blue; precision/optimality error), in R-separation, \cref{eq:r-sep}, between the hardest jet and photon (upper) and the transverse momentum of the hardest jet (lower).
        Refer to \incite{aylett-bullock:2021hmo} for details of cuts and for further distributions.
    }
    \label{fig:xs}
\end{figure}

\cref{fig:xs} shows excellent agreement between the distributions obtained from the model and \njet for two differential slices of phase space.
There are some fluctuations in the tails although they appear statistical rather than systematic and the model predictions mostly remain within the \njet \ac{MC} error bands.
Normalised \ac{NN} uncertainties are negligible compared to the \ac{MC} error.

In \incite{aylett-bullock:2021hmo}, we also demonstrate how agreement can be improved in $\mathcal{R}_{\textrm{div}}$ by reweighting event weights by the ratio of the emulated and true \acp{ME} at known points from the training data, as well as showing good performance for $gg\to\gamma\gamma g$ when relaxing cuts at inference compared to training.

Subsequent to this work, the authors of \incite{maitre:2021uaa} achieve improved per-point agreement at tree-level by exploiting the factorisation properties of \acp{ME} (\cref{sec:infrared-factorisation}).

\section{Summary}
\label{sec:nn-conc}

We extend previous work which pioneered the emulation of \acp{ME} with \acp{NN}, studying these techniques for the first time within a full hadron collider simulation.
We focus on loop-induced diphoton-plus-jets production via gluon fusion.
The difficulties introduced by \ac{IR} behaviour are tamed by partitioning the phase space as prescribed by \ac{FKS} subtraction.
We provide a general interface for trained models to \ac{SHERPA}.

We find that \ac{ME} \ac{NN} models provide an efficient general framework for optimising high-multiplicity observables at hadron colliders.
Agreement in differential distributions is excellent.
As the cost of inference on the model is negligible compared to the amplitude library call in training, the speedup factor in total simulation time (including training) compared to conventional methods is given by the ratio of the number of points used for inference and training,
\begin{align}
    \frac{N_{\mathrm{infer}}}{N_{\mathrm{train}}}\,.
\end{align}
For this study, this gave a factor of thirty, although for studies with higher statistics or coverage of multiple cut configurations, the factor would be much greater.

\chapter{Virtual \acs{QCD} corrections to the \texorpdfstring{$gg\to\gamma\gamma g$}{gluon-initiated diphoton-plus-jet production} amplitude}
\label{ch:yy-amp}

We present an analytic computation of the amplitudes for the gluon-initiated contribution to diphoton-plus-jet production at hadron colliders at up to two loops in \ac{QCD}.
We reconstruct the analytic form of the \acp{FR} from numerical evaluations over \acp{FF} including all colour contributions.
Compact expressions are found using the pentagon function basis~\cite{chicherin:2020oor}.
We provide a fast and stable implementation for the colour- and helicity-summed \acp{FR}, including the one-loop squared and the interference between one- and two-loop \acp{FR}, in \cpp as part of the version 3.0.0 release of the \njet library~\cite{njet}.

This chapter is organised as follows.
We first cover recent progress in two-to-three two-loop amplitude computations in \cref{sec:amp-back}.
Then, in \cref{sec:Kinematics}, we introduce the notation and describe the colour decomposition of the amplitudes.
We describe the methodology used to perform the \ac{IBP} reduction in \cref{sec:Reduction}.
In \cref{sec:momtwistors}, we include some details of the momentum twistor formalism used to provide a rational parametrisation of the kinematics.
Next, in \cref{sec:Reconstruction}, we discuss the reconstruction of the \acp{FR} over \acp{FF}.
In particular, we describe a method for performing a univariate partial fractioning of the rational coefficients of the special functions on the fly in \cref{sec:Apart}.
This approach can be used inside the \ac{FF} workflow, reducing significantly the number of sample points required to complete the analytic reconstruction and yielding compact analytic expressions.
In \cref{sec:AllPlus}, we show the simple analytic forms we obtained for the all-plus helicity amplitude, \ie~all external particles with positive helicity, which highlight its conformal properties.
Finally, in \cref{sec:performance}, we present the implementation in the \njet library and the performance of the code using a realistic set of phase-space points before concluding with a few remarks on future applications of the results and methods in \cref{sec:amp-conc}.

\section{Background}
\label{sec:amp-back}

As discussed in \cref{sec:yy}, diphoton-plus-jet production is an important experimental signature.
The Born-level amplitude for the gluon-initiated subprocess contains a closed quark loop coupling to both photons (\cref{sec:fixed-order}).
Consequently, this subprocess starts to contribute to the cross section only from \ac{NNLO} onwards (\cref{fig:pert-exp}).
Here, we derive the two-loop virtual amplitudes for $gg \to \gamma\gamma g$ that contribute to the \ac{NLO} corrections to the closed quark loop subprocess.
Curiously, the gluon channel has the opposite structure to the conventional expansion in the number of colour charges $\nc$.
The \ac{LC} contributions to the quark-initiated process, which are the dominant contributions, contain only planar diagrams~\cite{agarwal:2021vdh}.
However, in the gluon-initiated case, the \ac{LC} limit contains both planar and non-planar graphs at two loops.
Graphs with the highest complexity are thus unavoidable.

The last few years have seen rapid progress in our ability to compute two-loop two-to-three scattering processes in \ac{QCD} that had been intractable for a long time.
The analytic computation of the scattering amplitudes in a form suitable for phenomenological applications requires overcoming a number of major technical bottlenecks.
A basis of special functions must be identified that can be evaluated efficiently over the full phase space.
For massless five-point scattering, such a basis has been identified~\cite{chicherin:2017dob,papadopoulos:2015jft,gehrmann:2018yef,chicherin:2018mue,chicherin:2018old} and became recently available as a fast and stable implementation in \cpp valid in the physical scattering region~\cite{chicherin:2020oor}.
Secondly, the amplitude must be reduced from tensor Feynman integrals onto a basis of \acp{MI} that can subsequently be expanded in terms of special functions.
Currently, the only viable approach to this task is through the solution of enormous systems of \ac{IBP} identities~\cite{tkachov:1981wb,chetyrkin:1981qh,laporta:2001dd}, for which many public implementations now exist~\cite{anastasiou:2004vj,studerus:2009ye,vonmanteuffel:2012np,lee:2012cn,smirnov:2019qkx,klappert:2020nbg}.
There has been success in simplifying this problem using syzygy relations~\cite{gluza:2010ws,schabinger:2011dz,ita:2015tya,larsen:2015ped,boehm:2017wjc}, module intersection~\cite{boehm:2018fpv,boehm:2020ijp}, intersection theory~\cite{mastrolia:2018uzb,frellesvig:2019uqt,frellesvig:2019kgj,frellesvig:2020qot}, $\eta$ expansion~\cite{liu:2017jxz,liu:2018dmc,guan:2019bcx,zhang:2018mlo,wang:2019mnn}, direct solution of \acp{IBP} through recursive relations~\cite{kosower:2018obg}, multivariate partial fractioning~\cite{boehm:2020ijp}, and by-passing complicated algebraic steps through \ac{FF} arithmetic (\cref{sec:ff}).
The latter method can be applied more broadly~\cite{peraro:2016wsq,peraro:2019svx}, in particular to a complete reduction of the amplitudes into a representation using special functions.
New efficient reconstruction techniques~\cite{laurentis:2019bjh,delaurentis:2022otd,budge:2020oyl,campbell:2021mlr,campbell:2022qpq} allow compact expressions of the rational coefficients to be found.
In this chapter, we approach the problem through a direct analytic reconstruction of the amplitudes at the level of the pentagon functions, performing all intermediate steps numerically over \acp{FF}.
This technique has been applied successfully to \ac{LC} (planar) five-parton amplitudes first numerically~\cite{badger:2017jhb,abreu:2017hqn,badger:2018gip,abreu:2018jgq} and then analytically~\cite{gehrmann:2015bfy,badger:2018enw,abreu:2018zmy,abreu:2019odu,abreu:2021oya} (see \cref{ch:3j}).
\Ac{LC} triphoton production has also been completed and cross checked by two independent groups both at the level of the amplitudes~\cite{abreu:2020cwb,chawdhry:2020for} and of differential cross sections~\cite{chawdhry:2019bji,kallweit:2020gcp}.
Very recently, \ac{NNLO} \ac{QCD} predictions for a number of trijet observables and differential three-to-two jet ratios have been computed at \ac{LC} as well~\cite{czakon:2021mjy,chen:2022ktf}.
The process $gg\to g\gamma\gamma$ contains the most complicated non-planar topologies with up to rank five tensor numerators even at \ac{LC}.

We obtain compact analytic expressions for the complete set of helicity amplitudes for which the \ac{UV} and \ac{IR} poles have been subtracted, and implement them into an efficient and stable \cpp code as part of the \njet library~\cite{njet}.
These expressions take the form of rational coefficients multiplied by pentagon functions.
The code provides colour- and helicity-summed expressions for the interference between the one- and two-loop amplitudes, which can be used directly in phenomenological applications as in \cref{ch:yy-xs}.

\section{Kinematics and amplitude conventions}
\label{sec:Kinematics}

We consider the production of a pair of photons in association with a gluon from gluon fusion,
\begin{align}
    g(-p_1) + g(-p_2) \to g(p_3) + \gamma(p_4) + \gamma(p_5) \, ,
\end{align}
up to two-loop order in \ac{QCD}.
All particles are massless, \cref{eq:massless}, and we take all momenta as outgoing, \cref{eq:mom-cons}.
Without loss of generality, we assume that the external momenta $p_i$ live in a four-dimensional Minkowski spacetime, whereas the Feynman loop integrations are done in $d=4-2 \eps$ dimensions to regulate the divergences (\cref{sec:infinities}).
As discussed in \cref{sec:dof}, the kinematics are described by five \ac{LI} scalar invariants, which can be chosen as the set of momentum invariants $\{\s{12}, \s{23}, \s{34}, \s{45}, \s{15} \}$, and a pseudoscalar invariant $\trf$ defined in \cref{eq:tr5}.

We work in the $\s{12}$ physical scattering region, which is delimited by the requirements that all $s$-channel invariants are positive and all $t$-channel invariants are negative,
\begin{align}
    & \s{12}, \s{34}, \s{35}, \s{45} > 0 \,, \\
    & \s{13}, \s{14}, \s{15}, \s{23}, \s{24}, \s{25} < 0 \,,
\end{align}
together with the negativity of the Gram determinant defined in \cref{eq:gram}, $\Delta<0$, which follows from the real-valuedness of the momenta~\cite{gehrmann:2018yef}.

The scattering of gluons and photons is a one-loop process at \ac{LO}.
We decompose the scattering amplitude as
\begin{align}
    \am(1_g,2_g,3_g,4_\gamma,5_\gamma) = g_s e^2 \sum_{q=1}^{\nf} {Q_q}^2 f^{a_1 a_2 a_3}
    \sum_{\ell=1}^{\infty} \left(n_\eps \frac{\astr}{4\pi} \right)^{\ell} A^{(\ell)}(1_g,2_g,3_g,4_\gamma,5_\gamma) \,,
    \label{eq:colourdecomp}
\end{align}
where
\begin{align}
    n_{\eps} = \imi\br{\frac{4\pi}{{\mu_R}^2}}^\eps e^{-\eps \gamma_E} \,,
\end{align}
with the renormalisation scale $\mu_R$ and the Euler–Mascheroni constant $\gamma_E$.
The strong and \ac{EM} vertex couplings are denoted by $g_s$ and $e$ respectively, $\nf$ is the number of light quarks, $Q_q$ is the electric charge of quarks of flavour $q$ in units of the electron charge, $a_i$ is the adjoint \SUt{\nc} colour index of the $i^\mathrm{th}$ gluon, $f^{a_i,a_j,a_k}$ are the \SUt{\nc} structure constants, $\ell$ denotes the number of loops, and $\astr$ is the strong coupling parameter.
The one-loop diphoton amplitude can be obtained from permutations of pure-gluon scattering~\cite{dicus:1987fk,deflorian:1999tp}.

\begin{figure*}
    \centering
    \includegraphics[width=\textwidth]{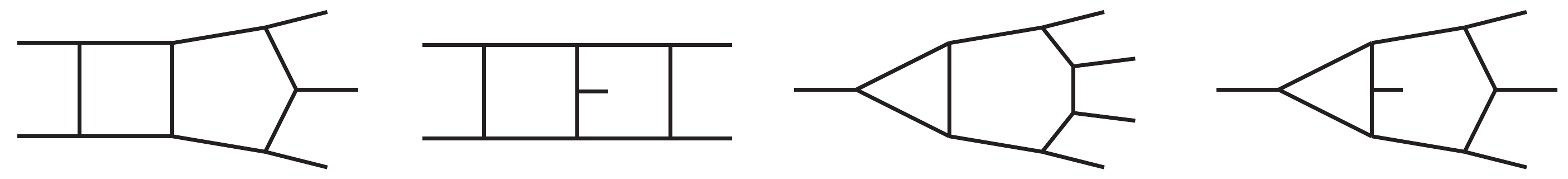}
    \caption{Independent integral families for the $gg \to g \gamma \gamma$ amplitude.
    The non-planar topologies (the second and fourth graphs) appear only in the \ac{LC} amplitude.}
    \label{fig:integral_topologies}
\end{figure*}

\begin{figure*}
    \centering
    \begin{subfigure}[t]{0.18\textwidth}
        \centering
        \includegraphics[height=1.2cm]{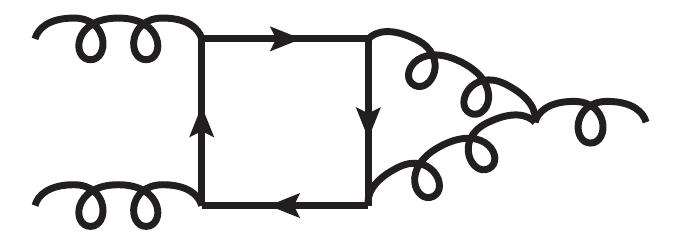}
        \caption{$\nc$}
    \end{subfigure}
    ~
    \begin{subfigure}[t]{0.18\textwidth}
        \centering
        \includegraphics[height=1.2cm]{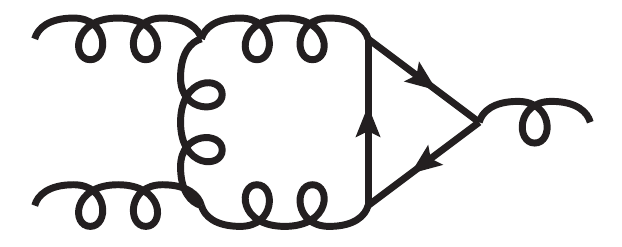}
        \caption{$\nc$}
    \end{subfigure}
    ~
    \begin{subfigure}[t]{0.18\textwidth}
        \centering
        \includegraphics[height=1.3cm]{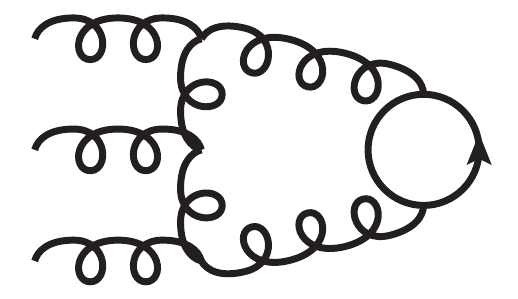}
        \caption{$\nc$}
    \end{subfigure}
    ~
    \begin{subfigure}[t]{0.18\textwidth}
        \centering
        \includegraphics[height=1.2cm]{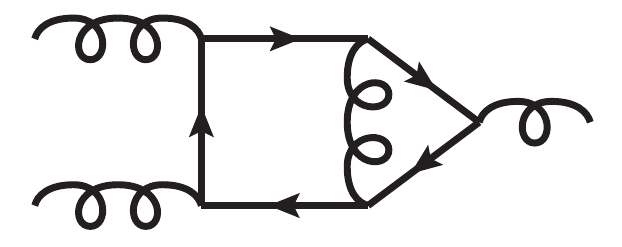}
        \caption{$\frac{1}{\nc}$}
    \end{subfigure}
    ~
    \begin{subfigure}[t]{0.18\textwidth}
        \centering
        \includegraphics[height=1.3cm]{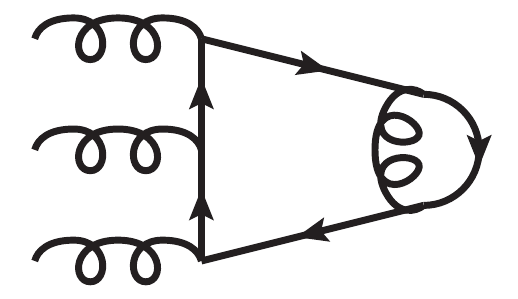}
        \caption{$\nc - \frac{1}{\nc}$}
    \end{subfigure}
    \caption{The colour factor of each diagram in the $gg \to g \gamma \gamma$ follows from the representative three-gluon two-loop diagrams with a closed fermion loop shown here.}
    \label{fig:diagram_colours}
\end{figure*}

We further expand the loop amplitudes in powers of $\nc$ and $\nf$,
\begin{align}
    \label{eq:Nc_exp}
    \begin{aligned}
        A^{(1)}(1_g,2_g,3_g,4_\gamma,5_\gamma) &= A^{(1)}_1(1_g,2_g,3_g,4_\gamma,5_\gamma) \,, \\\\
        A^{(2)}(1_g,2_g,3_g,4_\gamma,5_\gamma) &= \underbrace{\nc\,A^{(2)}_1(1_g,2_g,3_g,4_\gamma,5_\gamma)}_\text{\acs*{LC}} \\
        &+ \underbrace{\frac{1}{\nc} A^{(2)}_2(1_g,2_g,3_g,4_\gamma,5_\gamma) + \nf\,A^{(2)}_3(1_g,2_g,3_g,4_\gamma,5_\gamma)}_\text{\acs*{SLC}}
        \,.
    \end{aligned}
\end{align}
Surprisingly, the \ac{SLC} two-loop amplitudes contain only planar integrals, while the \ac{LC} contribution contains all of the four independent families shown in \cref{fig:integral_topologies}, including non-planar integrals.
This pattern is the opposite to that of the quark-initiated channels computed in \incites{agarwal:2021grm,chawdhry:2021mkw,agarwal:2021vdh}, for which the \ac{LC} contributions involve only the planar integrals and are therefore simpler to compute.
Providing a prediction for the gluon-initiated channel necessarily requires handling the most complicated integral families.
A simple analysis of the colour factors of each of the three-gluon vertex diagrams shown in \cref{fig:diagram_colours} illustrates how this pattern arises.
Photons couple to any of the fermion propagators, and the colour factors remain the same.
It can then be seen that non-planar contributions can come from the diagrams (a)--(c) only.
Diagrams (d)--(e), which contribute to \ac{SLC}, remain planar (allowing for permutations of the external momenta).

In our setup, we reduce directly to the \ac{FR} where the \ac{UV} and \ac{IR} poles have been subtracted analytically.
The poles take a particularly simple form since there is no tree-level process, the one-loop amplitudes are finite in $\eps$, and the two-loop amplitudes are only divergent to $\order{\eps^{-2}}$.
The one- and two-loop \acp{FR} are given in terms of the bare amplitudes~\cite{catani:1998bh,becher:2009qa,becher:2009cu,gardi:2009qi,gardi:2009zv} by
\begin{align}
    \label{eq:finremdef}
    \begin{aligned}
        F^{(1)} &= A^{(1)}(1_g,2_g,3_g,4_\gamma,5_\gamma) \,, \\
        F^{(2)} &= A^{(2)}(1_g,2_g,3_g,4_\gamma,5_\gamma) - \br{I^{(1)} + \frac{3}{2} \frac{\beta_0}{\eps} }A^{(1)}(1_g,2_g,3_g,4_\gamma,5_\gamma) \,,
    \end{aligned}
\end{align}
where
\begin{align}
    \begin{aligned}
        \beta_0 &= \frac{11}{3} \nc - \frac{2}{3} \nf \,, \\
        I^{(1)} &= - n_\Gamma(\eps) \, \cu{ \frac{\nc}{\eps^2} \sq{ \left(\frac{{\mu_R}^2}{-\s{12}} \right)^\eps+\br{\frac{{\mu_R}^2}{-\s{23}} }^\eps+\br{\frac{{\mu_R}^2}{-\s{13}} }^\eps } + 3 \frac{\gamma_g}{\eps} } \,,
        \label{eq:I1}
    \end{aligned}
\end{align}
with
\begin{align}
    n_\Gamma(\eps)&=\frac{e^{\eps \gamma_{E}}}{\Gamma(1-\eps)} = 1-\frac{(\pi\eps)^2}{12}+\order{\eps^3} \,, & \gamma_g &= \frac{\beta_0}{2} \,,
\end{align}
where $\Gamma(z)$ is the gamma function, in the \ac{tHV} scheme.
The logarithms arising from the $\eps$-expansion of $I^{(1)}$ can be analytically continued to the $\s{12}$ channel by adding a small positive imaginary part to each $\s{ij}$.
The $\beta_0$ term in the definition of the two-loop \ac{FR} is defined by \cref{eq:beta-fn} and accounts for the strong coupling renormalisation.
The \acp{FR} inherit from the amplitudes the decomposition in powers of $\nc$ and $\nf$ given by \cref{eq:Nc_exp},
\begin{align}
    \label{eq:F_Nc_exp}
    \begin{aligned}
        F^{(1)}(1_g,2_g,3_g,4_\gamma,5_\gamma) &= F^{(1)}_1(1_g,2_g,3_g,4_\gamma,5_\gamma) \,, \\
        F^{(2)}(1_g,2_g,3_g,4_\gamma,5_\gamma) &= \nc\,F^{(2)}_1(1_g,2_g,3_g,4_\gamma,5_\gamma) \\
        &+ \frac{1}{\nc} F^{(2)}_2(1_g,2_g,3_g,4_\gamma,5_\gamma)
        + \nf\,F^{(2)}_3(1_g,2_g,3_g,4_\gamma,5_\gamma)
        \,.
    \end{aligned}
\end{align}

Our final results are presented in the \ac{tHV} scheme, although we make the distinction between the dimension $d$ of the loop integration and the spin dimension $d_s$ arising from the numerator algebra (\cref{sec:amplitudes}).
We find that it is convenient to arrange terms by expanding the \ac{FR} around the physical degrees of freedom of the gluon, $d_s=2$.
The one-loop and $\nf$ two-loop \acp{FR} have only the $\br{d_s-2}^0$ component.
Thus, the expansions of all \acp{FR} are
\begin{align}
    \begin{aligned}
        \label{eq:ds_exp}
        F^{(1)}_1 &= F^{(1)}_{1;0} \,, \\
        F^{(2)}_k &= F^{(2)}_{k;0} + F^{(2)}_{k;1} \, (d_s-2) \qquad\forall k\in\cu{1,2} \,, \\
        F^{(2)}_3 &= F^{(2)}_{3;0} \,.
    \end{aligned}
\end{align}

\section{Computational setup and amplitude reduction}
\label{sec:Reduction}

\begin{figure}
    \begin{center}
        \includegraphics[width=0.8\textwidth]{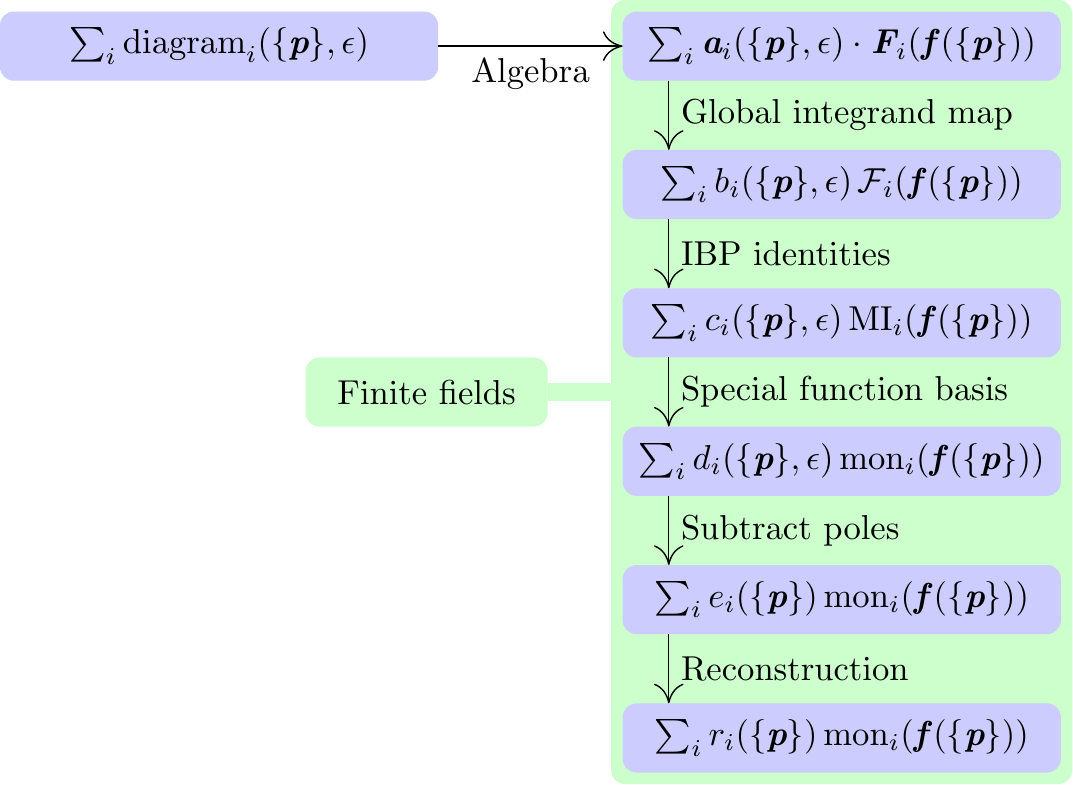}
        \caption{
            Schematic flow chart of the steps in our assembly of the \acsp*{FR}.
            We start by summing over all Feynman diagrams, then algebraically manipulating to a linear combination of tensor integrals $\textbi{F}_i$, which are composed of special functions $\textbi{f}\,(\cu{\textbi{p}})$.
            The coefficients $a_i$ are rational functions of the kinematics $\cu{\textbi{p}}$ and the dimensional regulator $\eps$.
            All computation from this point on is performed in \acp{FF}.
            We reduce to scalar integrals $\mathcal{F}_i(\cu{\textbi{p}})$, then to a set of \ac{LI} \acsp*{MI} using \acs*{IBP} identities.
            We project these onto a special function basis (the pentagon functions) to obtain a sum over \ac{LI} monomials of the special functions $\text{mon}_i(\textbi{f}\,(\cu{\textbi{p}}))$.
            We perform the \acs*{IR} subtraction to recover the \acs*{tHV} \acsp*{FR}.
            Finally, we reconstruct the coefficients to obtain compact analytic expressions $r_i(\cu{\textbi{p}})$ for them.
        }
        \label{fig:amp-pipeline}
    \end{center}
\end{figure}

We take a diagrammatic approach to the calculation of the amplitude along the lines of previous work~\cite{hartanto:2019uvl,badger:2021owl}.
Here we briefly summarise the steps, which are sketched in \cref{fig:amp-pipeline}, and refer the reader to \incite{badger:2021owl} for further details.
All Feynman diagrams are generated using
\qgraf~\cite{nogueira:1991ex} and subsequently processed using a combination of in-house \mathematica and
\form~\cite{kuipers:2012rf,ruijl:2017dtg} scripts.
In total, including contributions from ghost diagrams, we find 50 diagrams at one loop and 1527 at two loops.
Aided by the \spinney~\cite{cullen:2010jv} package to perform the 't Hooft algebra, the numerators are written for each independent helicity configuration.
From the loop denominator structure we assign an integral topology to each diagram.
At this point, the diagram numerators are linear combinations of monomials in loop-momentum dependent scalar and spinor products with coefficients depending only on external momenta.
These coefficients are loaded into a dataflow graph using \finiteflow~\cite{peraro:2019svx}.
This enables numerical sampling over \acp{FF}, thus sidestepping analytically complicated intermediate expressions in further steps.
We rewrite loop-momentum dependent monomials into inverse propagator denominators and a choice of \acp{ISP}.
The required mapping of the coefficients is performed numerically within the dataflow framework.
After summing all diagrams and dropping scaleless integrals, we arrive at an expression ready for \ac{IBP} reduction.

The reduction to \acp{MI} was obtained using an improved version of the Laporta algorithm~\cite{laporta:2001dd}.
For most integral families, we generated identities containing no higher power of propagators with respect to those appearing in the amplitude, following ideas proposed in \incites{gluza:2010ws,ita:2015tya,larsen:2015ped}.
These identities were found using the Baikov representation of loop integrals, for which identities (i) without higher powers of propagators and (ii) without dimension-shifted integrals can be found by solving polynomial equations called syzygy equations.
Closed-form solutions to both of these constraints are separately known.
Indeed, the solution of (i) is straightforward and the solution for (ii) has been found in \incite{boehm:2017wjc}.
The two syzygy solutions need to be combined for generating identities that satisfy both constraints.
For this purpose, we used a custom syzygy solver that implements the algorithm in \incite{schabinger:2011dz} using \finiteflow~\cite{peraro:2019svx}.
More details on this method can be found in \incites{gluza:2010ws,ita:2015tya,larsen:2015ped,boehm:2017wjc}.
The application of the syzygy technique leads to a substantial reduction in the size of the IBP system in the planar sector, which improves both the speed of solving the system and memory usage.
This improved performance in the planar sectors by almost a factor of ten, which was sufficient for the current calculation, without the need to extend it to the non-planar families---though we expect this will be necessary for future applications.

For each integral family, we generated integral identities for only one permutation of the external legs.
Numerical solutions for all the permutations contributing to an amplitude were found by solving the systems of equations several times, with different numerical inputs for the invariants.
Mappings between \acp{MI} with different permutations of external legs are applied afterwards to obtain a result in terms of a minimal set of \acp{MI}.

For each phase-space point evaluated on \iac{FF}, we reconstruct the full dependence on the dimensional regulator $\eps$ of the amplitude reduced onto \acp{MI} before substituting their expressions in terms of the basis of special functions and computing the Laurent expansion in $\eps$.
With this setup, fewer numerical solutions of the \ac{IBP} identities are needed in order to reconstruct analytic results for the amplitude.
This is due to the fact that if we were to expand the integrals into the pentagon functions before performing the Laurent expansion in $\eps$ of the final coefficients, it would complicate the dependence on $\eps$ of the result in this intermediate stage.

To make use of the \ac{FF} arithmetic we must have a rational parametrisation of the external kinematics.
As discussed in \cref{sec:mtv}, we parametrise the kinematics using \acp{MTV},
\begin{align}
    \label{eq:s2x}
    \begin{aligned}
        & \s{12} = x_1 \,, \\
        & \s{23} = x_1 x_4 \,, \\
        & \s{34} = \frac{x_1}{x_2} (x_4 + x_3 x_4 + x_2 x_3 x_5-x_2 x_3) \,, \\
        & \s{45} = x_1 x_5 \,, \\
        & \s{15} = x_1 x_3 (x_2-x_4+x_5)\,, \\
        & \trf = -\frac{x_1^2}{x_2}\left[ x_2 x_4 (1 + 2 x_3)  - x_4 (1 + x_3)  (x_4 - x_5) + x_2^2 x_3 (-1 + x_5)\right] \, .
    \end{aligned}
\end{align}
We stress that the pseudoscalar invariant $\trf$, and hence the square root of the Gram determinant $\Delta$ (see \cref{sec:dof}), is a rational function of the $x_i$ variables.
Moreover, since $x_1$ is the only dimensionful variable, we can set it equal to unity and recover the dependence on it after the reconstruction by dimensional analysis.
Further details on the momentum twistor parametrisation are presented in \cref{sec:momtwistors}.
In the following sections, we will consider all coefficients of the special functions to be rational functions of the variables $x_i$.

\section{Momentum twistor parametrisation}
\label{sec:momtwistors}

\index{Momentum!twistor!parametrisation}\Acp{MTV} were introduced in \cref{sec:mtv}.
This construction follows \incites{hodges:2009hk,badger:2013gxa,badger:2017jhb}.
It is possible to fix all but five of the entries of the momentum twistor matrix, \cref{eq:mom-twi-mat}, at five-point.
Explicitly, we choose the form,
\begin{align}
    Z = \begin{pmatrix} \lws_i \\ \mu_i \end{pmatrix}_{i\in\cu{1,\ldots,5}} \, =
        \begin{pmatrix}
            1 & 0 & \frac{1}{x_1} & \frac{1+x_2}{x_1 x_2} & \frac{1+x_3 (1+x_2)}{x_1 x_2 x_3} \\
            0 & 1 & 1 & 1 & 1 \\
            0 & 0 & 0 & \frac{x_4}{x_2} & 1 \\
            0 & 0 & 1 & 1 & \frac{x_4-x_5}{x_4}
        \end{pmatrix} \,.
        \label{eq:mtparam}
\end{align}
The parametrisation used in this work has some benefits: the only dimensionful quantity is $x_1$ (recall that we have set this to unity) and all holomorphic quantities are described using only $x_1,x_2,x_3$.
For real kinematics, only $x_2$ and $x_3$ are complex, while the other three are real.
Notice that the conversion between the \acp{MTV} and spinor-helicity expressions, \cref{eq:s2x}, is only invertible for phase-free quantities as the phase information is lost when translating to \acp{MTV}.
Thus, for an invertible mapping, we may use the following relations,
\begin{align}
    \label{eq:x2s}
    \begin{aligned}
        & x_1 = \s{12} \,, \\
        & x_2 =  - \frac{\ptrs{+}{p_1,p_2,p_3,p_4}}{\s{12} \s{34}} \,, \\
        & x_3 =  - \frac{\ptrs{+}{p_1,p_3,p_4,p_5}}{\s{13} \s{45}} \,, \\
        & x_4 = \frac{\s{23}}{\s{12}} \,, \\
        & x_5 = \frac{\s{45}}{\s{12}} \,,
    \end{aligned}
\end{align}
where $\tr_+(p_i,p_j,p_k,p_l)$ is defined in \cref{eq:tr+}.

In our work, we express the helicity amplitudes in terms of \acp{MTV} $x_i$.
The phase information can be restored by multiplying and dividing by a suitable phase factor,
\begin{align} \label{eq:phase_normalisation}
    \am = \Phi(\lws_i, \rws_i) \, \br{ \frac{\am(x_i)}{\Phi (x_i)} }\,,
\end{align}
where $\am$ is a helicity amplitude---or in general some object with a non-trivial phase---and $\Phi$ is an arbitrary factor with the same helicity weights (defined in \cref{sec:shf}) as $\am$.
The quantities $\am(x_i)$ and $\Phi(x_i)$ are both written in terms of \acp{MTV}.
Their ratio is phase-free and can thus be expressed in terms of the scalar and pseudoscalar invariants, $\s{ij}$ and $\trf$, for example through \cref{eq:x2s}, or evaluated directly in \acp{MTV}.
The phase $\Phi(\lws_i, \rws_i)$ can be constructed using \cref{eq:amp-hel-scaling}, is written in terms of the spinor-helicity variables, and carries all the phase information of $\am$.
The phase $\Phi(x_i)$ can be obtained from $\Phi(\lws_i, \rws_i)$ by applying the \ac{MTV} parametrisation, which in our case is given by the mapping \cref{eq:s2x}.

For the aid of comparisons with the data presented in this chapter, the specific choices of the amplitude phases $\Phi(\lws_i, \rws_i)$ were
\begin{align}
    \begin{aligned}
        \Phi(1_g^+,2_g^+,3_g^+,4_\gamma^+,5_\gamma^+) &= \frac{1}{\spA{1}{2}\spA{2}{3}\spA{3}{4}\spA{4}{5}\spA{5}{1}}, \\
        \Phi(1_g^-,2_g^+,3_g^+,4_\gamma^+,5_\gamma^+) &= \frac{\spB{2}{3}\spA{1}{3}}{\spA{2}{3}\spA{3}{4}\spA{4}{5}\spA{1}{5}}, \\
        \Phi(1_g^+,2_g^+,3_g^+,4_\gamma^-,5_\gamma^+) &= \frac{\spB{1}{5}\spA{1}{4}}{\spA{1}{2}\spA{2}{3}\spA{3}{4}\spA{1}{5}}, \\
        \Phi(1_g^-,2_g^-,3_g^+,4_\gamma^+,5_\gamma^+) &= \frac{\spA{1}{2}^3}{\spA{2}{3}\spA{3}{4}\spA{4}{5}\spA{5}{1}}, \\
        \Phi(1_g^+,2_g^+,3_g^+,4_\gamma^-,5_\gamma^-) &= \frac{\spA{4}{5}^3}{\spA{1}{2}\spA{2}{3}\spA{3}{4}\spA{5}{1}}, \\
        \Phi(1_g^-,2_g^+,3_g^+,4_\gamma^-,5_\gamma^+) &= \frac{\spA{1}{4}^4}{\spA{1}{2}\spA{2}{3}\spA{3}{4}\spA{4}{5}\spA{5}{1}}.
    \end{aligned}
    \label{eq:phaseconventions}
\end{align}
The \ac{MHV} trees were used along with contrived expressions for the \acp{UHV}.

\section{Analytic reconstruction over finite fields}
\label{sec:Reconstruction}

In this section, we present three general strategies to optimise the reconstruction over \acp{FF} of the rational coefficients in the \acp{FR}.
At this stage, each component $F(x)$ of the two-loop \ac{FR} is expressed as
\begin{align} \label{eq:F_r_pf}
    F(x) = \sum_{i} r_i(x) \, \mathrm{mon}_i\brf{f} \,,
\end{align}
where $r_i$ are rational functions of the \acp{MTV} $x$, and $\mathrm{mon}_{i}(f)$ are \ac{LI} monomials of the pentagon functions.
The entire chain of operations is implemented over \acp{FF} in the framework \finiteflow.
We therefore have a numerical algorithm which evaluates the rational coefficients $r_i(x)$ modulo some prime number.
The final step consists of reconstructing the analytic expression of the rational coefficients from a sufficient number of numerical evaluations.
We employ \finiteflow's multi-variate functional reconstruction algorithms, supplemented with three strategies to reduce the number of required sample points: we determine the linear relations among the rational coefficients and an ansatz, use univariate slices to identify the factors belonging to another ansatz, and perform a univariate \ac{PFD} on the fly.
In the following subsections we discuss thoroughly each of these procedures and their application to two-loop diphoton \acp{FR}.

\subsection{Linear relations among the rational coefficients}
\label{sec:LinearRelations}

The representation of the \acp{FR} in terms of rational coefficients and special function monomials given by \cref{eq:F_r_pf} is not optimal.
The special function monomials do not all appear independently.
They are present only in a number of \ac{LI} combinations that is typically much smaller than the total number of monomials.
As a result, the rational coefficients $r_i$ in the \acp{FR} are not \ac{LI}.
Expressing the \acp{FR} in terms of a set of \ac{LI} rational coefficients not only leads to more compact expressions, but may also simplify their reconstruction.

We can determine the linear relations among the rational coefficients $\{r_i(x)\}$ of the special function monomials by solving a linear fit problem,
\begin{align}
    \sum_i a_i \, r_i(x)  = 0 \, .
\end{align}
Since the coefficients of the linear relations $a_i$ are rational numbers, they require substantially fewer sample points to be reconstructed with respect to the rational coefficients themselves.
We can then use these relations to express the rational coefficients in terms of a set of \ac{LI} ones, which remain to be reconstructed.
Choosing the latter to be the simplest---\ie~those with the lowest polynomial degrees---may reduce the number of sample points required for the reconstruction.

This strategy can be further refined by supplying an ansatz for the rational coefficients.
We then fit the linear relations among the rational coefficients of the \acp{FR} and the coefficients of the ansatz, which we denote by $\{e_j(x)\}$, as
\begin{align}
    \sum_i a_i \, r_i(x) + \sum_j b_j \, e_j(x)  = 0 \, ,
\end{align}
with $a_i, b_j \in \rationals$.
In the best case scenario, all the rational coefficients $r_i$ can be expressed in terms of the ansatz coefficients $e_j$ and no further reconstruction needs to be performed.
Even when the ansatz does not entirely cover the rational coefficients, it may still lower the degrees of the \ac{LI} coefficients which have to be reconstructed.
The ansatz can be constructed from the tree-level amplitude and the rational coefficients of the one-loop amplitudes up to $\ord(\eps^2)$ from the analysis of the leading singularities~\cite{eden2002analytic,britto:2004nc,cachazo:2008vp,arkanihamed:2010gh} or from other related amplitudes.
In the diphoton case, we can use the two-loop five-gluon amplitudes.
At one loop, the $3g2\gamma$ amplitudes can be expressed in terms of permutations of the five-gluon ones~\cite{dicus:1987fk,deflorian:1999tp}.
While this is no longer true at two loops, we find there is an important overlap between the rational coefficients of the $3g2\gamma$ amplitudes and those of the five-gluon ones.
We use the rational coefficients of the \ac{LC} two-loop five-gluon amplitudes as ans\"atze in the linear relations; all two-loop five-parton amplitudes are available analytically at \ac{LC}~\cite{gehrmann:2015bfy,dunbar:2016aux,badger:2018enw,abreu:2018zmy,badger:2019djh,dunbar:2019fcq,abreu:2021oya,delaurentis:2020qle} and we made use of the independent results discussed in \cref{ch:3j}.

\subsection{Matching factors on univariate slices}
\label{sec:Factors}

The pole structure of the pentagon functions is determined by the letters of the pentagon alphabet~\cite{chicherin:2017dob}.
The pentagon functions (or their discontinuities) may in fact have logarithmic singularities in the phase-space points where one of the letters vanishes.
For this reason, it is natural to expect that the poles of the rational coefficients should be similarly linked to the pentagon alphabet.
Indeed, we observe that the denominators of the rational coefficients in front of the pentagon functions factorise into a product of letters of the pentagon alphabet.
In other words, each rational coefficient $r(x)$ has the form,
\begin{align} \label{eq:ansatz_factors}
    r(x) = \frac{n(x)}{\prod_{k} {\ell_k}^{e_k}(x)}\,,
\end{align}
where $e_k$ are integers, $n(x)$ is a polynomial in the variables $x$, and $\{\ell_k\}$ is an ansatz of factors from the pentagon alphabet.
The exponents $e_k$ in \cref{eq:ansatz_factors} may in general be negative, corresponding to factors in the numerator.
We use the following ansatz for the factors\footnote{Note that the ansatz in \cref{eq:coeffansatz} is independent of the rational parametrisation of the kinematics, \cref{eq:s2x}.
The list of independent polynomials used in the factor matching on the univariate slice is generated by evaluating this list using the specific parametrisation.},
\begin{align}
    \label{eq:coeffansatz}
    \begin{aligned}
        \cu{\ell_k(x)} = \bigl\{ & \spA{1}{2}, \spA{1}{3}, \spA{1}{4}, \spA{1}{5}, \spA{2}{3}, \spA{2}{4}, \spA{2}{5}, \spA{3}{4}, \spA{3}{5}, \spA{4}{5}, \spB{1}{2}, \spB{1}{3}, \spB{1}{4}, \\
        & \spB{1}{5}, \spB{2}{3}, \spB{2}{4}, \spB{2}{5}, \spB{3}{4}, \spB{3}{5}, \spB{4}{5}, \s{12}-\s{34}, \s{12}-\s{35}, \s{12}-\s{45}, \\
        & \s{13}-\s{24}, \s{13}-\s{25}, \s{13}-\s{45}, \s{14}-\s{23}, \s{14}-\s{25}, \s{14}-\s{35}, \s{15}-\s{23}, \\
        & \s{15}-\s{24}, \s{15}-\s{34}, \s{23}-\s{45}, \s{24}-\s{35}, \s{25}-\s{34},
        \trf \bigr\}\,.
    \end{aligned}
\end{align}
The exponents $e_k$ in the ansatz \cref{eq:ansatz_factors} can be determined by reconstructing $r(x)$ on a univariate slice modulo some prime number~\cite{abreu:2018zmy}.
The univariate slice is defined by parametrising the variables in terms of a single parameter $t$,
\begin{align}
    \{ x_i(t) = a_i + b_i t \} \,,
\end{align}
for constant $a_i$ and $b_i$.
The constants are chosen randomly in the \ac{FF} to avoid artificial simplifications.
The dependence on $t$ is chosen to be linear so that the degrees of the numerator and denominator of
\begin{align}
    r(t)\coloneqq r\brf{x(t)}
\end{align}
correspond to the total degrees of $r$ in $x$.
Matching the reconstructed $r(t)$ with the ansatz \cref{eq:ansatz_factors} evaluated on the same slice allows to determine the exponents $e_k$ straightforwardly.
With a univariate reconstruction on just one prime field we can thus infer a lot of information about the analytic form of the rational coefficients: the denominators are entirely fixed, and typically some factors of the numerators are determined as well.
What remains to be reconstructed therefore requires fewer sample points.

\subsection{Univariate partial fraction decomposition over finite fields}
\label{sec:Apart}

\Ac{PFD} is a standard and powerful tool for the simplification of rational functions.
The \ac{PFD} is however not unique in the multivariate case.
Its application to the multivariate rational functions in scattering amplitudes is therefore not straightforward.
The necessity to simplify the rational coefficients of two-loop five-particle scattering amplitudes has recently spurred several approaches to handle the multivariate case efficiently~\cite{abreu:2019odu,boehm:2020ijp,heller:2021qkz}, based upon Leinartas' algorithm~\cite{leinartas:1978,raichev:2012}.
These algorithms rely on algebraic geometry techniques, such as multivariate polynomial division and Gr\"obner bases, and require an arbitrary choice of a monomial ordering.

Our main goal in this work is to simplify the reconstruction of the rational coefficients over \acp{FF}.
In other words, we want to reconstruct the rational coefficients on the fly, directly in a form which is decomposed in partial fractions.
The simplification of the resulting analytic expressions comes as a welcome by-product.
We observe that a univariate \ac{PFD} is sufficient for this purpose.
The advantage is that it can be straightforwardly implemented over \acp{FF}, avoiding all algebraic geometry complications.
The only arbitrary choice that remains to be made is to choose which variable we will partial fraction with respect to.
This can be chosen by observing the impact of the \ac{PFD} with respect to each variable separately on the lower order amplitudes.
With the parametrisation of the kinematics in terms of momentum twistors, \cref{eq:mtparam}, we find it most convenient to decompose in partial fractions with respect to $x_4$.

We now discuss our algorithm to reconstruct the univariate \ac{PFD} of a multivariate rational function $r$ from its numerical evaluations over \acp{FF}.
The algorithm requires as input an ansatz for the factors which may appear in the denominator of $r$.
Only those factors which depend on the variable with respect to which the \ac{PFD} is being performed are strictly necessary.
Informed guesses of other factors may further simplify the reconstruction.
In the application to massless two-loop five-particle scattering amplitudes, the factor ansatz can be inferred from the letters of the pentagon alphabet~\cite{chicherin:2017dob}.
We use the factors in \cref{eq:coeffansatz}.

Let $r$ be a rational function of the set of variables,
\begin{align}
    x = \cu{x_i}_{i=1}^n\,.
\end{align}
In this work, the $x_i$ are the \acp{MTV} defined by \cref{eq:s2x} with $n=5$, but we outline the algorithm in general.
The goal is to decompose $r$ in partial fractions with respect to one of the variables, say $y\coloneqq x_k$.
We denote the remaining variables by
\begin{align}
    \bar{x}=\cu{x_i}_{i=1}^n \backslash \cu{y}\,.
\end{align}
We may not know the analytic expression of $r$, but we must be able to evaluate it numerically modulo some prime number.
Let
\begin{align}
    \label{eq:l-ansatz}
    \cu{ \ell_i\brf{\bar{x},y} }_{i=1}^m
\end{align}
be an ansatz for the $m$ factors which may appear in the denominator of $r$.
Without loss of generality, we assume that the $\ell_i$ are irreducible polynomials over $\rationals$.
In other words, we assume that $r$ has the form
\begin{align}
    \label{eq:Ansatz}
    r\brf{\bar{x},y} = \frac{N\brf{\bar{x},y}}{\prod_{i=1}^m {\ell_i}^{e_i}\brf{\bar{x},y}} \, ,
\end{align}
where the exponents $e_i\in \integers$, and $N\brf{\bar{x},y}$ is a function which depends polynomially on $y$ and rationally on $\bar{x}$.
The ansatz, \cref{eq:l-ansatz}, may catch some of the factors in the numerator of $r\brf{\bar{x},y}$, corresponding to negative values of $e_i$.
This lowers the total degree of $N\brf{\bar{x},y}$ and eventually simplifies its reconstruction, but is not necessary for the \ac{PFD} with respect to $y$.
Similarly, the ansatz may cover all the factors in the denominator of $r$, so that $N\brf{\bar{x},y}$ is a polynomial in $\bar{x}$ as well as $y$.
What is necessary for the \ac{PFD} algorithm to work is that the ansatz contains all the factors in the denominator of $r$ which depend on $y$.
We denote this subset by
\begin{align}
    \Lambda_y = \cu{i\in \cu{1,\ldots,m} : e_i>0 \land \mathrm{deg}_y\sqf{\ell_i\brf{\bar{x},y}}>0 } \, ,
\end{align}
where $\mathrm{deg}_y\sqf{\ell_i\brf{\bar{x},y}}$ is the degree in $y$ of the polynomial $\ell_i\brf{\bar{x},y}$.

The first step consists of fixing the exponents $e_i$ in the ansatz \cref{eq:Ansatz}.
We do this through the procedure discussed in \cref{sec:Factors}.
In the second step, we determine the degree in $y$ of $N\brf{\bar{x},y}$ in the ansatz \cref{eq:Ansatz},
\begin{align}
    d_N \coloneqq \mathrm{deg}_y\sqf{ N\brf{\bar{x},y} } \,.
\end{align}
We recall that $N\brf{\bar{x},y}$ is by construction a polynomial in $y$.
We compute its degree in $y$ by reconstructing it on another univariate slice, this time where only $y$ varies,
\begin{align}
    \label{eq:UnisliceY}
    \cu{\bar{x}_i(t) = \bar{a}_i \,, \quad y(t) = t } \,,
\end{align}
with $a_i$ chosen randomly from the \ac{FF}.
Clearly,
\begin{align}
    d_N &= \mathrm{deg}_t\sqf{N(t)}\,, &
    N(t)&\coloneqq N\brf{\bar{x}=\bar{a},y=t} \,.
\end{align}
We introduce the short-hand notation,
\begin{align}
    d_i \coloneqq \mathrm{deg}_y\sqf{ \ell_i\brf{\bar{x},y} } \,, \quad \qquad d_{\Lambda_y} \coloneqq \sum_{i\in\Lambda_y} e_i \, d_i \,,
\end{align}
for the degrees of the denominator factors $\ell_i\brf{\bar{x},y}$ in $y$.

Using the information about the factors in the denominator of $r$ and the degree in $y$ of its numerator, we construct the following ansatz for the \ac{PFD} of $r$ with respect to $y$,
\begin{align}
    \label{eq:AnsatzPartialFraction}
    r\brf{\bar{x},y} = \sum_{i\in \Lambda_y} \sum_{j=1}^{e_i} \sum_{k=0}^{d_i-1} \frac{U_{ijk}\brf{\bar{x}} \, y^k}{{\ell_i}^j\brf{\bar{x},y}} + R\brf{\bar{x}} + \sum_{h=1}^{d_N-d_{\Lambda_y}} V_h\brf{\bar{x}} \, y^h \,,
\end{align}
where $U_{ijk}\brf{\bar{x}}$, $R\brf{\bar{x}}$ and $V_h\brf{\bar{x}}$ are unknown rational functions of $\bar{x}$.
The rightmost term in \cref{eq:AnsatzPartialFraction} is required only if $d_N > d_{\Lambda_y}$, \ie~only if the numerator of $r$ has a higher degree in $y$ than the denominator.

The last step of the algorithm consists of reconstructing the analytic dependence on $\bar{x}$ of the unknown coefficients in the ansatz, \cref{eq:AnsatzPartialFraction}, from the numerical evaluations of $r\brf{\bar{x},y}$.
To solve this linear fit problem, we use the algorithm implemented in the \finiteflow~framework~\cite{peraro:2019svx}.
The solution comes in the form of an algorithm which numerically evaluates $U_{ijk}\brf{\bar{x}}$, $R\brf{\bar{x}}$ and $V_h\brf{\bar{x}}$.
The rational reconstruction may be simplified by first reconstructing the coefficients on a univariate slice where all the remaining variables $\bar{x}$ vary, and using that to match them with those factors in the ansatz, \cref{eq:l-ansatz}, which depend only on $\bar{x}$.
This may lower the total degree of the functions that need to be reconstructed.

In addition to the factors in the original ansatz, \cref{eq:l-ansatz}, the coefficients of the \ac{PFD}, \cref{eq:AnsatzPartialFraction}, may also contain spurious factors (\cref{sec:poles}).
For instance, consider the toy example,
\begin{align}
    \label{eq:example}
    \frac{1}{(y-a)(y-b)} = \frac{1}{(a-b)(y-a)} - \frac{1}{(a-b)(y-b)}  \, ,
\end{align}
where $a$ and $b$ are arbitrary constants such that $a\neq b$.
In this example, the inspection of the left-hand side indicates $\{y-a, y-b\}$ as ansatz for the irreducible denominator factors.
The \ac{PFD} however contains a factor of $a-b$ in the denominator, which arises from the residue of the function at the root of either of the denominator factors.
Clearly $a=b$ is a spurious singularity, manifestly absent on the left-hand side and produced by the \ac{PFD}.
In general, we can determine the potential spurious factors by evaluating the factors in the ansatz, \cref{eq:l-ansatz}, which depend on $y$ at their zeros,
\begin{align}
    \left\{ \ell_i\left(\bar{x},y^*_k\right)  \right\}_{i\in\Lambda_y, \, k\in\Lambda_{y}^1,\,i\neq k}\,,
\end{align}
where $y^*_k$ is the zero of $\ell_k\brf{\bar{x},y}$,
\begin{align}
    \ell_k\brf{\bar{x}, y^*_k} = 0\,,
\end{align}
and $\Lambda_{y}^1$ is the subset of factors which depend linearly on $y$,
\begin{align}
    \Lambda_{y}^1 = \cu{ i\in \Lambda_y : \mathrm{deg}_y\sqf{\ell_i\brf{\bar{x},y}}=1}\,.
\end{align}
The restriction to zeros of linear functions of $y$ is due to the facts that the $\ell_i$ are irreducible polynomials over $\rationals$ and that we are factoring over $\rationals$.
The zeros of higher-degree irreducible polynomials would introduce algebraic and/or complex dependence.

In practice, we observe that determining the spurious factors does not simplify the reconstruction.
The greatest part of the denominators of the coefficients in the \ac{PFD}, \cref{eq:AnsatzPartialFraction}, is in fact determined by the original ansatz, \cref{eq:l-ansatz}.
What remains after they are multiplied away has a total degree which is typically lower than that of the numerators, which therefore dominates the determination of the number of sample points required for the reconstruction.
While it is possible to determine entirely the denominators of the coefficients in \cref{eq:AnsatzPartialFraction}, it would not reduce the number of required sample points substantially, and for this reason we refrain from doing so.

Having determined as many factors as possible in the coefficients of the \ac{PFD}, we multiply them away and reconstruct the remainder using the multivariate rational reconstruction algorithms implemented in \finiteflow.
It is important to stress that the algorithm which evaluates the coefficients of the \ac{PFD} contains the solution of a linear fit.
For each numerical value of $\bar{x}$, \cref{eq:AnsatzPartialFraction} is sampled for several numerical values of $y$, roughly as many times as the number of unknowns.
This generates a linear system of equations for the unknowns evaluated at the chosen value of $\bar{x}$.
The redundant equations are removed after the learning phase.
Because it requires several evaluations of the original functions, the reconstruction on the univariate slices in the intermediate steps of the algorithm has a higher computational cost than directly evaluating $r$.
On the other hand, the coefficients of the \ac{PFD} depend on one fewer variable than the original function $r$, and may have substantially lower degrees.
As a result of all these aspects, the \ac{PFD} may be outperformed by a direct reconstruction for simple functions, but becomes increasingly convenient as the complexity of the functions rises.
It is well suited for application at two loops.

\subsection{Summary and impact of the reconstruction strategy}
\label{sec:reconstruction_summary}

The techniques discussed in the previous sections are general and can be applied to any rational reconstruction problem, in combination or separately.
In order to reconstruct the rational coefficients of the two-loop diphoton \acp{FR} we apply them consecutively as follows.
\begin{description}
    \item[Stage 1.] We fit the linear relations among the rational coefficients with an ansatz, as discussed in \cref{sec:LinearRelations}.
        We begin with the $(d_s-2)^1$ components and use the coefficients of the two-loop \ac{LC} five-gluon \acp{FR} as ans\"atze.
        For the $(d_s-2)^0$ components, which are more complicated, we add to the ansatz the $(d_s-2)^1$-coefficients already reconstructed.

    \item[Stage 2.] We guess the factors from the ansatz \cref{eq:coeffansatz} by reconstructing a univariate slice (\cref{sec:Factors}) and multiply them away.

    \item[Stage 3.] We partial fraction on the fly with respect to $x_4$, applying the algorithm presented in \cref{sec:Apart}.
        The coefficients to be reconstructed after this stage are those in the ansatz for the \ac{PFD} \cref{eq:AnsatzPartialFraction}, and depend on one fewer variable.

    \item[Stage 4.] We reconstruct another univariate slice and perform an additional factor guessing, as in Stage 2.
\end{description}

The drop in the complexity of the rational coefficients after each stage for the most complicated two-loop diphoton \acp{FR}, which are in the \ac{MHV} configurations, is illustrated in \cref{tab:degrees_MHV}.
As proxy for the complexity of the coefficients we use the maximal numerator/denominator polynomial degrees, which can be evaluated by reconstructing univariate slices as discussed in \cref{sec:Apart}.

\begin{table}
    \begin{center}
        \begin{tabular}[h!]{ccccccc}
            \hline
            Finite remainder & Original & Stage 1 & Stage 2 & Stage 3* & Stage 4* \\
            \hline
            $F^{(2)}_{1;1}(1_g^-,2_g^-,3_g^+,4_\gamma^+,5_\gamma^+)$ & $69/60$ & $28/20$ & $24/0$ & $19/10$ & $11/5$ \\
            $F^{(2)}_{1;0}(1_g^-,2_g^-,3_g^+,4_\gamma^+,5_\gamma^+)$ & $78/69$ & $44/35$ & $43/0$ & $21/10$ & $16/9$ \\
            $F^{(2)}_{1;1}(1_g^-,2_g^+,3_g^+,4_\gamma^-,5_\gamma^+)$ & $59/55$ & $30/27$ & $29/0$ & $18/15$ & $17/4$ \\
            $F^{(2)}_{1;0}(1_g^-,2_g^+,3_g^+,4_\gamma^-,5_\gamma^+)$ & $89/86$ & $38/36$ & $38/0$ & $20/16$ & $17/3$ \\
            $F^{(2)}_{1;1}(1_g^+,2_g^+,3_g^+,4_\gamma^-,5_\gamma^-)$ & $40/42$ & $25/27$ & $25/0$ & $15/18$ & $15/0$ \\
            $F^{(2)}_{1;0}(1_g^+,2_g^+,3_g^+,4_\gamma^-,5_\gamma^-)$ & $66/66$ & $32/33$ & $32/0$ & $13/13$ & $12/3$ \\
            \hline
        \end{tabular}
    \end{center}
    \caption{Maximal numerator/denominator polynomial degrees of the rational coefficients of the most complicated \acp{FR} at each stage of our reconstruction strategy.
    The column ``original'' refers to the rational coefficients prior to any optimisation.
    The asterisk highlights that, after the \ac{PFD} in Stage 3, the coefficients to be reconstructed depend on one fewer variable.
    }
    \label{tab:degrees_MHV}
\end{table}

Interestingly, we observe that the coefficients of the \ac{SLC} $3g2\gamma$ two-loop \acp{FR} $F^{(2)}_2$ can be expressed in terms of those of the \ac{LC} two-loop five-gluon \acp{FR}.
The coefficients of the \ac{LC} $3g2\gamma$ two-loop remainders $F^{(2)}_1$ instead are not entirely fixed by the five-gluon ones, but using the latter as ans\"atze in the linear relations reduces significantly the maximal polynomial degrees of the coefficients which remain to be reconstructed.

As can be appreciated in \cref{tab:degrees_MHV}, our strategy leads to a substantial drop in the polynomial degrees.
Furthermore, the coefficients to be reconstructed after the \ac{PFD} (Stage 3) depend on one fewer variable.
This makes the decrease in the number of sample points required for the reconstruction even more pronounced.
The price to pay for this is that performing the \ac{PFD} increases the evaluation time per point, as discussed at the end of \cref{sec:Apart}.
With our setup we observe that, for the most complicated \acp{FR}, the evaluation times grows roughly by one order of magnitude, while the number of sample points required for the reconstruction decreases by two orders of magnitude.
This leads to an overall gain of roughly one order of magnitude in the reconstruction time\footnote{To give a sense of the absolute scale of the improvement, we quote the explicit number of sample points required for the $-++-+$ helicity configuration for the \ac{LC} $(d_s-2)^0$ component, $F^{(2)}_{1;0}$.
Reconstructing after Stage 2 would have required \num{57291} sample points in four variables while reconstructing after Stage 4 requires 518 sample points in three variables.
Note that each of the 518 points requires a univariate fit in the additional variable and so the overall improvement is around a factor of 10.}.
We stress that the evaluation time relevant here is that of the algorithm which evaluates the rational coefficients over \acp{FF}, not the final evaluation time of the \acp{FR}.
Once the reconstruction is completed, in fact, the rational coefficients are evaluated from their analytic expressions.
We will discuss the evaluation time of the \acp{FR} in \cref{sec:performance}.

Our approach therefore leads to an important simplification in the reconstruction of the rational coefficients.
Moreover, the ensuing analytic expressions are dramatically more compact.
This makes them suitable for compilation in a \cpp library, an essential step for their phenomenological application, which we discuss in \cref{ch:yy-xs}.

\section{Compact analytic expressions for the all-plus configuration}
\label{sec:AllPlus}

Prior to discussing the numerical implementation of all two-loop helicity amplitudes, we would like to comment on the all-plus amplitude, which displays a particularly simple analytic form.
We find that the structures appearing are closely related to those appearing in the five-gluon all-plus amplitudes at one~\cite{bern:1993sx,mahlon:1993si,bern:1993qk,henn:2019mvc} and two loops~\cite{gehrmann:2015bfy,dunbar:2016aux,badger:2019djh,dunbar:2019fcq}.
We present the \acp{FR} in the expansion around $d_s=2$ (\cref{sec:Kinematics}).

The all-plus amplitude is finite and rational at one loop.
The \ac{FR} can be written as
\begin{align}
    F_{1;0}^{(1)}(1_g^+,2_g^+,3_g^+,4^+_\gamma,5^+_\gamma) = - 2 \frac{\spB{4}{5}^2}{\spA{1}{2}\spA{2}{3}\spA{3}{1}} \, .
    \label{eq:F1allplus}
\end{align}
Remarkably, this amplitude is invariant under conformal transformations, and the expression given here exhibits this property in a manifest way~\cite{henn:2019mvc}.
If all masses are neglected, the \ac{SM} Lagrangian is conformally invariant~\cite{braun:2003rp}.
This symmetry is obscured at loop level by the appearance of scales associated with the divergences and it is therefore rather surprising to observe it in a one-loop amplitude.
One might na\"ively suppose that this is a consequence of the finiteness of the all-plus one-loop amplitudes.
Yet, the single-minus one-loop amplitudes are equally finite, but they are not conformally invariant.
This phenomenon still calls for an explanation.
These properties are discussed in detail in \incite{henn:2019mvc}, where the authors prove that the $n$-gluon all-plus amplitudes in \ac{QCD} are conformally invariant at one loop.
Since the diphoton amplitudes can be expressed as permutations of pure-gluon scattering~\cite{dicus:1987fk,deflorian:1999tp} and the conformal generators commute with permutations, all considerations regarding conformal symmetry extend to the diphoton case.

At two-loop order, the \ac{LC} $(d_s-2)^0$ contribution is the only one involving transcendental functions (\cref{sec:loops}).
Its expression is remarkably simple,
\begin{align}
    F_{1;0}^{(2)}(1_g^+,2_g^+,3_g^+,4^+_\gamma,5^+_\gamma) = \frac{\spB{4}{5}^2}{\spA{1}{2}\spA{2}{3}\spA{3}{1}} \sum_{{\rm cyclic}(123)} F_{\rm box}(\s{12},\s{23};\s{45}) \, ,
    \label{eq:F2allplusA}
\end{align}
where the sum runs over the cyclic permutations of $(1,2,3)$, and
\begin{align}
    F_{\rm box}(\s{12},\s{23};\s{45}) = \li{2}{1-\frac{\s{12}}{\s{45}}} + \li{2}{1-\frac{\s{23}}{\s{45}}} + \ln^2\brf{\frac{\s{12}}{\s{23}}} + \frac{\pi^2}{6}
\end{align}
is the finite part of the one-loop box with an off-shell leg.
The dilogarithm is defined in \cref{eq:polylog}.
The analytic continuation of the box functions to any scattering region can be achieved by adding a small positive imaginary part to each two-particle momentum invariant,
\begin{align}
    \s{ij} \to \s{ij} + \imi0^+\,.
\end{align}
The other partial amplitudes at two loops are rational,
\begin{align}
    F_{1;1}^{(2)}(1_g^+,2_g^+,3_g^+,4^+_\gamma,5^+_\gamma) &=
    - \frac{\spB{4}{5}^2}{\spA{1}{2}\spA{2}{3}\spA{3}{1}}
    - \frac{1}{2} F_{3;0}^{(2)}(1_g^+,2_g^+,3_g^+,4^+_\gamma,5^+_\gamma) \,, \nonumber\\
    F_{2;0}^{(2)}(1_g^+,2_g^+,3_g^+,4^+_\gamma,5^+_\gamma) &= 0 \,, \nonumber\\
    F_{2;1}^{(2)}(1_g^+,2_g^+,3_g^+,4^+_\gamma,5^+_\gamma) &=
    - 3 \frac{\spB{4}{5}^2}{\spA{1}{2}\spA{2}{3}\spA{3}{1}}
    - \frac{1}{2} \frac{\ptrf{p_1,p_2,p_3,p_4-p_5}\spA{4}{5}}{\spA{1}{4}\spA{1}{5}\spA{2}{4}\spA{2}{5}\spA{3}{4}\spA{3}{5}} \,,  \nonumber\\
    F_{3;0}^{(2)}(1_g^+,2_g^+,3_g^+,4^+_\gamma,5^+_\gamma) &=
    \frac{1}{3} \ptrf{p_1,p_2,p_3,p_4-p_5} \sum_{{\rm cyclic}(123)} \frac{1}{\spA{2}{3}^2\spA{1}{4}\spA{1}{5} \spA{4}{5}}\,,
    \label{eq:F2allplusB}
\end{align}
where $\ptrf{p_i,p_j,p_k,p_l}$ is defined in \cref{eq:tr5g}.
The peculiar simplicity of this amplitude at two loops follows from the fact that it vanishes at tree level and it is rational in four dimensions at one loop.
The one-loop amplitude can in fact be used as an effective on-shell vertex in four-dimensional unitarity~\cite{bern:1994zx,bern:1994cg,dunbar:2016aux}.
In this way, the cuts of the two-loop amplitude become one-loop cuts with an insertion of the effective vertex.
The one- and two-loop all-plus \acp{FR} are thus treated as tree-level and one-loop objects, respectively.
As a result, the special functions appearing in the \ac{FR} at two loops can have at most transcendental weight two (up to $\ord(\eps^0)$).
Moreover, the rational coefficients of the transcendental functions can be shown through four-dimensional unitarity to be given by (permutations of) the one-loop all-plus \ac{FR}.
Thus, they inherit the symmetry under conformal transformations from the one-loop amplitude.
These properties are manifest in our explicit expressions, \cref{eq:F2allplusA,eq:F1allplus}.
Complementing four-dimensional unitarity with recursion relations for the rational terms allows us to compute the two-loop all-plus \acp{FR} in the purely gluonic case avoiding altogether the computation of the two-loop integrals~\cite{dunbar:2016aux,dunbar:2019fcq}.
Some results are available even for amplitudes involving more than five positive-helicity gluons~\cite{dunbar:2016cxp,dunbar:2016gjb,badger:2016ozq,dunbar:2017nfy,dunbar:2020wdh,dalgleish:2020mof}.

Amplitudes with a single minus helicity share some of the simplicity of the all-plus case.
They also vanish at tree level, and are finite and rational at one loop.
As a result, they also have maximum transcendental weight two at two loops.
In contrast to the all-plus amplitudes, however, they do not have the structure that $F_{1}^{(2)}$ exhibits; namely, $F_{1;0}^{(2)}$ has uniform transcendental weight two with all other contributions being rational.
For the amplitudes with two negative helicities, instead, the \acp{FR} have maximum weight two and four at one and two loops, respectively.

\section{Implementation and performance}
\label{sec:performance}

The \acp{FR} are implemented in the \njet \cpp library~\cite{njet}, which is linked to the \pentagonfunctions library~\cite{chicherin:2020oor} for the evaluation of the special functions.
The six independent helicity amplitudes (shown in \cref{tab:finremeval}) are permuted analytically onto the global basis of pentagon functions defined in the $12\to345$ scattering region to provide the complete list of 16 ``mostly-plus'' helicity amplitudes required for the sum.
This task is performed using the permuted coefficients from the six fully reconstructed amplitudes as an ansatz into the linear relations, so additional reconstruction time is avoided (see \cref{sec:LinearRelations}).
Having identified a global basis of pentagon functions for the complete colour and helicity sum, we formulate the partial amplitudes as
\begin{align}
    F^{h} = c^h_{i}\,M^h_{ij}\,f^h_j \,,
    \label{eq:partialrep}
\end{align}
where: $h$ is the helicity configuration;
$f^h_j$ is a list of integers corresponding to the global list of pentagon function monomials, which is evaluated once per phase-space point;
$M^h_{ij}$ are sparse matrices of rational numbers that are specific to each partial amplitude;
and $c^h_{i}$ are the independent rational coefficients for each helicity amplitude, written in terms of independent polynomials in the \acp{MTV} $x_i$.
The pentagon function monomials are split into parity-odd and -even components, which allows the remaining 16 ``mostly-minus'' helicities to be computed by simply flipping the parity of the special functions and applying complex conjugation to the coefficients.
The colour- and helicity-summed \ac{ME} is constructed numerically from these ingredients.
The sparse matrix multiplication is implemented using the \eigen library~\cite{eigenweb}.
Evaluation with \acp{f128} and \acp{f256} is provided via the \qd library~\cite{qdweb}.
The code is available through \njet~\cite{njet}, where we provide additional installation instructions and example programs demonstrating its usage.

\begin{figure}
    \begin{center}
        \includegraphics[width=\textwidth]{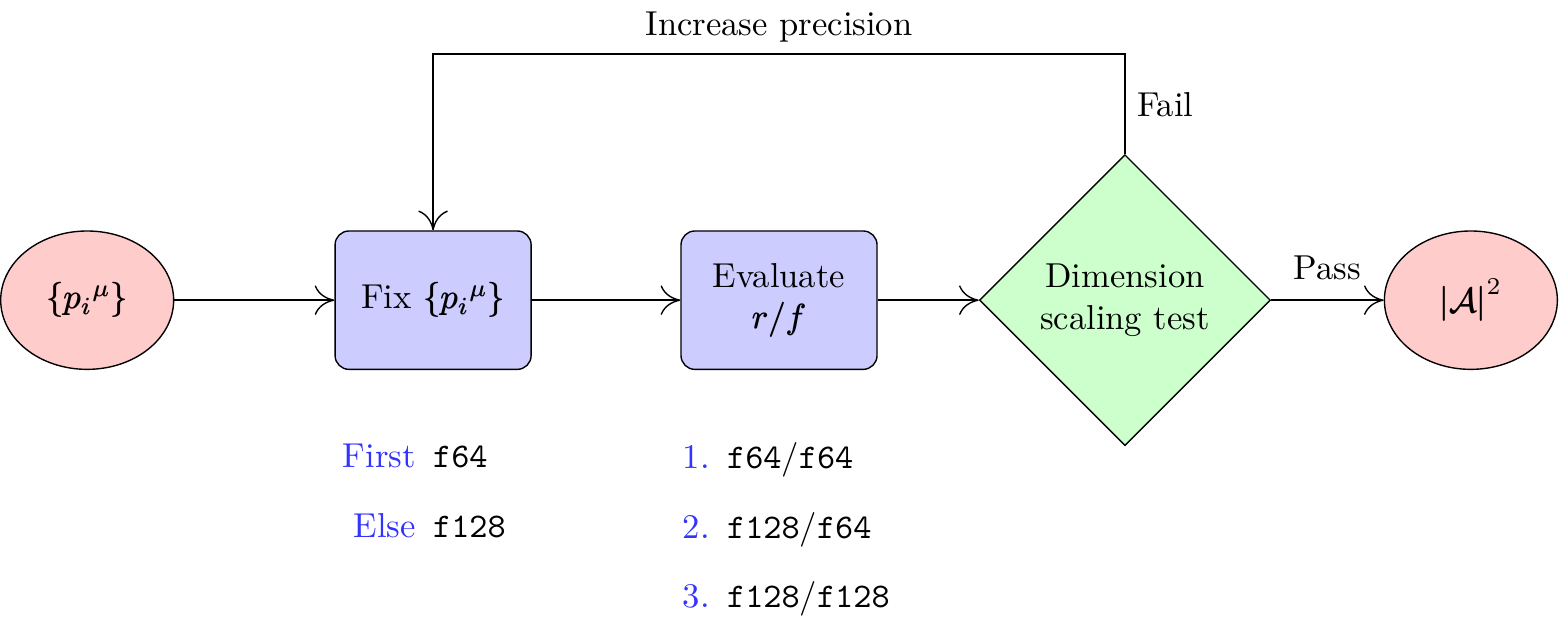}
        \caption{
            Flow chart of our evaluation strategy.
            $r/f$ indicates the rational coefficients and special functions respectively at the three precision levels.
        }
        \label{fig:eval-pipeline}
    \end{center}
\end{figure}

The \cpp code returns the values of the one- and two-loop hard functions, $\mathcal{H}^{(1)}$ and $\mathcal{H}^{(2)}$, obtained by squaring \cref{eq:colourdecomp}, substituting the decomposition in $\nc$ and
$\nf$ from \cref{eq:Nc_exp}, subtracting the \ac{IR} and \ac{UV} poles, and finally summing over colour and helicity,
\begin{align}
    \begin{aligned}
        \mathcal{H} &= \frac{\alpha^2 \astr^3}{(4\pi)^5} \left(
        \mathcal{H}^{(1)} + \frac{\astr}{4\pi} \, \mathcal{H}^{(2)}
        \right) + \order{\astr^5}\,,\\
        \mathcal{H}^{(2)} &=
        \nc \, \mathcal{H}^{(2)}_1
        + \frac{1}{\nc} \, \mathcal{H}^{(2)}_2
        + \nf \, \mathcal{H}^{(2)}_3\,.
        \label{eq:hardfunction}
    \end{aligned}
\end{align}
The sum over colours for each helicity can also be returned if required.
We find the evaluation time is dominated by the special functions,
particularly when higher precision is required.
In order to ensure fast and stable numerical evaluation, we adopt the following evaluation strategy, which is also depicted in \cref{fig:eval-pipeline}.
\begin{enumerate}
    \item The user-provided phase-space point is checked for the precision of the on-shell constraints.
        Points are adjusted in case the precision is not
        acceptable for the requested number of digits: \ac{f64}~$\sim \num{15}$ digits, \ac{f128}~$\sim \num{31}$ digits, and \ac{f256}~$\sim \num{62}$ digits~\cite{qdweb}.
    \item The colour- and helicity-summed amplitude is computed using \ac{f64} precision at two points which differ only by overall dimension scaling factor\index{Scale!dimension test}.
        After accounting for the overall dimension of the squared amplitude, and neglecting scale dependence arising due to the truncation of the perturbative series in $\astr$ to \ac{NLO}, the two evaluations only differ due to rounding errors at intermediate stages in the evaluation.
        This accuracy scaling test has been used extensively at one loop.
        Note that this is unrelated to the scale variation test discussed in \cref{sec:uncertainties}.
        We refer to this precision as \ac{f64}/\ac{f64} since both coefficients and special functions use \ac{f64} precision.
    \item If the estimated number of correct digits from the scaling tests falls below a user-defined threshold, the coefficients only are recomputed using \ac{f128} precision after the original point is corrected to \ac{f128} precision (as in step 1).
        We refer to this as \ac{f128}/\ac{f64} precision.
    \item The scaling test is performed again and if it fails the special functions are reevaluated in \ac{f128} precision.
        This is \ac{f128}/\ac{f128} precision.
\end{enumerate}

These steps can be repeated to obtain up to \ac{f256}/\ac{f256} precision.
In practice these steps are rather expensive and unnecessary for standard phenomenological applications, so they are omitted from our strategy.

While the dimension scaling test has been used successfully at one loop, we need to be more careful in our applications when linking the \pentagonfunctions library, which also makes use of the dimension rescaling internally.
To validate the reliability of the scaling test as an estimate of the error of the result, we evaluate both with a direct \ac{f128}/\ac{f128} computation and via a scaling test with an error cutoff of three digits at \ac{f64}/\ac{f64} for a set of \SI{60}{\kilo\relax} points.
To ensure a realistic validation, we use ``physical'' points with a phase-space sampling density determined by the one-loop process, obtained from \nnlojet.
We compare the estimated error provided by the \ac{f64}/\ac{f64} scaling test to the relative difference between the \ac{f64}/\ac{f64} and \ac{f128}/\ac{f128} evaluations, with the latter taken as the true error.
In the following, percentages are always stated with respect to the entire set of points.

\begin{figure}
    \centering
    \includegraphics[width=0.5\textwidth]{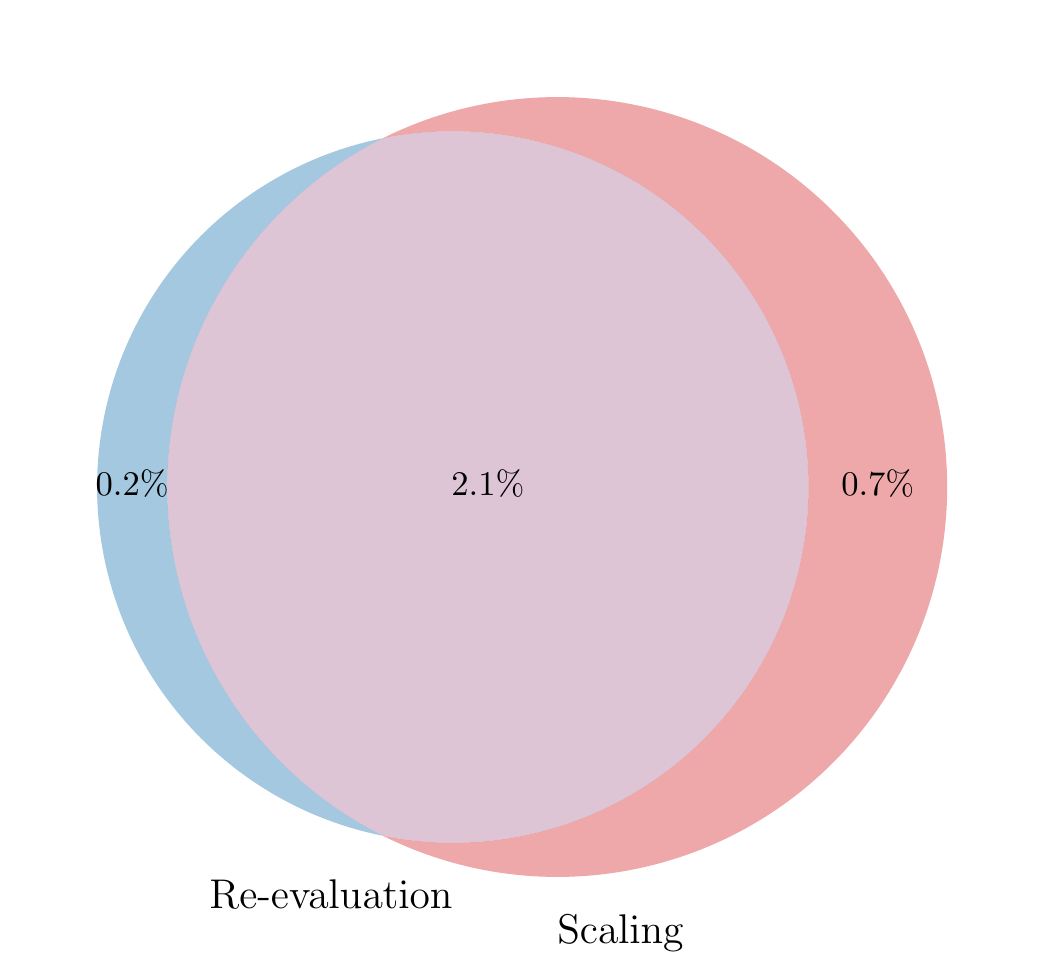}
    \caption{
        Venn diagram of evaluations failing an accuracy test with a cutoff of three digits as estimated by the \ac{f64}/\ac{f64} scaling test for the right circle (red), and the relative difference between the \ac{f64}/\ac{f64} and \ac{f128}/\ac{f128} evaluations for the left circle (blue).
        The reevaluation error is taken as truth in order to validate the scaling test as a reliable estimate of error.
        The sets are labelled by their percentage of the entire set of evaluated points in the validation.
        The leftmost set shows false positives, the centre set shows true negatives, and the rightmost set shows false negatives.
    }
    \label{fig:euler}
\end{figure}

As depicted in \cref{fig:euler}, the scaling test returns a negative for \SI{2.8}{\percent} of the points.
According to true error, an additional \SI{0.2}{\percent} of the points should be failed and are missed by the scaling test (false positive).
Of these points, almost all have true error of three digits, the remaining \SI{0.008}{\percent} with three digits, so the effect on stability is small.
The scaling test also fails some points unnecessarily (false negative), this subset comprising \SI{0.7}{\percent} of all points, which incurs a small performance penalty in the evaluation strategy.
The effects of the false estimates are considered to be allowably small.

We note that the dimension scaling test is statistical and therefore one will always find anomalies in a sufficiently large sample.
Care should be taken when integrating over extreme regions of phase space.

% https://gitlab.com/jetdynamics/njet-tools/-/blob/master/3g2a%402l/test/analyse-phys/main.py#L158
\begin{figure}
    \centering
    \includegraphics[width=0.8\textwidth]{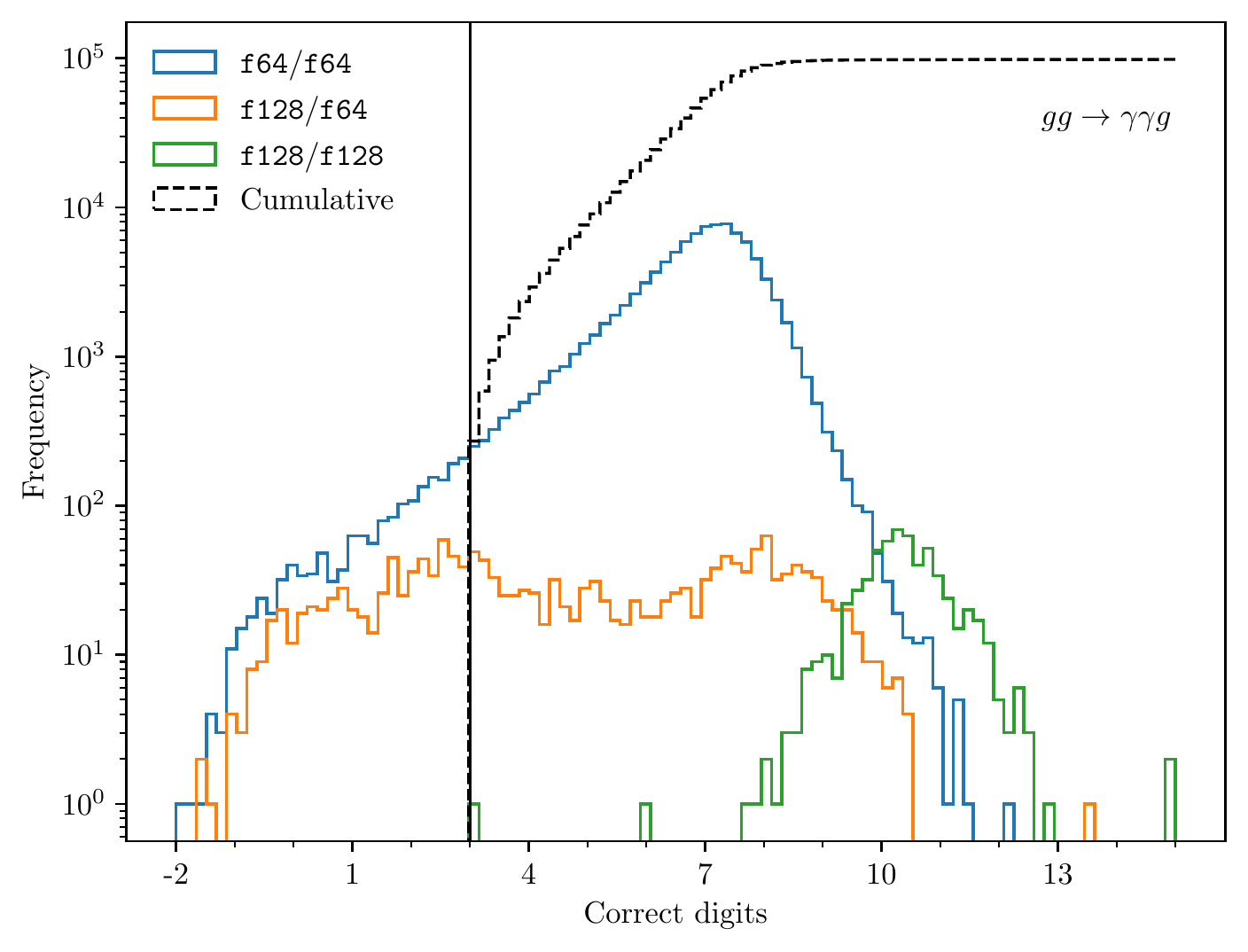}
    \caption{
        Histogram of the error estimate on the two-loop evaluations as given by the scaling test.
        We use the evaluation strategy with a target accuracy of three digits, denoted by the vertical black line, and show errors for all precision levels as well as the cumulative error on all passing points.
        A cumulative bin of height $h$ at $d$ digits indicates $h$ points have an accuracy of at best $d$ digits.
    }
    \label{fig:stability}
\end{figure}

To assess the stability of our implementation (\cref{fig:stability}) and measure timings, we evaluate the amplitude squared over \SI{100}{\kilo\relax} points of the physical phase space.
We see \SI{1.8}{\percent} of points failing \ac{f64}/\ac{f64} evaluation, with \SI{1.2}{\percent} passing at \ac{f128}/\ac{f64} and \SI{0.6}{\percent} passing at \ac{f128}/\ac{f128}.
The evaluation strategy achieves target accuracy for all of the \SI{100}{\kilo\relax} physical phase-space points tested.
We find a single \ac{f64}/\ac{f64} call has a mean time of \SI{9}{\second}, with \SI{99}{\percent} of that time spent evaluating the pentagon functions.
Using the full evaluation strategy with a target minimum accuracy of three digits, we obtain a mean timing per phase-space point of \SI{26}{\second}.

We present a benchmark evaluation at a point taken from the physical phase space.
We choose a generic configuration where the momentum invariants $\s{ij}$ ($\si{GeV}^2$) and pseudoscalar $\trf$ ($\si{GeV}^4$) take the values, quoted to four significant figures,
\sisetup{round-precision = 4, round-mode = figures, scientific-notation = fixed,}
\begin{align}
    \begin{aligned}
        \s{12} &= \num{1.4116251163350877377236204264908242298667471893246336180766098423e+04} \,, &
        \s{23} &= \num{-1.4046834737972321415660577595055706351599646008939620136601159902e+03} \,, &
        \s{34} &= \num{7.6669799448946075091609646313550363122719368647199176601634259772e+03} \,, \\
        \s{45} &= \num{5.4932450565561162032281304796011000981320172481749213347830057155e+03} \,, &
        \s{15} &= \num{-4.4044289245917486109857445869899023397535714857496713286053642996e+03} \,, &
        \trf &= \num{-1.7599755750687916479576367229182851347753538557888213932493870876e+07}\imi\,.
        \label{eq:benchmarkpoint}
    \end{aligned}
\end{align}
\sisetup{round-precision = 1, scientific-notation = false,}\noindent
High precision \ac{f128}/\ac{f128} evaluations are given in the ancillary files of \incite{badger:2021imn}.
The values for the \acp{FR} and the two-loop hard function, normalised by the \ac{LO}, are shown in \cref{tab:finremeval,tab:hardfuneval} respectively.
The \ac{SLC} corrections are \num{645} times smaller than the \ac{LC} at the point \cref{eq:benchmarkpoint}, while the closed fermion loop corrections are \num{133} times smaller.
These ratios do change as we sample different points.
Averaging over \num{100} physical points, the ratio is
\begin{align*}
    \ab{ \nc\,\frac{\mathcal{H}^{(2)}_1}{\mathcal{H}^{(1)}}}:\ab{\frac{1}{\nc}\,\frac{\mathcal{H}^{(2)}_2}{\mathcal{H}^{(1)}}}:\ab{ \nf\,\frac{\mathcal{H}^{(2)}_3}{\mathcal{H}^{(1)}}}
    &= \num{2061}:\num{1}:\num{14}.
\end{align*}

While the evaluation is considerably more difficult than the massless planar five-gluon scattering (\cref{ch:3j}), owing to the more complicated set of pentagon functions arising from the non-planar integral topologies, our tests show the amplitudes are clearly ready for phenomenological applications (\cref{ch:yy-xs}).
In addition, recent improvements to \pentagonfunctions~\cite{chicherin:2021dyp} would improve the timing and stability, as found in \cref{ch:3j}.

\sisetup{
    round-precision = 4,
    scientific-notation = fixed,}

\begin{table}
    \centering
    \begin{tabular}[h]{cccc}
        \hline
        Helicity & $\nc F^{(2)}_1/F^{(1)}$ & $\frac{1}{\nc} F^{(2)}_2/F^{(1)}$ & $\nf F^{(2)}_3/F^{(1)}$ \\
        \hline
        $+++++$ &
        $\num{-2.77582993046583427605142515235285e+01} - \num{1.01745076290502519208036374477010e+01}$i &
        $\num{-1.67327484248907239815459032713580e+00} - \num{2.39649153322601693283371805022356e-01}$i &
        $\num{-5.22837426806977838213461692255251e+00} - \num{4.03428197735324580204112002553437e+00}$i \\
        $-++++$ &
        $\num{-2.57580532623280503092540771294086e+01} + \num{2.78347349175690890510559905150441e+01}$i &
        $\num{3.57081631633550999314235864904020e-01} - \num{3.21338918788382341155048018442021e-01}$i &
        $\num{3.36268118497271552633054996889330e-01} - \num{4.42401378002389410311387525774025e+00}$i \\
        $+++-+$ &
        $\num{-2.41566460358473862557597565808062e+01} + \num{1.45880275249122297539666099129978e+01}$i &
        $\num{3.69756140499054359520282727639744e-01} - \num{5.53932445125758408476522017564549e-01}$i &
        $\num{-4.95123441825520347867841272615952e+00} + \num{6.67167689442155278362227939920601e-01}$i \\
        $--+++$ &
        $\num{-2.02342531062579671222860854129219e+01} + \num{8.20384686558306263565400583003510e-01}$i &
        $\num{-4.05487410679291377838146170351151e-01} - \num{3.54867788156361451381190048399781e-01}$i &
        $\num{5.35518188040192554893963855293205e-02} + \num{2.47830230193615584923072985322998e-04}$i \\
        $-++-+$ &
        $\num{-2.85844161095369508004626750797306e+01} + \num{3.29007964718656854030254373989609e+01}$i &
        $\num{3.91732905857814948144244362925973e-01} - \num{5.48869824722050041995728687006486e-04}$i &
        $\num{3.02156400028686127902143151309653e+00} + \num{1.47549772885303042900035952799006e+00}$i \\
        $+++--$ &
        $\num{-2.09417707770365479210244429535450e+01} - \num{1.53427303145404829662973631186851e+01}$i &
        $\num{-3.07968164685809980488634602515262e-01} - \num{4.55762937816688321755049623598366e-01}$i &
        $\num{-4.87984856920121200597547643690781e+00} - \num{5.86222628879890415882513441668661e-03}$i \\
        \hline
    \end{tabular}
    \caption{Numerical values of the partial amplitudes for the six independent helicities at the benchmark point in \cref{eq:benchmarkpoint}.
    Values are quoted with $\nc=3$ and $\nf=5$, to four significant figures.}
    \label{tab:finremeval}
\end{table}

\begin{table}
    \centering
    \begin{tabular}[h]{ccc}
        \hline
        $\nc\mathcal{H}^{(2)}_1/\mathcal{H}^{(1)}$ &
        $\frac{1}{\nc}\mathcal{H}^{(2)}_2/\mathcal{H}^{(1)}$ &
        $\nf\mathcal{H}^{(2)}_3/\mathcal{H}^{(1)}$ \\
        \hline
        $\num{52.74593955908158767014772455004}$ &
        $\num{0.08176280908057119497025187026506}$ &
        $\num{0.39564191164271056748420506863181}$ \\
        \hline
    \end{tabular}
    \caption{
        Numerical values for the components of the two-loop hard function normalised to the one-loop hard function defined in
        \cref{eq:hardfunction} at the benchmark point of \cref{eq:benchmarkpoint}.
        Values are quoted with $\nc=3$ and $\nf=5$, to four significant figures.}
    \label{tab:hardfuneval}
\end{table}

\sisetup{
    round-mode = off,
    scientific-notation = false,}

\section{Summary}
\label{sec:amp-conc}

In this chapter, we have presented a complete \ac{FC} five-point amplitude at two loops in \ac{QCD}.
All helicity configurations have been implemented in the \njet \cpp library, which provides efficient and stable evaluation over the physical scattering region.
Although the algebraic complexity of the amplitude is considerable, the direct analytic reconstruction of the \acp{FR} was possible by making use of linear relations amongst the coefficients, and univariate partial fractioning that could be done without any analytic knowledge of the intermediate steps in the reduction.
We expect these techniques will have applications to other important high-multiplicity two-loop calculations with more external scales, such as five-particle scattering with an off-shell leg, for which there has also been recent progress~\cite{hartanto:2019uvl,papadopoulos:2015jft,papadopoulos:2019iam,abreu:2020jxa,canko:2020ylt,syrrakos:2020kba,badger:2021nhg,krys:2022gby,badger:2022ncb,abreu:2021smk,kardos:2022tpo,chicherin:2021dyp,badger:2022mrb,bevilacqua:2022nrm}.
We have found a form that is suitable for phenomenological applications.
New high-precision predictions for diphoton-plus-jet production at the \ac{LHC}, which include the dominant \ac{N3LO} corrections we have computed here, are presented in \cref{ch:yy-xs}.

\chapter{\acs{NLO} \acs{QCD} corrections to \texorpdfstring{$gg\to\gamma\gamma g$}{diphoton-plus-jet production through gluon fusion} distributions}
\label{ch:yy-xs}

Having obtained the virtual \ac{QCD} corrections to the amplitude in \cref{ch:yy-amp}, we now compute the \ac{NLO} \ac{QCD} corrections to the cross section of the gluon-fusion contribution to diphoton-plus-jet production at the \ac{LHC}.
We produce fully differential distributions by combining the two-loop virtual corrections with the one-loop real corrections using antenna subtraction to cancel \ac{IR} divergences.
We observe significant corrections at \ac{NLO} which demonstrate the importance of combining these corrections with the quark-induced diphoton-plus-jet channel at \ac{NNLO}.

This chapter is organised as follows.
We first discuss recent developments in the field that motivated this study in \cref{sec:xs-back}.
Next, in \cref{sec:setup}, we review the computational setup, discussing the amplitude-level ingredients and antenna subtraction method used to cancel \ac{IR} divergences.
We then present results for the \ac{NLO} corrections to differential cross sections at the \SI{13}{\TeV} \ac{LHC} in \cref{sec:results}.
We study the perturbative convergence in both transverse momentum and mass variables as well angular distributions in rapidity and the Collins-Soper angle.
We comment on the computational cost of the calculation in \cref{sec:cost}, before drawing our conclusions in \cref{sec:concl}.

\section{Background}
\label{sec:xs-back}

As discussed in \cref{sec:yy}, diphoton-plus-jet production is of high phenomenological relevance.
The recently computed \ac{NNLO} corrections of diphoton-plus-jet production~\cite{chawdhry:2021hkp} display a good perturbative convergence, except in regions where the loop-mediated gluon-fusion subprocess (which contributes to the cross section only from \ac{NNLO} onwards) is numerically sizeable compared to other contributions.
In order to capture the full effects of the \ac{QCD} corrections,
it is important to include loop-induced gluon-fusion channels from at least one order higher in the perturbative series.
These corrections, which are \ac{N3LO} in the full proton-initiated process, but \ac{NLO} in the gluon-fusion channel (\cref{sec:fixed-order}), are the subject of this chapter.

As discussed in \cref{sec:subtraction-methods}, combining and integrating the amplitudes into differential cross sections requires the subtraction of \ac{IR} divergences.
To achieve this in a stable and efficient way is an extremely difficult problem and many solutions have been proposed and applied in calculations up to \ac{NNLO}.
Such subtraction schemes often scale poorly with the number of external particles and only a handful of examples for high-multiplicity processes at \ac{NNLO} currently exist~\cite{czakon:2021mjy,chawdhry:2019bji,kallweit:2020gcp,chawdhry:2021hkp}.

For the process considered in this chapter, the \ac{IR} divergences are only at \ac{NLO}.
However, since the real radiation involves two-to-four one-loop squared amplitudes, the automated numerical algorithms for the one-loop amplitudes are tested in extreme phase-space regions.
The \ac{LO} \ac{QCD} contributions to the gluonic subprocess were first considered in \incite{deflorian:1999tp}, based on the compact one-loop five-gluon amplitudes~\cite{bern:1993mq}.

\section{Computational setup}
\label{sec:setup}

We consider the inclusive scattering process,
\begin{align}
    g g \to \gamma \gamma g + X \,,
    \label{eq:process}
\end{align}
at a hadron collider.
As the process is loop induced, the \ac{LO} contribution is at $\ord(\astr^3)$ and involves the integration of a one-loop amplitude squared.
The \ac{NLO} \ac{QCD} corrections are computed by combining the two-loop virtual corrections to the $2\to3$ process $g g \to \gamma \gamma g$ with the $2\to 4$ processes with an additional unresolved parton: $gg\to \gamma \gamma gg$ and $gg\to \gamma \gamma q\anti{q}$.
Pictorially, we can represent the parton level cross sections up to \ac{NLO} in \ac{QCD} as
\begin{align}
    \begin{aligned}
        \sigma_{gg\to \gamma\gamma g + X}^\text{NLO} =&
        \bigint \dd\Phi_3 \msq{
            \raisebox{-6mm}{\includegraphics[width=17mm]{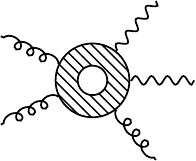}}
        } +
        2 \, \bigint \dd\Phi_3 \, \Re\brf{
            \raisebox{-6mm}{\includegraphics[width=17mm]{ggyyg_1l}}
            ^\dagger\cdot
            \raisebox{-6mm}{\includegraphics[width=17mm]{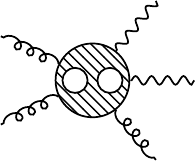}}
        } \\
        +&\bigint \dd\Phi_4 \msq{
            \raisebox{-7.5mm}{\includegraphics[width=17mm]{ggyygg_1l}}
        } +
        \bigint \dd\Phi_4 \msq{
            \raisebox{-7.5mm}{\includegraphics[width=17mm]{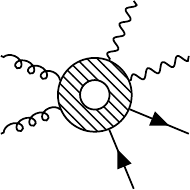}}
        }
        + \order{\astr^5} \,,
        \label{eq:xsdef}
    \end{aligned}
\end{align}
where $\dd\Phi_N$ represents the on-shell phase-space measure for $N$ massless final state particles, \cref{eq:ndNpLips}.
The one-loop amplitude for $gg\to \gamma \gamma q\anti{q}$ indicates the loop contribution in which the photons couple to an internal fermion loop.
The observable process $pp \to \gamma\gamma j$ also includes channels where the photons couple to an external quark pair, which are not included in the gluon-fusion subprocess.
The expansion up to the \ac{NNLO} of $pp \to \gamma\gamma j$ includes terms up to $\ord(\astr^3)$ and so the contributions coming from \cref{eq:xsdef} are technically \ac{N3LO}, as depicted in \cref{fig:pert-exp}.
However, due to the large gluon flux at high-energy hadron colliders, such contributions can be significant (\cref{sec:yy}).

The one-loop amplitudes for the \ac{LO} process and the real correction are finite, since the corresponding tree-level processes vanish.
The renormalised two-loop five-particle amplitude contains explicit \ac{IR} divergences generated by the integration over the loop-momenta, while the one-loop six-particle amplitudes exhibit a divergent behaviour when a final-state parton becomes unresolved.
The divergences cancel in the final result, as established by the \ac{KLN} theorem, and \iac{FR} of the virtual amplitudes can be defined using \ac{QCD} factorisation~\cite{catani:1998bh}.
In our calculation, this cancellation is performed using the antenna subtraction method (\cref{sec:subtraction-methods}).
The method extracts the \ac{IR} singular contributions from the real radiation subprocess, and combines its integrated form with the virtual subprocess, thus enabling their numerical integration using \ac{MC} methods, performed here in the \nnlojet framework.
The \ac{QCD} structure of the process under consideration is very similar to Higgs-plus-jet production in gluon fusion, which has been computed previously~\cite{chen:2014gva,chen:2016zka} using antenna subtraction,  and identical antenna subtraction terms are applied here.

The \acp{FR} of the two-loop amplitudes were computed in \cref{ch:yy-amp}, providing the \ac{FC} colour- and helicity-summed expressions through the \njet amplitude library.
We set a three-digit accuracy threshold for the dimension scaling test (\cref{sec:performance}), which guarantees a stable result without significantly affecting the performance.

The one-loop six-particle amplitudes are obtained using a combination of implementations from the \openloops~\cite{buccioni:2017yxi,buccioni:2019sur} library and from the generalised unitarity approach~\cite{bern:1994zx,bern:1994cg,britto:2004nc} within \njet (\cref{sec:loops}).
We use an improved version of \openloops in combination with the new extension \otter~\cite{otter:2020}, which is a tensor integral library based on the ``on-the-fly reduction''~\cite{buccioni:2017yxi} of \openloops and on various stability improvements~\cite{buccioni:2019sur}.
This new version of \openloops allows for a stable computation of the needed one-loop squared amplitudes in deep \ac{IR} regions.
Internally, \otter uses \ac{f64} scalar integrals that are provided by \collier~\cite{denner:2014gla,denner:2016kdg}, as well as \ac{f128} scalar integrals provided by \oneloop~\cite{vanhameren:2010cp}.
Minor modifications were made in \njet to avoid de-symmetrisation over the two photons and allow for a pointwise correspondence with the subtraction terms.
To compute the one-loop amplitude $gg\to\gamma\gamma gg$, the \openloops implementation was generally more efficient, but for exceptional phase-space points it was necessary to use the \ac{f128} implementation within \njet.
For the $gg\to\gamma\gamma\anti{q}q$ channel, we used
\njet, which allowed for a straightforward selection of the required loop contribution.
We note that this amplitude is also available within \openloops and we checked that the two implementations agree.

The amplitude-level ingredients have been validated in all relevant collinear and soft limits by checking their convergence towards the respective antenna subtraction terms.

\section{Results}
\label{sec:results}

For the numerical evaluation of our \ac{NLO} results on the gluon-induced diphoton-plus-jet process, we apply the same kinematic cuts  as were used for the \ac{NNLO} calculation of the quark-induced processes~\cite{chawdhry:2021hkp}.
These represent a realistic setup relevant for physics studies at the \SI{13}{\TeV} \ac{LHC}.
Using the quantities defined in \cref{sec:cuts} and with $\gamma_i$ as the $i^\text{th}$ hardest photon, the cuts are as follows.
\begin{itemize}
    \item Minimum photon transverse momenta:
        \begin{itemize}
            \item $p_T(\gamma_1) > \SI{30}{\GeV}$,
            \item $p_T(\gamma_2) > \SI{18}{\GeV}$.
        \end{itemize}
    \item Smooth photon isolation criterion with:
        \begin{itemize}
            \item $\Delta R = 0.4$,
            \item $E_{T,\gamma} = \SI{10}{\GeV}$,
            \item $\epsilon_\gamma=1$.
        \end{itemize}
    \item Minimum pseudorapidity of the photon pair: $|\eta(\gamma\gamma)|<2.4$.
    \item Minimum invariant mass of the photon pair: $m(\gamma\gamma) \geq \SI{90}{\GeV}$.
    \item Minimum separation of the photons: $\Delta R\brf{\gamma\gamma} > 0.4$.
    \item Minimum transverse momentum of the photon pair: $p_{T}(\gamma\gamma) > \SI{20}{\GeV}$.
\end{itemize}
We consider kinematic distributions in the following diphoton variables.
\begin{itemize}
    \item Transverse momentum of the diphoton system $p_{T}(\gamma\gamma)$.
    \item Diphoton invariant mass $m\brf{\gamma\gamma}$.
    \item Diphoton total rapidity $|y(\gamma\gamma)|$.
    \item Diphoton rapidity difference $\Delta y(\gamma\gamma)$.
    \item Collins-Soper angle\footnote{The Collins-Soper angle~\cite{collins:1977iv,richter-was:2016mal} $\phi_{CS}\brf{a b}$ for particles $a$ and $b$ is defined by
        \begin{align}
            \cos(\phi_{CS}\brf{a b}) &= \text{sign}\brf{p_z\brf{ab}}\,2\frac{{p_a}^+{p_b}^--{p_a}^-{p_b}^+}{m\brf{ab}\sqrt{m^2\brf{ab}+{p_T}^2\brf{ab}}} \,, &
            {p_i}^\pm &= \frac{1}{\sqrt{2}}\br{E_i\pm (p_i)_z} \,.
        \end{align}} $\left|\phi_{CS}(\gamma\gamma)\right|$.
    \item Diphoton azimuthal decorrelation $\Delta\phi (\gamma\gamma)$.
\end{itemize}
For these distributions, no jet requirement is applied since the transverse momentum cut on the diphoton system---which the jet(s) recoil against---is already sufficient to avoid \ac{NNLO}-like configurations where all final-state \ac{QCD} partons become unresolved.
This treatment follows that of \incite{chawdhry:2021hkp}.

Our numerical results use the \ac{NNLO} set of the \nnpdf \acp{PDF}~\cite{nnpdf:2017mvq} throughout, thus allowing a straightforward comparison with the existing \ac{NNLO} results~\cite{chawdhry:2021hkp} in the quark-initiated channels.
The strong coupling is evaluated using \lhapdf~\cite{buckley:2014ana}, with $\astr\brf{m_Z}=0.118$.
The \ac{EM} coupling is set to $\alpha=1/\num{137.035999139}$.
The choices of couplings are also made to match \incite{chawdhry:2021hkp}.
\Ac{MC} integration errors are below one percent on average and not displayed in the plots.

The uncertainty on our theory predictions is estimated by a seven-point variation of the renormalisation and factorisation scales around a central value (\cref{sec:uncertainties}), chosen in a dynamical manner on an event-by-event basis to be
\begin{align}
    \label{eq:mu}
    \mu_F = \mu_R = \frac{m_{T}}{2} \coloneqq \frac{1}{2} \sqrt{m^2(\gamma\gamma)+{p_T}^2(\gamma\gamma)} \,,
\end{align}
which is typical for diphoton studies~\cite{delduca:2003uz,gehrmann:2013aga,chawdhry:2021hkp}.

\begin{figure}
    \begin{center}
        \includegraphics[width=0.7\textwidth]{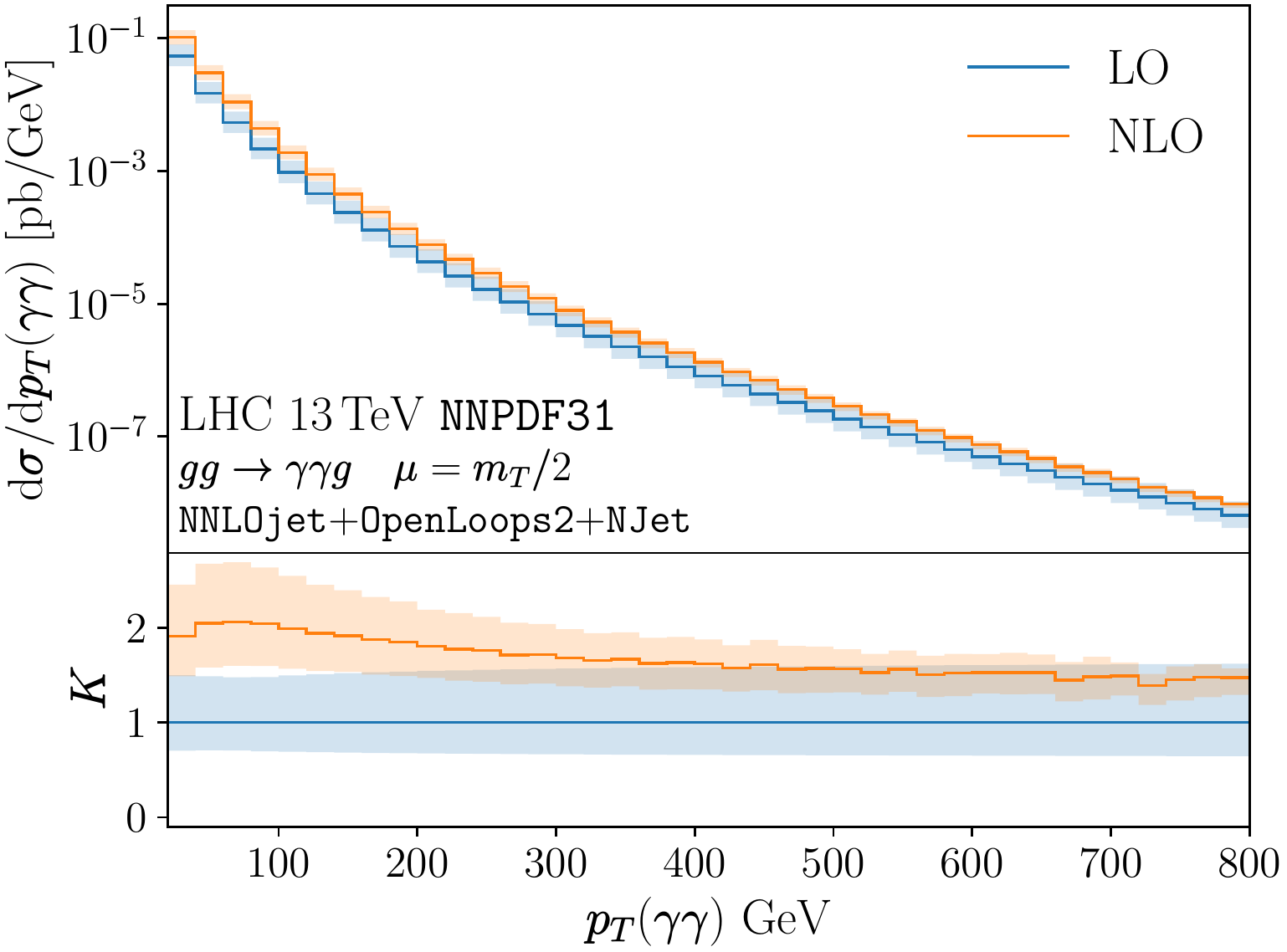}
    \end{center}
    \caption{
        Differential distributions in the transverse momentum $p_T(\gamma\gamma)$ of the diphoton system.
    }
    \label{fig:pt_maadista}
\end{figure}

\begin{figure}
    \begin{center}
        \includegraphics[width=0.7\textwidth]{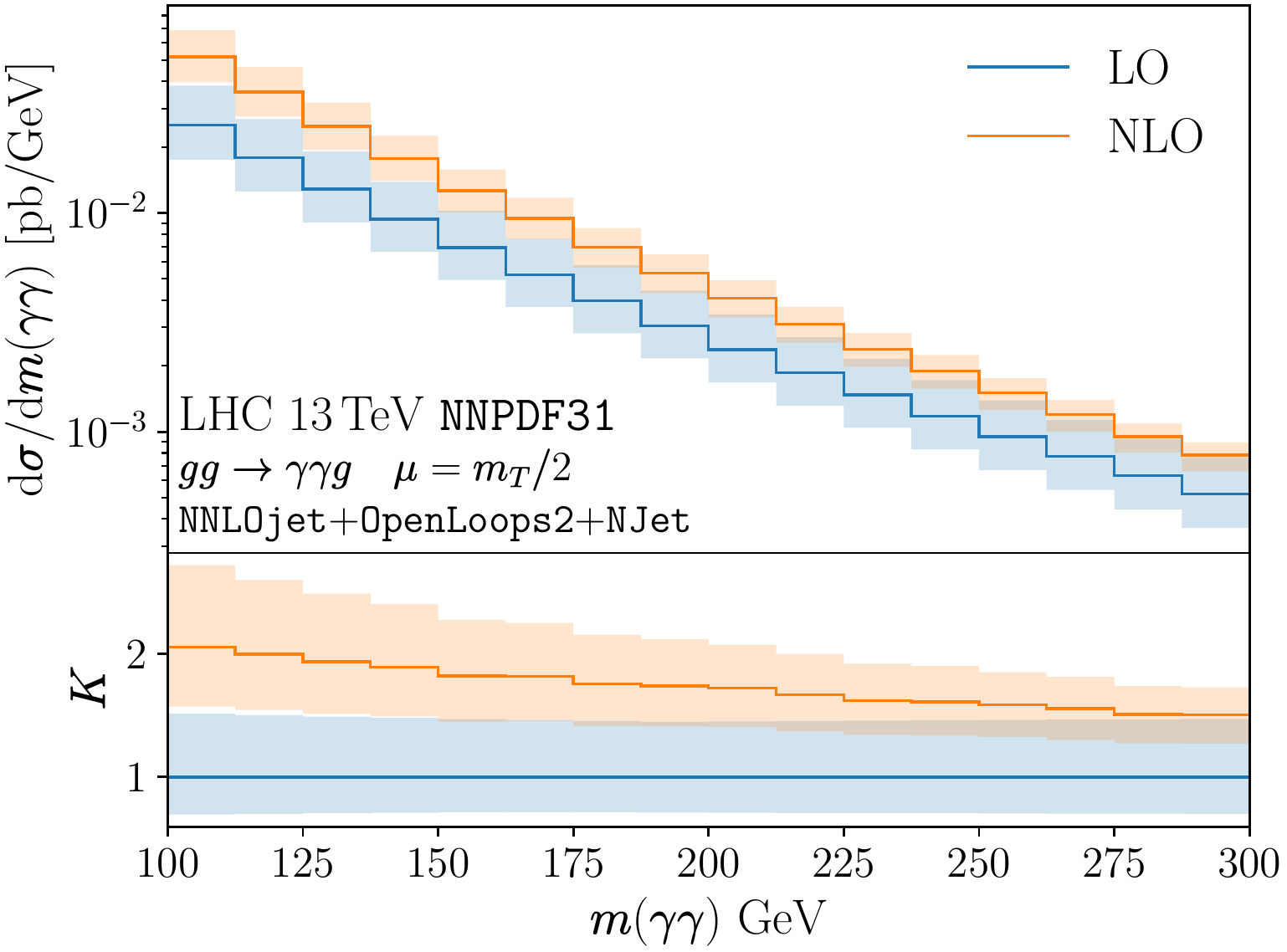}
    \end{center}
    \caption{
        Differential distributions in the invariant mass $m(\gamma\gamma)$ of the diphoton system.
    }
    \label{fig:pt_maadistb}
\end{figure}

\begin{figure}
    \begin{center}
        \includegraphics[width=0.7\textwidth]{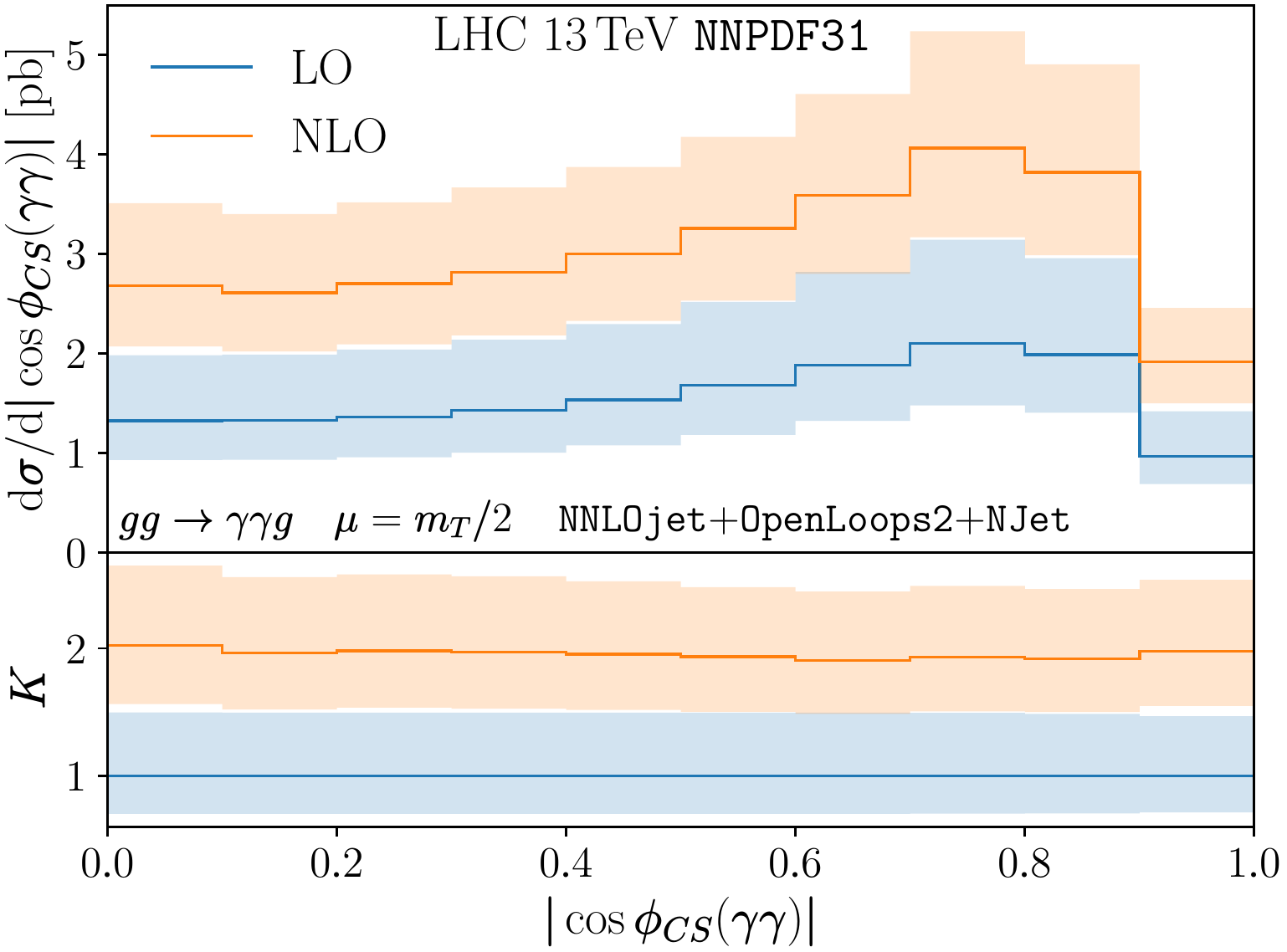}
    \end{center}
    \caption{
        Differential distribution in the Collins-Soper angle $\left|\cos \phi_{CS}(\gamma\gamma)\right|$ of the diphoton system.
    }
    \label{fig:phidista}
\end{figure}

\begin{figure}
    \begin{center}
        \includegraphics[width=0.7\textwidth]{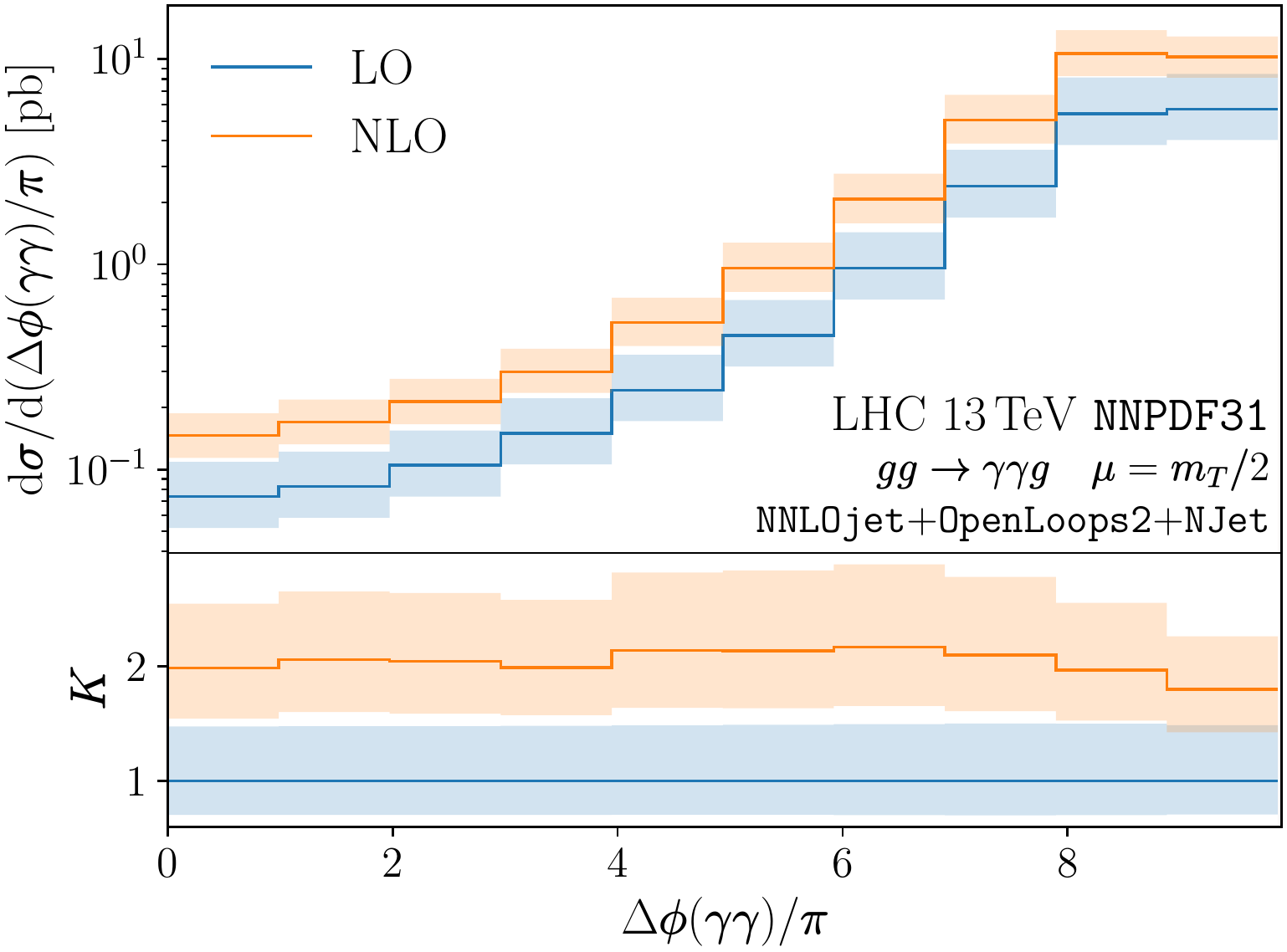}
    \end{center}
    \caption{
        Differential distribution in the azimuthal decorrelation $\Delta \phi(\gamma\gamma)$ of the diphoton system.
    }
    \label{fig:phidistb}
\end{figure}

\begin{figure}
    \begin{center}
        \includegraphics[width=0.7\textwidth]{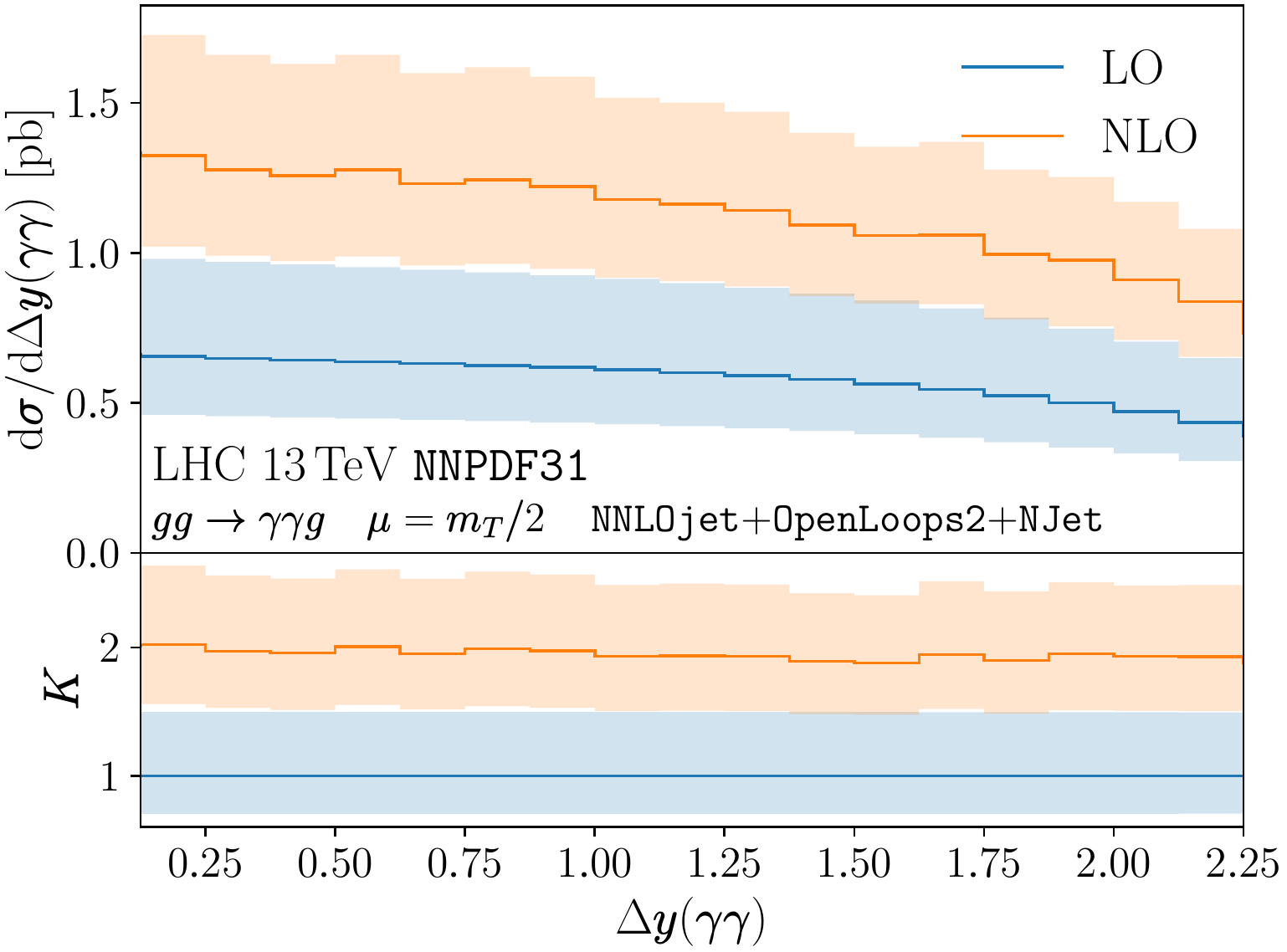}
    \end{center}
    \caption{
        Differential distribution in the diphoton rapidity difference $\Delta y(\gamma\gamma)$.
    }
    \label{fig:ydista}
\end{figure}

\begin{figure}
    \begin{center}
        \includegraphics[width=0.7\textwidth]{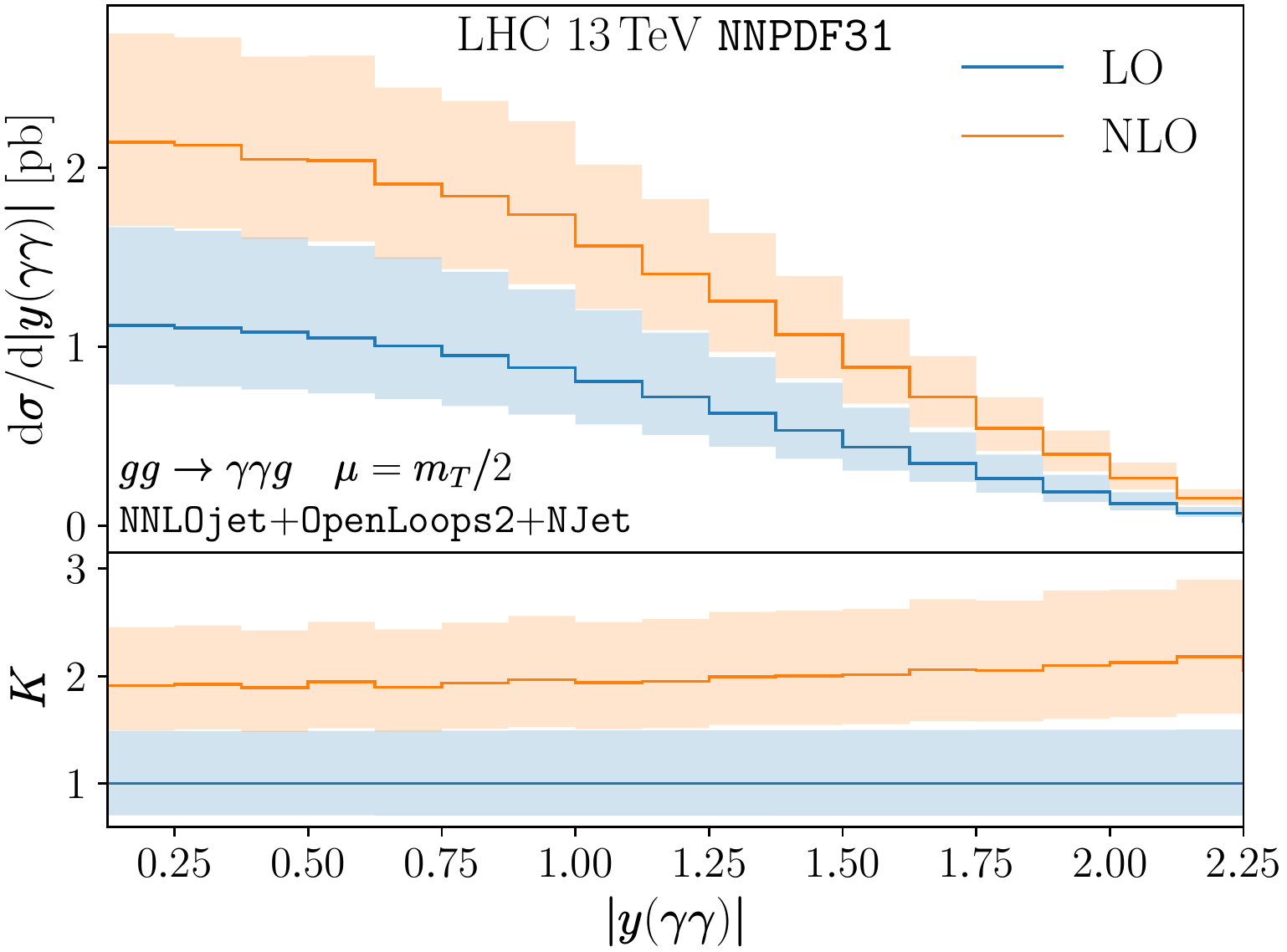}
    \end{center}
    \caption{
        Differential distribution in the diphoton total rapidity $|y(\gamma\gamma)|$.
    }
    \label{fig:ydistb}
\end{figure}

\Cref{fig:pt_maadista,fig:pt_maadistb,fig:phidista,fig:phidistb,fig:ydista,fig:ydistb} display the theory predictions for the different single-differential distributions in the diphoton variables.
We observe the \ac{NLO} corrections to be sizeable, often being comparable in size to the \ac{LO} predictions.
We define the $K$-factor as the ratio \ac{NLO}/\ac{LO}.
The corrections are largest at low $p_T(\gamma\gamma)$ (\cref{fig:pt_maadista}) or $m(\gamma\gamma)$ (\cref{fig:pt_maadistb}) where $K\sim2$ and \ac{NLO} and \ac{LO} uncertainties fail to overlap.
The ratio smoothly decreases towards $K\sim1.5$ for large $p_T(\gamma\gamma)$ or $m(\gamma\gamma)$, with overlapping scale uncertainty bands above $p_T(\gamma\gamma)=\SI{200}{\GeV}$ or $m(\gamma\gamma)=\SI{175}{\GeV}$.

The integrated cross section is dominated by the region of low $p_T(\gamma\gamma)$ or low $m(\gamma\gamma)$, such that distributions that are differential only in geometrical photon variables (\cref{fig:phidista,fig:phidistb,fig:ydista,fig:ydistb}) display typically near-uniform $K\sim2$, and no overlap of the \ac{LO} and \ac{NLO} scale uncertainty bands.
Visually, the scale uncertainty bands at \ac{NLO} and \ac{LO} appear to be of comparable width in all distributions.
However, owing to the large size of the \ac{NLO} corrections, the relative scale uncertainty is reduced from about \SI{50}{\percent} at \ac{LO} to \SI{30}{\percent} at \ac{NLO}.

\begin{figure}
    \begin{center}
        \includegraphics[width=0.7\textwidth]{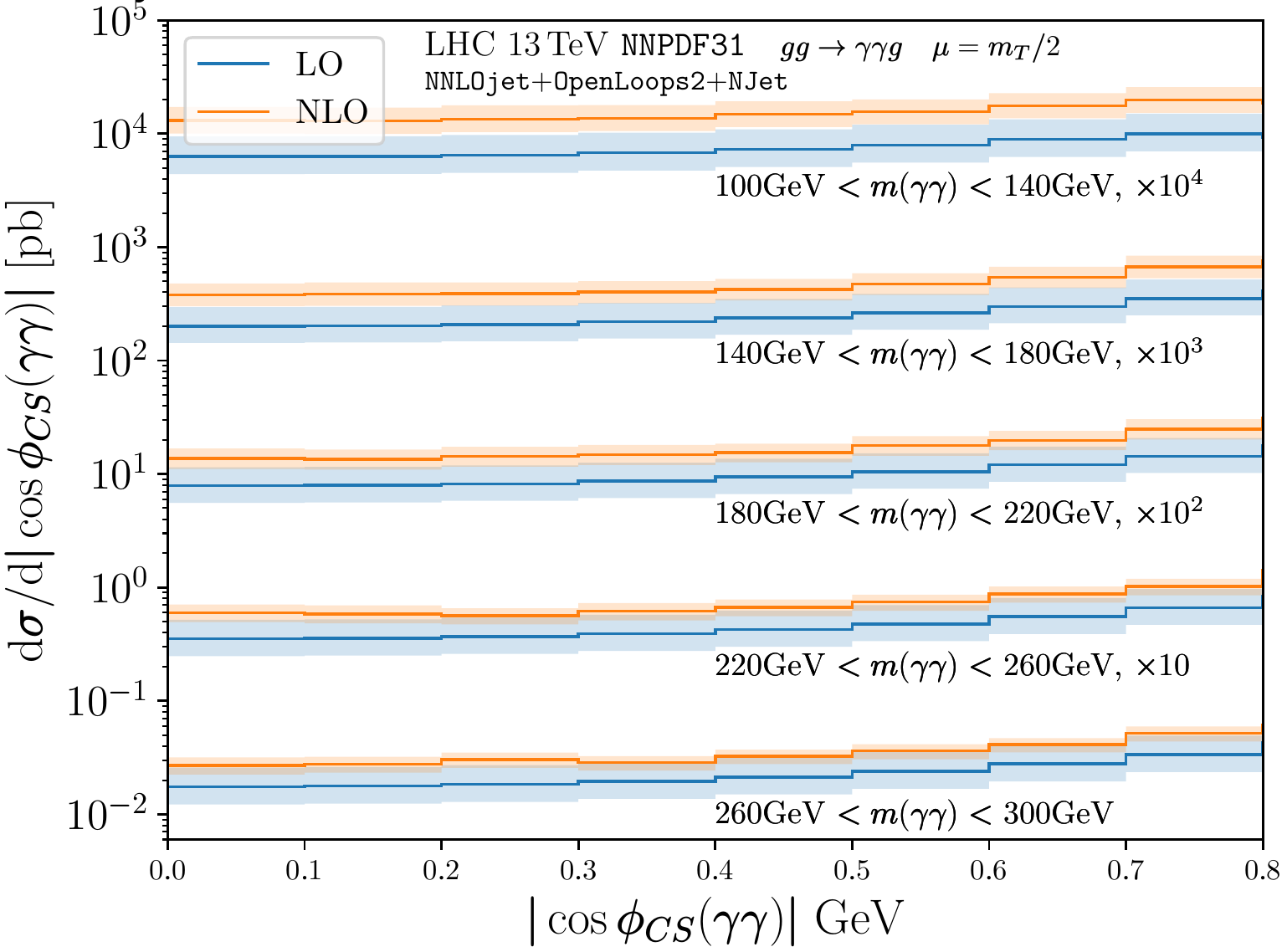}
    \end{center}
    \caption{
        Two-dimensional differential distribution in the diphoton invariant mass $m(\gamma\gamma)$ and Collins-Soper angle $\left|\phi_{CS}(\gamma\gamma)\right|$.
    }
    \label{fig:2Ddista}
\end{figure}

\begin{figure}
    \begin{center}
        \includegraphics[width=0.7\textwidth]{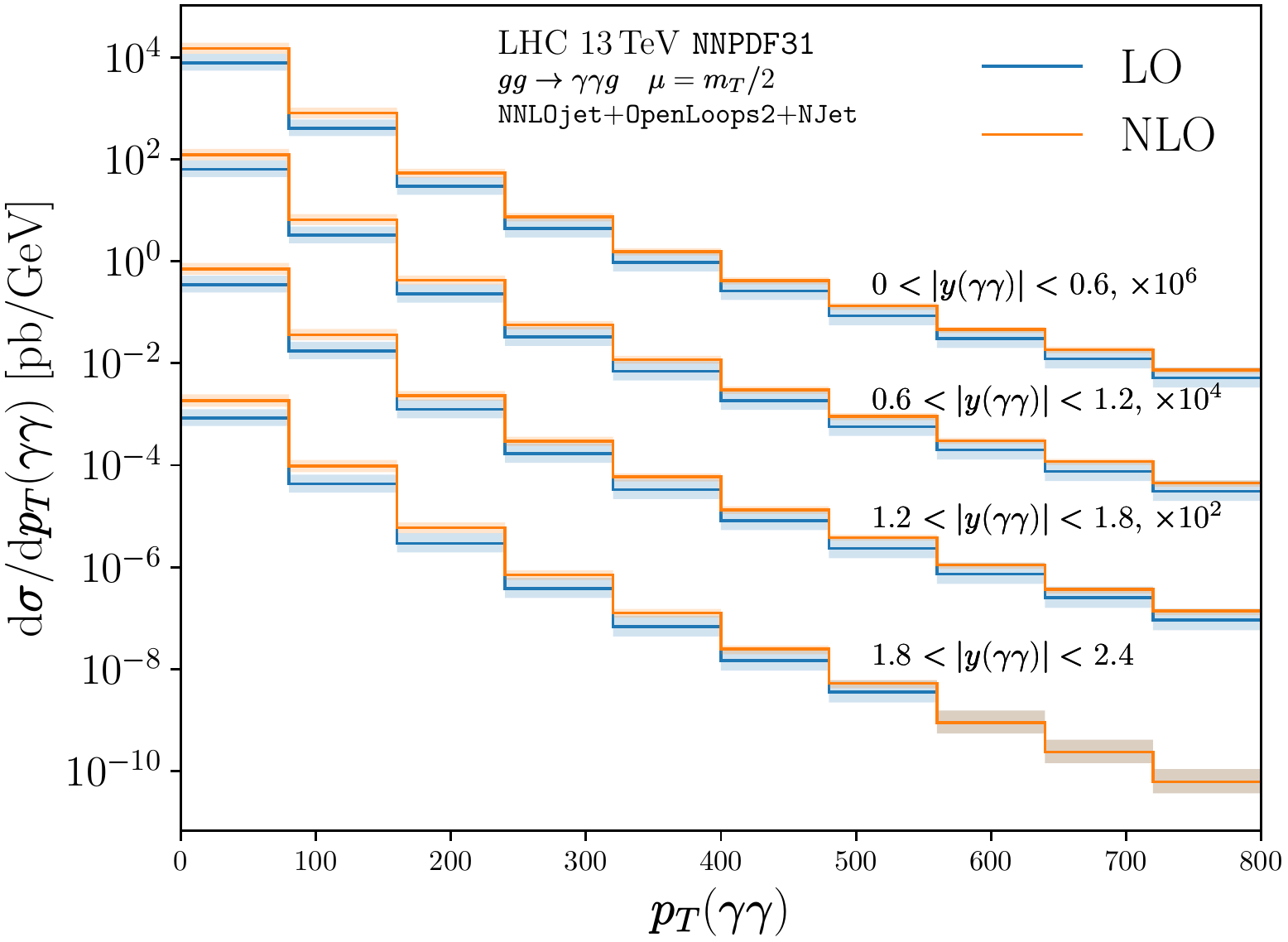}
    \end{center}
    \caption{
        Two-dimensional differential distribution in the diphoton rapidity $|y(\gamma\gamma)|$  and transverse momentum $p_T(\gamma\gamma)$.
    }
    \label{fig:2Ddistb}
\end{figure}

\begin{figure}
    \begin{center}
        \includegraphics[width=0.7\textwidth]{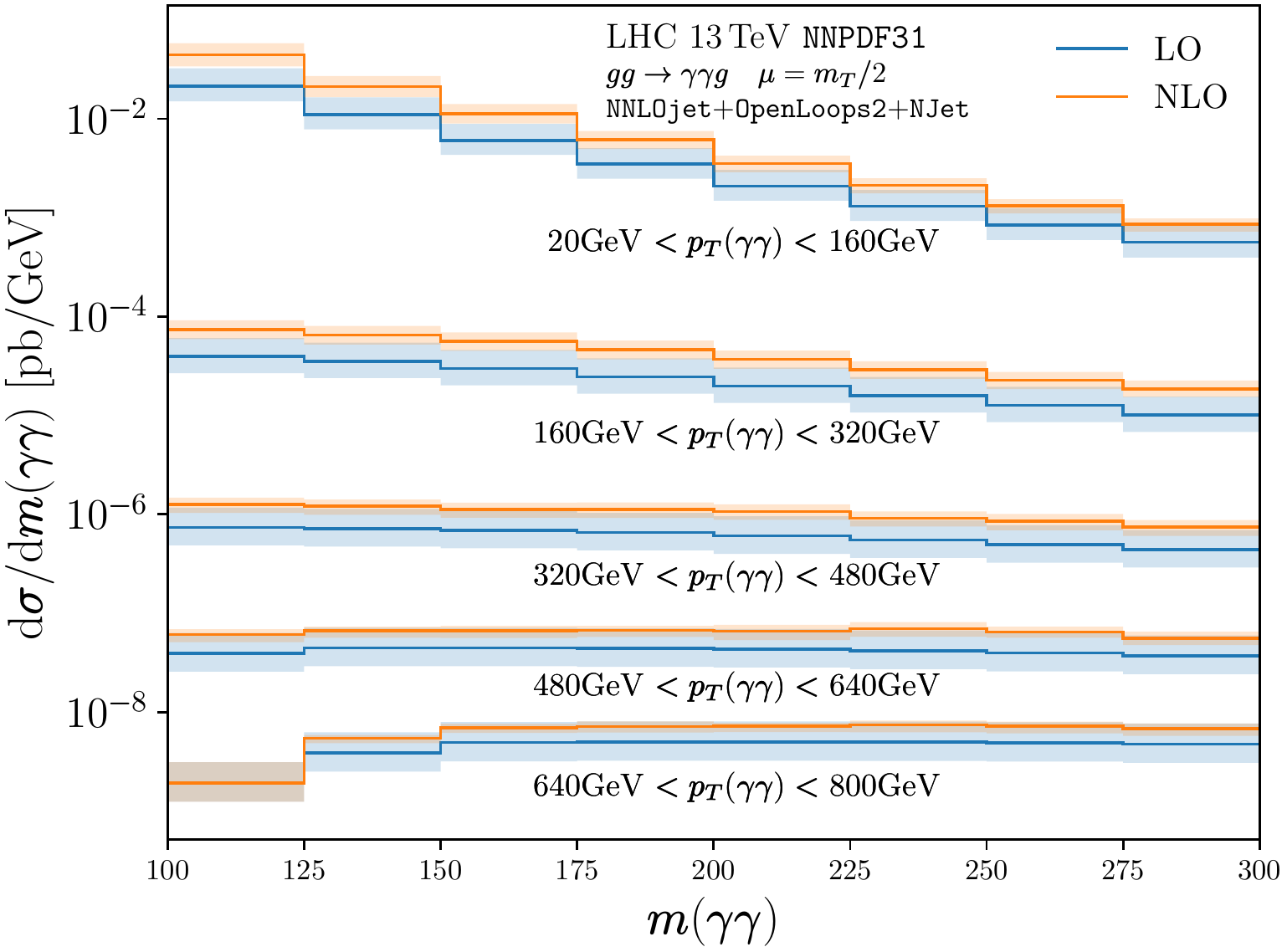}
    \end{center}
    \caption{
        Two-dimensional differential distribution in the diphoton transverse momentum $p_T(\gamma\gamma)$ and invariant mass $m(\gamma\gamma)$, in bins in $p_T(\gamma\gamma)$.
    }
    \label{fig:2Ddist2a}
\end{figure}

\begin{figure}
    \begin{center}
        \includegraphics[width=0.7\textwidth]{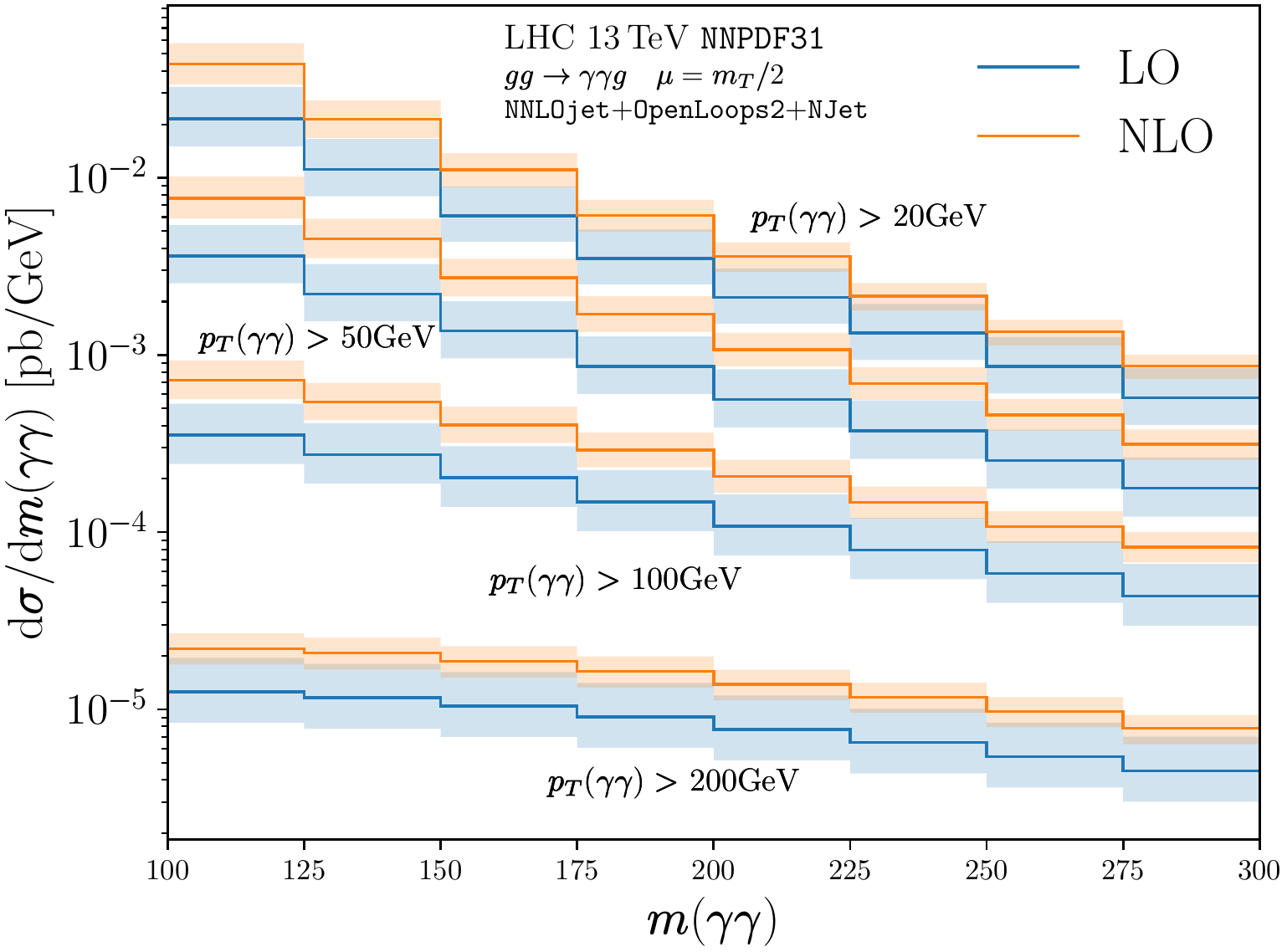}
    \end{center}
    \caption{
        Two-dimensional differential distribution in the diphoton transverse momentum $p_T(\gamma\gamma)$ and invariant mass $m(\gamma\gamma)$, for varying lower $p_T(\gamma\gamma)$ cut.
    }
    \label{fig:2Ddist2b}
\end{figure}

By inspecting the two-dimensional differential distribution in $m(\gamma\gamma)$ and $\left|\phi_{CS}(\gamma\gamma)\right|$ (\cref{fig:2Ddista}) we observe that the relative magnitude of the \ac{NLO} corrections decreases with increasing $m(\gamma\gamma)$, while the corrections remain uniform in $\left|\phi_{CS}(\gamma\gamma)\right|$ for all bins in $m(\gamma\gamma)$.
The two-dimensional differential distribution in $|y(\gamma\gamma)|$ and $p_T(\gamma\gamma)$ (\cref{fig:2Ddistb}) also shows the decrease of the corrections towards larger $p_T(\gamma\gamma)$.
The decrease is more pronounced at forward rapidity (large $|y(\gamma\gamma)|$) than at central rapidity (small $|y(\gamma\gamma)|$).

The two-dimensional distributions in $p_T(\gamma\gamma)$ and $m(\gamma\gamma)$ (\cref{fig:2Ddist2a,fig:2Ddist2b}) largely reproduce the features of the one-dimensional distributions of \cref{fig:pt_maadista,fig:pt_maadistb}, both for distributions in bins of $p_T(\gamma\gamma)$ (\cref{fig:2Ddist2a}) or for varying lower cut in $p_T(\gamma\gamma)$ (\cref{fig:2Ddist2b}).
The only novel feature is a non-uniform shape in $m(\gamma\gamma)$ for the highest bin in $p_T(\gamma\gamma)$ (lowest curve in \cref{fig:2Ddist2a}), which is indicative of the onset of large logarithmic corrections in $\log(m(\gamma\gamma)/p_T(\gamma\gamma))$ in this range.

The numerical size of the \ac{NLO} corrections and the scale uncertainties at \ac{LO} and \ac{NLO} are comparable to what was observed in inclusive Higgs boson production in gluon fusion~\cite{djouadi:1991tka} or in the Higgs boson transverse momentum distribution in gluon fusion~\cite{deflorian:1999zd,ravindran:2002dc}.
These processes are mediated through a heavy top quark loop and are very similar to the diphoton-plus-jet production considered here in terms of kinematics and initial-state parton momentum range.
The pathology of the \ac{NLO} corrections observed here is thus not that surprising after all; it does, however, indicate the potential numerical importance of corrections beyond \ac{NLO}.

The Born-level $gg\to \gamma\gamma g$ subprocess (corresponding to the \ac{LO} in our results)
contributes to the full diphoton-plus-jet production as part of the \ac{NNLO} corrections.
Corrections to this order were recently computed~\cite{chawdhry:2021hkp}.
They were observed to be moderate and within the scale uncertainty of the previously known \ac{NLO} results~\cite{delduca:2003uz} for most of the kinematic range, where they also led to a substantial reduction of the scale uncertainty at \ac{NNLO}.
At low $p_T(\gamma\gamma)$ or low $m(\gamma\gamma)$, larger positive corrections and an increased scale uncertainty were observed~\cite{chawdhry:2021hkp}.
These effects could be identified to be entirely due to the contribution of the $gg\to \gamma\gamma g$, which only starts to contribute from \ac{NNLO} onwards, and it was anticipated in
\incite{chawdhry:2021hkp} that \ac{NLO} corrections to the  $gg\to \gamma\gamma g$ (which form a subset of the \ac{N3LO} corrections to the full diphoton-plus-jet process) could help to stabilise the predictions in the relevant kinematic ranges.

Our results demonstrate that this is not the case.
The absolute scale uncertainty on the gluon-induced process does not decrease from \ac{LO} to \ac{NLO}, and the
\Ac{NLO} correction is of about the size of the \ac{LO} contribution.
Consequently, inclusion of the \ac{NLO} corrections to the $gg\to \gamma\gamma g$
into the full \ac{NNLO} diphoton-plus-jet process will further enhance the predictions at low $p_T(\gamma\gamma)$ or low $m(\gamma\gamma)$, thereby further elongating them from the previously known order, and will leave the scale uncertainty band largely unchanged.
This is an indication that the seven-point scale variation technique may not be providing a reliable estimate of the uncertainty for \ac{NNLO} diphoton-plus-jet production.

\section{Computational cost}
\label{sec:cost}

State-of-the-art calculations in precision phenomenology require high-performance computing resources~\cite{cordero:2022gsh}.
In this section, we comment on the computational cost of our calculation.

To reach an average \ac{MC} error of at most one percent in the total cross section and all differential distributions, we find that approximately \SI{5}{\mega\relax} points in the virtual corrections and \SI{30}{\mega\relax} points in the real corrections are required after cuts are applied.
The real corrections require more evaluations due to the higher dimensionality of the phase space and the presence of \ac{IR} limits, so a factor of six is unsurprising.

To estimate the relative time per \ac{ME} call, we time the evaluation of approximately \SI{1}{\kilo\relax} points after cuts.
We find a mean time per point in the virtual corrections of \SI{16}{\s}, with some of the points in \ac{f128} precision taking up to 5 minutes.
This is a much shorter mean time than the \SI{26}{\s} found in \cref{sec:performance}, which uses the same target accuracy of three digits.
Most of this time is in the evaluation of the pentagon functions, the timing of which at fixed precision is sensitive to the degree of \ac{IR} divergence of the phase-space point.
The proportion of points requiring \ac{f128} evaluation, which requires significantly more time than \ac{f64} evaluation, is the dominant contribution to the mean time.
Therefore, the timing is highly dependent on the choice of phase space.
Both of these benchmarks use version 1.0 of \pentagonfunctions~\cite{chicherin:2020oor}; using the recently released version 2.0~\cite{chicherin:2021dyp} would reduce these times, as found in \cref{ch:3j}.

For the real contributions, we find an mean time of \SI{11}{\s} for the $gg\to \gamma\gamma gg$ channel and \SI{14}{\s} for the $gg\to\gamma\gamma q \anti{q}$ channel.
The combined average time per point in the real contributions is \SI{25}{\s}.
We stress that the real integration includes extreme regions of phase space, requiring evaluations in \ac{f128} precision.

We find a cost of around \SI{50}{\kilo\relax} \ac{CPU} hours in the virtual corrections and \SI{500}{\kilo\relax} \ac{CPU} hours in the real corrections for our simulation.
The order of magnitude increase in the time cost of the real corrections is good motivation for the \ac{NN} approximation techniques for high-multiplicity scattering explored in \cref{ch:ann}.

\section{Summary}
\label{sec:concl}

In this chapter, we have presented the \ac{NLO} \ac{QCD} corrections to the diphoton-plus-jet production in the gluon-fusion channel for the first time.
The loop-induced process requires the evaluation of six-point one-loop real emission amplitudes and \ac{FC} five-point two-loop virtual amplitudes.
To the best of our knowledge it is the first time that five-point two-loop \ac{FC} amplitudes have been integrated to provide fully differential cross section predictions relevant for the \ac{LHC} experiments.

Using a realistic set of kinematic cuts and simulation parameters, we find significant corrections at \ac{NLO}.
This is particularly relevant at low values of $p_T(\gamma\gamma)$ and $m(\gamma\gamma)$.
Since observables that are differential only in angles, such as rapidity and the Collins-Soper angle, are inclusive over the energy variables, one observes significant \ac{NLO} corrections across the full parameter range.
Double-differential distributions further highlight this feature, which is reminiscent of the perturbative convergence observed in other gluon-induced processes such as inclusive Higgs production and the Higgs boson transverse momentum distribution.
The relative scale uncertainty is reduced by the higher-order corrections, although in absolute terms the scale uncertainty does not decrease from \ac{LO} to \ac{NLO} in the low $p_T(\gamma\gamma)$ and $m(\gamma\gamma)$ regions.

This work demonstrates the importance of a combined prediction for quark-induced and gluon-induced diphoton-plus-jet signatures for future precision studies at the \ac{LHC}.

\chapter{Leading-colour double-virtual \acs{QCD} corrections to \texorpdfstring{$pp\to3j$}{trijet production at hadron colliders}}
\label{ch:3j}

We present an analytic computation of the helicity amplitudes for trijet production at hadron colliders up to two loops in \ac{QCD}, providing the virtual and \ac{VV} contributions for two-to-three \ac{NNLO} predictions.
We provide a fast and stable implementation of the colour- and helicity-summed \acp{FR} in \cpp as part of the version 3.1.0 release of the \njet library~\cite{njet}.

The conventions, including the decompositions in channels, helicities, and the ratio $\nf/\nc$, and the \ac{IR} subtraction to \ac{FR}, follow those of \incite{abreu:2021oya}.
The methods of construction and implementation are the same as those described in \cref{ch:yy-amp}.
We cross-checked our implementation against the benchmark point in \incite{abreu:2021oya}, finding exact agreement.

We express the amplitudes with the scale dependence factorised out.
This allows for efficient computation of the amplitudes at a single phase-space point with many different scales, as is used for scale variation uncertainty estimates (\cref{sec:uncertainties}), with only a single evaluation of the scale-independent part of the amplitudes.

We will now proceed with an analysis of the implementation, looking at the numerical stability and evaluation timing over a sample phase space in \cref{sec:3j-stab}, and the performance in \ac{IR} limits in \cref{sec:3j-ir}.

\section{Stability and timing}
\label{sec:3j-stab}

% https://gitlab.com/jetdynamics/njet-tools/-/blob/master/5p2l-checks/stability/stability.py
\begin{figure}
    \begin{center}
        \includegraphics[width=\textwidth]{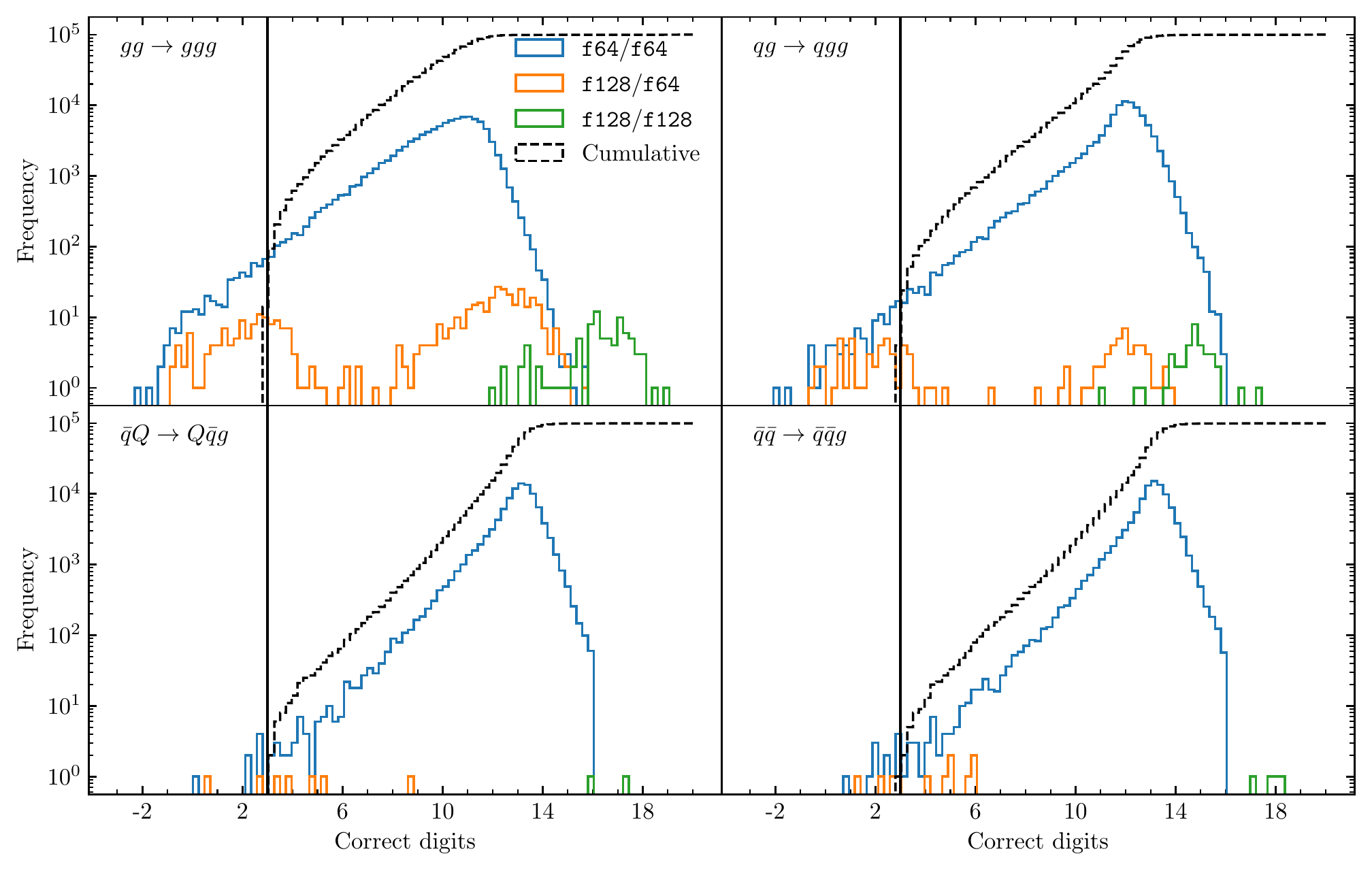}
        \caption{
            Stability plots for a single channel in each of the $5g$, $2q3g$, $2q2Q1g$, and $4q1g$ two-loop trijet subprocesses.
            Other channels within each of the subprocesses perform similarly.
            Run of \SI{100}{\kilo\relax} points per channel over a randomly sampled uniform five-point phase space.
            The evaluation strategy is used with a target accuracy of three digits, denoted on the plots by a black vertical line.
        }
        \label{fig:stability-3j}
    \end{center}
\end{figure}

% https://gitlab.com/jetdynamics/njet-tools/-/blob/master/5p2l-checks/stability/timing_paper.py
\begin{table}
    \begin{center}
        \begin{tabular}{| l | r r | r r |}
            \hline
            \multicolumn{1}{|c|}{\multirow{2}*{Channel}} & \multicolumn{2}{c|}{\ac{f64}/\ac{f64}} & \multicolumn{2}{c|}{Precision rescue system} \\
            \cline{2-5}
            & Time (s) & Pentagons (\%) & Time (s) & Pentagons (\%) \\
            \hline
            $gg \to ggg$                             & 1.39 & 69 & 1.89 & 77 \\
            $gg \to \anti{q}qg$                      & 1.35 & 91 & 1.37 & 91 \\
            $qg \to qgg$                             & 1.34 & 92 & 1.57 & 93 \\
            $q\anti{q} \to ggg$                      & 1.34 & 93 & 1.38 & 93 \\
            $\anti{q}Q \to Q\anti{q}g$               & 1.14 & 99 & 1.16 & 99 \\
            $\anti{q}\anti{Q} \to \anti{q}\anti{Q}g$ & 1.36 & 99 & 1.39 & 99 \\
            $\anti{q}g \to \anti{q}Q\anti{Q}$        & 1.36 & 99 & 1.39 & 99 \\
            $\anti{q}q \to Q\anti{Q}g$               & 1.14 & 99 & 1.14 & 99 \\
            $\anti{q}g \to \anti{q}q\anti{q}$        & 1.84 & 99 & 1.90 & 99 \\
            $\anti{q}\anti{q} \to \anti{q}\anti{q}g$ & 1.82 & 99 & 1.94 & 99 \\
            $\anti{q}q \to q\anti{q}g$               & 1.71 & 99 & 1.77 & 99 \\
            \hline
            % $gg\to\gamma\gamma g$   & 9    & 99 & 26   & 99 \\ % these numbers are for the old pentagon functions
            % \hline
        \end{tabular}
        \caption{
            Mean timing per phase-space point for combined evaluation of Born, virtual, and \ac{VV} contributions in \ac{f64}/\ac{f64} and evaluation strategy with target accuracy of three digits.
        }
        \label{tab:timing}
    \end{center}
\end{table}

We perform similar stability tests on the trijet production channels to those in \cref{sec:performance}.
We generate \SI{100}{\kilo\relax} phase-space points isotropically with the algorithm from~\incite{byckling:1971vca}, which is provided in the \njet library.
We use the cuts $p_T > \SI{3}{\MeV}$, $\eta < 2.8$, and $\Delta R > 0.4$ for all jets, which is likely to be at least as inclusive over phase space as any phenomenological application.
We use arbitrary scales for testing, choosing centre-of-mass energy $\sqrt{s_{12}}=\SI{1}{\GeV}$ and renormalisation scale $\mu_R=m_Z$ with $Z$-boson mass $m_Z=\SI{91.188}{\GeV}$\index{Boson!$Z$!mass}.
We evaluate all channels over this phase space using the evaluation strategy with a target accuracy of three digits, generating the stability plots shown in \cref{fig:stability-3j} and mean timings of \cref{tab:timing}.

The stability and evaluation times improve for the simpler channels, which are those with more quarks and less gluons.
In \cref{fig:stability-3j}, the channels with four quarks even show an \ac{f64} bin with around \SI{1}{\percent} of the points at the maximum possible precision of 16 digits.
The $4q1g$ channels are slower as they are each constructed from a pair of independent $2q2Q1g$ channels, although they show similar stability to the $2q2Q1g$ channels.

Comparing the trijet stability plots to that for $gg\to\gamma\gamma g$, \cref{fig:stability}, which includes non-planar integrals, fewer points are required to be reevaluated with higher precision, as expected.
The evaluation time is also significantly less for the trijet channels than for $gg\to\gamma\gamma g$, owing to the \ac{LC} trijet computations containing only planar diagrams.

Compared to \incite{abreu:2021oya}, we find a reduced evaluation time.
This is most pronounced in the most difficult channel, pure-gluon scattering $gg\to ggg$, where the time is halved.
This is because the former was evaluated with version 1.0 of \pentagonfunctions~\cite{chicherin:2020oor}, while our setup uses version 2.0, which additionally supports one massive leg~\cite{chicherin:2021dyp}.
While we should be careful when comparing to the stability plots of \incite{chicherin:2021dyp} as they use a different phase space, it also appears that the tail of failing points is shorter in our results, suggesting \pentagonfunctions version 2.0 also improves stability.

\section{Infrared performance}
\label{sec:3j-ir}

% https://gitlab.com/jetdynamics/njet-tools/-/blob/master/collinear-phase-space/collinear.py
\begin{figure}
    \begin{center}
        \includegraphics[width=\textwidth]{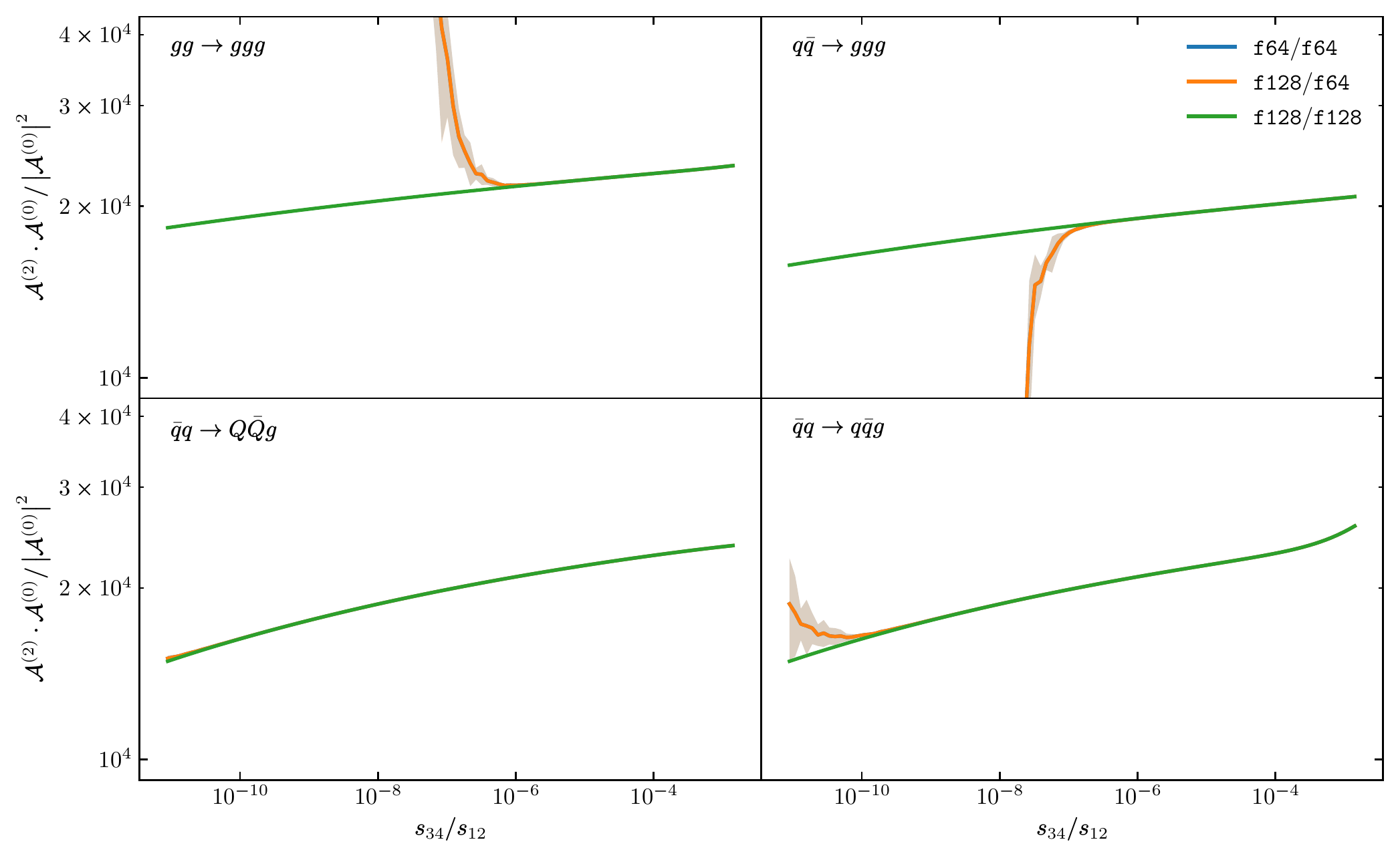}
        \caption{
            The two-loop \ac{ME} for a selection of channels over a slice of phase space which drives into \iac{IR} configuration that is collinear in the first two outgoing legs.
            The error band is given by the dimension scaling test.
            Where it is not visible, the \ac{f64}/\ac{f64} (blue) line coincides with the \ac{f128}/\ac{f64} (orange) line, which similarly coincides with the \ac{f128}/\ac{f128} (green) line.
        }
        \label{fig:collinear}
    \end{center}
\end{figure}

We prepare another phase space using the method described in \cref{sec:valid}, generating a \num{100}-point slice that approaches \iac{IR} limit.
The four-point seed phase space is generated randomly using the generator discussed in \cref{sec:3j-stab}.

We perform this for ten different four-point seeds and plot the mean in \cref{fig:collinear} to avoid any irregularities that may arise when approaching the limit in an exceptional direction.
The lower precision evaluations diverge when they get too close to the collinear limit; for example in the five-gluon channel, this occurs at around $\s{34}/\s{12}=10^{-7}$.
The origin of this numerical divergence lies in the evaluation of the pentagon functions as it is not fixed by the \ac{f128}/\ac{f64} evaluation.
Evaluation in \ac{f128}/\ac{f128} remains unproblematic deep into the limit.
This demonstrates that these amplitudes are suitable not only for integrating over two-to-three \ac{VV} phase spaces at \ac{NNLO}, but also the more difficult two-to-two \ac{RVV} phase space at \ac{N3LO}.

\chapter{Conclusion}
\label{ch:conc}

To tackle the challenges that face precision tests of the \ac{SM}, we must leverage novel techniques in mathematics and computation.
The first step for bringing theoretical predictions into the realm of percent precision is the further mapping of the phase space of \ac{NNLO} processes within fixed-order perturbation theory.
In particular, in order for these to be applicable to phenomenology, it is vital that they are implemented in an optimally efficient and stable way.

Gluon-initiated diphoton-plus-jet production offers an attractive testing ground for new technology as it is loop-induced, presenting challenges for conventional event generator techniques.
It also exhibits a complicated two-to-three two-loop structure including non-planar integrals, while having simpler poles, being at \ac{NLO}, and colour than similar partonic processes.
Moreover, it is a relevant background for interesting phenomenology such as probing the Higgs coupling.

After building up the technology necessary to perform precision \ac{QCD} calculations, we discussed the \ac{IR} behaviour of \ac{QCD} before presenting a library of \ac{IR} \ac{QCD} functions up to at most \ac{NNLO}, which have a variety of uses in the field.

Next, we reviewed the feasibility of using \acp{NN} to optimise the evaluation of the \ac{ME} for cross section calculations for hadron colliders.
We used the $gg\to\gamma\gamma+n\times g$ amplitudes from \njet to train \iac{NN} ensemble, and fed it into the \ac{MC} event generator \ac{SHERPA} to investigate its behaviour within a full hadronic simulation.
At six-point, we found that the total simulation time was sped up by a factor of thirty.
This method offers a performant way to run high-multiplicity radiative contributions for event generator simulations where conventional techniques are prohibitively slow.

We then computed the \ac{FC} two-loop amplitudes for $gg\to\gamma\gamma g$, with an efficient public implementation released in \njet.
We found that their \ac{LC} contribution contains non-planar integrals, which are those with the highest complexity.
Such ``industrialised'' analytical computations of two-loop five-point \ac{QCD} processes present huge technical challenges.
We first needed a basis of special functions offering fast and stable evaluation over the physical phase space; for massless and single-mass scattering, the pentagon functions library has recently made this possible.
Generating the amplitude via colour-ordered diagrams, we reduced tensor integrals to \iac{MI} basis which could be expressed in terms of these special functions.
This required constructing and solving a very large system of \ac{IBP} identities.
Finally, we obtained the coefficients in an efficient form using \ac{FF} reconstruction techniques.

Following the virtual $gg\to\gamma\gamma g$ amplitude calculation, we used the antenna subtraction scheme to combine the virtual result with the one-loop real corrections to obtain \ac{NLO} \ac{QCD} differential cross sections.
Considering the full process $pp\to\gamma\gamma j$, the gluon-fusion subprocess enters at \ac{N3LO}, but the large gluonic \ac{PDF} means it provides a dominant correction to \ac{NNLO}.
We computed various observable distributions, finding significant corrections that highlight the importance of including the gluon-initiated channels in these predictions.

We also presented an efficient computation of the \ac{LC} two-loop amplitudes for hadronic trijet production and tested their public implementation within \njet.

This thesis has focused on the calculation of \ac{QCD} amplitudes at the precision frontier, in particular, the process of diphoton-plus-jets production through gluon fusion.
This presented the challenge of handling the algebraic and analytic complexity of higher-order perturbative expressions.
Furthermore, assembly of observables from these contributions required careful regularisation of \ac{IR} behaviour to cancel poles at each order.
All amplitudes were implemented as analytical expressions into the public \cpp library \njet; such distribution is vital for the collaborative progress of the whole theoretical community.
They provide a vital ingredient for theoretical predictions of cross sections at the \ac{LHC} in the search for deviations from the \ac{SM}.

\bibliographystyle{JHEP}
\bibliography{bibliography}

\printindex

\end{document}